\documentclass[12pt]{article}
 \pdfoutput=1 
\usepackage{geometry}
\usepackage{amsthm,amsmath,amssymb}
\usepackage{natbib}
\usepackage{multirow}
\usepackage{graphicx}
\usepackage{makecell}
\usepackage{booktabs}
\usepackage{array}
\usepackage{fullpage}
\usepackage{url}
\usepackage{algorithm}
\usepackage{algorithmic}
\usepackage{bm}
\usepackage{mathtools}
\usepackage{wrapfig}
\usepackage{lipsum}
\usepackage{mathrsfs}
\usepackage{dsfont}
\usepackage{verbatim}
\usepackage{graphicx}
\usepackage{enumitem}
\usepackage{hyperref}
\usepackage[utf8]{inputenc}
\usepackage[english]{babel}
\usepackage{multirow}
\usepackage{rotating}
\usepackage[toc,page]{appendix}
\usepackage{xcolor}
\usepackage{caption}
\usepackage{subcaption}
\usepackage{longtable}
\usepackage{appendix}
\allowdisplaybreaks

\title{Accounting for Smoking in Forecasting Mortality and Life Expectancy\thanks{Yicheng Li is Graduate 
Research Assistant and Adrian E. Raftery is Boeing International Professor
of Statistics and Sociology, both at the Department of Statistics,
Box 354322, University of Washington, Seattle, WA 98195-4322.
This research was supported by NIH grants R01 HD054511 and R01 HD070936,
and by the Center for Advanced Research in the Behavioral Sciences at
Stanford University. The authors are grateful to John Bongaarts for 
helpful discussions.} }
\date{\today}
\author{Yicheng Li,  Adrian E. Raftery \\
University of Washington}

\begin{document}
	\maketitle
	
\begin{abstract}
	Smoking is one of the main risk factors that has affected human mortality and life expectancy over the past century. 
	Smoking accounts for a large part of the nonlinearities in the growth of life expectancy and of the geographic and sex differences in mortality. As \citet{bongaarts2006long} and \citet{janssen2018mort} suggested, accounting for smoking could improve the quality of mortality forecasts due to the predictable nature of the smoking epidemic. We propose a new Bayesian hierarchical model to forecast life expectancy at birth for both sexes and for 69 countries with good data on smoking-related mortality. The main idea is to convert the forecast of the non-smoking life expectancy at birth (i.e., life expectancy at birth removing the smoking effect) into life expectancy forecast through the use of the age-specific smoking attributable fraction (ASSAF). We introduce a new age-cohort model for the ASSAF and a Bayesian hierarchical model for non-smoking life expectancy at birth. The forecast performance of the proposed method is evaluated by out-of-sample validation compared with four other commonly used methods for life expectancy forecasting. Improvements in forecast accuracy and model calibration based on the new method are observed. 
\end{abstract}

\section{Introduction}
Forecasting human mortality and life expectancy is of considerable 
importance for public health policy, planning social security systems,
life insurance,  and other areas, particularly as the world's population
continues to age.
It is also a major component of population projections, as it impacts
the number of people alive and their distribution by age and sex.
Population projection are themselves a major input to 
government planning at all levels, as well as private sector planning,
monitoring international development and environmental goals, and research in the
health and social sciences.

Many methods for forecasting mortality have been developed. 
The Lee-Carter method \citep{lee1992modeling} for forecasting age-specific
mortality rates was a milestone and has developed rapidly since it was
proposed. \citet{lee2001evaluating} modified the Lee-Carter method by matching estimated life expectancy  with the observed value. Other variations of the Lee-Carter method include adding a cohort effect \citep{renshaw2006cohort}, applying a functional data approach \citep{ hyndman2007robust, shang2016mortality}, and incorporating biomedical information \citep{janssen2013including}.
Bayesian Lee-Carter methods have also been proposed
\citep{pedroza2006bayesian, king2011future, wisniowski2015bayesian}.
See \citet{booth2006lee} for a review.

The main organization that produces regularly updated mortality and population
forecasts for all countries is the United Nations, which publishes these 
forecasts every two years in the {\it World Population Prospects} 
\citep{WPP2017}. 
Traditionally since the 1940s, 
population projections have been done
using deterministic methods that do not primarily use statistical estimation
methods or assess uncertainty in a statistical way \citep{Whelpton1936,Preston&2000}. In 2015, in a major advance, the UN changed the method for 
producing their official mortality and population forecasts
from the traditional deterministic method
to a Bayesian approach that estimates and assesses uncertainty about 
future trends in a principled statistical way using Bayesian hierarchical
models for life expectancy and fertility 
\citep{Raftery&2012PNAS,raftery2013bayesian,Raftery&2014,WPP2015}.

The basic approach of these methods is to extrapolate past trends in
observed mortality rates, which have been dominated by a monotone increasing
trend in life expectancy for over a century. 
However, it may also be helpful to include risk factors that can impact health,
and hence mortality \citep{janssen2018mort}. 
This has been done, for example, for the HIV/AIDS epidemic \citep{godwin2017bayesian}, alcohol consumption \citep{trias2019gender}, 
and the obesity epidemic \citep{vidra2017impact}.
Another major factor is smoking, which is mainly responsible for lung cancer  and is a risk factor for many other fatal diseases, and causes about 6 million deaths per year \citep{britton2017death}. 
Smoking can account for some nonlinear trends, cohort effects, 
and between-country and between-sex differentials observed in mortality,
suggesting that it could be used to improve mortality and life expectancy 
projections \citep{bongaarts2014trends}.

Here we propose a Bayesian method for doing this for both sexes and
multiple countries jointly. It uses the 
smoking attributable fraction (SAF) of mortality, estimated by the
Peto-Lopez method \citep{peto1992mortality,bongaarts2006long,janssen2013including,stoeldraijer2015future}. The proposed method consists of two main components,
one to forecast the age-specific SAF (ASSAF), and the other to forecast
non-smoking life expectancy. Our method develops male and female forecasts
jointly, since the female smoking epidemic tends to resemble the male one,
but with a lag, and possibly a different maximum level,  
a fact that can be used to improve forecasts.
The female advantage in life expectancy is partly due to smoking effects,
and our method quantifies this and uses it to forecast the future
life expectancy gap between females and males.
We apply our method to 69 countries with high quality data on the 
historical impact of smoking on mortality.

The paper is organized as follows. The methodology is described in Section \ref{sec:med}. Section \ref{subsec:assaf} describes the method for estimating and forecasting the ASSAF. Section \ref{subsec:nonsmke0} presents the estimation and forecasting method for non-smoking life expectancy. Section \ref{subsec:mfjoint} 
describes our model for the gap between male and female life expectancy to complete the coherent projection. 
An out-of-sample validation experiment is reported in Section \ref{sec:res} to evaluate and compare the projection accuracy and calibration of our model with several benchmark methods. We then study the details of the forecast results for four selected countries in Section \ref{sec:case}. We conclude with a discussion in Section \ref{sec:disc}.

\section{Method}\label{sec:med}
\subsection{Notation}
We use indices $\ell$ for country (always as a superscript unless otherwise indicated), $s$ for sex, $t$ for time (usually in terms of the year), and $c$ for cohort (usually in terms of the year of birth). We use $x$ to denote the left end of an age group, i.e., $x$ represents the $a$-year age group $[x,x+a)$, and $x+$ represents the age group $[x, +\infty)$. 

A key general concept in our approach is the smoking attributable fraction (SAF) of mortality for a population of interest. This is  defined as the proportion by which mortality would be reduced if the population were not exposed to smoking. 
We focus on the age-specific SAF (ASSAF) of mortality for age group $x$
in country $\ell$ and time period $t$, denoted by $y_{x,t}^\ell$. 
The all-age smoking attributable fraction (ASAF) of mortality 
is defined as a weighted average of the ASSAF over all age groups, where the weights are the age-specific mortality rates.  We use the symbols $d$, $e_0$, and $e_0^{NS}$ to denote the mortality rate, the life expectancy at birth, and the non-smoking life expectancy at birth, respectively.

We denote by $\mathcal{N}_{[u, v]}(\lambda, \kappa)$ the truncated normal distribution with mean $\lambda$ and variance $\kappa$ on the support $[u, v]$ (the subscript $[u, v]$ is omitted if supported on the whole real line), by $\mathcal{G}(\lambda, \kappa)$ the Gamma distribution with mean $\lambda/\kappa$ and shape parameter $\kappa$, by $\mathcal{IG}(\lambda, \kappa)$ the inverse-Gamma distribution with mean $\kappa/(\lambda-1)$ and shape parameter $\kappa$, and by $\mathcal{U}_{[u,v]}$ the continuous uniform distribution on the support $[u,v]$. We denote the cardinality of a set $\mathcal{A}$ by $|\mathcal{A}|$ and the absolute value of a number $b$ by $|b|$. A truncated function is written as $b_+ := \max\{b, 0\}$.

\subsection{Data}\label{subsec:data}
To calculate the ASAF and ASSAF, we need annual death counts by country, age group, sex, and cause of death from the WHO Mortality Database \citep{WHO2017}, which covers data from 1950 to 2015 for more than 130 countries and regions around the world. This dataset comprises death counts registered in national vital registration systems and is coded under the rules of the International Classification of Diseases (ICD). Quinquennial population,  mortality rates, and life expectancy at birth were obtained from the 2017 Revision of the \textit{World Population Prospects} \citep{WPP2017} for each country, sex, and age group.

\subsection{Age-specific Smoking Attributable Fraction}\label{subsec:assaf}
\begin{figure}[!ht]
	\begin{center}
		\includegraphics[scale=0.35]{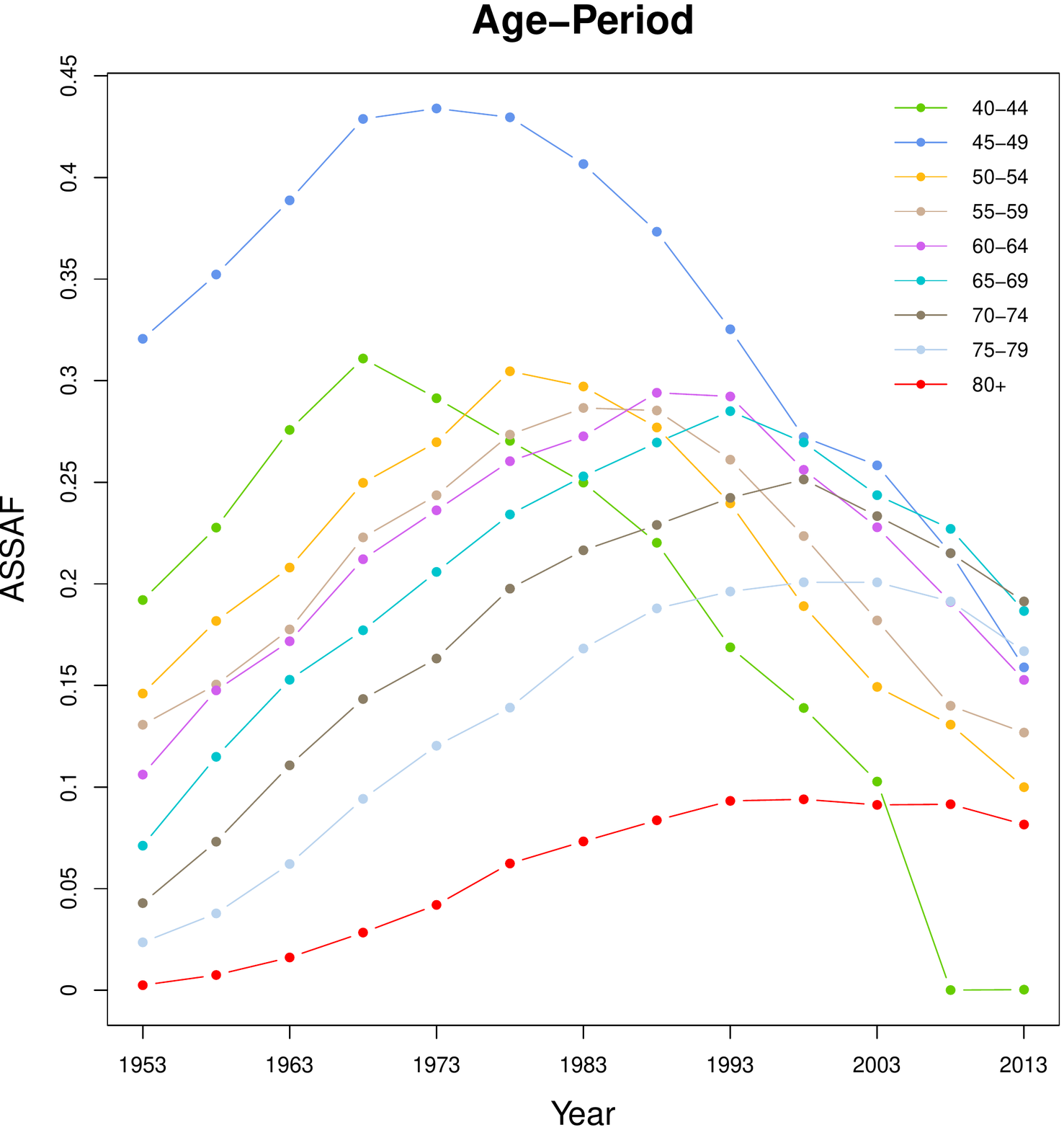}
		\includegraphics[scale=0.35]{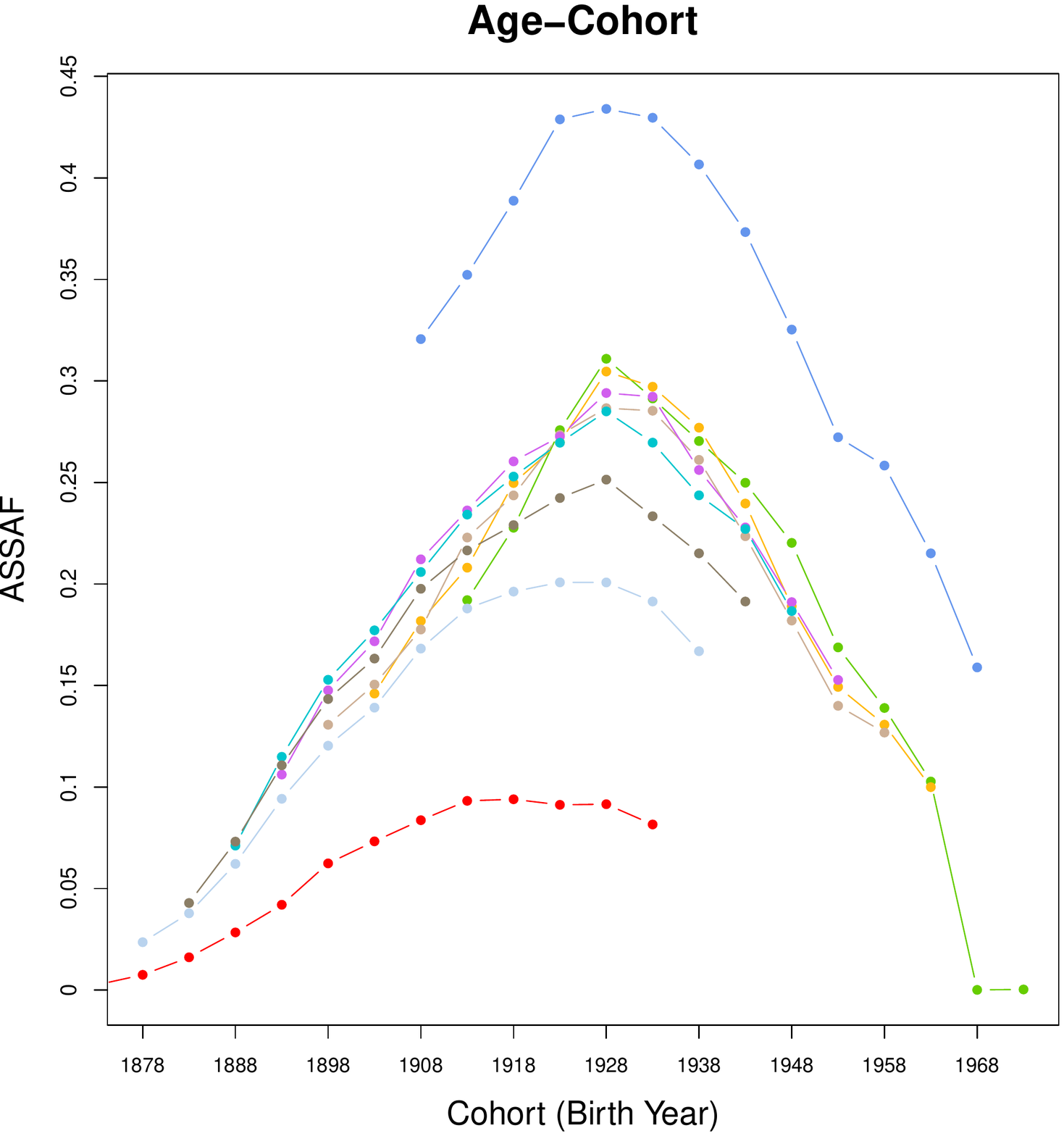}
	\end{center}	
	\caption{Age-specific smoking attributable fractions (ASSAF) for the male population in the United States from 1950-2015. Left: Age-period plot. The horizontal axis is the year of observation and colors differentiate age groups. Right: Age-cohort plot. The horizontal axis is the year of birth for all cohorts, where the values for each age group are shown by a different color. }\label{fg:asaf1}
\end{figure}

We use estimates of the smoking attributable fraction (SAF) obtained with the Peto-Lopez method, an indirect method based on the
observed lung cancer count data \citep{peto1992mortality,kong2016comparison,li2019estimating}.
Here we use a modified version of the Peto-Lopez method proposed by \citet{rostron2011estimating} to estimate the ASSAF. The modified method calculates the ASSAF for all 5-year age groups from 35 to 100, which is finer than the original Peto-Lopez method. Also, the reference lung cancer mortality rates used in the original Peto-Lopez method were underestimated because of selection bias, and the modified method addresses this by introducing an inflation factor. 
Because of data quality issues, we set ASSAF for age groups less than 40 to 0, and ASSAF for age groups 85 and older to the same value as that for the 80--84 age group. These rules follow the guidelines in \citet{peto1992mortality} and \citet{rostron2011estimating} with minor modifications, and result in nine  age groups with non-zero ASSAF. The left panel of Figure \ref{fg:asaf1} shows the estimated quinquennial  ASSAF of US males for all nine age groups (shown in different colors) from 1953 to 2013.

\subsubsection{Estimation and Forecasting: Age-cohort Modeling}\label{subsubsec:acm}
We propose a probabilistic age-cohort approach to estimate and forecast the ASSAF for the male population. The age-cohort plot of the US male ASSAF (right panel) in Figure \ref{fg:asaf1} has two main features that lead to our modeling. First, the ASSAF can be well approximated by the product of an age effect and a cohort effect. The ASSAF of age group $80+$ tends to shift horizontally from other age groups for most of the countries (e.g., see the red dashed line in the age-cohort plot of Figure \ref{fg:asaf1} for the case of US males). 
Hence, we apply a cohort effect $\tau$ for all age groups less than 80, and a separate cohort effect $\tilde{\tau}$ for the $80+$ age group. 

\begin{figure}[!ht]
	\begin{center}
		\includegraphics[scale=0.4]{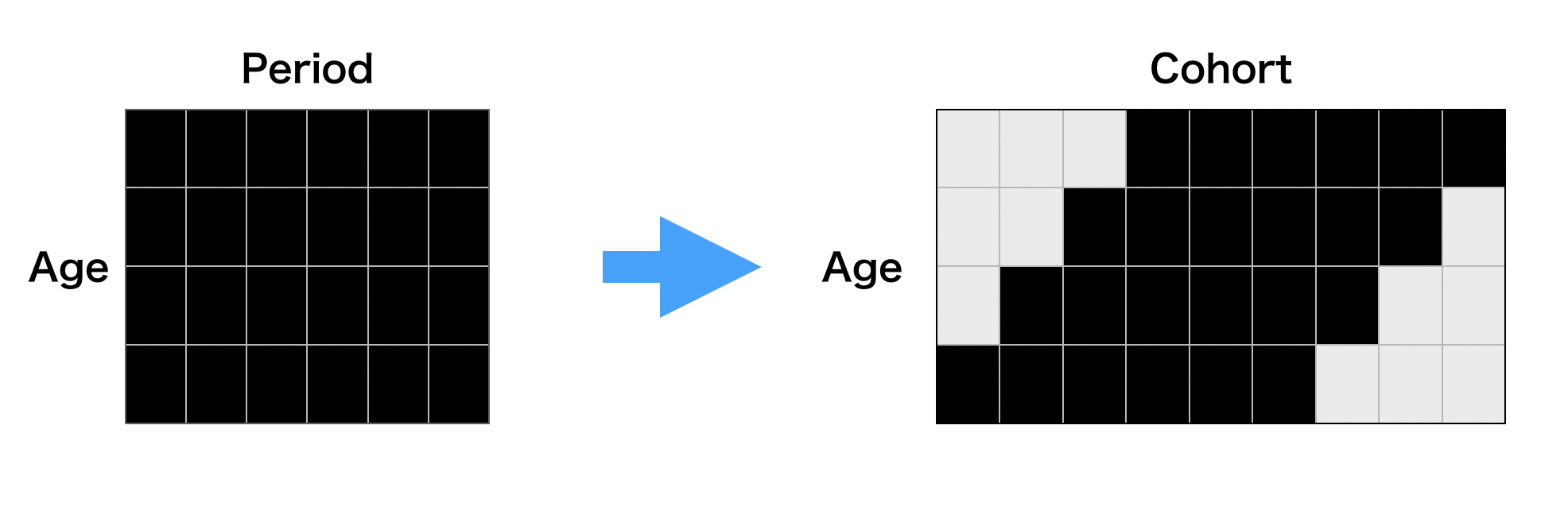}
	\end{center}	
	\caption{Transformation from age-period matrix (left) to age-cohort matrix (right). Black and grey cells represent observed and missing values, respectively.} \label{fg:cohorttb}
\end{figure}

The probabilistic model of ASSAF in country $\ell$ is 
\begin{align}\label{eq:assaf1}
y_{x,t}^\ell  &\stackrel{\rm ind}{\sim} \mathcal{N}(\xi_{x}^\ell \tau_{t-x}^\ell \mathbf{1}_{x \neq 80}+\xi_{x}^\ell \tilde{\tau}_{t-x}^\ell \mathbf{1}_{x = 80},\ \sigma^2_{\ell}),
\end{align}
where $x$ takes values in $\{40, 45, 50, 55, 60, 65, 70, 75, 80\}$. To ensure identifiability, we set $\xi_{40}^\ell = 1$ for all countries. Eq. (\ref{eq:assaf1}) is also closely related to a low-rank matrix completion method. The age-cohort matrix based on the observed period ASSAF inevitably contains missing values since we do not observe the ASSAF of early cohorts at young ages or that of late cohorts at old ages (see Figure \ref{fg:cohorttb}).

Second, the cohort pattern of the male ASSAF has a strong increasing-peaking-declining pattern. 
This trend can be well captured by a five-parameter double logistic function \citep{meyer1994bi}:
\begin{align}\label{eq:dl}
g(c|\theta) &:= \frac{k}{1+\exp\{-\triangle_1(c-1873-\triangle_2)\}} - \frac{k}{1+\exp\{-\triangle_3(c-1873-\triangle_2- \triangle_4)\}},
\end{align}
where $\theta := (\triangle_1, \triangle_2, \triangle_3, \triangle_4,  k)$. The double logistic curve is a flexible parametric curve, which has been used in many scientific fields such as hematology, phenology, and agricultural science. Due to its scientific interpretability, it is often used to describe social change, diffusion, and substitution processes \citep{grubler1999dynamics, fokas2007growth, kucharavy2011logistic}. Examples of the use of a double logistic curve to describe dynamics in human demography include mortality rates \citep{marchetti1996human}, life expectancy at birth \citep{raftery2013bayesian}, and total fertility rates \citep{alkema2011probabilistic}.

Most developed countries have already entered the declining stage of the smoking epidemic. The epidemic started in the early 1900s with a steady increase until the 1950s-60s when the adverse impact of smoking became widely known and anti-smoking measures started to be put in place. Since then, the smoking epidemic has continued to decline.
Thus the cohort effect of smoking exhibits a similar increasing-peaking-decreasing trend, which can be captured naturally by the double logistic curve. 

The cohort effect $\tilde{\tau}$ for ages $80+$ is just a horizontal shift of the cohort effect $\tau$ for younger ages, so we use two related double logistic curves to bridge them:
\begin{align}\label{eq:assaftau}
\tau_c^\ell|\theta^\ell, \sigma^{2[\tau]}  \stackrel{\rm ind}{\sim} \mathcal{N}(g(c|\theta^\ell),\ \sigma^{2[\tau]}),\hspace{1cm}
\tilde{\tau}_c^\ell|\tilde{\theta}^\ell, \sigma^{2[\tau]}\stackrel{\rm ind}{\sim} \mathcal{N}(g(c|\tilde{\theta}^\ell), \sigma^{2[\tau]}) ,
\end{align}
where $c := t-x$, $\theta^\ell := (\triangle_1^\ell, \triangle_2^\ell, \triangle_3^\ell, \triangle_4^\ell, k^\ell)$, and $\tilde{\theta}^\ell := (\triangle_1^\ell, \triangle_2^\ell, \triangle_3^\ell, \triangle_4^\ell + \delta^\ell, k^\ell)$. Here $\delta^\ell$ is a shift parameter controlling the amount of horizontal translation $\tilde{\tau}$ can make with respect to $\tau$. 


We use a three-level Bayesian hierarchical model (BHM) to estimate and forecast male ASSAF for all countries of interest jointly. Level 1 models the observed male ASSAF in terms of the tensor product of the age effect and the cohort effect (i.e., Eq. (\ref{eq:assaf1})). Level 2 models the distributions (conditioning on the global parameters) of the country-specific age effect $\xi_x^\ell$, the country-specific cohort effects $\tau^\ell_c$ and $\tilde{\tau}^\ell_c$ in Eq. (\ref{eq:assaftau}), the country-specific parameters $\theta^\ell$ and $\tilde{\theta}^\ell$ of the double logistic function, and the country-specific measurement variance $\sigma_\ell^2$.
Level 3 sets hyperpriors on the global parameters \\
$\psi := (\{\mu_{x}^{[\beta]}\}_{x \neq 40},\ \{\sigma^{2[\beta]}_{x}\}_{x \neq 40},\ \sigma^2,\ \sigma^{2[\tau]},\ \mu_{\triangle_1},\ \mu_{\triangle_2},\ \sigma^2_{\triangle_2},\ \mu_{\triangle_3},\ \mu_{\triangle_4},\ \sigma^2_{\triangle_4},\ \mu_{k},\ \sigma^2_{k},\ \mu_\delta,\ \sigma^2_\delta)$. More details of the specification of the full model are given in the Appendix \ref{app1}.

\begin{figure}[!ht]
	\begin{center}
		\includegraphics[scale=0.35]{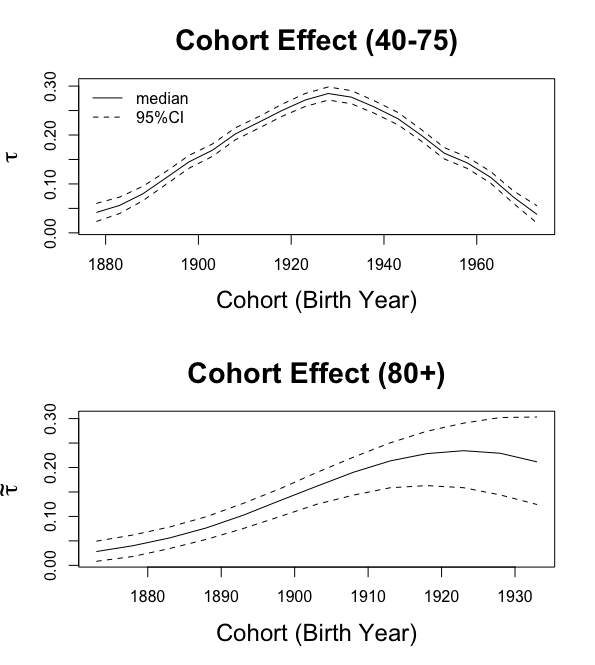}
		\includegraphics[scale=0.35]{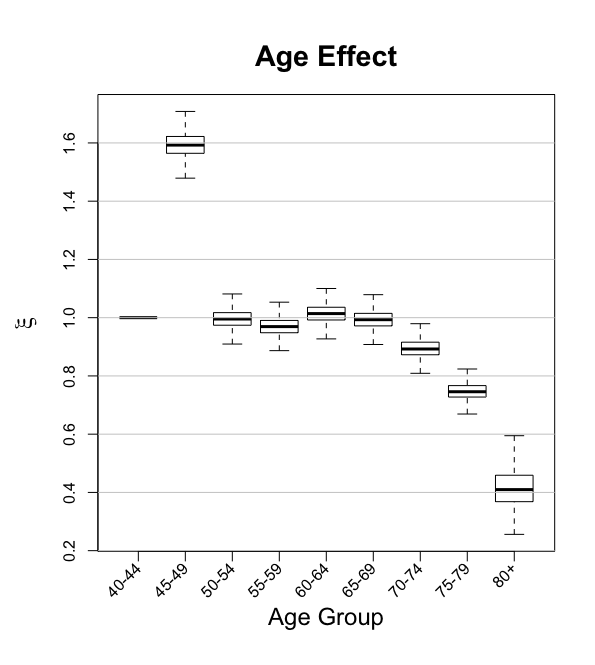}
	\end{center}	
	\caption{Posterior distributions of cohort and age effects of United States male ASSAF. Top Left: posterior median and $95\%$ credible intervals of the cohort effects $\tau$ for the 40--79 age groups. Bottom Left: posterior median and $95\%$ credible intervals of the cohort effect $\tilde{\tau}$ for the $80+$ age groups. Right: boxplot of posterior distribution of the age effect.} \label{fg:usaasaf1}
\end{figure}

\begin{figure}[!ht]
	\begin{center}
		\includegraphics[scale=0.3]{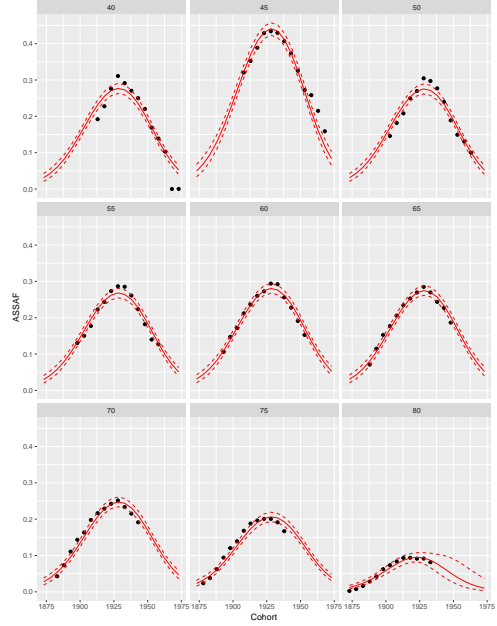}
	\end{center}	
	\caption{Posterior distributions of the means of US male ASSAF for all 9 age groups. The observed ASSAF is shown by black dots. The posterior median and $95\%$ credible intervals  of the means are shown by solid and dashed red lines, respectively.}\label{fg:usaasaf2}
\end{figure}

The left and right panels of Figure \ref{fg:usaasaf1} plot the cohort effects and age effect of US male ASSAF, respectively. The estimated cohort effect $\tau$ for the age groups 45-79 shows a clear increasing-peaking-decreasing trend as observed in Figure \ref{eq:assaf1}. The estimated cohort effect $\tilde{\tau}$ for the $80+$ age group shows the same trend for the 13 cohorts reaching age 80  by 2015. We could forecast any cohort effects based on the posterior distribution of the double logistic function. 
The estimated age effect indicates that the smoking-attributed fraction of mortality is higher among middle-aged males (aged 40--69) in the US than among older males (70 and over). Figure \ref{fg:usaasaf2} plots the posterior distributions of the means of the US male ASSAF for all 9 age groups and all 21 cohorts.

To project the future ASSAF, we first generate future cohort effects by plugging samples drawn from the posterior distributions of country-specific parameters $\theta^\ell$ and $\tilde{\theta}^\ell$ in Eq. (\ref{eq:dl}) and (\ref{eq:assaftau}). Then, we apply Eq. (\ref{eq:assaf1}) using samples drawn from posterior distributions of the future cohort effects, age effect, and country-specific variance $\sigma^2_\ell$ to get projections of ASSAF.  

\subsection{Non-smoking Life Expectancy}\label{subsec:nonsmke0}
The non-smoking life expectancy at birth, $e_0^{NS}$, is the life expectancy at birth that a population would have if no one smoked, but all mortality risks were otherwise the same \citep{bongaarts2006long}. 
To estimate $e_0^{NS}$, we need the age-specific mortality rates $d_x$ and the ASSAF $y_x$ described in Section \ref{subsec:assaf}. As in the last section, all quantities described in this section are specific to the male population, and the sex index $s$ is omitted unless otherwise specified. 

\begin{figure}[!ht]
	\begin{center}
		\includegraphics[scale=0.4]{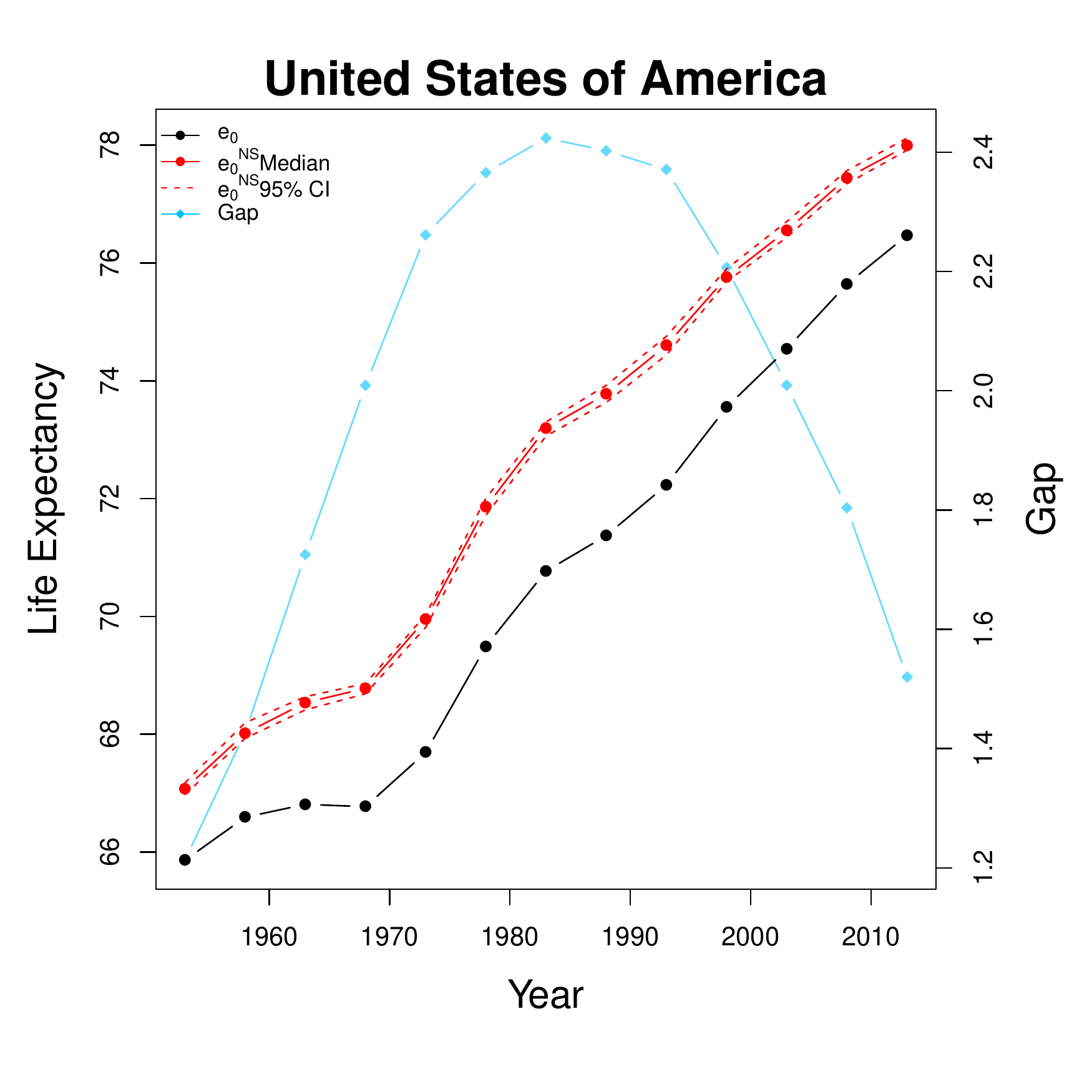}
		\includegraphics[scale=0.4]{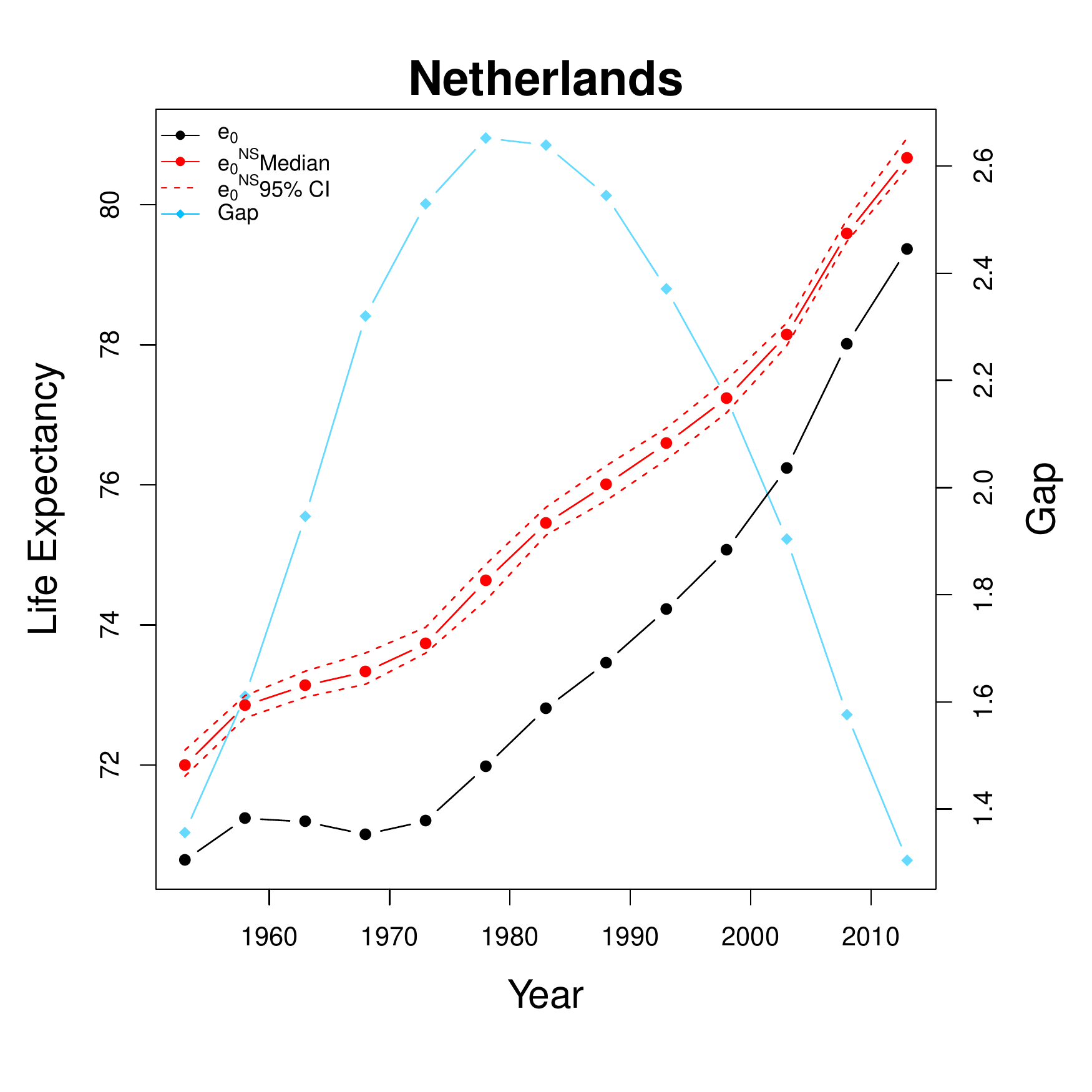}
	\end{center}	
	\caption{Male life expectancy at birth, $e_0$, and male non-smoking life expectancy at birth, $e_0^{NS}$, for the United States (left) and the Netherlands (right).  The black line shows $e_0$. The solid red line and the dashed red lines show the posterior median and the $95\%$ credible interval  of $e_0^{NS}$. The blue line represents the gap between $e_0$ and $e_0^{NS}$. }\label{fg:nonsmke0}
\end{figure}

The calculation of $e_0^{NS}$ consists of two steps. First, the age-specific non-smoking attributable mortality rate for a given country $\ell$, age group $x$, and period $t$ (denoted by $d^{NS}_{\ell, x, t}$) is calculated as 
\begin{align}\label{eq:dns}
d^{NS}_{\ell, x,t} := (1-y_{x,t}^\ell)\cdot d_{x,t}^\ell. 
\end{align}
Second, we convert the set of $d^{NS}_{\ell, x,t}$ to $e_0^{NS}$ using the standard period life table method \citep[Chapter 3]{Preston&2000}, as implemented in the \textit{life.table} function in the R package \texttt{MortCast} \citep{sevcikova2019mortcast}. Figure \ref{fg:nonsmke0} shows the relationship between quinquennial $e_0$ and $e_0^{NS}$ for US males and Netherlands males from 1950 to 2015, respectively. The vertical gap between $e_0$ and $e_0^{NS}$ at each time point presents the years of life expectancy lost due to smoking. The changes in the gaps also follow a similar increasing-peaking-decreasing trend over the period 1950 to 2015.

\subsubsection{Estimation and Forecasting: Non-linear Life Expectancy Gain Model}
We forecast $e_0^{NS}$ by investigating the nonlinear five-year gains of $e_0^{NS}$. As discussed by \citet{raftery2013bayesian}, the improvement of gains on $e_0$ for most of the countries has experienced a  slow-rapid-slow increasing pattern and a six-parameter double logistic function is used to capture the non-linearity of five-year gains of $e_0$:
\begin{align}\label{eq:dl2}
\tilde{g}(e_0|\zeta) := \frac{w}{1 + \exp\{-\frac{4.4}{a_2}(e_0 -a_1 - 0.5a_2 )\}  } + \frac{z-w}{1 + \exp\{-\frac{4.4}{a_4}(e_0 -\sum_{i = 1}^3a_i - 0.5a_4 )\} },
\end{align}
where $\zeta := (a_1, a_2, a_3, a_4, w, z)$
and $z$ is the asymptotic average rate of increase in $e_0$. 
We assume that $z$ is nonnegative, implying that life expectancy will continue to increase on average \citep{oeppen2002enhanced, bongaarts2006long}.

\begin{figure}[!ht]
	\begin{center}
		\includegraphics[scale=0.30]{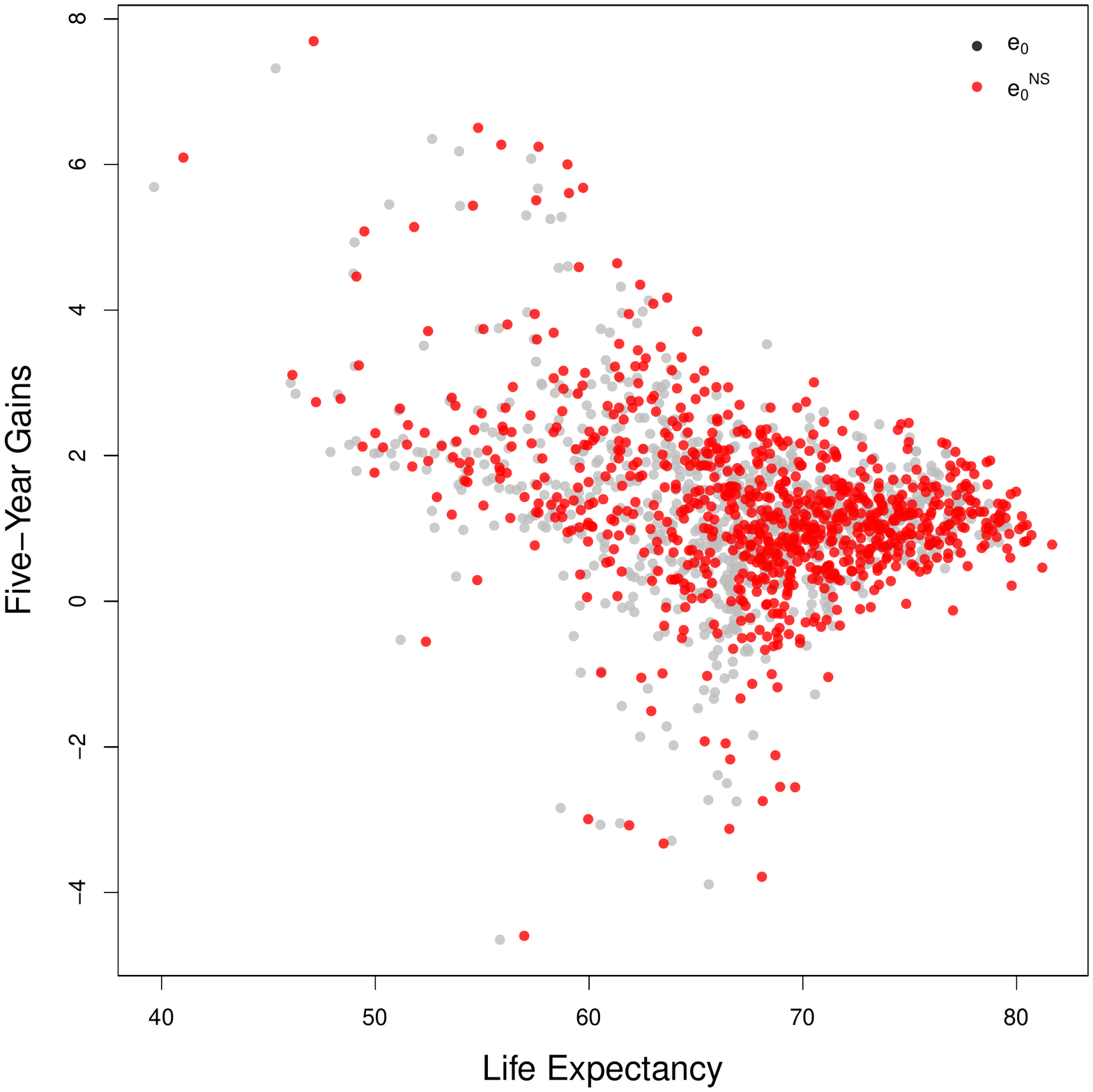}
		\includegraphics[scale=0.30]{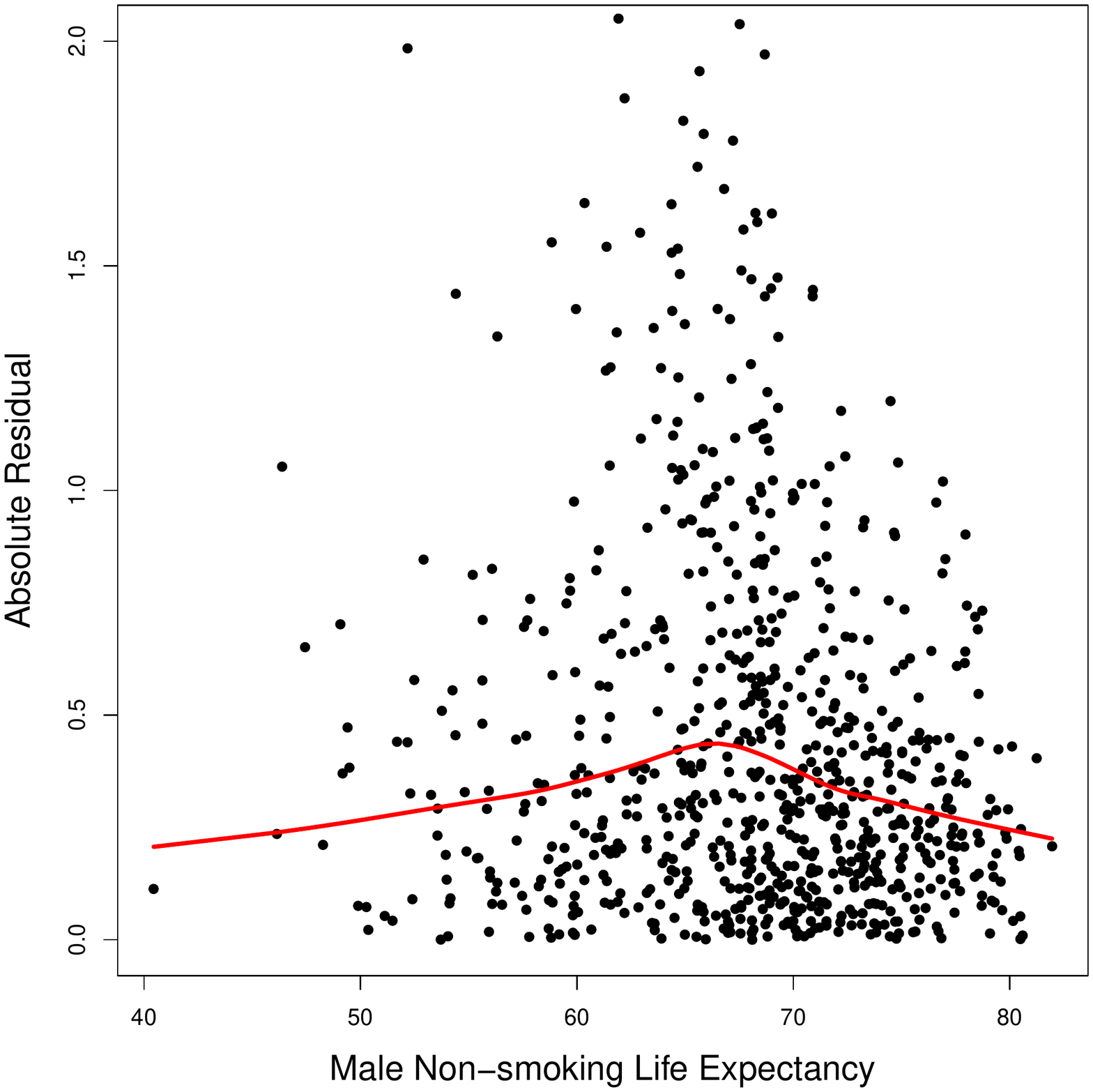}
	\end{center}	
	\caption{Left: Five-year gains of $e_0$ and $e_0^{NS}$ for 69 countries from 1950 to 2015. The gains in $e_0$ and $e_0^{NS}$ are represented using grey and red dots, respectively. Right: Plot of absolute residuals estimated from the constant variance model against life expectancy shown by black dots, with fitted regression spline shown by the red line. }\label{fg:nonsmke0gain}
\end{figure}

The five-year gains in $e_0^{NS}$ exhibit this nonlinear pattern as well. The left panel of  Figure \ref{fg:nonsmke0gain} plots the observed five-year gains of $e_0$ (in grey dots) and $e_0^{NS}$ (in red dots) for the 69 countries with data of high enough quality from 1950 to 2015. The five-year gains in $e_0^{NS}$ have nearly the same shape as the five-year gains in $e_0$, which supports using the same double logistic function to model the gains. Also, $e_0^{NS}$ has almost the same five-year gain at the highest age as $e_0$, suggesting that the asymptotic average rate of increase $z$ for $e_0^{NS}$ should be similar to that of $e_0$. 
Further, the variability of the five-year gains of $e_0^{NS}$ changes from a low level to a high level of $e_0^{NS}$, which suggests including a nonconstant variance component in the model.

We use a three-level Bayesian hierarchical model for $e_0^{NS}$. Level 1 models $e_{0, \ell, t}^{NS}$ for country $\ell$ and period $t$ by
\begin{align}\label{eq:bl1}
e_{0, \ell, t}^{NS}  &\stackrel{\rm ind}{\sim} \mathcal{N} (e_{0, \ell, t-1}^{NS} + \tilde{g}(e_{0, \ell, t-1}^{NS} |\zeta^{\ell}),\ (\omega^\ell \cdot \phi (e_{0, \ell, t-1}^{NS}))^2),
\end{align}
with country-specific parameters $\zeta^\ell := (a_1^\ell, a_2^\ell, a_3^\ell, a_4^\ell, w^\ell, z^\ell)$. Here $\phi (\cdot)$ is a regression spline fitted to the absolute residuals resulting from the model with constant variance in Eq. (\ref{eq:bl1}) with the same estimation method described later. The regression spline is used to account for the changing variability of the observed data. The right panel of Figure \ref{fg:nonsmke0gain} illustrates the varying absolute residuals with the fitted spline in red. Level 2 specifies the conditional distribution for all country-specific parameters including $\zeta^\ell$ and $\omega^{\ell}$.
Level 3 sets the hyperpriors for the global parameters $\tilde{\psi} := (\{\mu_{a_i}\}_{i = 1}^4, \{\sigma^2_{a_i}\}_{i = 1}^4, \mu_w, \sigma^2_w, \mu_z, \sigma^2_z)$. The full specification of the model is given in the Appendix \ref{app2}. 

To produce a probabilistic forecast, we sample from the joint posterior distributions of the country-specific parameters $\zeta^\ell$ to calculate the five-year gains $\tilde{g}(e_0^{NS})$ together with the posterior distributions of $\omega^\ell$. For the variance component, we evaluate $ \phi (e_{0, \ell, t-1}^{NS})$ if $e_{0, \ell, t-1}^{NS}$ is within the range of the fitted data; otherwise, it is set equal to the spline value evaluated at the largest observed $e_{0}^{NS}$. We then use Eq. (\ref{eq:dl2}) and (\ref{eq:bl1}) to generate samples from the posterior predictive distribution for future country-specific $e^{NS}_{0,\ell, t}$. The set of samples approximates the posterior predictive distribution.

\subsection{Male-Female Joint Forecast}\label{subsec:mfjoint}
\subsubsection{Male $e_0$ Forecast}\label{subsubsec:e0male}
\begin{figure}[!ht]
	\begin{center}
		\includegraphics[scale=0.35]{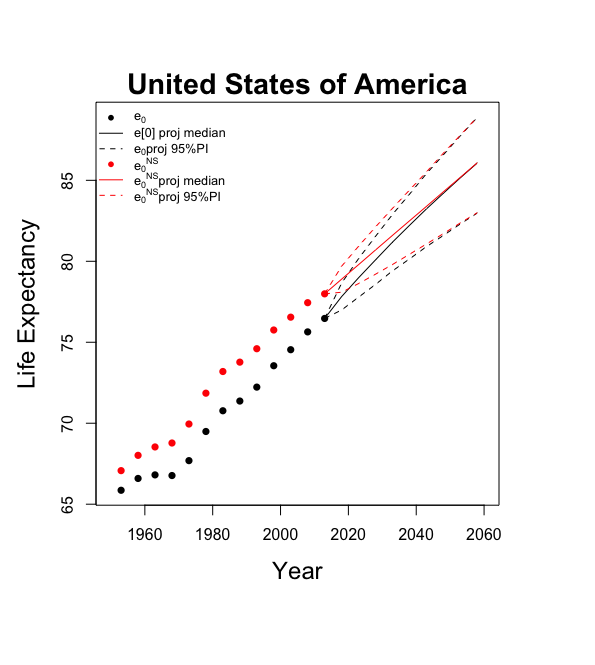}
		\includegraphics[scale=0.35]{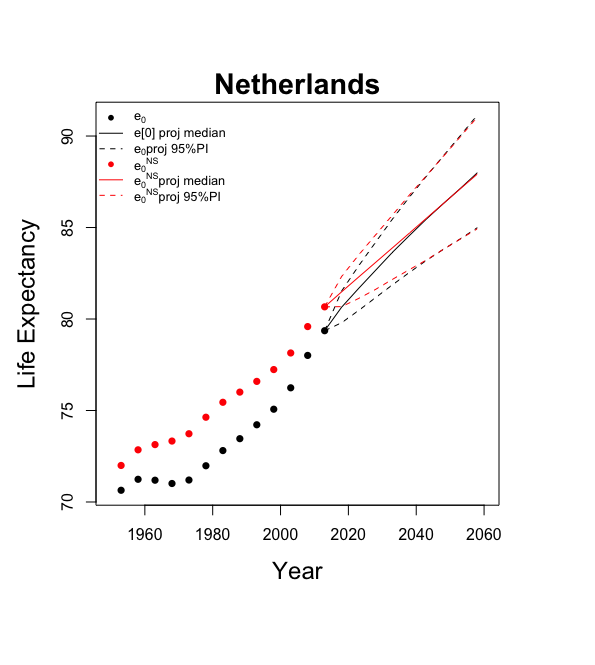}
	\end{center}	
	\caption{Projections of $e_0^{NS}$ and $e_0$ of US (left) and the Netherlands (right) males to 2060. The posterior medians and the $95\%$ predictive intervals of projected $e_0^{NS}$ are shown by solid and dashed red lines, respectively. The posterior medians and the $95\%$ predictive intervals of projected $e_0$ are shown by solid and dashed black lines, respectively. }\label{fg:usae0}
\end{figure}

First, we use the coherent Lee-Carter method \citep{LiLee2005,vsevvcikova2016age} to convert the projected $e_{0, \ell, t}^{NS}$ back to $d^{NS}_{\ell, x, t}$ for all age groups $x$ at period $t$ of country $\ell$. Then, we invert Eq. (\ref{eq:dns}) to get the projected age-specific all-cause mortality, i.e., $d_{x, t}^\ell = d^{NS}_{\ell, x, t}/(1- y_{x,t}^\ell)$ for any age groups $x$, period $t$, and country $\ell$. Finally, applying the same life table method described in Section \ref{subsec:nonsmke0} to the forecast $d_{x, t}^\ell$, we obtain the forecast life expectancy at birth for period $t$ and country $\ell$. Figure \ref{fg:usae0} illustrates the projections of $e_0^{NS}$ and $e_0$ for US and the Netherlands males to 2060. The projected $e_0$ converges to the projected $e_0^{NS}$ as ASSAF decreases towards 0 for all age groups of US and the Netherlands males. 

\subsubsection{Female $e_0$ Forecast: Gap Model}\label{subsubsec:gap}
We propose a gap model similar to that of \citet{raftery2014joint} to produce a coherent projection of male-female life expectancy at birth.
It has been argued that  differences in smoking largely account for the life expectancy gap between males and females \citep{preston2006sex,wang2009forecasting}. 
Here we explore the relationship between the between-sex gap in life expectancy and the between-sex gap in the all-age smoking attributable fraction (ASAF). The ASAF is a single statistic summarizing the smoking effect on mortality and is defined as a weighted average of the ASSAF values as calculated in Section \ref{subsec:assaf}, where the weights are the age-specific mortality rates.  \citet{li2019estimating} describe 
the estimation of ASAF, as well as a method for forecasting it
using a four-level Bayesian hierarchical model. 


We modify the gap model of \citet{raftery2014joint} by adding the country-specific between-sex ASAF gap as a covariate. The proposed gap model is as follows:
\begin{align}\label{eq:gap}
G_{t}^\ell &:= \min\{\max\{\tilde{G}_{t}^{\ell}, L  \}, U\}\\\nonumber
\tilde{G}_{t}^\ell  &\stackrel{\rm ind}{\sim} 
\mathcal{N}(\beta_0 + \beta_1e_{0,m, 1953}^\ell + \beta_2G_{t-1}^\ell + \beta_3e_{0,m,t}^\ell +
\beta_4(e_{0,m,t}^\ell - \varpi)_+ +  \beta_5h_{t}^\ell,\ \sigma^{2}_G),
\end{align}
where $U$ and $L$ are the observed historical maximum and minimum of the between-sex gap in $e_0$,
$\varpi$ is the level of male $e_0$ at which the gap is expected to stop widening, and $h_t$ is the between-sex gap (male minus female) of the posterior median of ASAF in period $t$. 

The estimated parameters of the model based on the data for 69 countries for 1950--2015 are reported in Table \ref{tb:gap}. Our estimates indicate that the $e_0$ sex gap has a strong positive association with the ASAF gap after adjusting for other factors ($\hat{\beta}_5 =  1.180$ with p-value $<0.01$). Since the estimated lower bound of the life expectancy gap $L$ is positive, our model guarantees that no crossover of male and female life expectancy forecasts will happen for all trajectories. The other coefficients have similar estimates and significance as in \citet{raftery2014joint}, which accounts for the remaining variability in the between-sex life expectancy gap, possibly due to biological and other social factors \citep{janssen2015adoption}. 


When performing projection, we forecast all terms in Eq. (\ref{eq:gap}) forward. Instead of using a random walk as in \citet{raftery2014joint}, we make use of the ASAF gap to guide our projection. However, we constrain the quantity $(e_{0,m,t}^\ell - \varpi)_+$ to be 20 when $e_{0,m,t}^\ell $ is greater than 81 years, which is the largest male $e_0$ observed in countries of interest up to 2015, since there is not enough information to determine whether the gap will continue to shrink for higher $e_0$. After the gender gap has been forecast, we add the gap to each posterior trajectory of the forecast male $e_0$ to get the full posterior predictive distribution of female $e_0$.

\begin{table}[tbp]
	\caption{Estimated gap model coefficients with standard errors in parentheses, if available.} \label{tb:gap}
	\begin{center}
		\begin{tabular}{ccc|ccc}
			Variable&	Parameter&  Estimate & Variable & Parameter & Estimate\\
			\hline
			Intercept &	$\beta_0$& -2.173 (0.627) & $h_t^\ell$ &	$\beta_5$	& 1.180 (0.384)\\
			$e_{0,m,1953}^\ell$ &	$\beta_1$& 0.012 (0.003) &  &$\sigma_G$& 0.496  \\
			$G_{t-1}^\ell$	&$\beta_2$&0.901  (0.010) & & $\varpi$ & 61  \\
			$e_{0,m,t}^\ell$&	$\beta_3$&0.043 (0.011) && $L$ & 0.03  \\
			$(e_{0,m,t}^\ell - \varpi)_+$&	$\beta_4$& -0.107 (0.012) &	&$U$ & 13.35  \\
			\hline
			\hline
			\multicolumn{3}{c}{$R^2$ } & \multicolumn{3}{c}{0.933}
		\end{tabular}
	\end{center}
\end{table}

\subsection{Estimation and Projection of the Full Model}

We use data from 69 countries for which the data on the male smoking-attributable mortality was of good enough quality. The precise data quality criteria and thresholds used are described in \citet{li2019estimating}.
Of these 69 countries, two are in Africa, 16 are in the Americas, nine are in Asia, 40 are in Europe and two in Oceania.
Estimation of the full model makes uses of male ASSAF, male age-specific mortality rates, both sexes $e_0$, and both sexes ASAF of all 69 clear-pattern countries over 13 five-year periods during 1950--2015. Future $e_0$ of the same set of countries over 9 five-year periods from 2015 to 2060 is projected based on the joint posterior predictive distribution of the full model. The full procedure is described in the Appendix \ref{app1}.

We use Markov Chain Monte Carlo (MCMC) to sample from the joint posterior distributions of the parameters of interest. For the BHM of the ASSAF, we ran three chains, each of length 100,000 iterations thinned by 20 iterations with a burn-in of 2,000. This yielded a final, approximately independent sample of size 3,000 for each chain.
For the BHM of each of the 30 samples of $e_0^{NS}$, we ran one chain with length 100,000 iterations thinned by 50 with a burn-in of 1,000. This yielded a final, approximately independent sample of size 1,000 for each chain. 
We monitored convergence by inspecting trace plots and using standard convergence diagnostics, details of which are given in the Appendix \ref{app2}. We include the plots of $e_0$ projections for all 69 countries and both sexes in the Appendix \ref{app3}.

\section{Results}\label{sec:res}
We assess the predictive performance of our model using out-of-sample predictive validation.
\subsection{Study design}\label{subsec:des}
The data we used for out-of-sample validation cover the period 1950--2015,  
dividing it into an earlier training period and a later test period.
We fit the model using only data from the training period, and then generated
probabilistic forecasts for the training period. We finally compared the 
probabilistic forecasts with the observations for the training period.
We used two different choices of test period: 2000--2015, and 2010--2015.
The former allows us to assess longer-term forecasts, while the latter
focuses on shorter-term forecasts.

To assess the accuracy of the probabilistic forecasts, we define the sex-specific mean absolute error (MAE) as
\begin{align}\label{eq:mae}
\mbox{MAE}_s = \frac{1}{|\mathcal{L}||\mathcal{T}|}\sum_{\ell \in \mathcal{L}} \sum_{t \in \mathcal{T}} |\hat{e}_{0,s,t}^\ell - e_{0,s,t}^\ell|,
\end{align}
where $\mathcal{L}$ is the set of countries considered in the validation, $\mathcal{T}$ is the set of training periods, and $\hat{e}_{0,s,t}^\ell $ is the posterior median of the predictive distribution of life expectancy at birth at year $t$ for country $\ell$ and sex $s$. To assess the calibration and sharpness of the model, we calculated the average empirical coverage of the prediction interval over the validation period, which we hope to be close to its nominal level with as short a halfwidth of the interval as possible \citep{gneiting2007strictly}. 

\subsection{Out-of-sample validation}\label{subsec:osv}
We evaluated and compared the performance of the proposed model with four commonly used methods for forecasting $e_0$: the Lee-Carter method \citep{lee1992modeling}, the Lee-Miller method \citep{lee2001evaluating}, the Hyndman-Ullah functional data method \citep{hyndman2007robust}, and the Bayesian hierarchical model as implemented in the {\tt bayesLife} R package \citep{raftery2013bayesian}. We refer to the last as the bayesLife method. The first three methods were implemented using the corresponding functions with default settings in the \texttt{demography} R package \citep{booth2006lee, hyndman2019demo}. The bayesLife method was implemented under default settings using the R package \texttt{bayesLife}  \citep{raftery2013bayesian, raftery2014joint, sevcikova2019bayeslife}.

Table \ref{tb:osv} gives the out-of-sample validation results for the four methods described above as well as our proposed method. Our method had the smallest MAE for both sexes and both choices of test period among the five methods. For predicting one five-year period ahead, our method improved accuracy over the Lee-Carter method by $70\%$ ($60\%$), and over the bayesLife method by $24\%$ ($11\%$) for males (females).  For predicting three five-year periods ahead, the new method improved accuracy over the Lee-Carter method by $53\%$ ($40\%$), and over the bayesLife method by $24\%$ ($17\%$) for males (females). 

For model calibration, the Lee-Carter-type models produced predictive intervals that are too narrow, thus underestimating the predictive uncertainty in the testing period. The bayesLife method and the new method produced predictive intervals with coverage close to the nominal level. We assess the sharpness of the forecast method using the $80\%$ predictive interval halfwidth. 
For male data under the three five-year periods prediction, the $80\%$ predictive interval of the new method was $30\%$ shorter on average, but yielded the same empirical coverage as the bayesLife method. Under the one five-year out-of-sample predictions, the $80\%$ predictive interval of the new method was $30\%$ shorter on average but yielded even higher empirical coverage than the bayesLife method. For female data, the predictive intervals of our method overcovered the observations slightly for each choice of test period, but their median halfwidths were not much wider than those of the bayesLife method (e.g., the largest increment was less than $18\%$). The major source of variability in the female projections of the new method comes from the gap model.

\begin{table}[!hp]
	\caption{Out-of-sample validation results for forecasting life expectancy at birth of males and females one and three five-year periods ahead. ``Num" is the number of countries used in the validation. In the ``Method" column, 
		``H-U FDA" is the Hyndman-Ullah functional data analysis method, ``bayesLife" represents the method described in \citet{raftery2013bayesian}, and ``smokeLife" is the our proposed method.
		``Halfwidth" represents the median of the halfwidth of the prediction interval.}\label{tb:osv}
	\centering
	\begin{tabular}{c|c|c|c|c|cc|cc}
		\hline
		\multirow{2}{*}{Period} & 	\multirow{2}{*}{Num}	&	\multirow{2}{*}{Sex} &	\multirow{2}{*}{Method} & 	\multirow{2}{*}{MAE}   & \multicolumn{2}{c|}{Coverage}&  \multicolumn{2}{c}{Halfwidth}  \\
		&  & & & & $80\%$ & $95\%$ & $80\%$ & $95\%$\\
		\hline
		\multirow{5}{*}{Train:1950--2000} & \multirow{10}{*}{67} &
		\multirow{5}{*}{M} &
		Lee-Carter &2.043 & 0.144& 0.199 & 0.368 & 0.568 \\
		&	& & Lee-Miller &  1.536 & 0.318& 0.418& 0.831&1.239  \\
		&	 & & H-U FDA & 2.206 & 0.189 & 0.274& 0.808& 1.259 \\
		&	 & & bayesLife  &1.273 &\textbf{0.741}  & \textbf{0.950}& 1.722 & 2.714\\
		&   & & smokeLife &\textbf{0.962} &\textbf{0.741 }& 0.896 & 1.197& 1.943\\
		\cline{3-9}
		\multirow{5}{*}{Test: 2000--2015}& &	\multirow{5}{*}{F}
		& Lee-Carter &1.210 & 0.199 & 0.294 & 0.391 & 0.599\\
		&	& & Lee-Miller & 0.748 & 0.602 & 0.756& 0.612& 0.940\\
		&	 & & H-U FDA &  1.430 & 0.114 & 0.299 & 0.412 & 0.633\\
		&	 & & bayesLife &0.876  &\textbf{0.816} &\textbf{0.955} &1.312 & 1.985\\
		&    & & smokeLife &  \textbf{0.718}& 0.891  & 1.000 & 1.380 & 2.173 \\
		\hline
		\multirow{5}{*}{Train:1950--2010} &\multirow{10}{*}{68} & \multirow{5}{*}{M} 
		& Lee-Carter &  1.741 & 0.103 & 0.118 & 0.306 & 0.448 \\
		&	 & & Lee-Miller &  0.853 & 0.544& 0.721& 0.581&0.931  \\
		& & & H-U FDA &1.364 & 0.191& 0.324& 0.548& 0.791 \\
		&	 & & bayesLife &0.688 &\textbf{0.824}  & 0.897& 1.098 & 1.748\\
		&    & & smokeLife&\textbf{0.523} &0.912& \textbf{0.985} & 0.773 & 1.250 \\
		\cline{3-9}
		\multirow{5}{*}{Test: 2010--2015} & &	\multirow{5}{*}{F}
		& Lee-Carter & 1.025& 0.118 & 0.221& 0.279& 0.436\\
		&	 & & Lee-Miller &  0.486& 0.662& 0.779& 0.476& 0.708\\
		&	 & & H-U FDA & 0.895 & 0.250 & 0.368& 0.373 & 0.573\\
		&	 & & bayesLife  &0.464  &\textbf{0.868}&\textbf{0.941} &0.853 & 1.291\\
		&    & & smokeLife  &\textbf{0.413 }& 0.971  & 1.000 & 0.974& 1.517 \\
		\hline
	\end{tabular}
	
\end{table}

\section{Case studies}\label{sec:case}
On average, smoking results in 1.4 years lost of male life expectancy at birth for the 69 countries over 1950-2015. The trend in years lost due to smoking also follows the pattern of the smoking epidemic. The average years lost due to smoking  among males increased from 0.9 in 1953 to a maximum of 1.7 in 1993, and decreased to 1.3 in 2013. 

For male populations of most countries, the ASSAF 
has already passed the peak for most age groups. 
When this is the case, accounting for the smoking effect 
leads to higher forecasts of life expectancy at birth.
On average, our proposed method gives forecasts of male life expectancy at birth that are 1.1 years higher than the bayesLife method used by the UN for the 69 countries over the period 2015--2060.

Most female populations are still at the increasing or peaking stage of the smoking epidemic. However, for 2055-2060, we expect to see an increment of 1.0 in female life expectancy compared to the forecast result from the bayesLife method, since the female smoking epidemic will be following the same decreasing trend as that of males by then.

We now study four countries in detail, representing different patterns of the smoking epidemic.

\subsection{United States}
The United States of America has one of the best vital registration systems in the world and also high quality data on cause of death. It thus has high quality data on the SAF. The smoking epidemic started in the early 1900s among the male population and rose to the historical maximum of around $60\%$ in the 1950s. 
At that point, government programs and social movements against smoking began to develop, and the US public became increasingly aware of the adverse impacts of smoking.
Since then, there has been a substantial decrease in smoking prevalence, going down to about $20\%$ in the 1990s, and $17.5\%$ in 2016 \citep{burns1997cigarette, islami2015global}. 

\begin{figure}[!hp]
	\centering
	\begin{subfigure}[tbp]{.4\textwidth}
		\centering	
		\includegraphics[width=\textwidth]{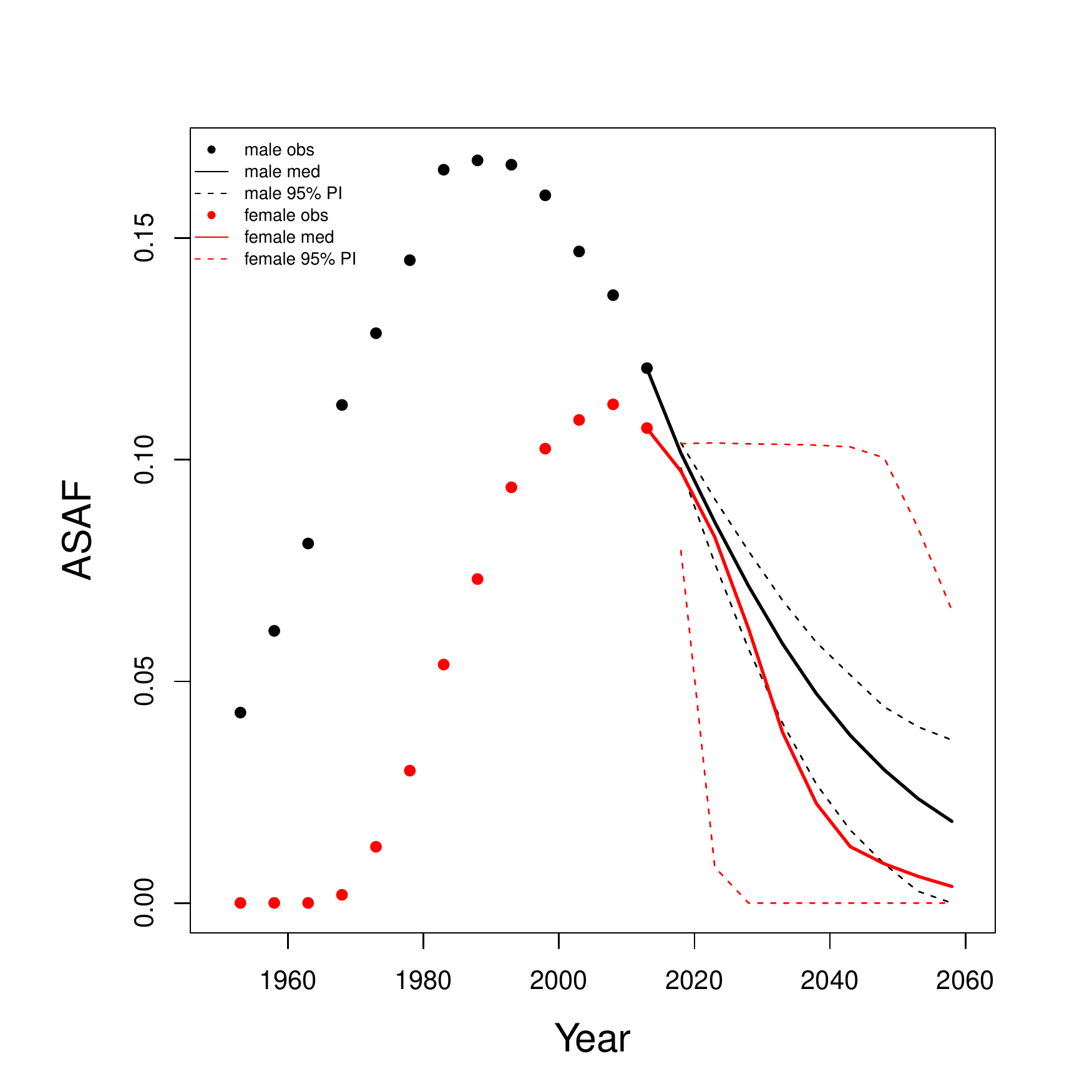}
		\caption{}
	\end{subfigure}
	\begin{subfigure}[tbp]{.4\textwidth}
		\centering
		\includegraphics[width=\linewidth]{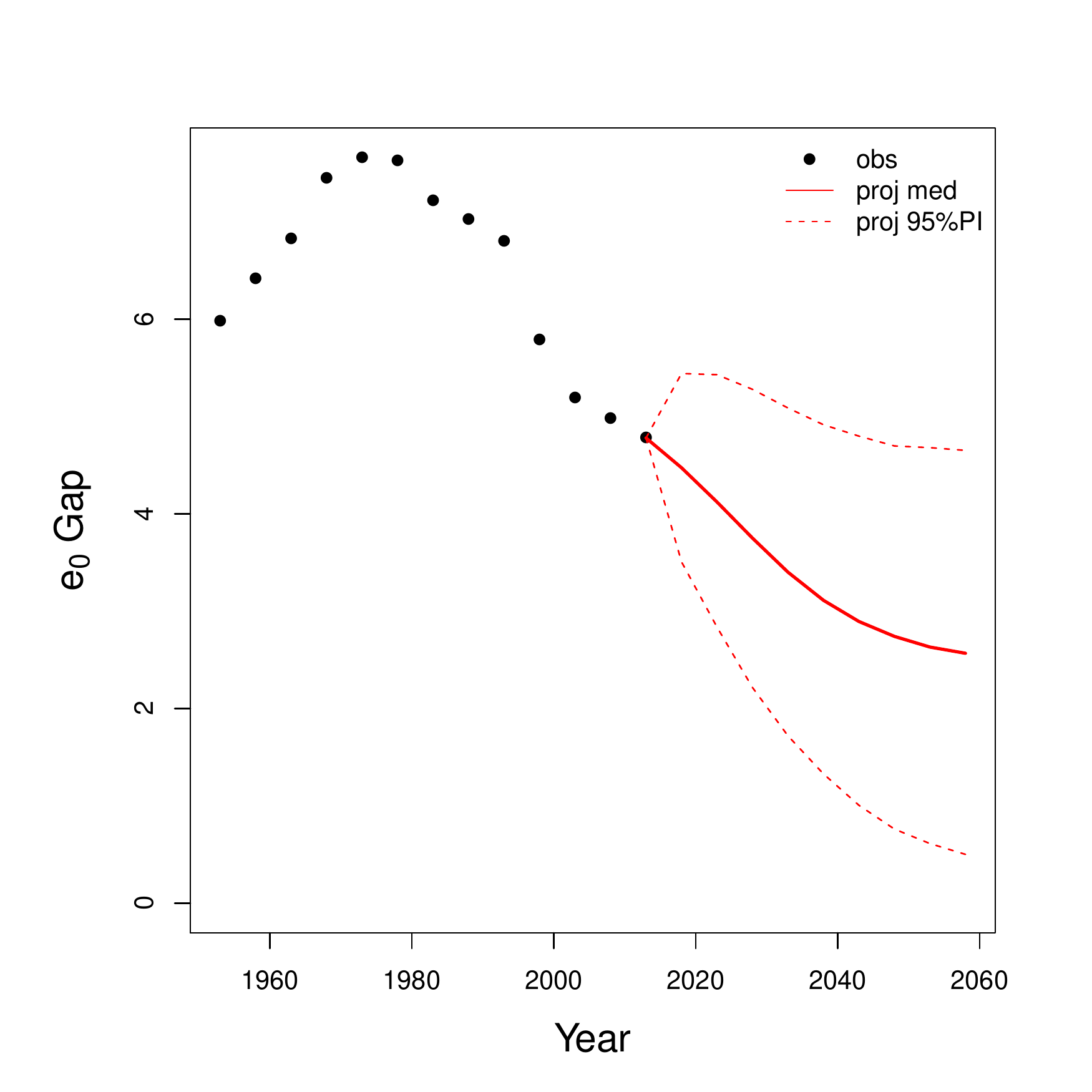}
		\caption{}
	\end{subfigure}	
	\begin{subfigure}[tbp]{.4\textwidth}
		\centering
		\includegraphics[width=\linewidth]{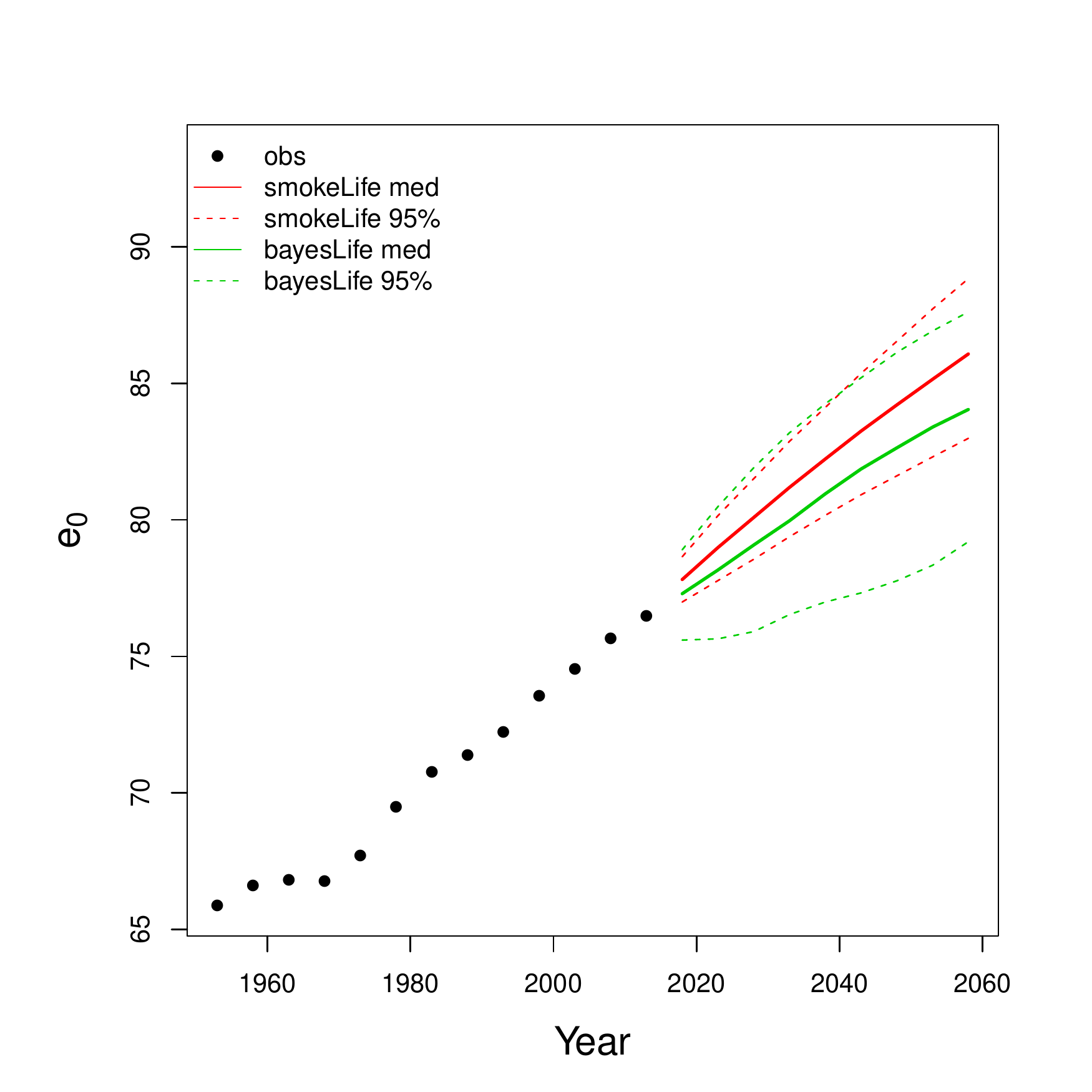}
		\caption{}
	\end{subfigure}		
	\begin{subfigure}[tbp]{.4\textwidth}
		\centering
		\includegraphics[width=\linewidth]{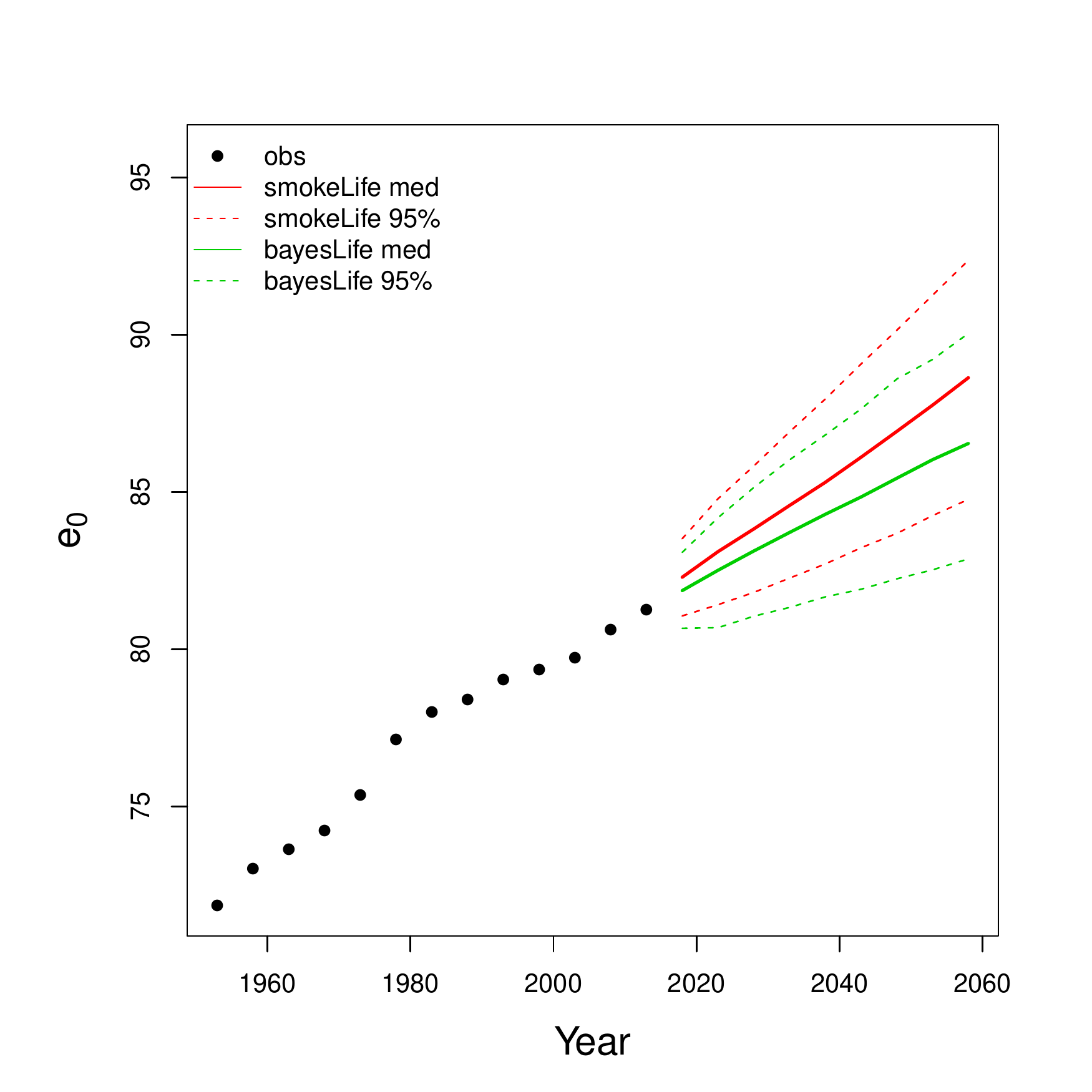}
		\caption{}
	\end{subfigure}	
	\caption{United States of America. (a) All-age smoking attributable fraction (ASAF) for male (black) and female (red) with median and $95\%$ PI of posterior predictive distributions. (b) Between-sex gap of life expectancy at birth with posterior predictive median (red solid) and $95\%$ PI  (red dotted). (c) Forecasts of male life expectancy at birth to 2060 using bayesLife method (green) and our proposed method (red) with posterior predictive medians (dashed) and $95\%$ PI (dotted).  (d) Forecasts of female life expectancy at birth to 2060 using bayesLife method (green) and our proposed method (red) with posterior predictive medians (dashed) and $95\%$ PI (dotted).}\label{fg:usa}
\end{figure}

The female smoking epidemic started two decades later than the male one with a maximum prevalence of around $30\%$ in the 1960s. Female smoking prevalence decline to about $20\%$ in 1990s and $13.5\%$ in 2016 \citep{burns1997cigarette, islami2015global}. Figure \ref{fg:usa}a shows projections of the US male and female ASAF to 2060. Figure \ref{fg:usa}b predicts a continuously narrowing gap of the between-sex life expectancy due to the shrinking gap between male and female ASAF up to 2060.

Figures \ref{fg:usa}c and \ref{fg:usa}d show projections of male and female life expectancy for the period 2015-2060. The bayesLife method projects male life expectancy in 2055-2060 to be 84.0 years, with $95\%$ predictive interval (79.2, 87.6). We project male life expectancy to be 86.1 in 2060, with $95\%$ predictive interval (83.0, 88.9). 
The bayesLife method projects US female life expectancy for 2055-2060 to be 
86.5 with $95\%$ predictive interval (82.9, 90.0). We project female life expectancy to be 88.6 with interval (84.8, 92.4). 

Our method gives forecasts of life expectancy that are about two years higher than those from the bayesLife method for both males and females, because of accounting for the smoking effect.
Our predictive interval for male life expectancy at birth is $29\%$ shorter than the bayesLife one, while our female interval is comparable with that of the bayesLife method. Both of our $95\%$ predictive intervals cover the posterior medians from the bayesLife method. 

\subsection{The Netherlands}
The Netherlands is a western European country with a long history of the smoking epidemic, which can be dated back to the 1880s when the cigarette industry began there. Male smoking prevalence reached $90\%$ in most age groups in the 1950s, but dropped rapidly to $30\%$ in the 2010s. In contrast, smoking was more prevalent among females in the 1970s, when about $40\%$ of female smoked, and after 1975 there was a sustained drop to $24\%$ in the 2010s \citep{stoeldraijer2015future}. 

\begin{figure}[!hp]
	\centering
	\begin{subfigure}[tbp]{.4\textwidth}
		\centering	
		\includegraphics[width=\textwidth]{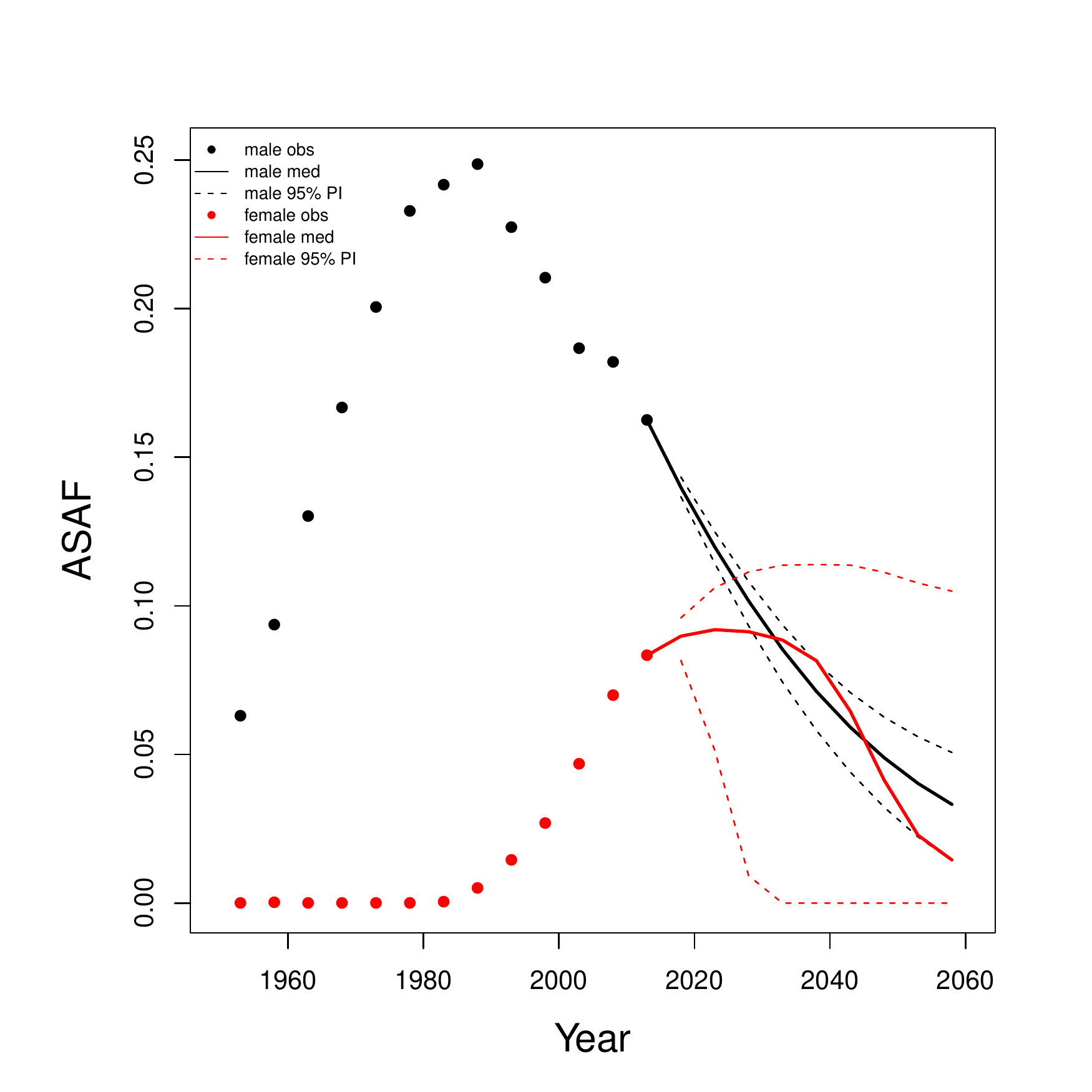}
		\caption{}
	\end{subfigure}
	\begin{subfigure}[tbp]{.4\textwidth}
		\centering
		\includegraphics[width=\linewidth]{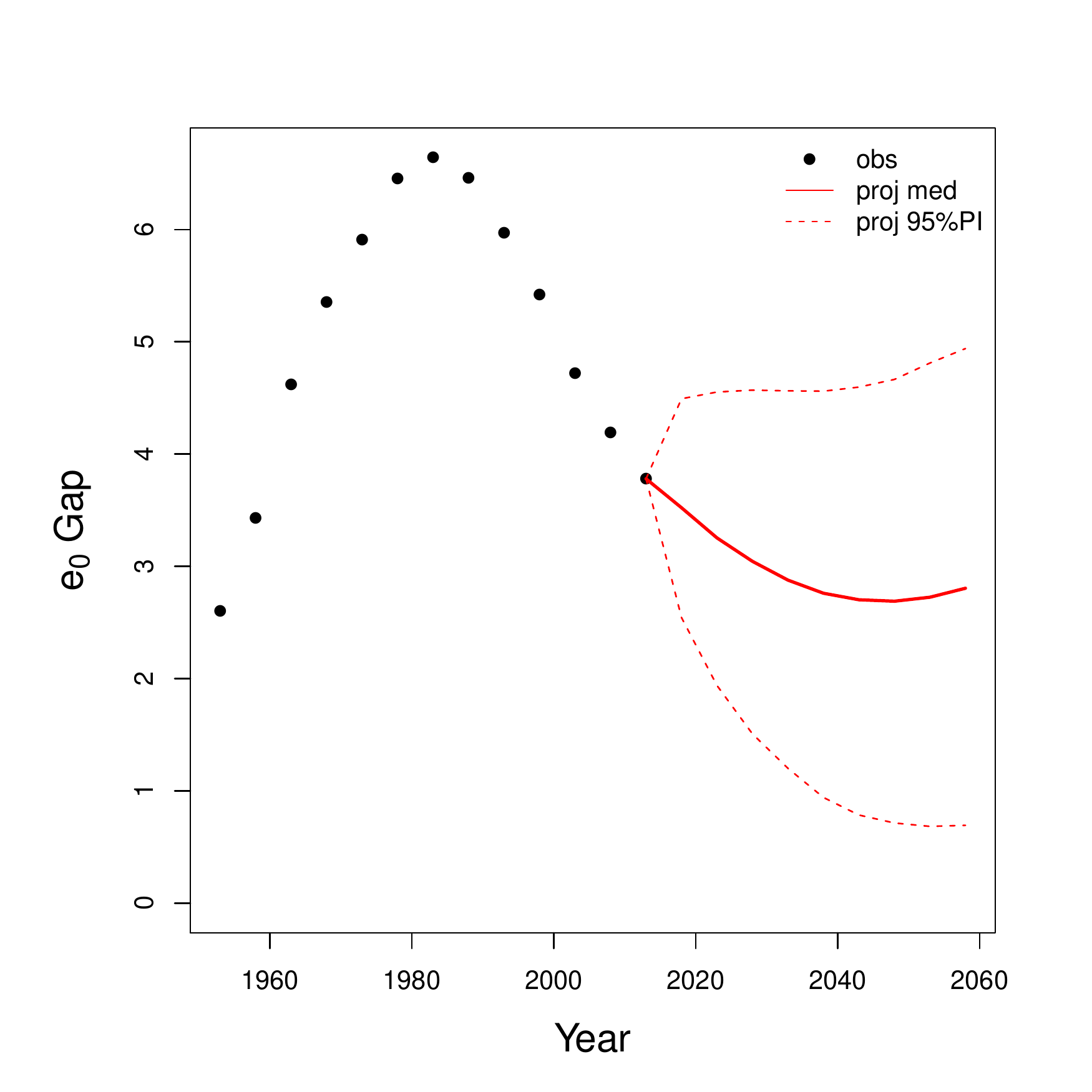}
		\caption{}
	\end{subfigure}	
	\begin{subfigure}[tbp]{.4\textwidth}
		\centering
		\includegraphics[width=\linewidth]{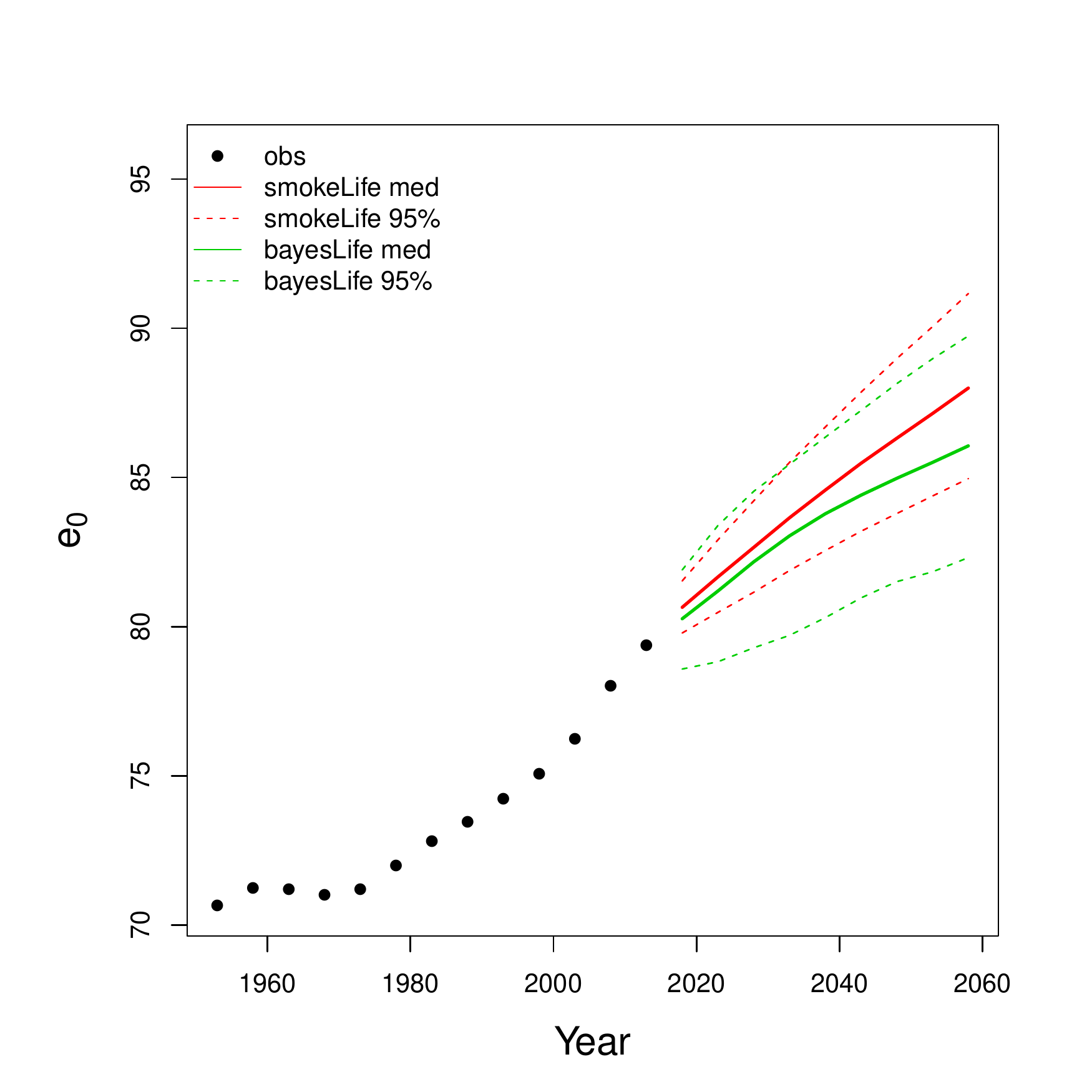}
		\caption{}
	\end{subfigure}		
	\begin{subfigure}[tbp]{.4\textwidth}
		\centering
		\includegraphics[width=\linewidth]{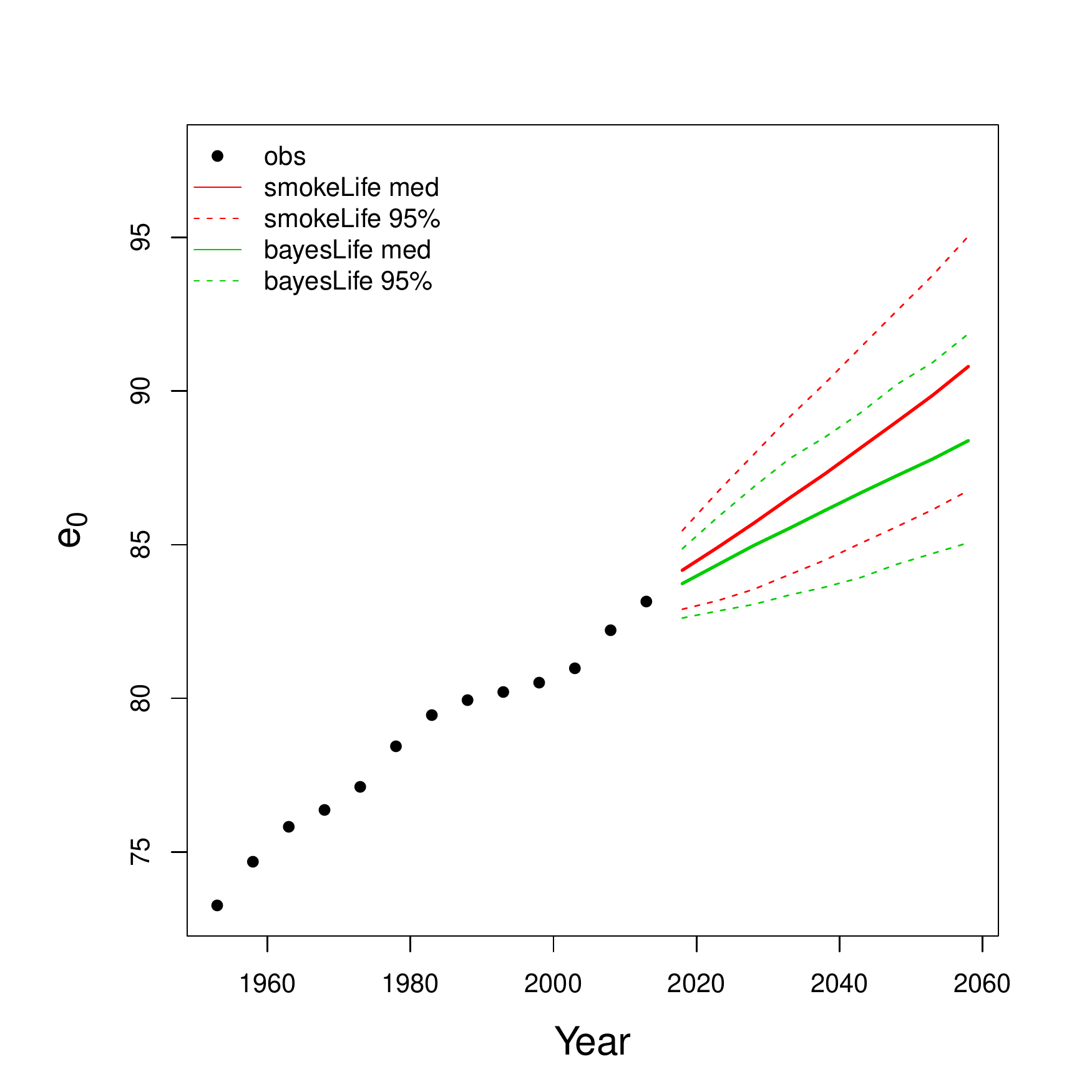}
		\caption{}
	\end{subfigure}	
	\caption{The Netherlands. (a) All-age smoking attributable fraction (ASAF) for male (black) and female (red) with median and $95\%$ PI of posterior predictive distributions. (b) Between-sex gap of life expectancy at birth with posterior predictive median (red solid) and $95\%$ PI  (red dotted). (c) Forecasts of male life expectancy at birth to 2060 using bayesLife method (green) and our proposed method (red) with posterior predictive medians (dashed) and $95\%$ PI (dotted).  (d) Forecasts of female life expectancy at birth to 2060 using bayesLife method (green) and our proposed method (red) with posterior predictive medians (dashed) and $95\%$ PI (dotted).}\label{fg:netherlands}
\end{figure}

Figure  \ref{fg:netherlands}a shows that 
the female ASAF is forecast to surpass the male ASAF for the next two decades and by 2060, both male and female ASAF will be at about the same level. 
Figure \ref{fg:netherlands}b shows that the turning point in the between-sex gap of life expectancy happened around the 1990s, when the male ASAF had passed its peak and the female ASAF started to climb. With the shrinking of the ASAF gap, the projected life expectancy gap is forecast to continue to shrink and plateau around 2.8, due to biological and social factors \citep{janssen2015adoption}. 

Both Dutch males and females experienced a period of stagnation in life expectancy gains---in the 1960s for males and the 1990s for females. Smoking is a major reason for this stagnation. The right panel of Figure \ref{fg:nonsmke0} indicates that the forecast Dutch male life expectancy gain is more linear and sustained after removing the smoking effect. Figures \ref{fg:netherlands}c and \ref{fg:netherlands}d show projections of male and female life expectancy for 2015--2060. We project male life expectancy for the period 2055-2060 to be 88.0 years, with a $95\%$ prediction interval of (85.0, 91.1), while the bayesLife method projects 86.1, with interval of (82.3, 89.7). We project female life expectancy for the period 2055-2060 to be 90.8, with a $95\%$ prediction interval of (86.6, 95.0), while the bayesLife method projects 88.4 years, with interval of (85.1, 91.9). 

Similarly to the US, our forecast of life expectancy in 2060 is about two years higher than a forecast that does not take account of smoking. By considering the decreasing trend of the smoking epidemic, our forecast is 1.9 years higher for males and 2.3 years higher for females expectancy compared with the bayesLife method. 
\citet{janssen2013including} forecast the Dutch male and female life expectancy in 2040 to be 84.6 years and 87.2  years respectively, taking account of smoking.  This agrees well with our forecasts ---85.0 for males and 87.3 for female---in 2040.

\subsection{Chile}
Chile is a South American country where the smoking epidemic had a late start, and it is currently one of the countries with the highest smoking prevalence in the Americas.
Smoking prevalence decreased from $50\%$ in 2000 to $40\%$ in 2016 among males, and from $44\%$ to $36\%$ among females. This decline is modest compared to that in the United States \citep{islami2015global}. 

\begin{figure}[!hp]
	\centering
	\begin{subfigure}[tbp]{0.4\textwidth}
		\centering	
		\includegraphics[width=\textwidth]{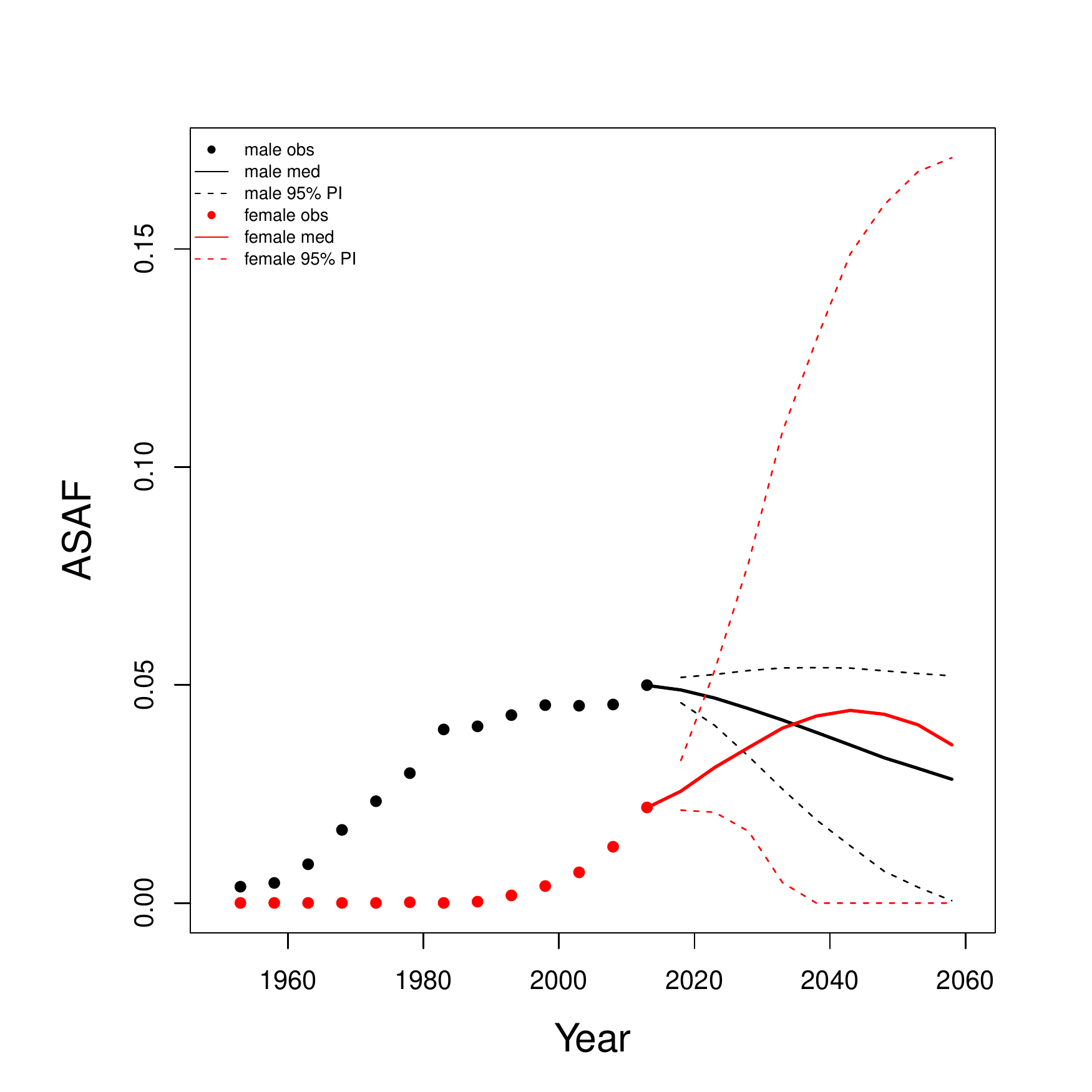}
		\caption{}
	\end{subfigure}
	\begin{subfigure}[tbp]{.4\textwidth}
		\centering
		\includegraphics[width=\linewidth]{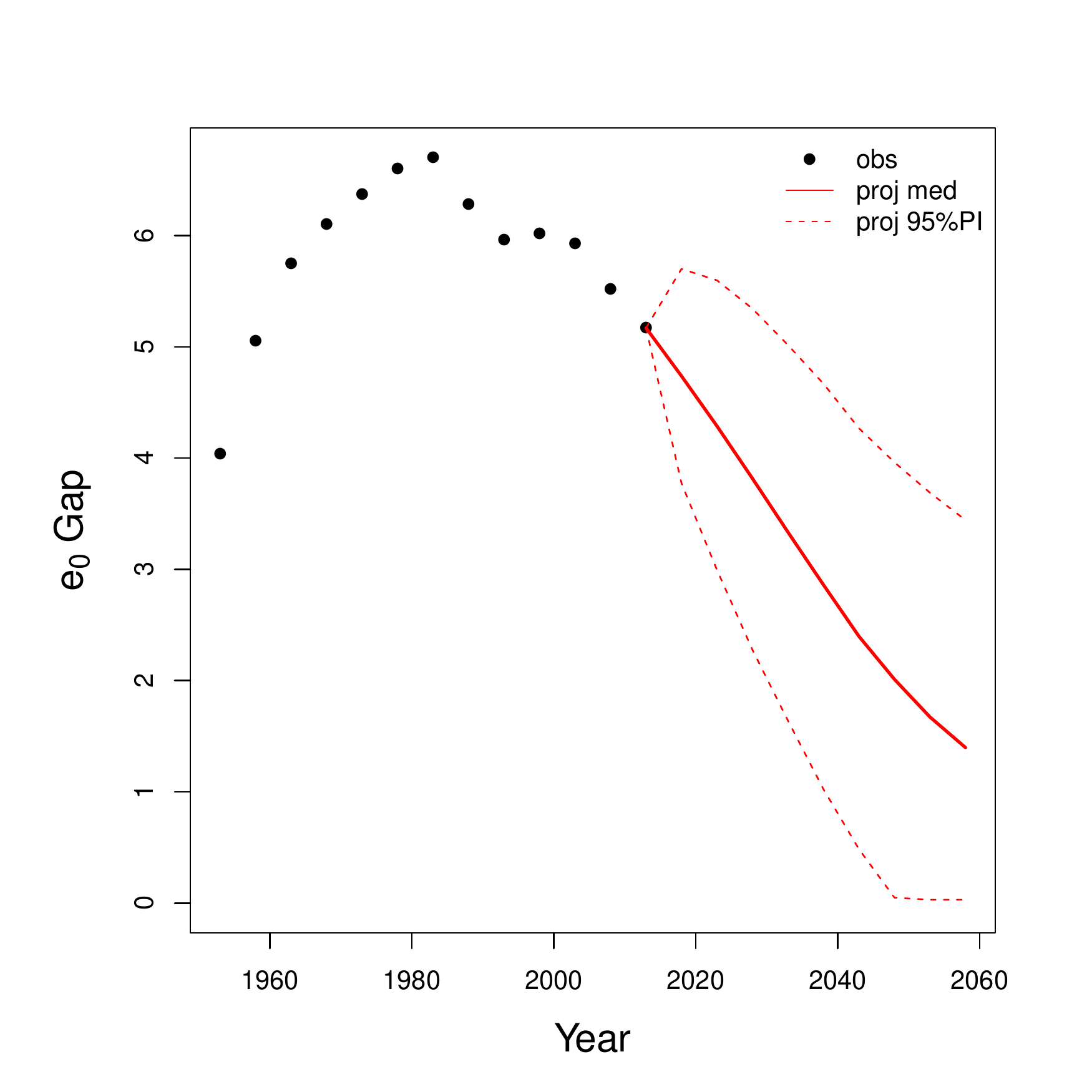}
		\caption{}
	\end{subfigure}	
	\begin{subfigure}[tbp]{.4\textwidth}
		\centering
		\includegraphics[width=\linewidth]{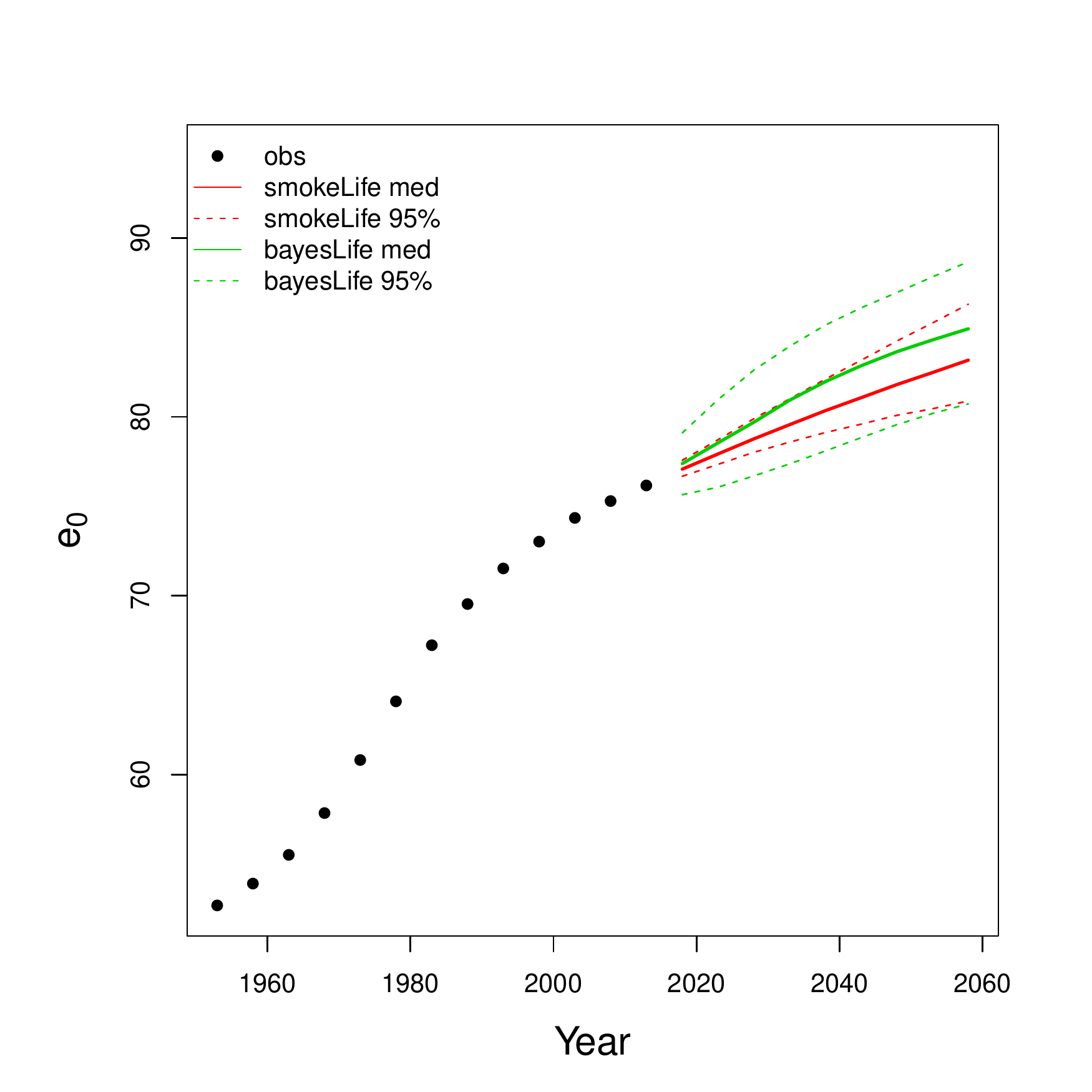}
		\caption{}
	\end{subfigure}		
	\begin{subfigure}[tbp]{.4\textwidth}
		\centering
		\includegraphics[width=\linewidth]{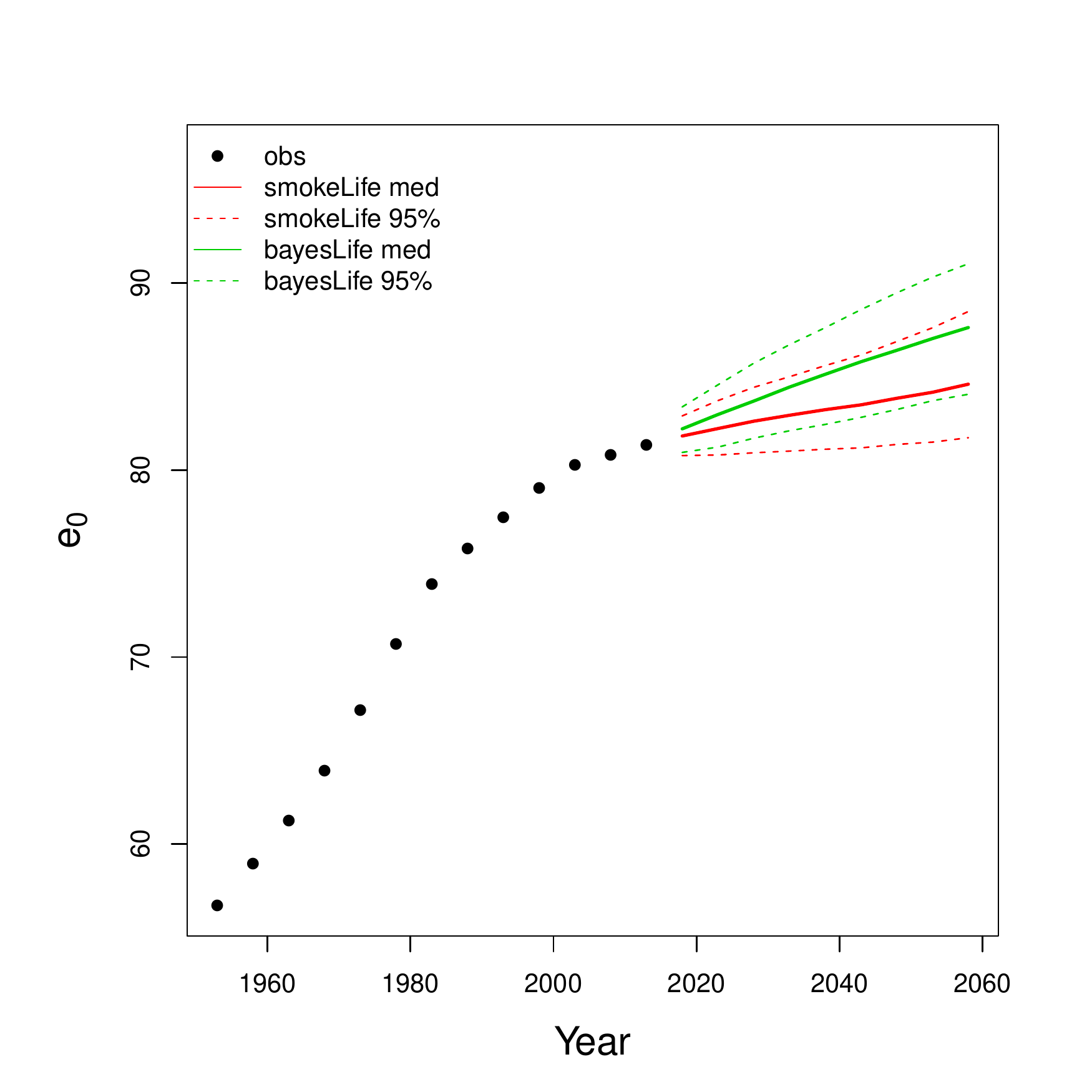}
		\caption{}
	\end{subfigure}	
	\caption{Chile. (a) All-age smoking attributable fraction (ASAF) for male (black) and female (red) with median and $95\%$ PI of posterior predictive distributions. (b) Between-sex gap of life expectancy at birth with posterior predictive median (red solid) and $95\%$ PI  (red dotted). (c) Forecasts of male life expectancy at birth to 2060 using bayesLife method (green) and our proposed method (red) with posterior predictive medians (dashed) and $95\%$ PI (dotted).  (d) Forecasts of female life expectancy at birth to 2060 using bayesLife method (green) and our proposed method (red) with posterior predictive medians (dashed) and $95\%$ PI (dotted).}\label{fg:chile}
\end{figure}

Figure \ref{fg:chile}a shows the projections of male and female ASAF. Chilean male ASAF has been at the peaking stage for a long time, with high prevalence and no sign of a decline. Female ASAF is predicted to grow to approach the male level. The narrowing of the ASAF gap is forecast to lead to a sustained closing of the life expectancy between-sex gap (Figure \ref{fg:chile}b).

Figures \ref{fg:chile}c and \ref{fg:chile}d show projections of  male and female life expectancy for 2015--2060. We project male life expectancy for the period 2055-2060 be 83.2, with a $95\%$ predictive interval of (80.9, 86.3). In contrast with the USA and the Netherlands, our median projection is 1.8 years {\it less} than that from bayesLife method. 
This is due to the fact that the epidemic has not yet clearly peaked.
We project female life expectancy to be 84.5, with a $95\%$ predictive interval of (81.7, 88.5), which is again substantially smaller than that from the bayesLife method with forecast median 87.6 years and $95\%$ prediction interval (84.1, 91.0). This is due to the increasing impact of smoking on the Chilean female population.

\subsection{Japan}
\begin{figure}[!hp]
	\centering
	\begin{subfigure}[tbp]{.4\textwidth}
		\centering	
		\includegraphics[width=\textwidth]{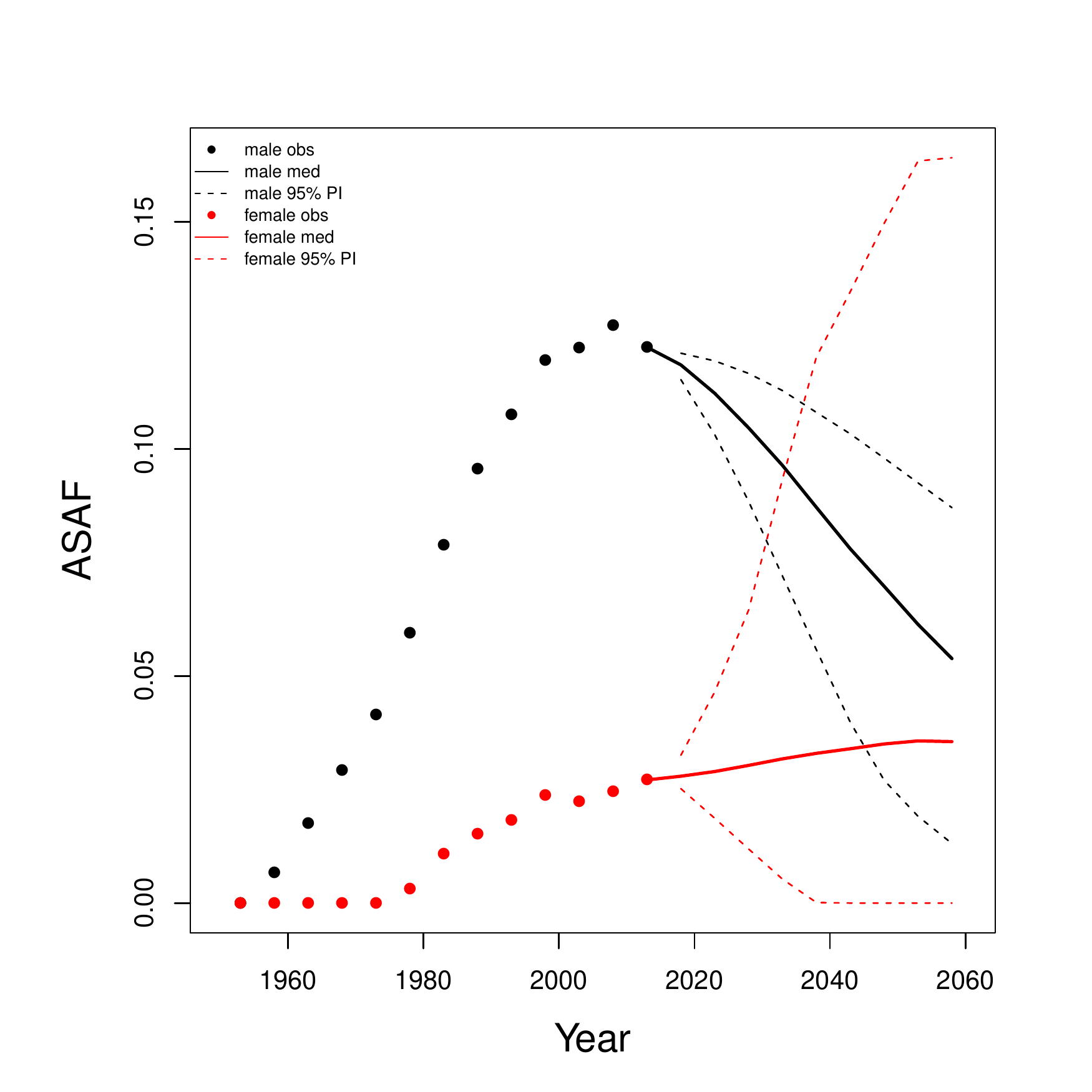}
		\caption{}
	\end{subfigure}
	\begin{subfigure}[tbp]{.4\textwidth}
		\centering
		\includegraphics[width=\linewidth]{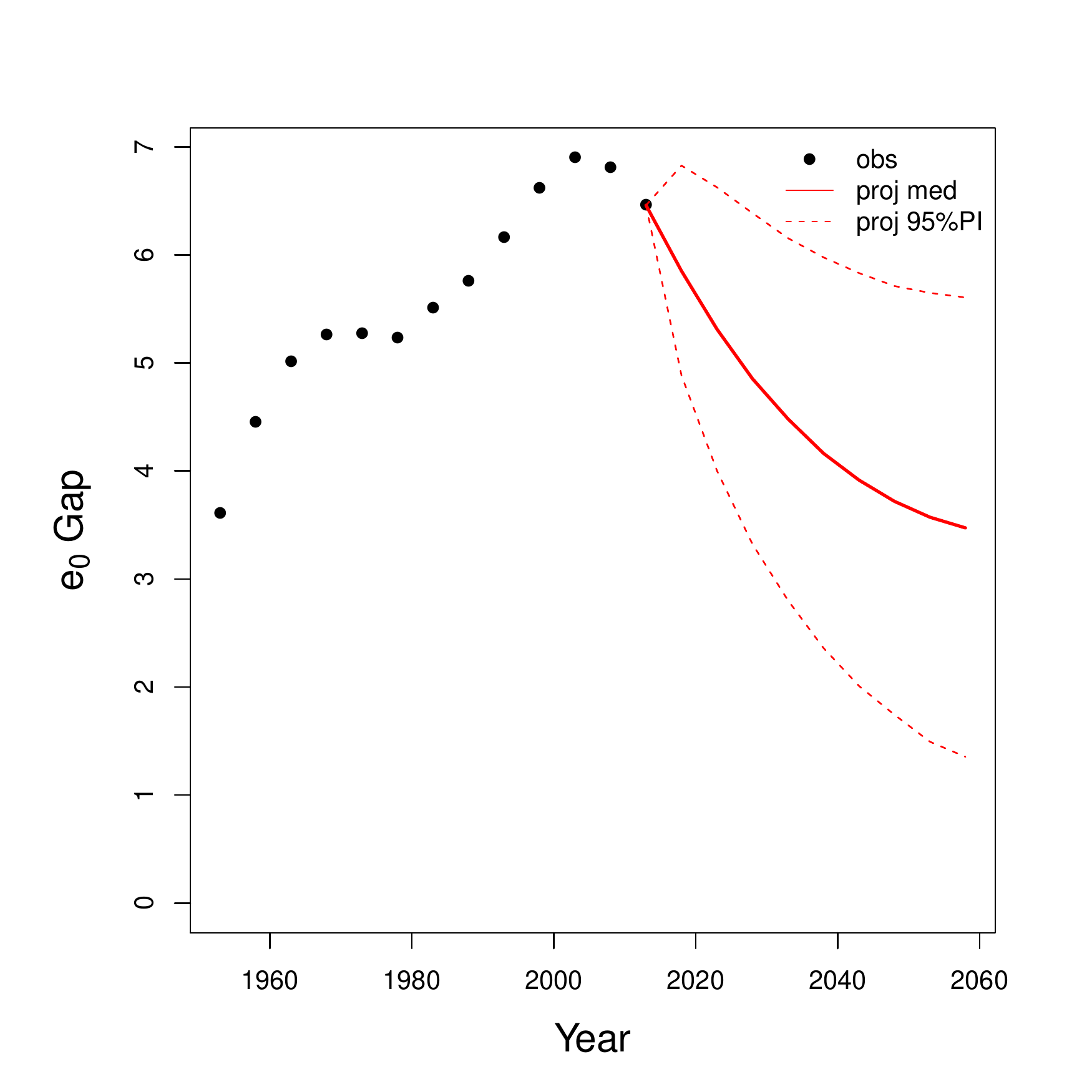}
		\caption{}
	\end{subfigure}	
	\begin{subfigure}[tbp]{.4\textwidth}
		\centering
		\includegraphics[width=\linewidth]{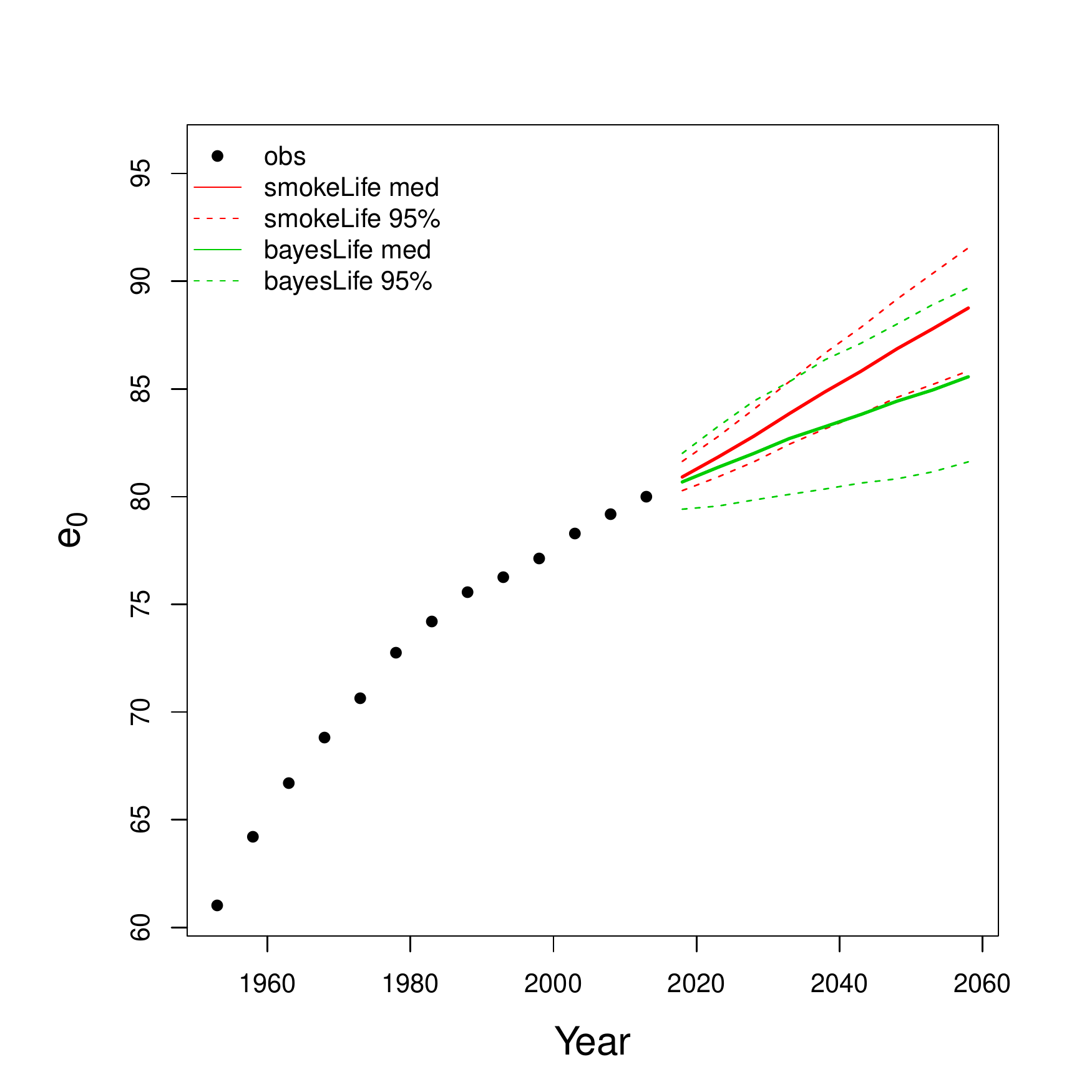}
		\caption{}
	\end{subfigure}		
	\begin{subfigure}[tbp]{.4\textwidth}
		\centering
		\includegraphics[width=\linewidth]{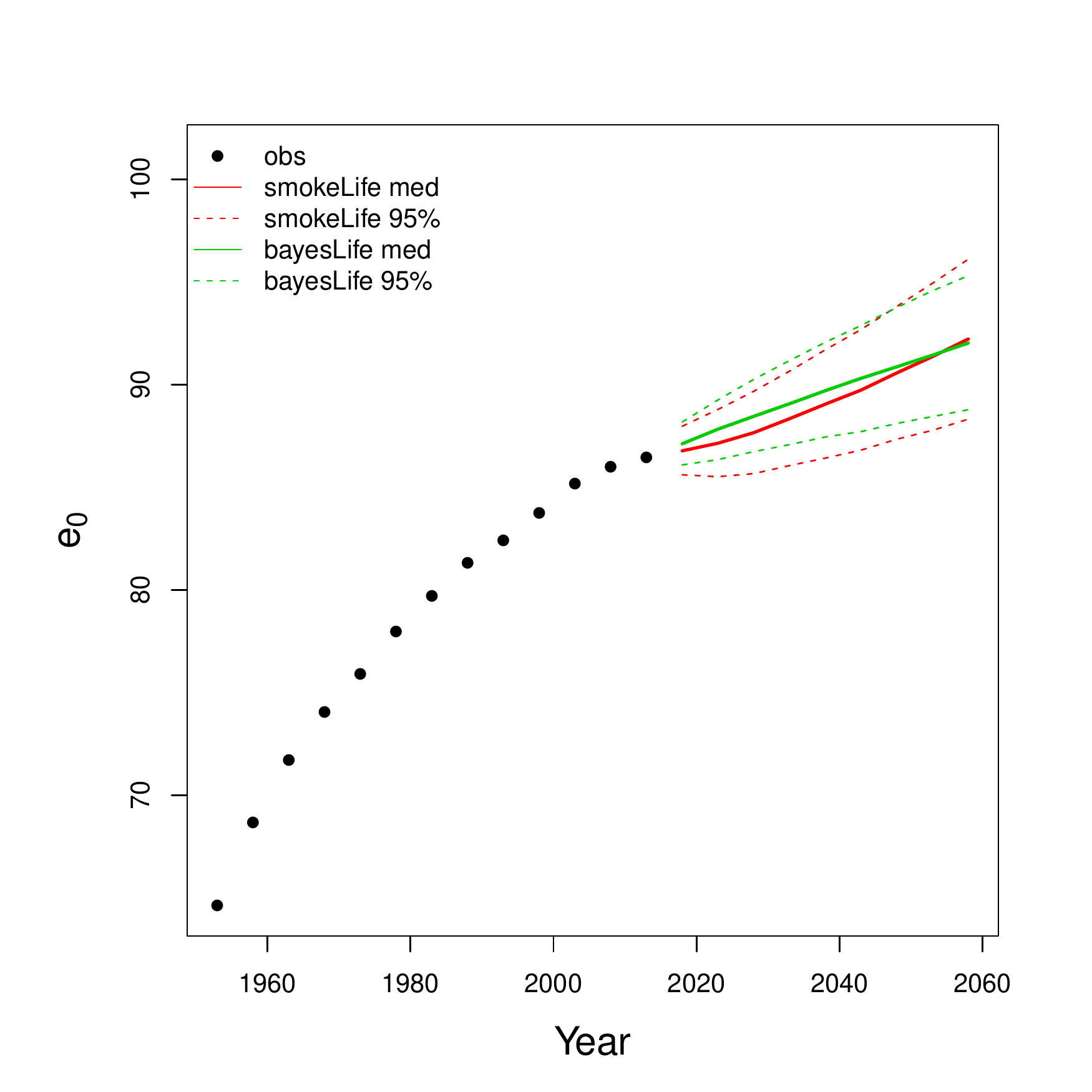}
		\caption{}
	\end{subfigure}	
	\caption{Japan. (a) All-age smoking attributable fraction (ASAF) for male (black) and female (red) with median and $95\%$ PI of posterior predictive distributions. (b) Between-sex gap of life expectancy at birth with posterior predictive median (red solid) and $95\%$ PI  (red dotted). (c) Forecasts of male life expectancy at birth to 2060 using bayesLife method (green) and our proposed method (red) with posterior predictive medians (dashed) and $95\%$ PI (dotted).  (d) Forecasts of female life expectancy at birth to 2060 using bayesLife method (green) and our proposed method (red) with posterior predictive medians (dashed) and $95\%$ PI (dotted).}\label{fg:japan}
\end{figure}

Japan has been a leading country in life expectancy for a long period, while it also has a long history of smoking and is one of the largest tobacco consumers.
Male smoking prevalence reached $83.7\%$ in 1966. That number dropped to $36\%$ in the 1990s and halved again by 2018. Female smoking prevalence is far lower and changes less dramatically than that of males. Female smoking prevalence reached $16\%$ in the 1970s and decreased to $9.7\%$ in 2015. The significant changes result mainly from government regulations and anti-smoking movements starting in the 1980s. Figure \ref{fg:japan}a shows the forecast male and female ASAF. Figure \ref{fg:japan}b shows the narrowing of the life expectancy gap as a result. 

Figures \ref{fg:japan}c and  \ref{fg:japan}d show projections of life expectancy for males and females. We project male life expectancy for the period 2055-2060 to be 88.8, with a $95\%$ predictive interval of (85.8, 91.5). The bayesLife method forecasts 85.6, with a projection interval (81.6, 89.7). Notice that 
our median forecast is 3.2 years higher than that of bayesLife, while its interval is 1.4 years narrower.
We project female life expectancy to be 92.2 with a $95\%$ prediction interval of (88.3, 96.1). Our forecast shows a noticeable slowdown of the growth of female life expectancy due to the smoking effect. The bayesLife method projects 92.0 years with interval (88.8, 95.3). Though both methods produce comparable forecast results for 2055-2060, the bayesLife method forecasts a more linear increase while ours reflects the nonlinear smoking effect on the life expectancy forecast.

\section{Discussion}\label{sec:disc}
We have proposed a method for probabilistic forecasting of mortality and life expectancy that takes account of the smoking epidemic. The method is based on the idea of the smoking attributable fraction of mortality, as estimated by the Peto-Lopez method using data on lung cancer mortality. The age-specific smoking attiributable fraction (ASSAF) of mortality is estimated and used to infer the non-smoking life expectancy at birth, $e_0^{NS}$. Both the ASSAF and $e_0^{NS}$ are then forecast using a Bayesian hierarchical models for all countries with sufficiently good data. This in turn yields posterior predictive distributions of 
mortality rates and life expectancy at birth.
The method performed well in an out-of-sample validation study. 

The strength of the method derives from the fact that the smoking attributable
fraction of mortality follows a very strong increasing-peaking-decreasing trend over time in all countries where the smoking epidemic has been going for long enough. This pattern is strong, broadly the same across countries, is to a large extent socially determined, and is also not highly correlated over time with the life expectancy at birth itself, which follows a broadly increasing pattern over time. However, smoking does impact mortality. Thus smoking mortality can be predicted with considerable accuracy, and accurate predictions improve mortality forecasts. 

Another strength of the method is its use of a hierarchical model, 
which greatly facilitates forecasting, particularly  for countries where 
the smoking epidemic is at an early stage. This allows forecasts for such countries to be informed by information from other countries, especially those where the epidemic is more advanced. It also makes it easier to incorporate all major sources of uncertainty.

The results indicate that for country-sex combinations where the smoking epidemic is advanced enough that we can expect it to be declining by 2060, incorporating smoking increases forecasts of life expectancy by about two years. When the epidemic is at an earlier stage, though, incorporating smoking tends to reduce
forecasts of life expectancy. The results also indicate that much of change over time in the female-male gap in life expectancy is due to relative changes in
smoking related mortality.

The biggest limitation of our method is that it
relies on the availability of high-quality data on cause of death,
particularly lung cancer, which are available for only 69 countries of the
201 or so countries in the world. Thus the biggest improvement in the method
would come from improvements in data quality. In particular, China and India
are missing from our study, because national data on cause of death of 
high enough quality are not available. Producing such data should be a focus
of future data collection and research. This is very important because, not
only are China and India the two most populous countries in the world,
but they also have high smoking rates and are likely to experience high
smoking mortality in the coming decades.

Several other approaches to the problem have been proposed. 
\citet{bongaarts2006long} introduced the concept of non-smoking life
expectancy, and proposed modeling it in a linear way. 
However, the time evolution of non-smoking life expectancy appears generally
to follow a nonlinear pattern, with gains that broadly follow a 
non-monotonic increasing-peaking-declining patter. This is modeled in our
method by a random walk with a the double logistic drift.

\citet{janssen2013including} proposed directly modeling the ASSAF and the age-specific non-smoking attributable mortality rates. They observed that non-smoking mortality rates decline more linearly than overall mortality rates,
making the data fit a Lee-Carter model better. 
They conducted an age-period-cohort analysis, while we found an age-cohort 
model to be sufficient. There are well-known identifiability issues with
age-period-cohort analysis that our approach avoids. They used a coherent 
Lee-Carter method. This assumes linear progress in log mortality rates, while
in fact progress tends to be nonlinear, and also tends to be more linear
on the scale of life expectancy than of log mortality rates, which our 
double logistic random walk attempts to represent.

The mortality component of the UN's population projections for all countries 
is based on the Bayesian hierarchical model of \citet{raftery2013bayesian},
which does not take account of smoking. We have shown that this could be 
improved significantly by taking account of smoking. However, the data to do this are available for only 69 countries currently, and the UN aims to use 
a unified approach for all the 230 countries and territories that they analyze.
Thus extending the UN's method to take account of smoking in this way might not
be feasible in the short term. To do this would likely require a major improvement in data availability for many countries.
However, it could be useful for national population and mortality projections for individual countries, for example for planning health services, and also for the private sector, for example for actuarial and insurance analyses.

\bibliographystyle{apalike}
\bibliography{smokingbib}

\begin{thebibliography}{}

\bibitem[Alkema et~al., 2011]{alkema2011probabilistic}
Alkema, L., Raftery, A.~E., Gerland, P., Clark, S.~J., Pelletier, F., Buettner,
  T., and Heilig, G.~K. (2011).
\newblock Probabilistic projections of the total fertility rate for all
  countries.
\newblock {\em Demography}, 48(3):815--839.

\bibitem[Bongaarts, 2006]{bongaarts2006long}
Bongaarts, J. (2006).
\newblock How long will we live?
\newblock {\em Population and Development Review}, 32(4):605--628.

\bibitem[Bongaarts, 2014]{bongaarts2014trends}
Bongaarts, J. (2014).
\newblock Trends in causes of death in low-mortality countries: implications
  for mortality projections.
\newblock {\em Population and Development Review}, 40(2):189--212.

\bibitem[Booth et~al., 2006]{booth2006lee}
Booth, H., Hyndman, R.~J., Tickle, L., and De~Jong, P. (2006).
\newblock {L}ee-{C}arter mortality forecasting: a multi-country comparison of
  variants and extensions.
\newblock {\em Demographic Research}, 15:289--310.

\bibitem[Britton, 2017]{britton2017death}
Britton, J. (2017).
\newblock Death, disease, and tobacco.
\newblock {\em The Lancet}, 389(10082):1861--1862.

\bibitem[Burns et~al., 1997]{burns1997cigarette}
Burns, D.~M., Lee, L., Shen, L.~Z., Gilpin, E., Tolley, H.~D., Vaughn, J.,
  Shanks, T.~G., et~al. (1997).
\newblock Cigarette smoking behavior in the {United States}.
\newblock {\em Changes in cigarette-related disease risks and their implication
  for prevention and control. Smoking and Tobacco Control Monograph}, 8:13--42.

\bibitem[Fokas, 2007]{fokas2007growth}
Fokas, N. (2007).
\newblock Growth functions, social diffusion, and social change.
\newblock {\em Review of Sociology}, 13(1):5--30.

\bibitem[Gneiting and Raftery, 2007]{gneiting2007strictly}
Gneiting, T. and Raftery, A.~E. (2007).
\newblock Strictly proper scoring rules, prediction, and estimation.
\newblock {\em Journal of the American Statistical Association},
  102(477):359--378.

\bibitem[Godwin and Raftery, 2017]{godwin2017bayesian}
Godwin, J. and Raftery, A.~E. (2017).
\newblock Bayesian projection of life expectancy accounting for the {HIV/AIDS}
  epidemic.
\newblock {\em Demographic Research}, 37:1549.

\bibitem[Gr{\"u}bler et~al., 1999]{grubler1999dynamics}
Gr{\"u}bler, A., Naki{\'c}enovi{\'c}, N., and Victor, D.~G. (1999).
\newblock Dynamics of energy technologies and global change.
\newblock {\em Energy Policy}, 27(5):247--280.

\bibitem[Hyndman et~al., 2019]{hyndman2019demo}
Hyndman, R.~J., Booth, H., Tickle, L., and Maindonald., J. (2019).
\newblock {\em Demography: {F}orecasting {M}ortality, {F}ertility, {M}igration
  and {P}opulation {D}ata}.
\newblock R package version 1.22.
  \url{https://CRAN.R-project.org/package=demography}.

\bibitem[Hyndman and Ullah, 2007]{hyndman2007robust}
Hyndman, R.~J. and Ullah, M.~S. (2007).
\newblock Robust forecasting of mortality and fertility rates: a functional
  data approach.
\newblock {\em Computational Statistics \& Data Analysis}, 51(10):4942--4956.

\bibitem[Islami et~al., 2015]{islami2015global}
Islami, F., Torre, L.~A., and Jemal, A. (2015).
\newblock Global trends of lung cancer mortality and smoking prevalence.
\newblock {\em Translational Lung Cancer Research}, 4(4):327.

\bibitem[Janssen, 2018]{janssen2018mort}
Janssen, F. (2018).
\newblock Advances in mortality forecasting: {I}ntroduction.
\newblock {\em Genus}, 74(1):21.

\bibitem[Janssen and van Poppel, 2015]{janssen2015adoption}
Janssen, F. and van Poppel, F. (2015).
\newblock The adoption of smoking and its effect on the mortality gender gap in
  {N}etherlands: a historical perspective.
\newblock {\em BioMed Research International}, 2015.

\bibitem[Janssen et~al., 2013]{janssen2013including}
Janssen, F., van Wissen, L.~J., and Kunst, A.~E. (2013).
\newblock Including the smoking epidemic in internationally coherent mortality
  projections.
\newblock {\em Demography}, 50(4):1341--1362.

\bibitem[King and Soneji, 2011]{king2011future}
King, G. and Soneji, S. (2011).
\newblock The future of death in {A}merica.
\newblock {\em Demographic Research}, 25:1--38.

\bibitem[Kong et~al., 2016]{kong2016comparison}
Kong, K.~A., Jung-Choi, K.-H., Lim, D., Lee, H.~A., Lee, W.~K., Baik, S.~J.,
  Park, S.~H., and Park, H. (2016).
\newblock Comparison of prevalence-and smoking impact ratio-based methods of
  estimating smoking-attributable fractions of deaths.
\newblock {\em Journal of Epidemiology}, 26(3):145--154.

\bibitem[Kucharavy and De~Guio, 2011]{kucharavy2011logistic}
Kucharavy, D. and De~Guio, R. (2011).
\newblock Logistic substitution model and technological forecasting.
\newblock {\em Procedia Engineering}, 9:402--416.

\bibitem[Lee and Carter, 1992]{lee1992modeling}
Lee, R.~D. and Carter, L.~R. (1992).
\newblock Modeling and forecasting {US} mortality.
\newblock {\em Journal of the American Statistical Association},
  87(419):659--671.

\bibitem[Lee and Miller, 2001]{lee2001evaluating}
Lee, R.~D. and Miller, T. (2001).
\newblock Evaluating the performance of the {L}ee-{C}arter method for
  forecasting mortality.
\newblock {\em Demography}, 38(4):537--549.

\bibitem[Li and Lee, 2005]{LiLee2005}
Li, N. and Lee, R.~D. (2005).
\newblock Coherent mortality forecasts for a group of populations: An extension
  of the {L}ee-{C}arter method.
\newblock {\em Demography}, 42:575--594.

\bibitem[Li and Raftery, 2019]{li2019estimating}
Li, Y. and Raftery, A.~E. (2019).
\newblock Estimating and forecasting the smoking-attributable mortality
  fraction for both sexes jointly in 69 countries.
\newblock {\em arXiv preprint arXiv:1902.07791}.

\bibitem[Marchetti et~al., 1996]{marchetti1996human}
Marchetti, C., Meyer, P.~S., and Ausubel, J.~H. (1996).
\newblock Human population dynamics revisited with the logistic model: how much
  can be modeled and predicted?
\newblock {\em Technological Forecasting and Social Change}, 52(1):1--30.

\bibitem[Meyer, 1994]{meyer1994bi}
Meyer, P. (1994).
\newblock Bi-logistic growth.
\newblock {\em Technological Forecasting and Social Change}, 47(1):89--102.

\bibitem[Oeppen and Vaupel, 2002]{oeppen2002enhanced}
Oeppen, J. and Vaupel, J.~W. (2002).
\newblock Broken limits to life expectancy.
\newblock {\em Science}, 296(5570):1029--1031.

\bibitem[Pedroza, 2006]{pedroza2006bayesian}
Pedroza, C. (2006).
\newblock A {B}ayesian forecasting model: predicting {US} male mortality.
\newblock {\em Biostatistics}, 7(4):530--550.

\bibitem[Peto et~al., 1992]{peto1992mortality}
Peto, R., Boreham, J., Lopez, A.~D., Thun, M., and Heath, C. (1992).
\newblock Mortality from tobacco in developed countries: indirect estimation
  from national vital statistics.
\newblock {\em The Lancet}, 339(8804):1268--1278.

\bibitem[Preston et~al., 2000]{Preston&2000}
Preston, S., Heuveline, P., and Guillot, M. (2000).
\newblock {\em Demography: Measuring and Modeling Population Processes}.
\newblock Wiley-Blackwell.

\bibitem[Preston and Wang, 2006]{preston2006sex}
Preston, S.~H. and Wang, H. (2006).
\newblock Sex mortality differences in the {United States}: The role of cohort
  smoking patterns.
\newblock {\em Demography}, 43(4):631--646.

\bibitem[Raftery et~al., 2014a]{Raftery&2014}
Raftery, A.~E., Alkema, L., and Gerland, P. (2014a).
\newblock Bayesian population projections for the {U}nited {N}ations.
\newblock {\em Statistical Science}, 29:58--68.

\bibitem[Raftery et~al., 2013]{raftery2013bayesian}
Raftery, A.~E., Chunn, J.~L., Gerland, P., and {\v{S}}ev{\v{c}}{\'\i}kov{\'a},
  H. (2013).
\newblock Bayesian probabilistic projections of life expectancy for all
  countries.
\newblock {\em Demography}, 50(3):777--801.

\bibitem[Raftery et~al., 2014b]{raftery2014joint}
Raftery, A.~E., Lalic, N., and Gerland, P. (2014b).
\newblock Joint probabilistic projection of female and male life expectancy.
\newblock {\em Demographic Research}, 30:795--822.

\bibitem[Raftery and Lewis, 1992]{raftery1992mcmc}
Raftery, A.~E. and Lewis, S.~M. (1992).
\newblock One long run with diagnostics: Implementation strategies for {M}arkov
  chain {M}onte {C}arlo.
\newblock {\em Statistical Science}, 7(4):493--497.

\bibitem[Raftery et~al., 2012]{Raftery&2012PNAS}
Raftery, A.~E., Li, N., {\v{S}}ev{\v{c}}{\'\i}kov{\'a}, H., Gerland, P., and
  Heilig, G.~K. (2012).
\newblock Bayesian probabilistic population projections for all countries.
\newblock {\em Proceedings of the National Academy of Sciences},
  109:13915--13921.

\bibitem[Renshaw and Haberman, 2006]{renshaw2006cohort}
Renshaw, A.~E. and Haberman, S. (2006).
\newblock A cohort-based extension to the {L}ee-{C}arter model for mortality
  reduction factors.
\newblock {\em Insurance: Mathematics and Economics}, 38(3):556--570.

\bibitem[Rostron and Wilmoth, 2011]{rostron2011estimating}
Rostron, B.~L. and Wilmoth, J.~R. (2011).
\newblock Estimating the effect of smoking on slowdowns in mortality declines
  in developed countries.
\newblock {\em Demography}, 48(2):461--479.

\bibitem[{\v{S}}ev{\v{c}}{\'\i}kov{\'a} et~al., 2019a]{sevcikova2019mortcast}
{\v{S}}ev{\v{c}}{\'\i}kov{\'a}, H., Li, N., and Gerland, P. (2019a).
\newblock {\em MortCast: Estimation and Projection of Age-Specific Mortality
  Rates}.
\newblock R package version 2.1-1.
  \url{https://CRAN.R-project.org/package=MortCast}.

\bibitem[{\v{S}}ev{\v{c}}{\'\i}kov{\'a} et~al., 2016]{vsevvcikova2016age}
{\v{S}}ev{\v{c}}{\'\i}kov{\'a}, H., Li, N., Kantorov{\'a}, V., Gerland, P., and
  Raftery, A.~E. (2016).
\newblock Age-specific mortality and fertility rates for probabilistic
  population projections.
\newblock In {\em Dynamic Demographic Analysis}, pages 285--310. Springer.

\bibitem[{\v{S}}ev{\v{c}}{\'\i}kov{\'a} et~al., 2019b]{sevcikova2019bayeslife}
{\v{S}}ev{\v{c}}{\'\i}kov{\'a}, H., Raftery, A., and Chunn, J. (2019b).
\newblock {\em bayes{L}ife: {B}ayesian {P}rojection of {L}ife {E}xpectancy}.
\newblock R package version 4.0-2.
  \url{https://CRAN.R-project.org/package=bayesLife}.

\bibitem[Shang, 2016]{shang2016mortality}
Shang, H.~L. (2016).
\newblock Mortality and life expectancy forecasting for a group of populations
  in developed countries: a multilevel functional data method.
\newblock {\em The Annals of Applied Statistics}, 10(3):1639--1672.

\bibitem[Stoeldraijer et~al., 2015]{stoeldraijer2015future}
Stoeldraijer, L., Bonneux, L., van Duin, C., van Wissen, L., and Janssen, F.
  (2015).
\newblock The future of smoking-attributable mortality: the case of {England}
  \& {Wales}, {Denmark} and the {Netherlands}.
\newblock {\em Addiction}, 110(2):336--345.

\bibitem[Trias~Llim{\'o}s and Janssen, 2019]{trias2019gender}
Trias~Llim{\'o}s, S. and Janssen, F. (2019).
\newblock Gender gaps in life expectancy and alcohol consumption in {E}astern
  {E}urope.
\newblock {\em N-IUSSP}.

\bibitem[{United Nations}, 2015]{WPP2015}
{United Nations} (2015).
\newblock {\em World Population Prospects 2015}.
\newblock United Nations, New York, N.Y.

\bibitem[{United Nations}, 2017]{WPP2017}
{United Nations} (2017).
\newblock {\em World Population Prospects 2017}.
\newblock United Nations, New York, N.Y.
\newblock Accessed: Oct. 15, 2018 \url{
  http://population.un.org/wpp/Download/Standard/Population/}.

\bibitem[Vidra et~al., 2017]{vidra2017impact}
Vidra, N., Trias-Llim{\'o}s, S., and Jansse, F. (2017).
\newblock Impact of obesity on trends in life expectancy among different
  {E}uropean countries, 1975-2012.
\newblock {\em European Journal of Public Health}, 27(suppl\_3).

\bibitem[Wang and Preston, 2009]{wang2009forecasting}
Wang, H. and Preston, S.~H. (2009).
\newblock Forecasting {United States} mortality using cohort smoking histories.
\newblock {\em Proceedings of the National Academy of Sciences},
  106(2):393--398.

\bibitem[Whelpton, 1936]{Whelpton1936}
Whelpton, P.~K. (1936).
\newblock An empirical method of calculating future population.
\newblock {\em Journal of the American Statistical Association},
  31(195):457--473.

\bibitem[Wi{\'s}niowski et~al., 2015]{wisniowski2015bayesian}
Wi{\'s}niowski, A., Smith, P.~W., Bijak, J., Raymer, J., and Forster, J.~J.
  (2015).
\newblock Bayesian population forecasting: extending the {L}ee-{C}arter method.
\newblock {\em Demography}, 52(3):1035--1059.

\bibitem[{World Health Organization}, 2017]{WHO2017}
{World Health Organization} (2017).
\newblock Mortality database.
\newblock Last accessed: Oct. 15, 2018 \url{
  http://www.who.int/healthinfo/statistics/mortality_rawdata/en/}.

\end{thebibliography}

\pagebreak

\begin{appendices}
	
\section{Full Bayesian Hierarchical Model} \label{app1}
	We first describe the estimating and projection of the full model. 
	
	\begin{enumerate}
		\item Estimate and forecast the male ASSAF using the 3-level Bayesian hierarchical model described in Section \ref{subsec:assaf}, and generate 30 samples from the posterior distributions of the mean of ASSAF of all 69 clear-pattern countries for all 13 five-year estimation periods and all 9 five-year periods forecast period;
		\item For each country, generate 30 samples of male $e_0^{NS}$ based on the ASSAF samples drawn in Step 2 for all 13 five-year estimation periods, and for each of the 30 samples, forecast male $e_0^{NS}$ of all 69 countries for all 9 five-year periods using the 3-level Bayesian hierarchical model described in Section \ref{subsec:nonsmke0};
		\item For each country, forecast male $e_0$ based on the method described in Section \ref{subsubsec:e0male} for each of the 30 samples, and combine trajectories from all 30 samples to get the full posterior predictive distribution of male $e_0$;
		\item For each country, apply the gap model described in Section \ref{subsubsec:gap} to the combined trajectories of male $e_0$ to get the full posterior predictive distribution of female $e_0$.
	\end{enumerate}

	The Bayesian hierarchical model for modeling age-specific smoking attributable fraction (ASSAF) described in Section \ref{subsec:assaf} is specified as follows.
	\begin{alignat*}{2}
	\mbox{Level 1:}\ &
	y_{x,t}^\ell \stackrel{\rm ind}{\sim} \mathcal{N}(\xi_{x}^\ell \tau_{t-x}^\ell \mathbf{1}_{x \neq 80}+\xi_{x}^\ell \tilde{\tau}_{t-x}^\ell \mathbf{1}_{x = 80},\ \sigma^2_{\ell});\\
	\mbox{Level 2:}\ &
	\beta_{40}^\ell = 1,\
	&&\beta_x^\ell|\mu_x^{[\beta]}, \sigma^{2[\beta]}_x \stackrel{\rm i.i.d}{\sim} \mathcal{N}(\mu_x^{[\beta]}, \sigma^{2[\beta]}_x)\ \mbox{for all $x$ except 40},\\
	&\tau_c^\ell|\theta^\ell, \sigma^{2[\tau]}  \stackrel{\rm ind}{\sim} \mathcal{N}(g(c|\theta^\ell),\ \sigma^{2[\tau]}),\
	&&\tilde{\tau}_c^\ell|\tilde{\theta}^\ell, \sigma^{2[\tau]}\stackrel{\rm ind}{\sim} \mathcal{N}(g(c|\tilde{\theta}^\ell), \sigma^{2[\tau]})\ \mbox{for $c = t-x$},\\
	&\triangle_1^\ell|\mu_{\triangle_1} \stackrel{\rm i.i.d}{\sim}  \mathcal{G}(2, 2/\mu_{\triangle_1}),\ 
	&&\triangle_2^\ell|\mu_{\triangle_2}, \sigma^2_{\triangle_2} \stackrel{\rm i.i.d}{\sim}  \mathcal{N}(\mu_{\triangle_2}, \sigma^2_{\triangle_2}),\\
	& \triangle_3^\ell|\mu_{\triangle_3} \stackrel{\rm i.i.d}{\sim}  \mathcal{G}(2, 2/\mu_{\triangle_3}),\ 
	&&\triangle_4^\ell |\mu_{\triangle_4}, \sigma^2_{\triangle_4}    \stackrel{\rm i.i.d}{\sim} \mathcal{N}(\mu_{\triangle_4}, \sigma^2_{\triangle_4}),\\
	& k^\ell|\mu_k, \sigma^2_k \stackrel{\rm i.i.d}{\sim}  \mathcal{N}(\mu_k, \sigma^2_k),\ 
	&&\delta^\ell|\mu_\delta, \sigma^2_\delta \stackrel{\rm i.i.d}{\sim}  \mathcal{N}(\mu_\delta, \sigma^2_\delta),\\
	& \sigma^2_{\ell}|\sigma^2 \stackrel{\rm i.i.d}{\sim}  \mathcal{IG}(2, \sigma^2);\\
	\mbox{Level 3:}\ &
	\mu_x^{[\beta]} \stackrel{\rm i.i.d}{\sim} \mathcal{N}(1, 5),\ &&\sigma_x^{2[\beta]} \stackrel{\rm i.i.d}{\sim} \mathcal{IG}(2, 5),\\
	&\sigma^2 \sim \mathcal{IG}(2, 0.01),\ &&\sigma^{2[\tau]} \sim \mathcal{IG}(2, 0.01),\\
	&\mu_{\triangle_1} \sim \mathcal{G}(2, 0.1 ),\ &&\mu_{\triangle_2} \sim \mathcal{N}(20, 1000),\\
	&\mu_{\triangle_3} \sim \mathcal{G}(2, 0.1),\ &&\mu_{\triangle_4} \sim \mathcal{N}(20, 1000),\\
	&\mu_k \sim \mathcal{N}(0.3, 0.25),\ &&\mu_\delta \sim \mathcal{N}(0, 100),\\
	&\sigma^2_{\triangle_2} \sim \mathcal{IG}(2, 1000),\ \sigma^2_{\triangle_4} \sim \mathcal{IG}(2, 1000),\\
	&\sigma^2_{k} \sim \mathcal{IG}(2, 0.25),\ \sigma^2_{\delta} \sim \mathcal{IG}(2, 100),
	\end{alignat*}
	where 
	$ \theta^\ell := (\triangle_1^\ell, \triangle_2^\ell, \triangle_3^\ell, \triangle_4^\ell, k^\ell),\ \tilde{\theta}^\ell := (\triangle_1^\ell, \triangle_2^\ell, \triangle_3^\ell, \triangle_4^\ell + \delta^\ell, k^\ell),$ and
	$$
	g(c|\theta) =  \frac{k}{1+\exp\{-\triangle_1(c-1873-\triangle_2)\}} -  \frac{k}{1+\exp\{-\triangle_3(c-1873-\triangle_2- \triangle_4)\}}.$$
	\linebreak
	
	The Bayesian hierarchical model for modeling non-smoking life expectancy ($e_0^{NS}$) described in Section \ref{subsec:nonsmke0} is specified as follows.
	\begin{alignat*}{2}
	\mbox{Level 1:}\ &
	e_{0, \ell, t}^{NS}  \stackrel{\rm ind}{\sim} \mathcal{N} (e_{0, \ell, t-1}^{NS} + \tilde{g}(e_{0, \ell, t-1}^{NS} |\zeta^{\ell}),\ (\omega^\ell \cdot \phi (e_{0, \ell, t-1}^{NS}))^2);\\
	\mbox{Level 2:}\ &
	a_i^\ell|\mu_{a_i}, \sigma_{a_i}^2 \stackrel{\rm i.i.d}{\sim}  \mathcal{N}_{[0,100]}(\mu_{a_i}, \sigma_{a_i}^2), &&i = 1,\ \cdots,\ 4,\\
	&w^\ell|\mu_w, \sigma^2_w \stackrel{\rm i.i.d}{\sim}  \mathcal{N}_{[0,15]}(\mu_w, \sigma^2_w),\ && z^\ell|\mu_z, \sigma^2_z \stackrel{\rm i.i.d}{\sim}  \mathcal{N}_{[0,1.15]}(\mu_z, \sigma^2_z),\\
	&\omega^{\ell} \stackrel{\rm i.i.d}{\sim}  \mathcal{U}_{[0,10]};\\
	\mbox{Level 3:}\ 
	&\mu_{a_1} \sim \mathcal{N}(15.77, 15.6^2),\ && \mu_{a_2} \sim \mathcal{N}(40.97, 23.5^2),\\
	&\mu_{a_3} \sim \mathcal{N}(0.21, 14.5^2),\ && \mu_{a_4} \sim \mathcal{N}(19.82, 14.7^2),\\
	&\mu_{w} \sim \mathcal{N}(2.93, 3.5^2),\ && \mu_{z} \sim \mathcal{N}(0.40, 0.6^2),\\
	&\sigma^2_{a_1} \sim \mathcal{IG}(2, 15.6^2),\ && \sigma^2_{a_2} \sim \mathcal{IG}(2, 14.5^2),\\
	&\sigma^2_{a_3} \sim \mathcal{IG}(2, 14.7^2),\ && \sigma^2_{a_4} \sim \mathcal{IG}(2, 3.5^2),\\
	&\sigma^2_{w} \sim \mathcal{IG}(2, 0.6^2),\ && \sigma^2_{z} \sim \mathcal{IG}(2, 0.6^2),
	\end{alignat*}
	where $\zeta := (a_1, a_2, a_3, a_4, w, z)$ and
	\begin{align*}
	\tilde{g}(e_{0}^{NS}|\zeta) := \frac{w}{1 + \exp\{-\frac{4.4}{a_2}(e_{0}^{NS} -a_1 - 0.5a_2 )\}  } + \frac{z-w}{1 + \exp\{-\frac{4.4}{a_4}(e_{0}^{NS} -\sum_{i = 1}^3a_i - 0.5a_4 )\} }.
	\end{align*}
	
		\section{MCMC Diagnostics}\label{app2}
\begin{figure}[hpb]
		\begin{center}
			\includegraphics[scale=0.35]{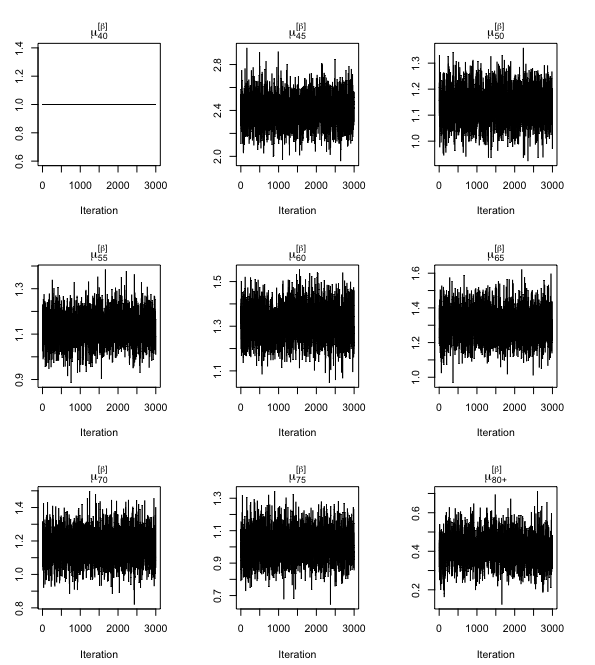}
			\includegraphics[scale=0.35]{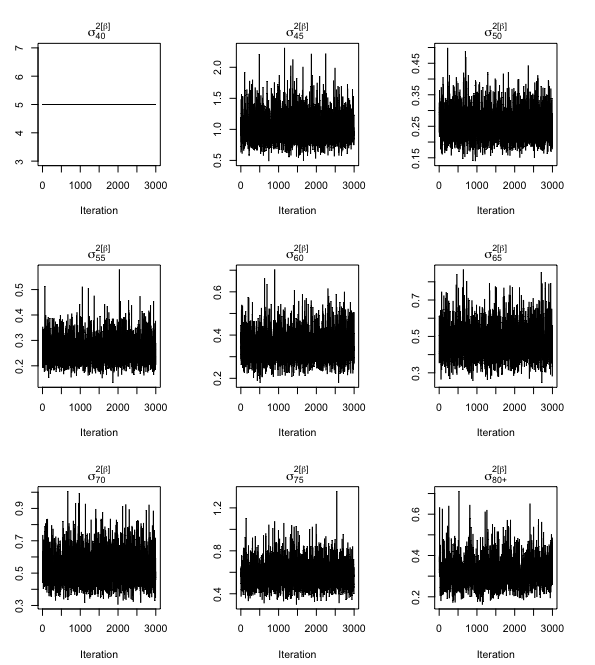}
			\includegraphics[scale=0.35]{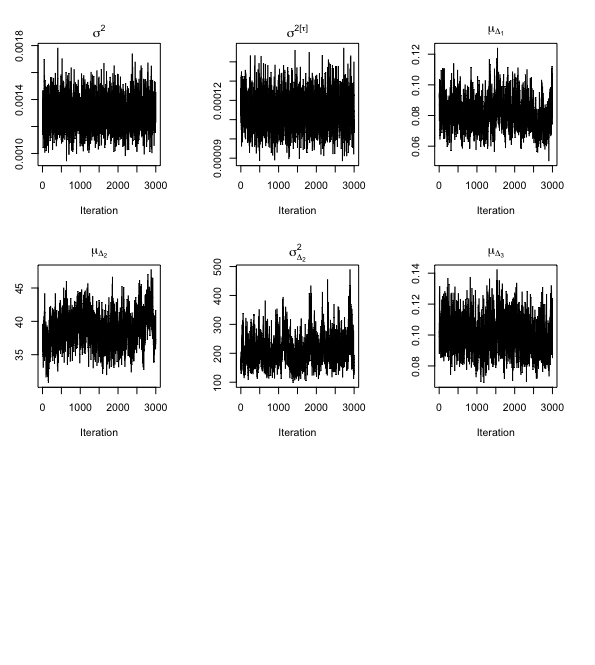}
			\includegraphics[scale=0.35]{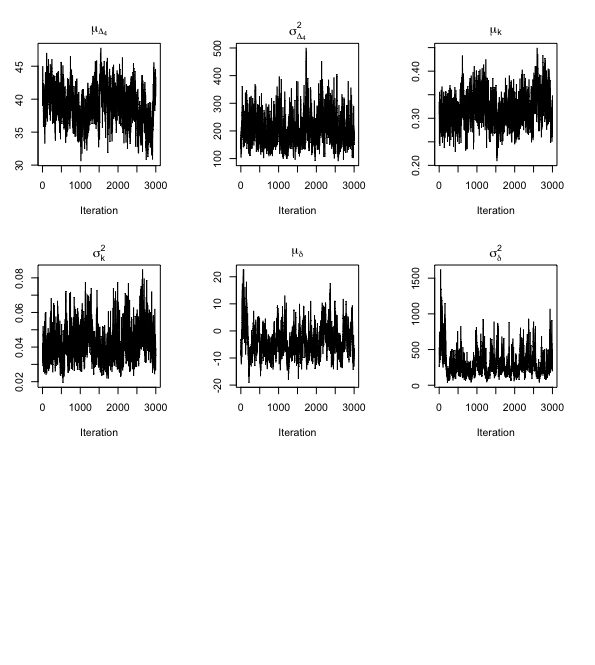}
		\end{center}
		\caption{Traceplots for the hyperparameters in BHM for ASSAF.} \label{fg:assaftr}
	\end{figure}

	We check the convergence of BHM for ASSAF based on trace plots and diagnostics \citep{raftery1992mcmc} for the global parameters in Level 3. We check one chain with 2,000 burn in iterations and 100,000 samples,
	with a thinning period of 20 iterations. Table \ref{tb:assafraf} shows the summary statistics of the diagnostics. Figure \ref{fg:assaftr} shows the trace plots of all 3,000 samples of the global parameters. 
	
\begin{figure}[!h]
		\begin{center}
			\includegraphics[scale=0.35]{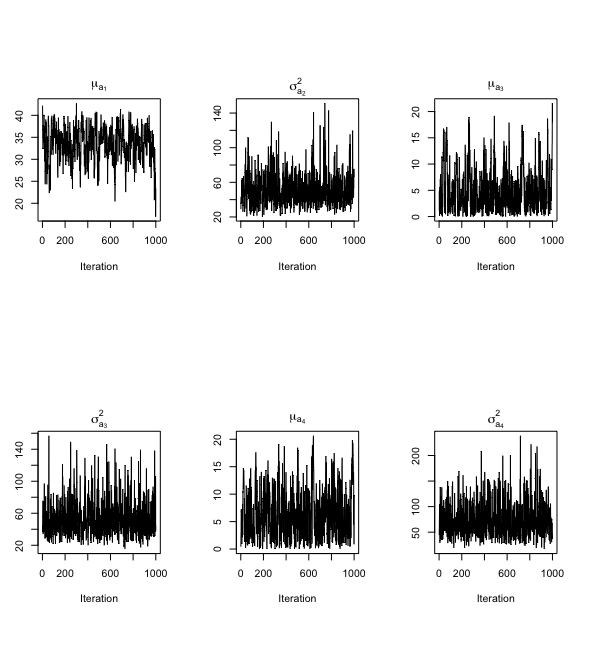}
			\includegraphics[scale=0.35]{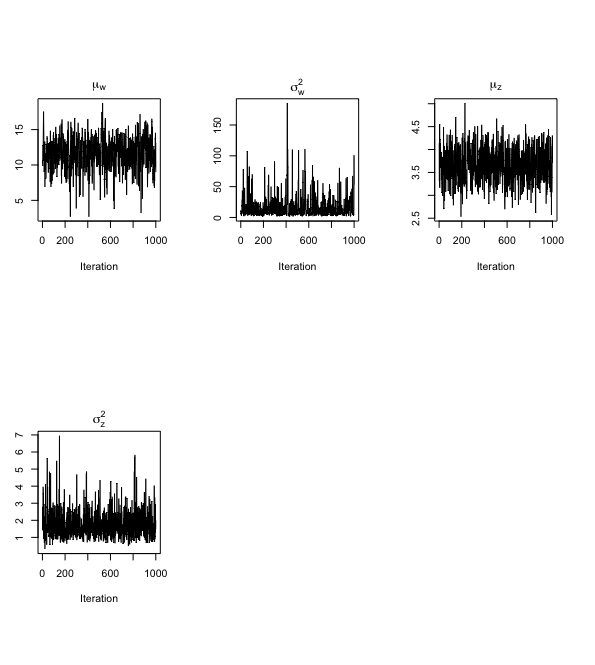}
		\end{center}
		\caption{Traceplots for the hyperparameters in BHM for $e_0^{NS}$.} \label{fg:e0nstr}
	\end{figure}

	We do the same for $e_0^{NS}$.
	We check one of the 30 samples with 1,000 burn in iterations, 100,000 samples
	and a thinning period of 50 iterations. Table \ref{tb:e0nsraf} shows the summarizing statistics of the diagnostics. Figure \ref{fg:e0nstr} shows the trace plots of all 1,000 samples of the global parameters.
	
	\begin{longtable}{|c|c|c|c|c|c|c|}
		\caption{Diagnostic statistics for global parameters in BHM for ASSAF.  Burn1, Size1, and
			DF1 are the length of burn-in, required sample size, and dependence factor of the Raftery-Lewis diagnostics with parameters $q = 0.025$, $r = 0.0125$, and $s= 0.95$. Burn2, Size2, and DF2 are the same, but for $q = 0.975$.}
		\label{tb:assafraf} \\
		\hline
		Parameters& Burn1& Size1 & DF1 & Burn2& Size2& DF2\\ 
		\hline \endhead
		\hline
		\endfoot
		
		\hline
		\endlastfoot
		$\mu_{40}^{2[\beta]}$ & - & - &- & - & - &-  \\ 
		$\mu_{45}^{2[\beta]}$ & 2 & 606  & 1.01 & 2 & 631& 1.05 \\ 
		$\mu_{50}^{2[\beta]}$ & 2 & 641 & 1.07 & 2 & 581 & 0.97 \\ 
		$\mu_{55}^{2[\beta]}$ & 2 & 577 & 0.96 & 2 & 631 & 1.05 \\ 
		$\mu_{60}^{2[\beta]}$ & 2 & 616 & 1.03 & 2 & 591& 0.98 \\ 
		$\mu_{65}^{2[\beta]}$ & 2 & 601& 1.00 & 2 & 621 & 1.03 \\ 
		$\mu_{70}^{2[\beta]}$ & 2 & 616 & 1.03 & 2 & 621 & 1.03 \\ 
		$\mu_{75}^{2[\beta]}$ & 2 & 587 & 0.98 & 2 & 611 & 1.02 \\ 
		$\mu_{80}^{2[\beta]}$ & 2 & 626& 1.04 & 3 & 664 & 1.11 \\  
		$\sigma_{40}^{2[\beta]}$ & - &- &-  & - &- &-\\
		$\sigma_{45}^{2[\beta]}$ & 3 & 669& 1.12 & 3 & 653 & 1.09 \\  
		$\sigma_{50}^{2[\beta]}$ & 2 & 601 & 1.00 & 2 & 601 & 1.00 \\ 
		$\sigma_{55}^{2[\beta]}$ & 2 & 591& 0.98 & 2 & 621 & 1.03 \\ 
		$\sigma_{60}^{2[\beta]}$ & 2 & 606 & 1.01 & 2& 572 & 0.95 \\ 
		$\sigma_{65}^{2[\beta]}$ & 2 & 611 & 1.02 & 2& 611 & 1.02 \\ 
		$\sigma_{70}^{2[\beta]}$ & 1 & 595 & 0.99 & 2 & 591 & 0.98 \\ 
		$\sigma_{75}^{2[\beta]}$ & 3 & 648 & 1.08 & 2 & 641 & 1.07 \\ 
		$\sigma_{80}^{2[\beta]}$ & 2 & 621 & 1.03 & 3 & 676 & 1.13 \\ 
		$\sigma$ & 2 & 591 & 0.98 & 3 & 658 & 1.10 \\
		$\sigma^{2[\tau]}$ & 2 & 591& 0.98 & 3 & 648 & 1.08 \\ 
		$\mu_{\triangle_1}$ & 6 & 1308& 2.18 & 9& 2040 & 3.40 \\ 
		$\mu_{\triangle_2}$ & 8& 1610& 2.68 & 2 & 641 & 1.07 \\ 
		$\sigma_{\triangle_2}^2$ & 12 & 2160 & 3.60 & 8 & 1586 & 2.64 \\ 
		$\mu_{\triangle_3}$& 4 & 790 & 1.32 & 3 & 686 & 1.14 \\ 
		$\mu_{\triangle_4}$ & 12& 1980 & 3.30 & 15 & 2211& 3.68 \\  
		$\sigma_{\triangle_4}^2$ & 6 & 1701 & 2.84 & 8& 1656 & 2.76 \\ 
		$\mu_{k}$ & 6 & 1432 & 2.39 & 8 & 1456 & 2.43 \\ 
		$\sigma_{k}^2$ & 4 & 1236 & 2.06 & 4 & 771 & 1.28 \\ 
		$\mu_{\delta}$ & 18 & 3090& 5.15  & 24 & 4912 & 8.19 \\  
		$\sigma_{\delta}^2$ & 15 & 2499& 4.16 & 30 & 6955 & 11.60 
	\end{longtable}

	\begin{table}
		\caption{Diagnostic statistics for global parameters in BHM for $e_0^{NS}$.  The quantities shown are defined as in Table \protect\ref{tb:assafraf}.}
		\label{tb:e0nsraf} 
		\centering
		\begin{tabular}{|c|ccc|ccc|}
		\hline
		Parameters& Burn1& Size1 & DF1 & Burn2& Size2& DF2\\ 
		\hline
		$\mu_{a_1}$ & 164 & 14734& 24.60 & 3& 703 & 1.17 \\ 
		$\sigma^2_{a_1}$ & 2 & 596 & 0.99 & 3 & 648& 1.08 \\ 
		$\mu_{a_2}$& 2 & 572 & 0.95 & 81 & 10795 & 18.00 \\ 
		$\sigma^2_{a_2}$& 3 & 648& 1.08 & 3 & 648 & 1.08 \\ 
		$\mu_{a_3}$ & 2 & 572 & 0.95 & 7 & 1179& 1.96 \\ 
		$\sigma^2_{a_3}$ & 2 & 596 & 0.99 & 2 & 621 & 1.03 \\ 
		$\mu_{a_4}$ & 3 & 703 & 1.17 & 6& 1005& 1.68 \\ 
		$\sigma^2_{a_4}$ & 3& 675 & 1.12 & 5 & 867& 1.44 \\ 
		$\mu_{w}$& 3 & 648 & 1.08 & 2& 596 & 0.99 \\ 
		$\sigma^2_{w}$ & 2& 572 & 0.95 & 2 & 596 & 0.99 \\ 
		$\mu_{z}$ & 3& 662 & 1.10 & 2& 572 & 0.95 \\ 
		$\sigma^2_{z}$ & 2 & 596 & 0.99 & 3 & 689 & 1.15 \\ 
		\hline
		\end{tabular}
	\end{table}
	
	\section{Life Expectancy Forecast Plots of 69 Countries}\label{app3}
	\begin{figure}[H]
		\begin{center}
			\includegraphics[scale=0.45]{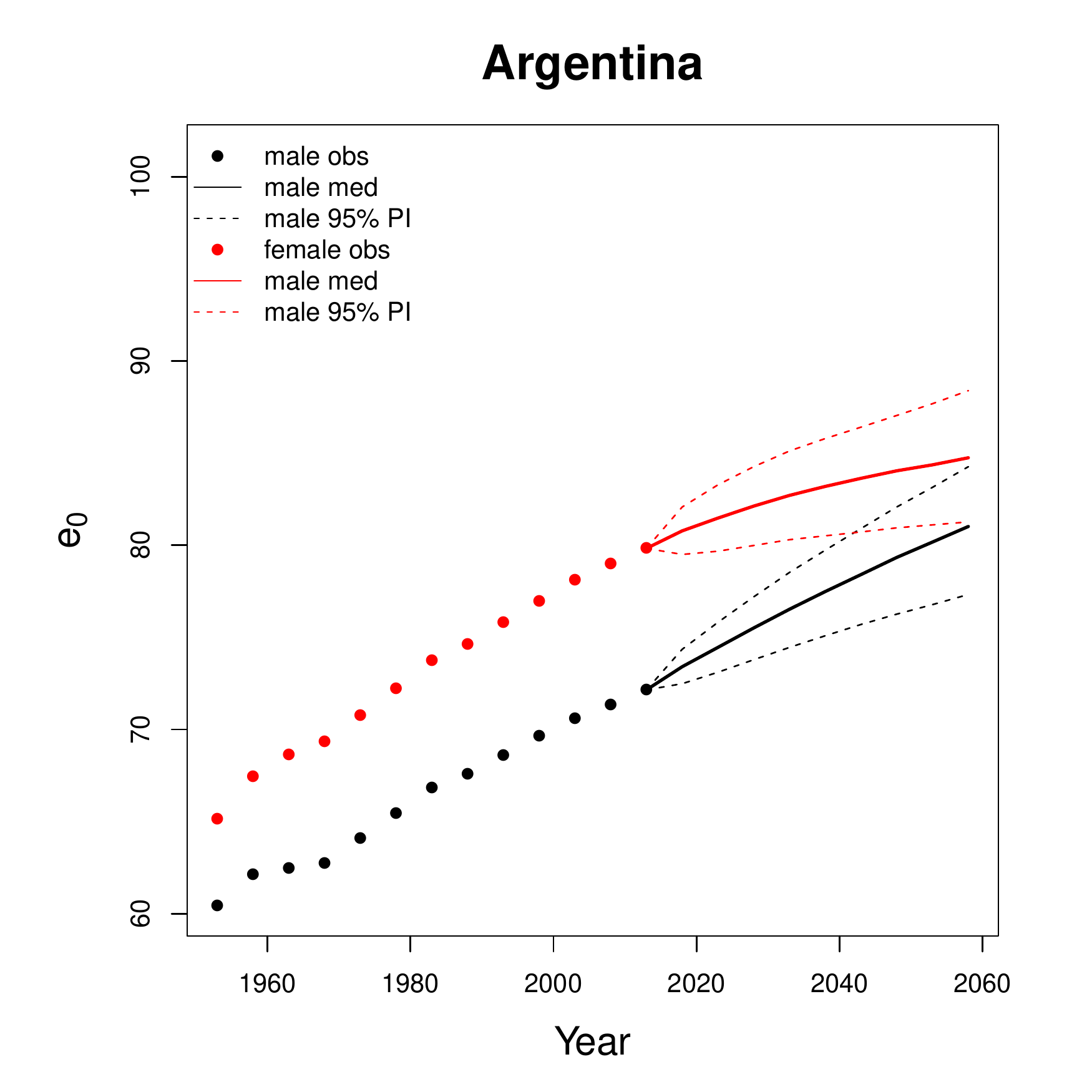}
			\includegraphics[scale=0.45]{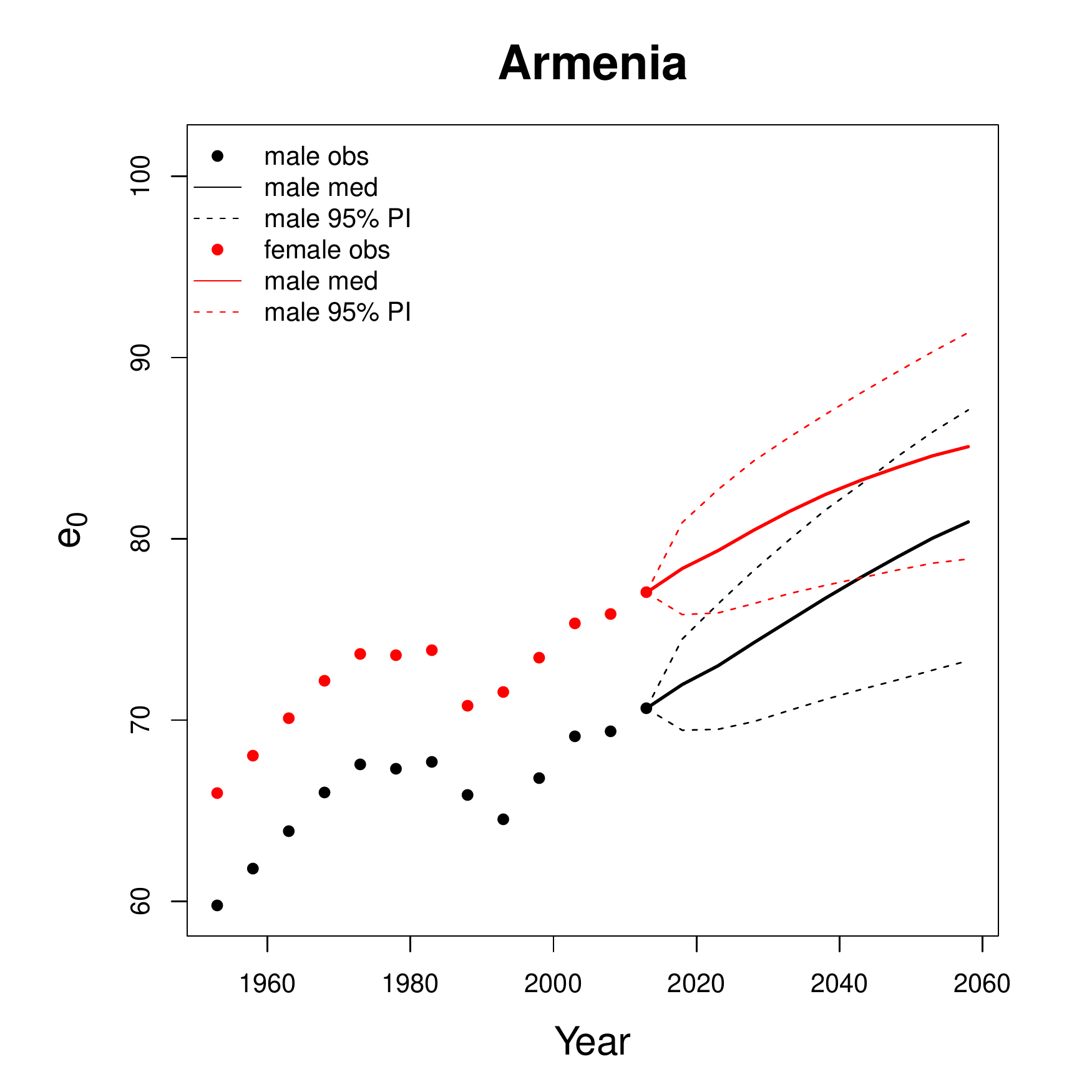}
			\includegraphics[scale=0.43]{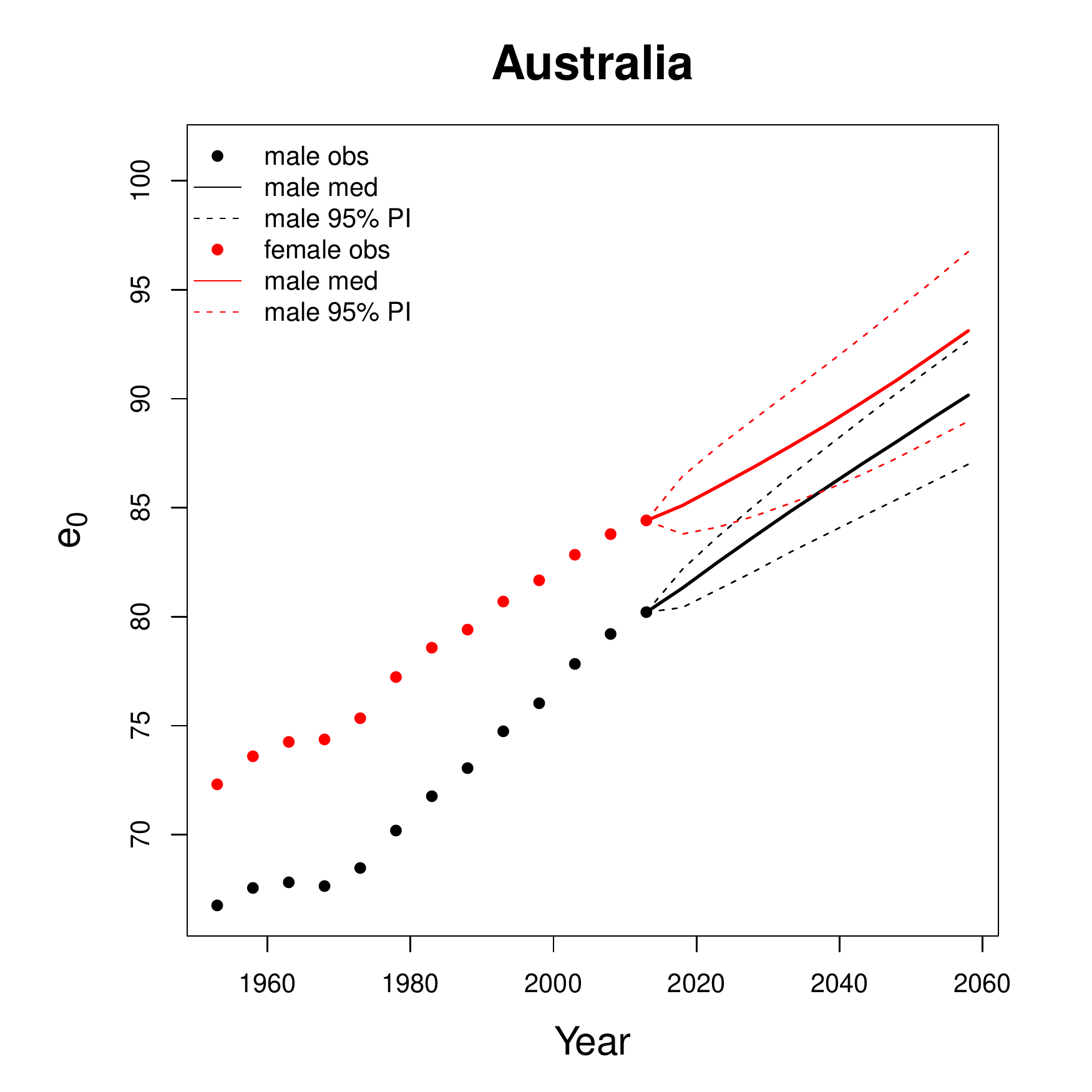}
			\includegraphics[scale=0.43]{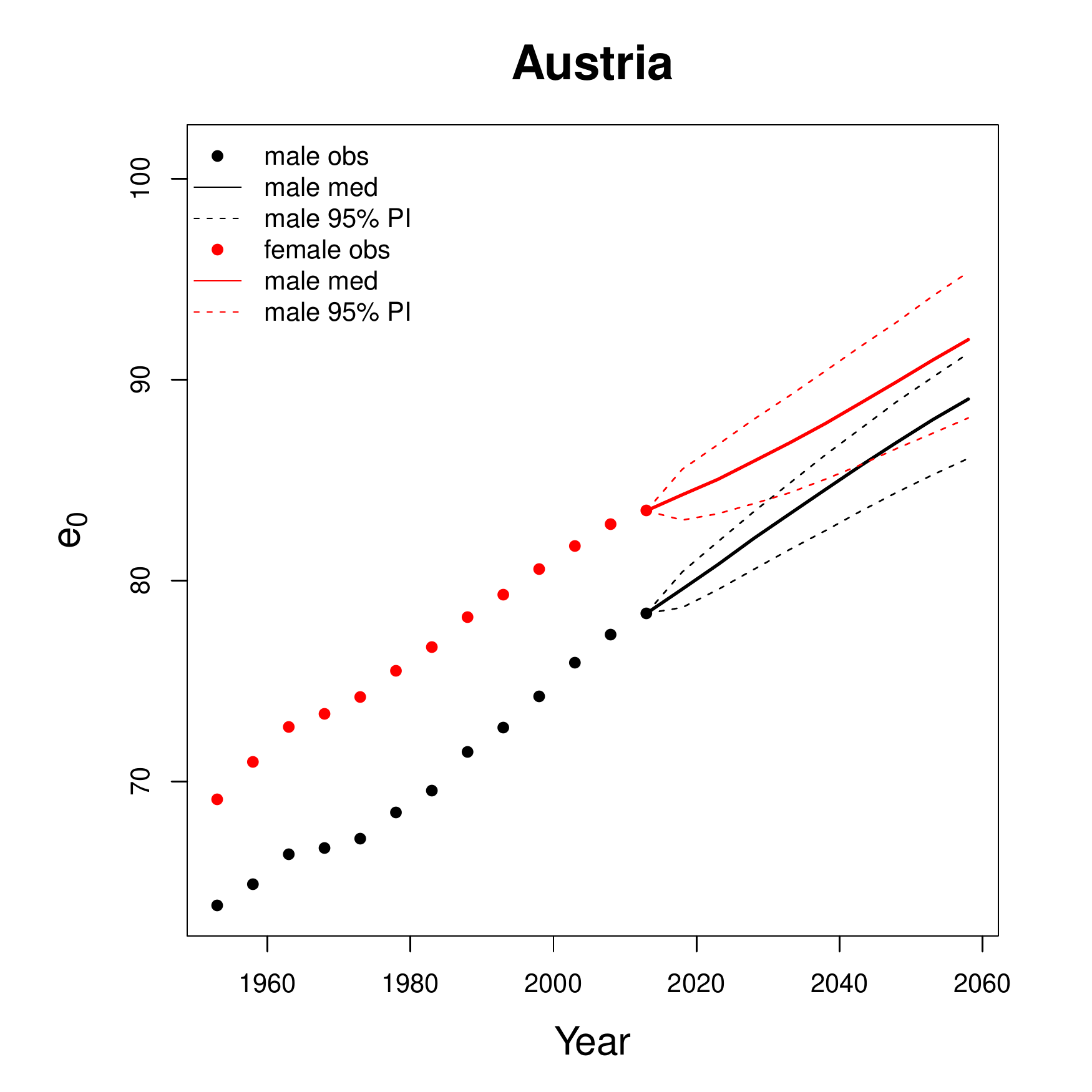}
			\includegraphics[scale=0.43]{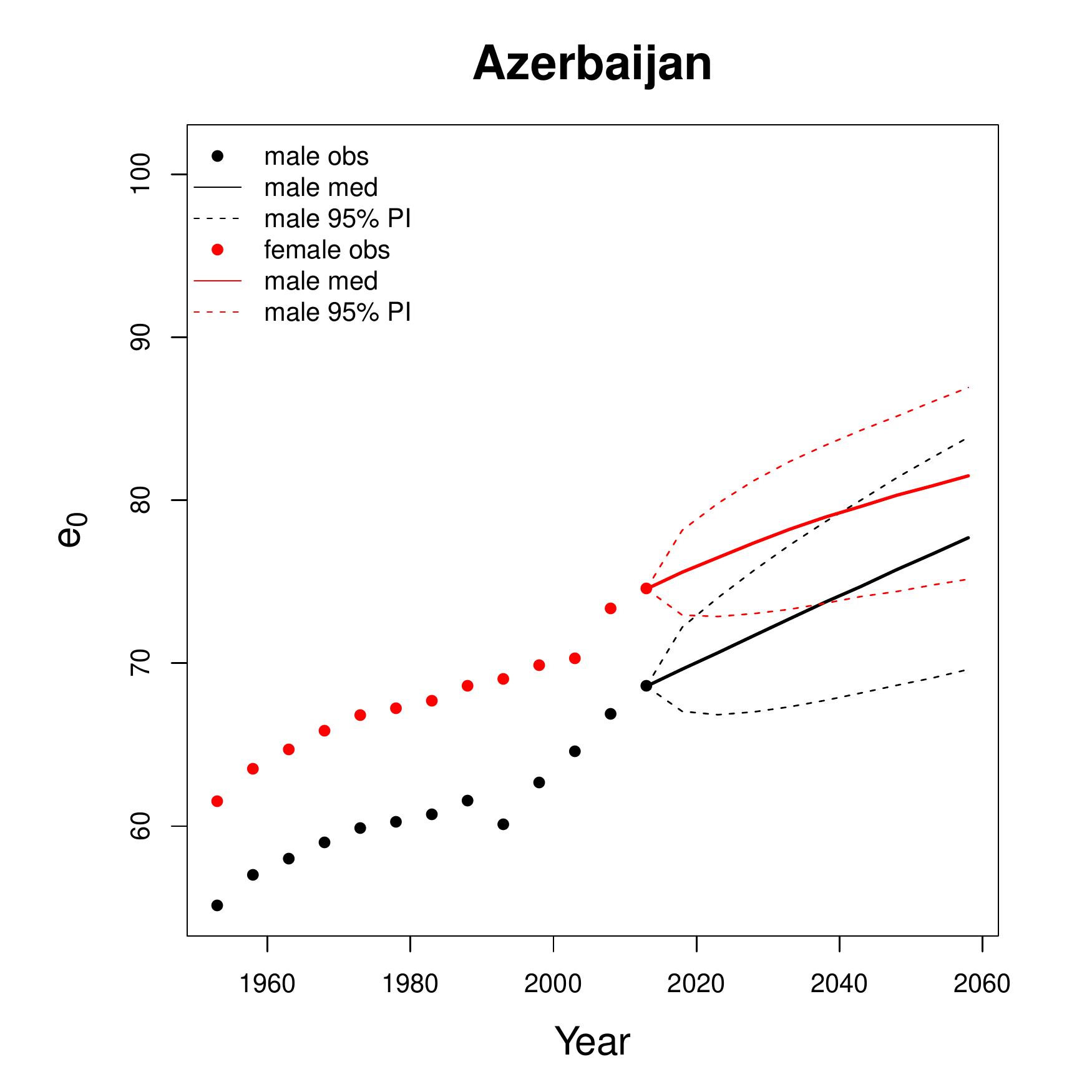}
			\includegraphics[scale=0.43]{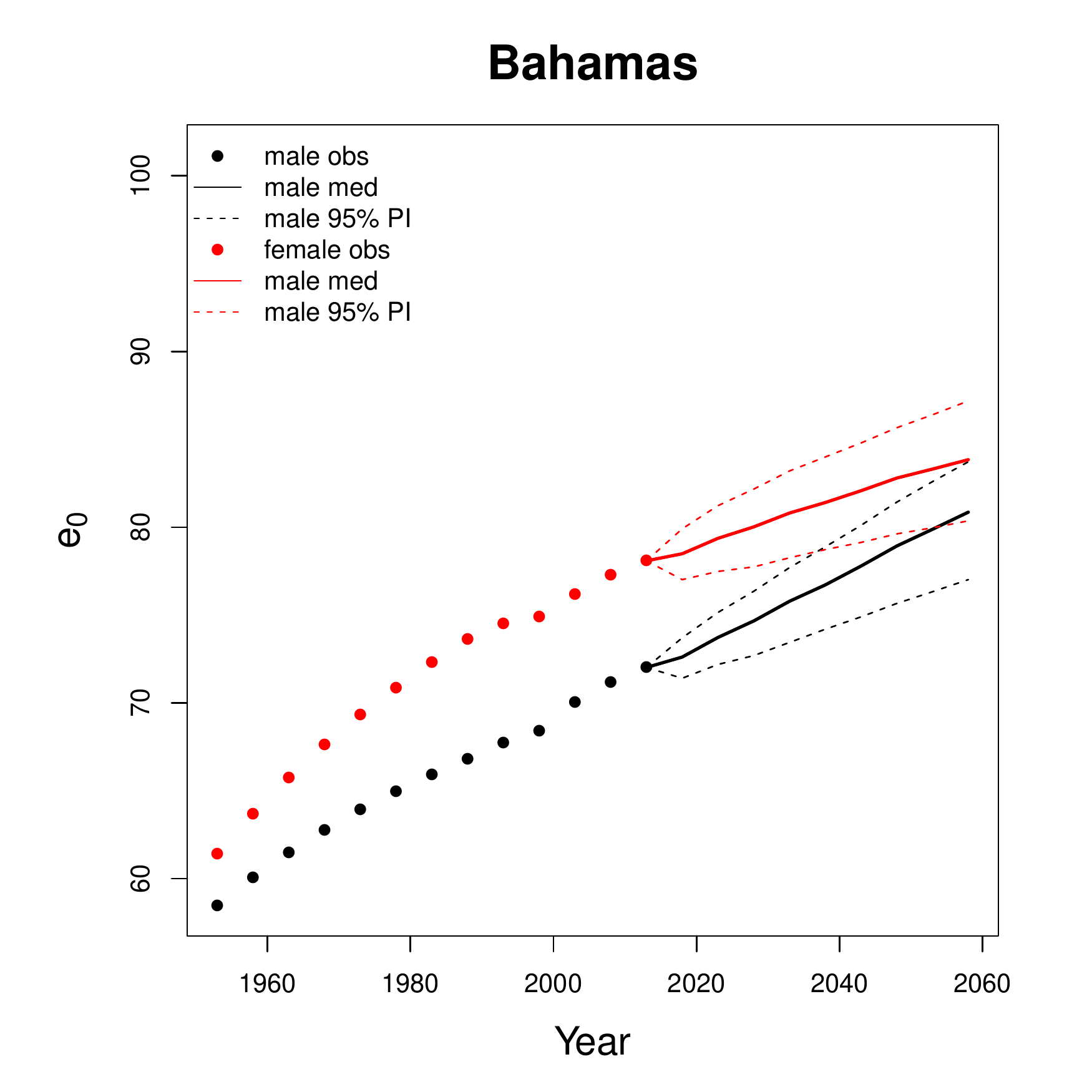}
		\end{center}
	\end{figure}

	\begin{figure}[H]
		\begin{center}
			\includegraphics[scale=0.43]{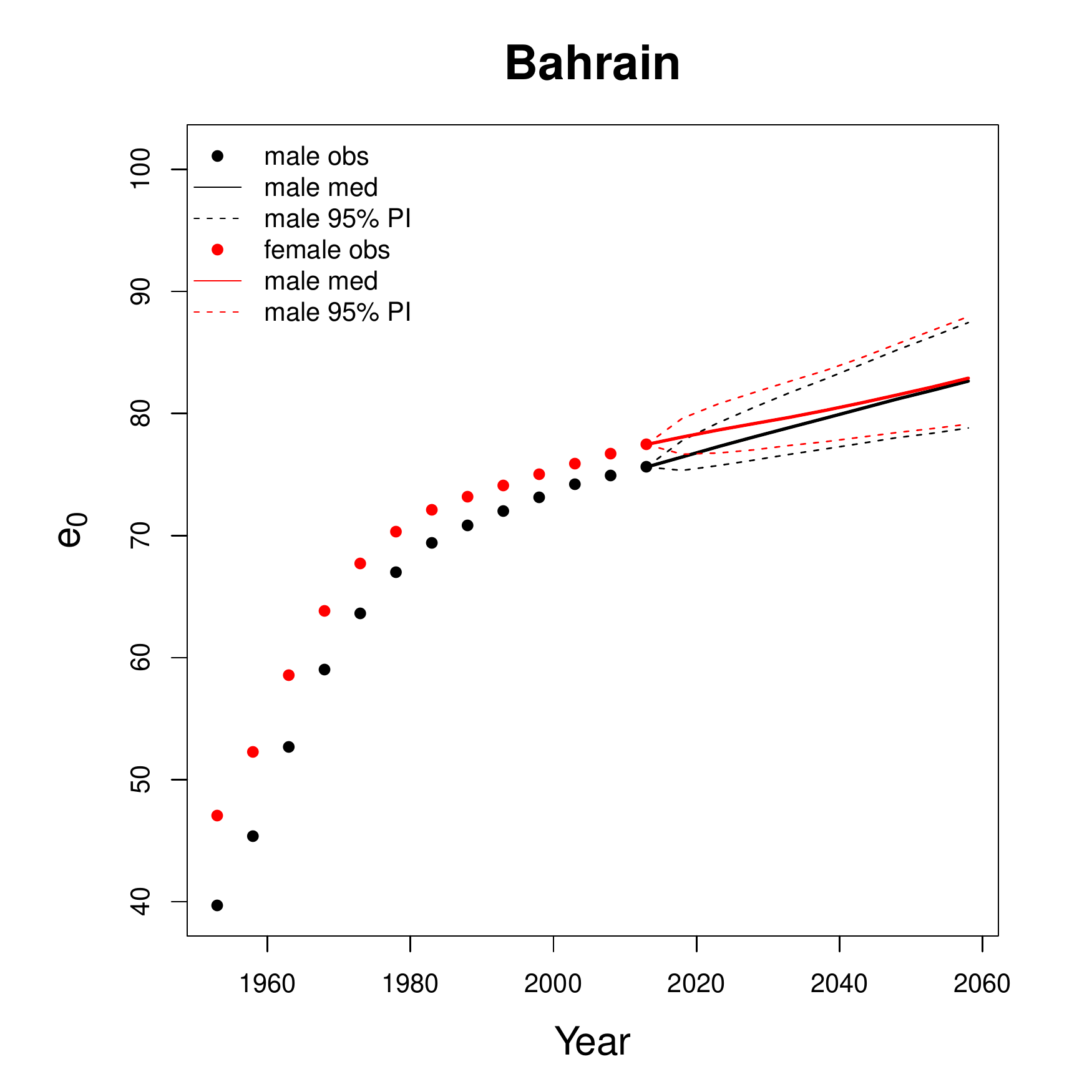}
			\includegraphics[scale=0.43]{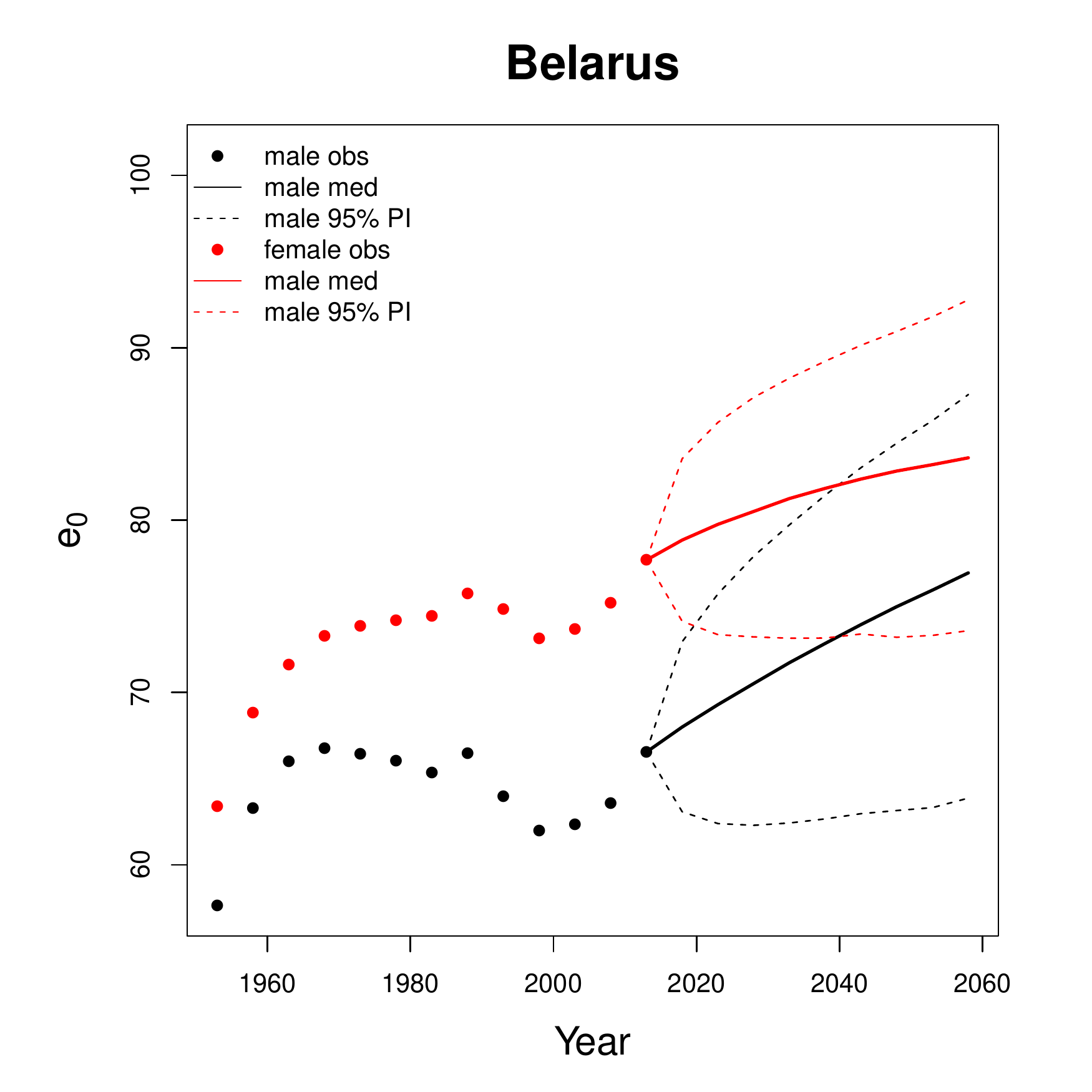}	\includegraphics[scale=0.43]{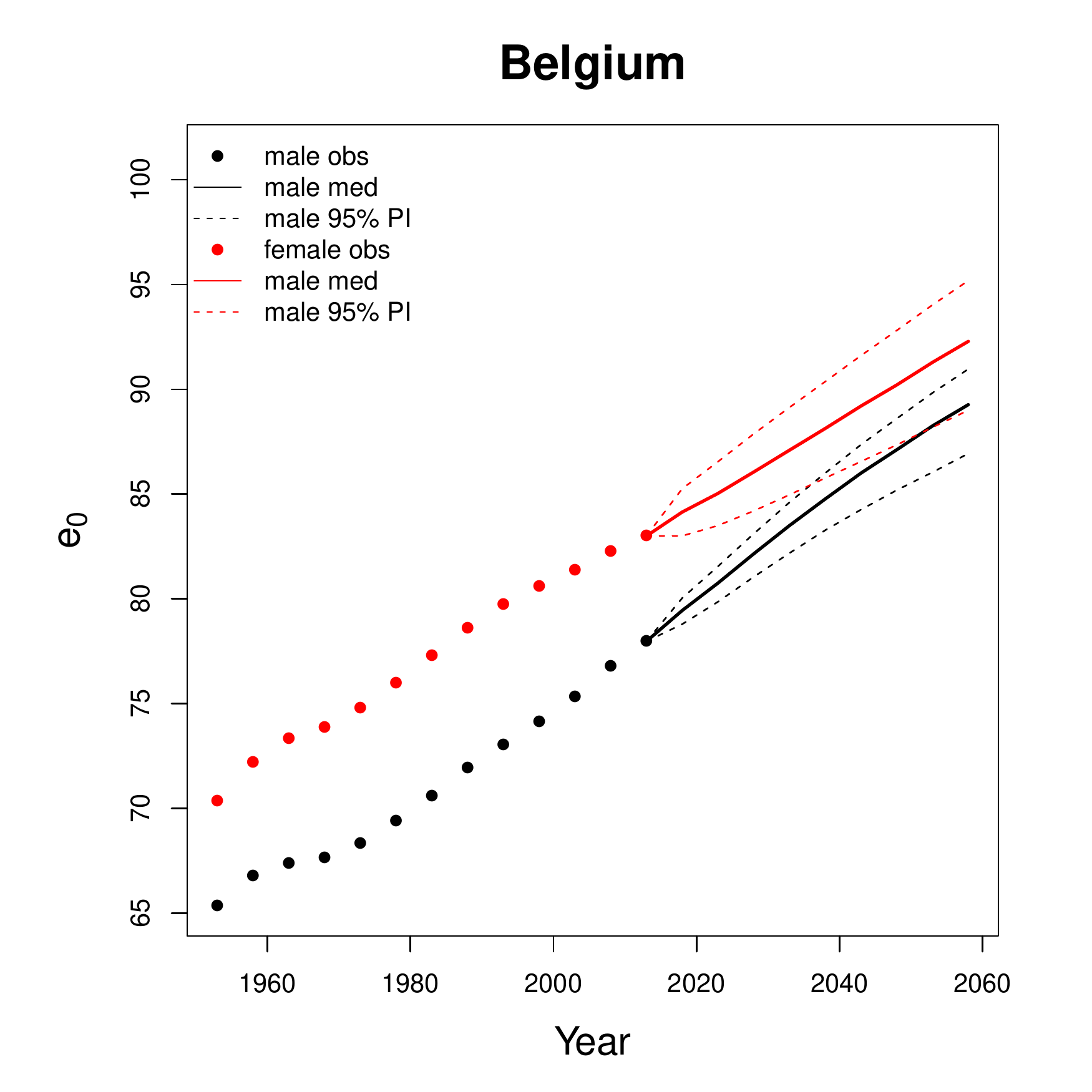}
			\includegraphics[scale=0.43]{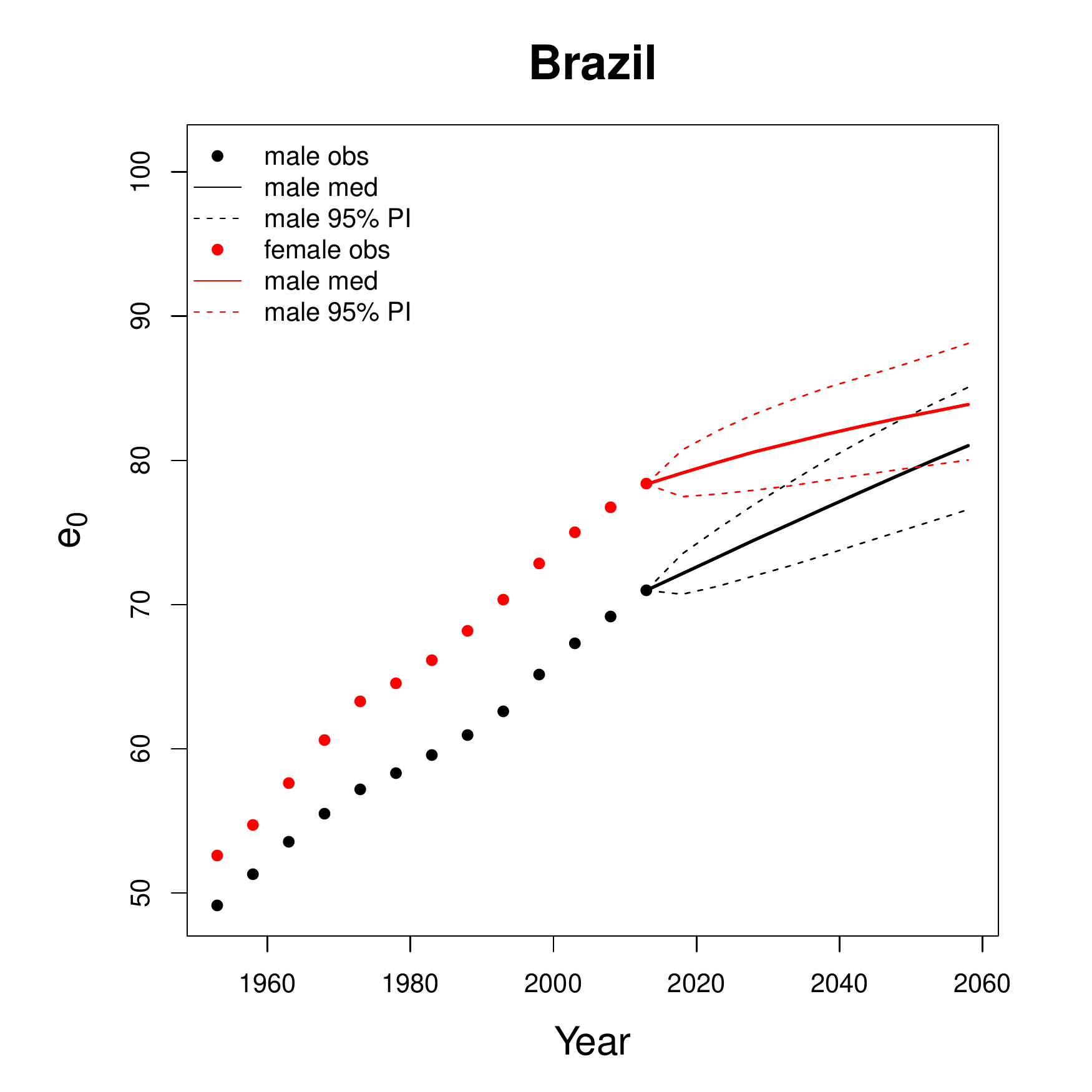}
			\includegraphics[scale=0.43]{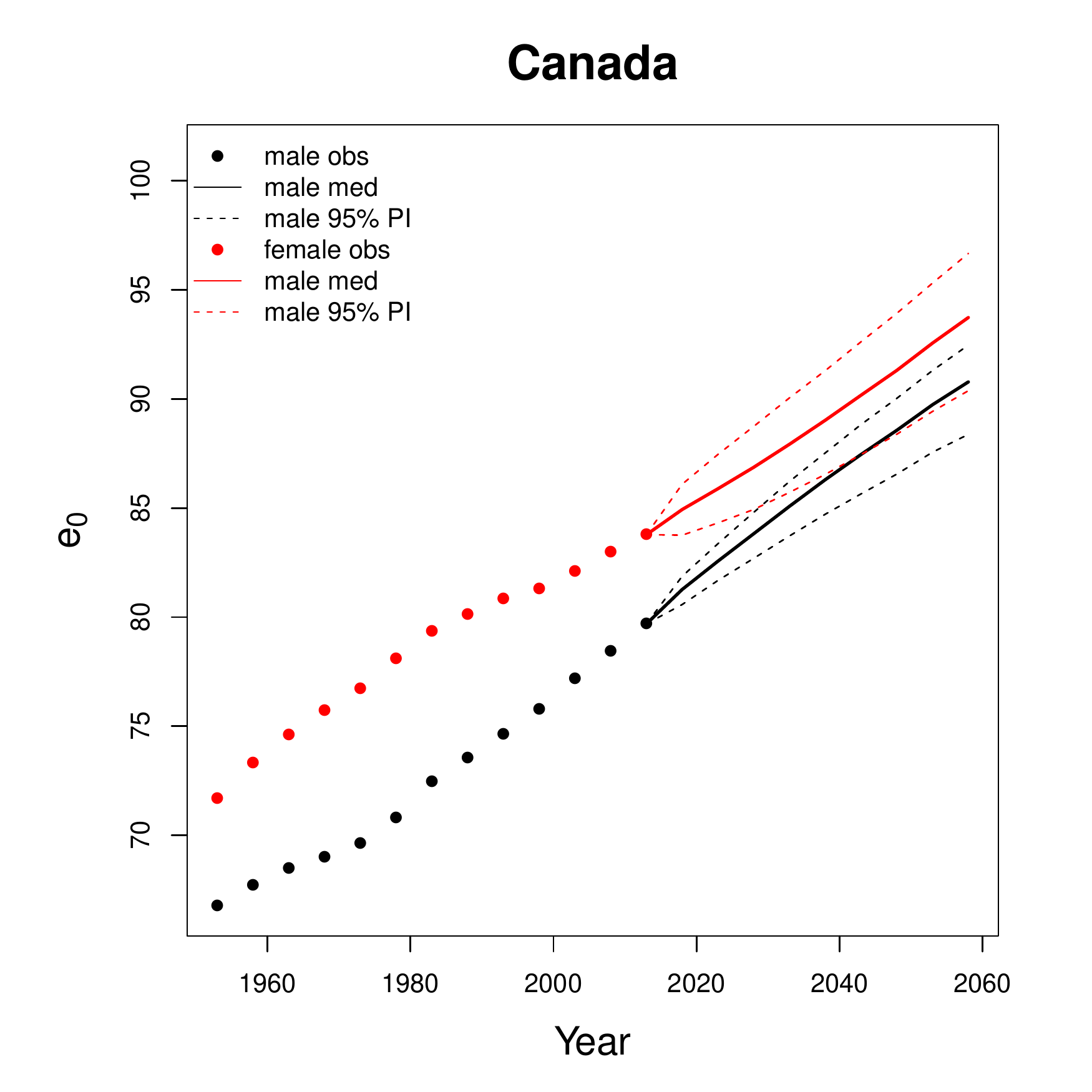}	
			\includegraphics[scale=0.43]{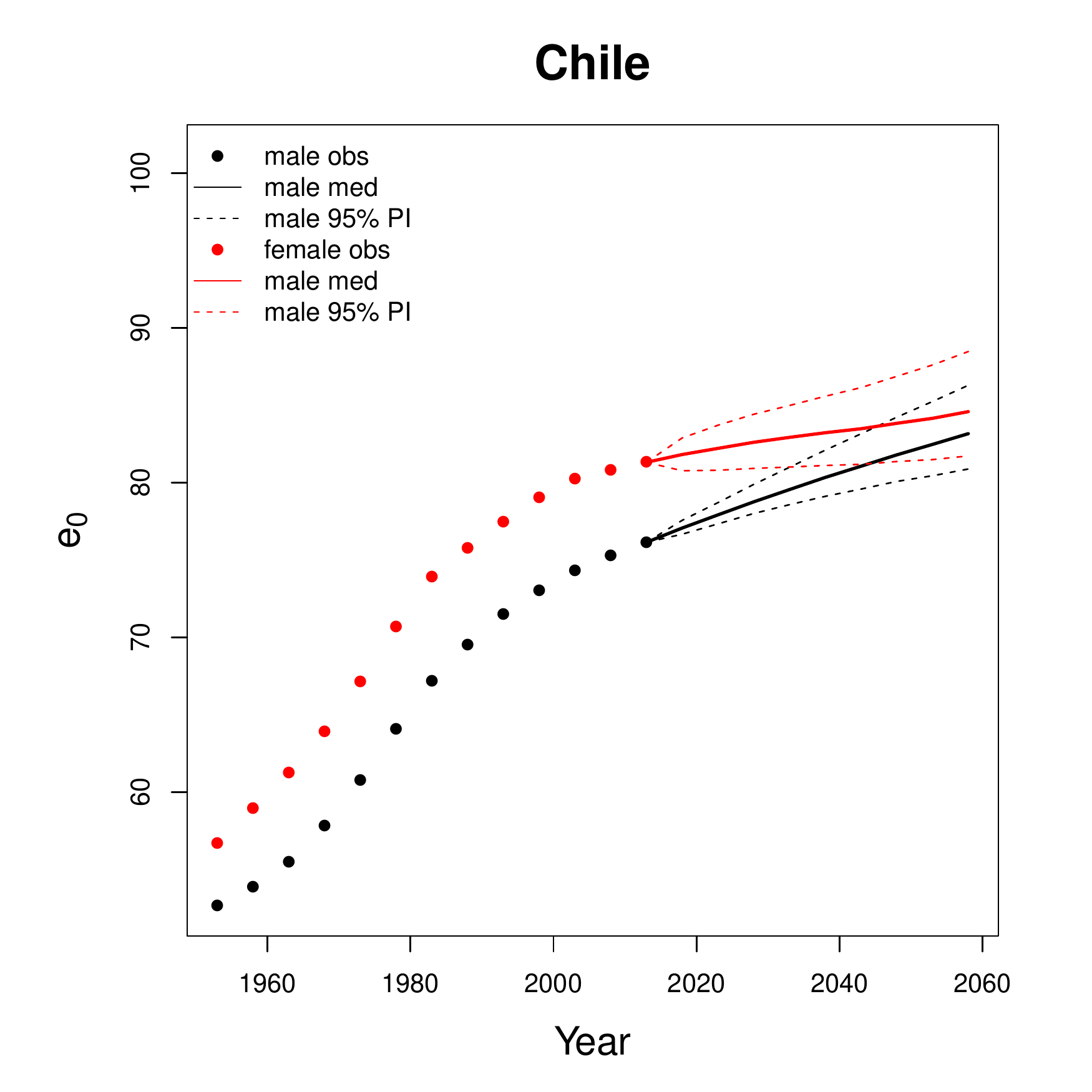}		
			
		\end{center}
	\end{figure}

	\begin{figure}[H]
		\begin{center}
			\includegraphics[scale=0.43]{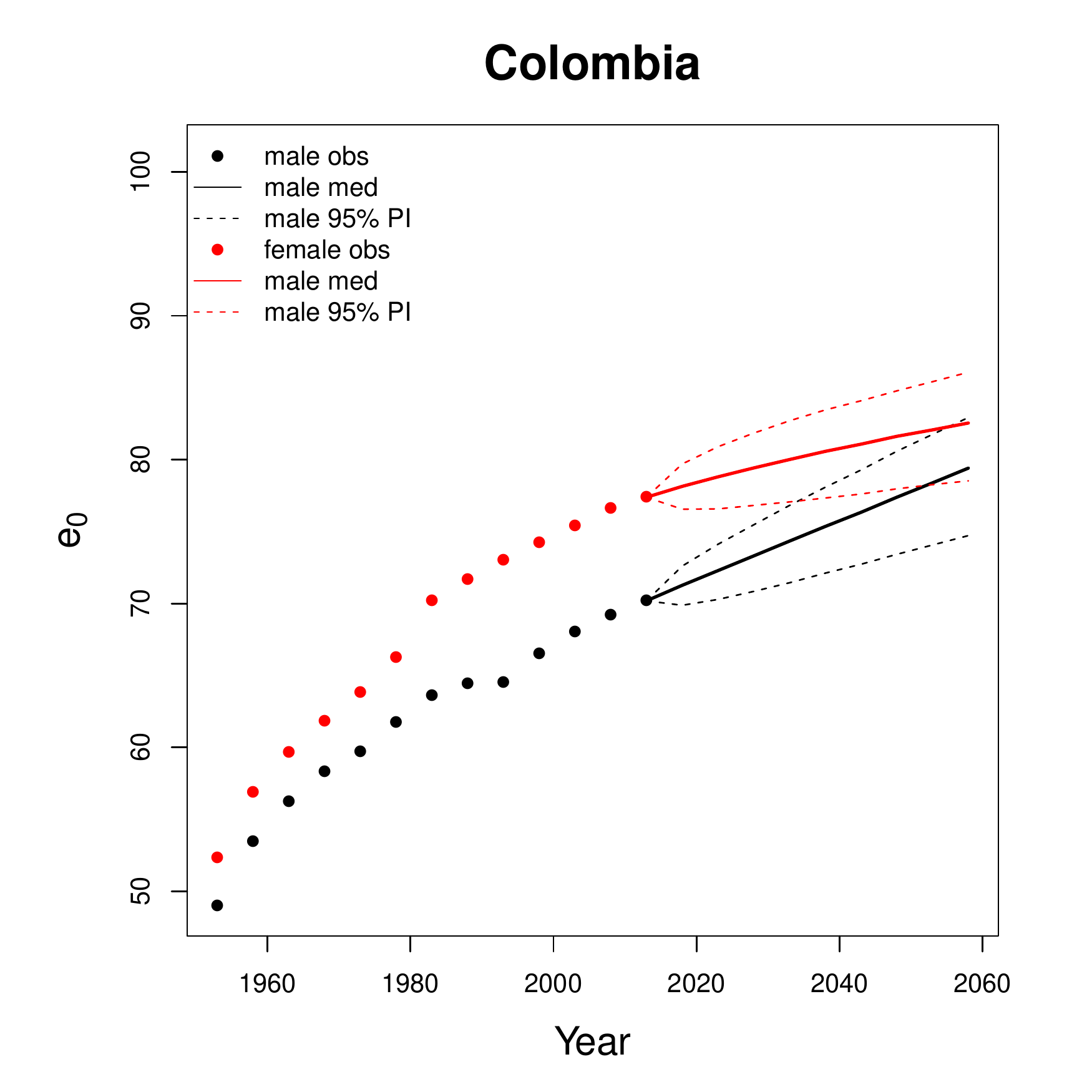}
			\includegraphics[scale=0.43]{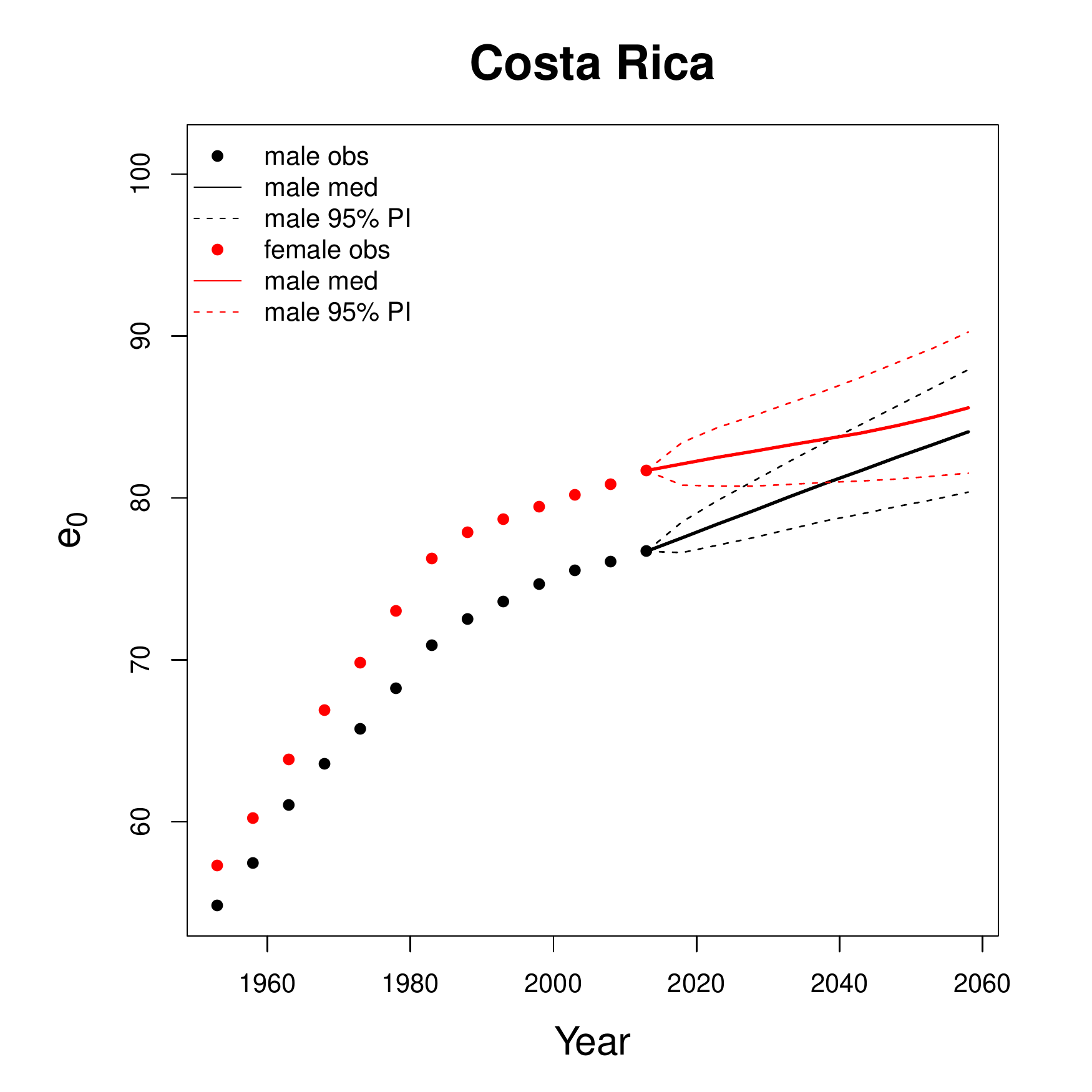}
			\includegraphics[scale=0.43]{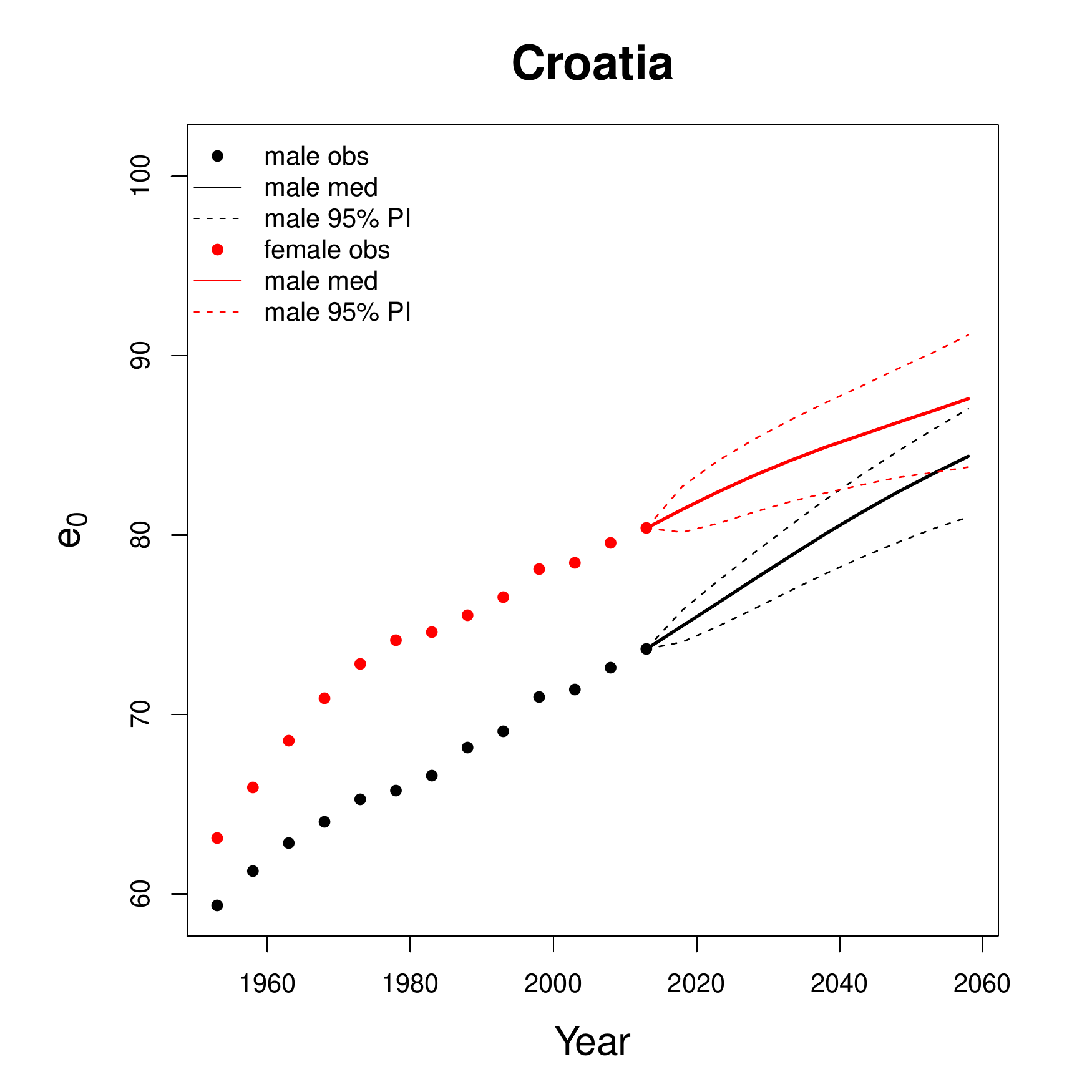}	
			\includegraphics[scale=0.43]{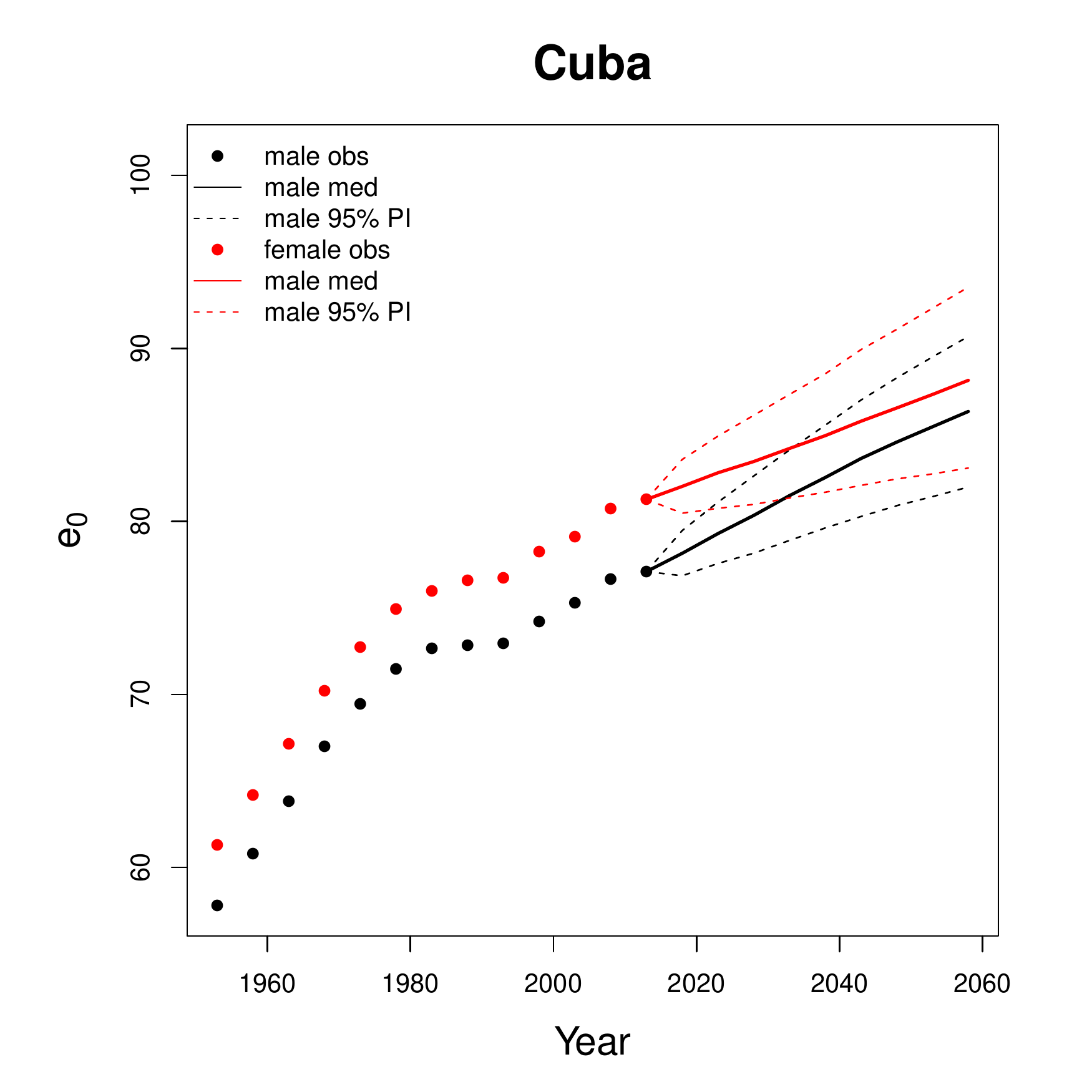}
			\includegraphics[scale=0.43]{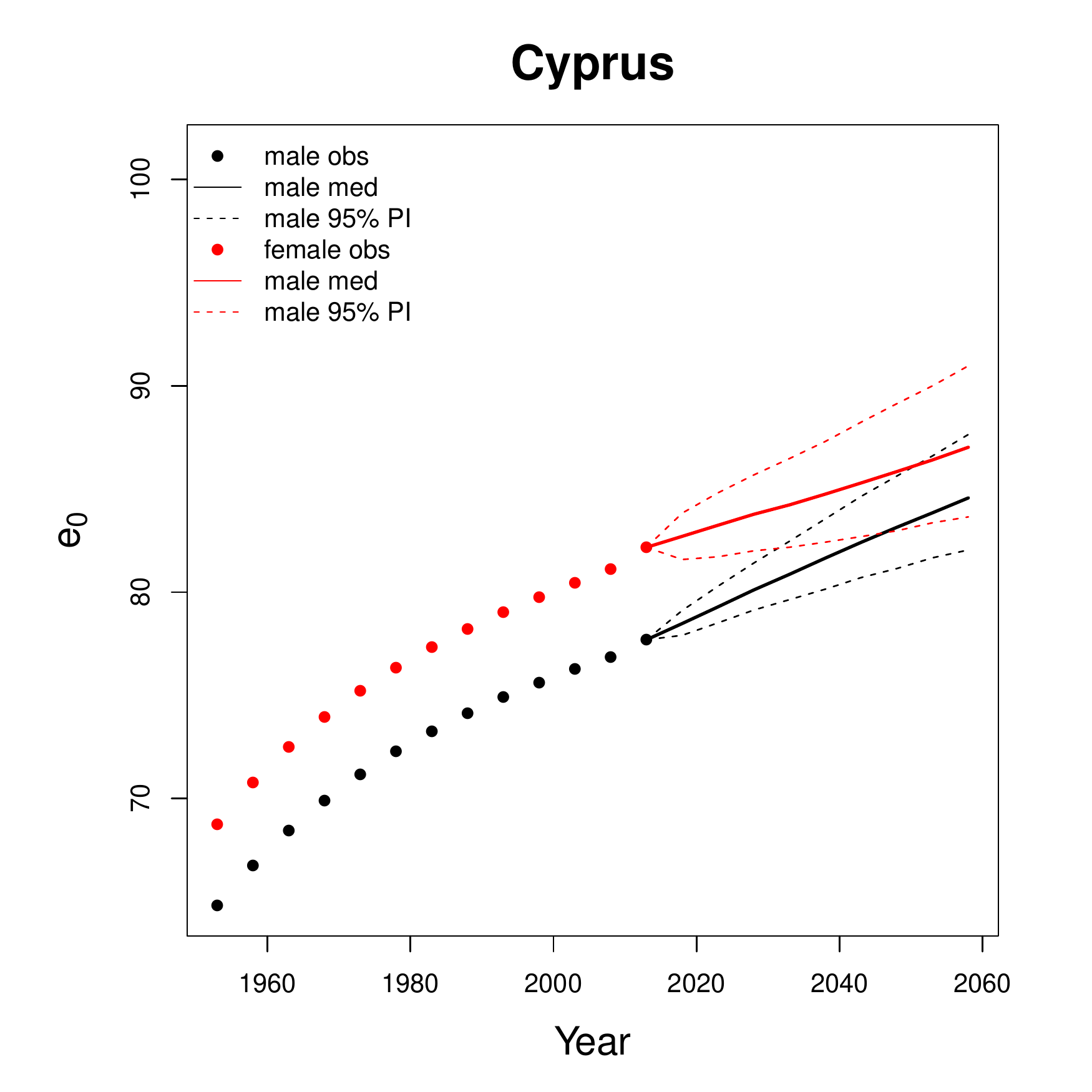}			\includegraphics[scale=0.43]{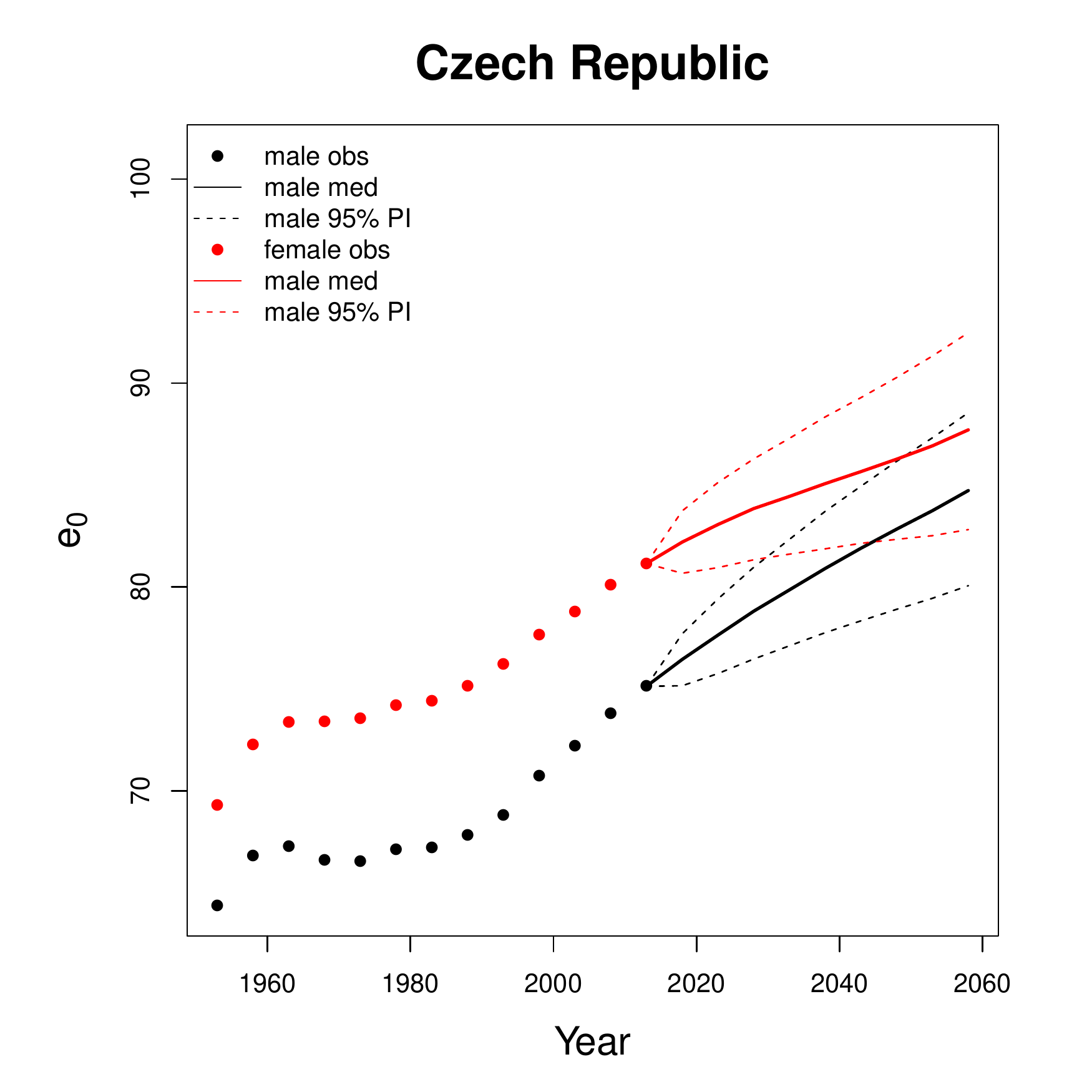}
		\end{center}
	\end{figure}

	\begin{figure}[H]
		\begin{center}
			\includegraphics[scale=0.43]{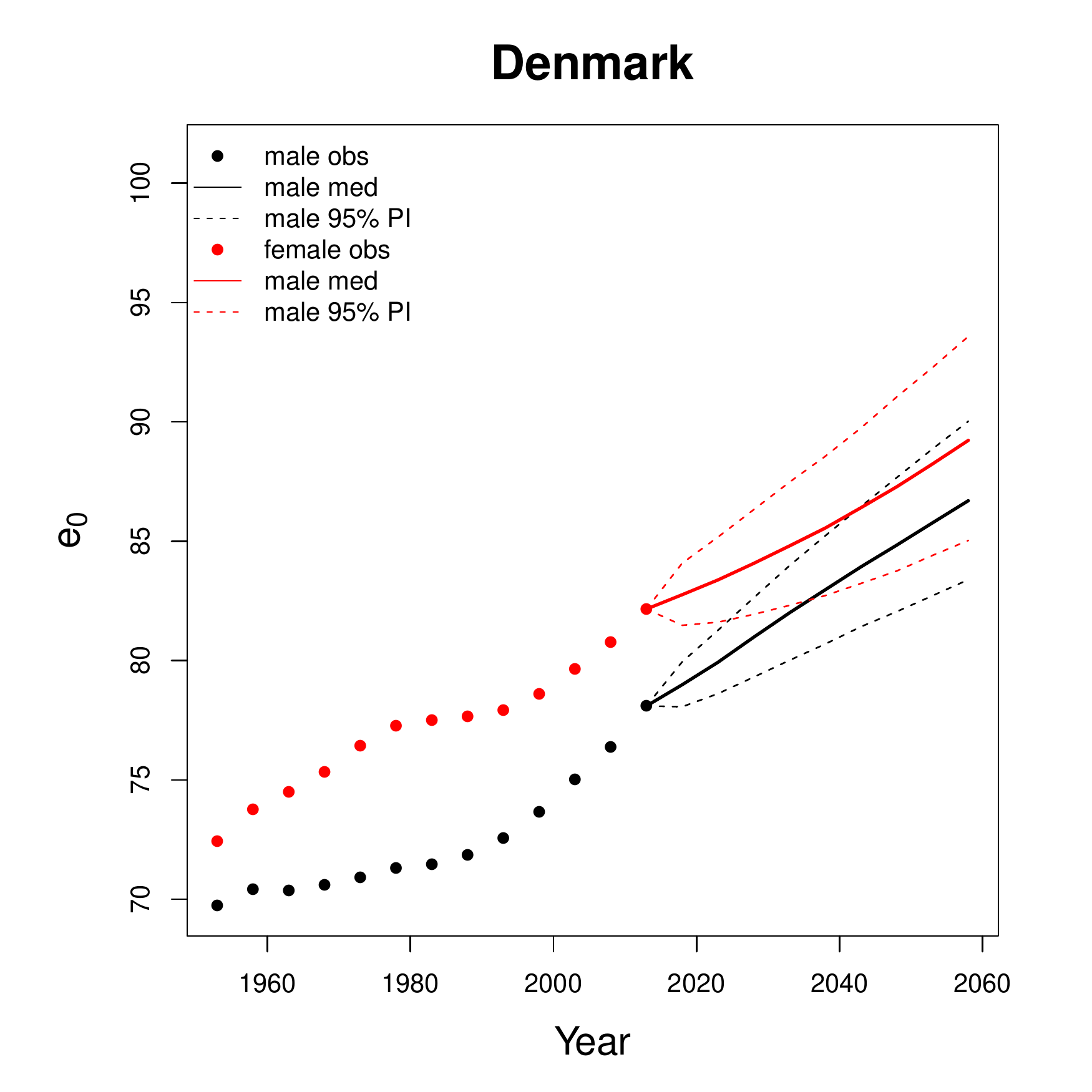}	
			\includegraphics[scale=0.43]{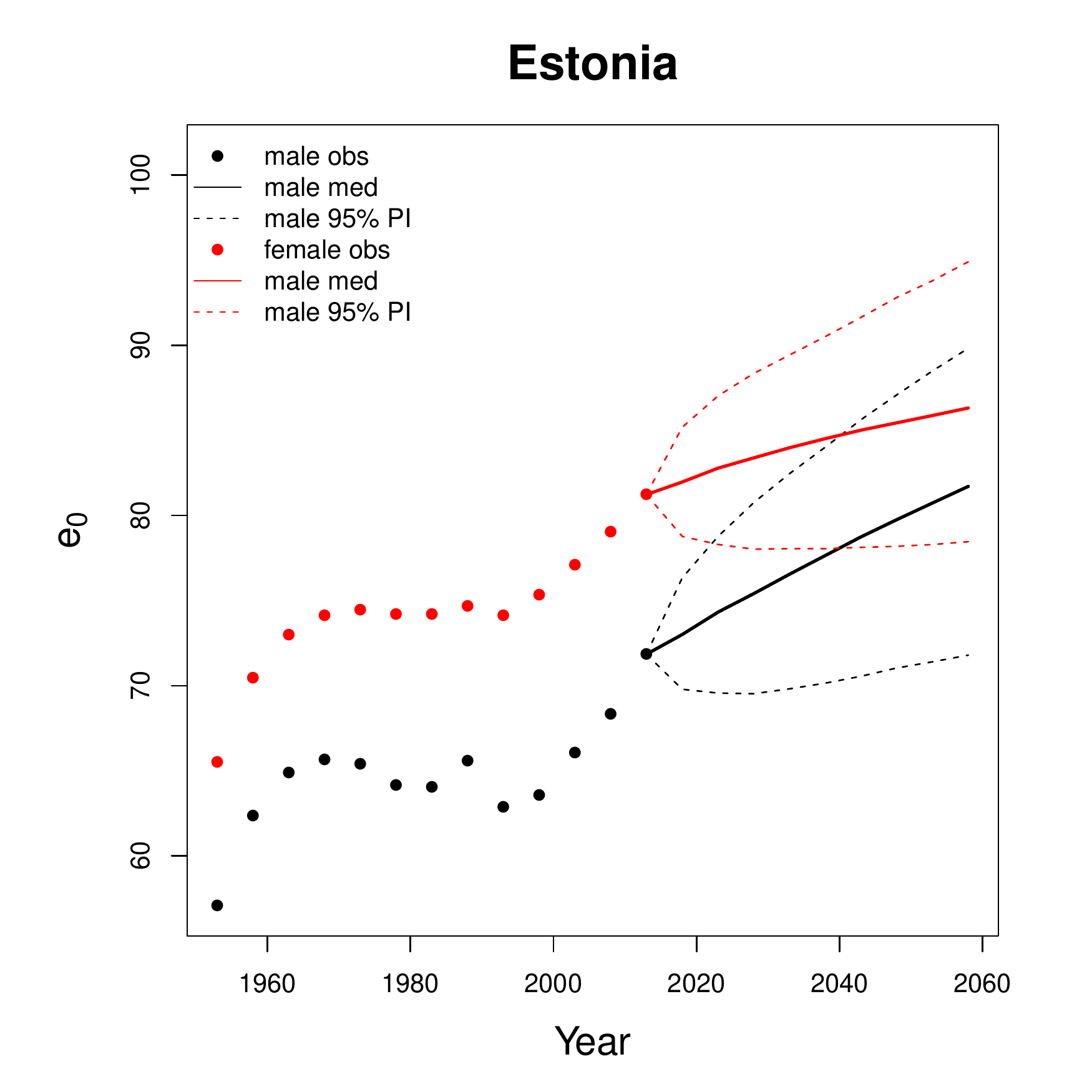}
			\includegraphics[scale=0.43]{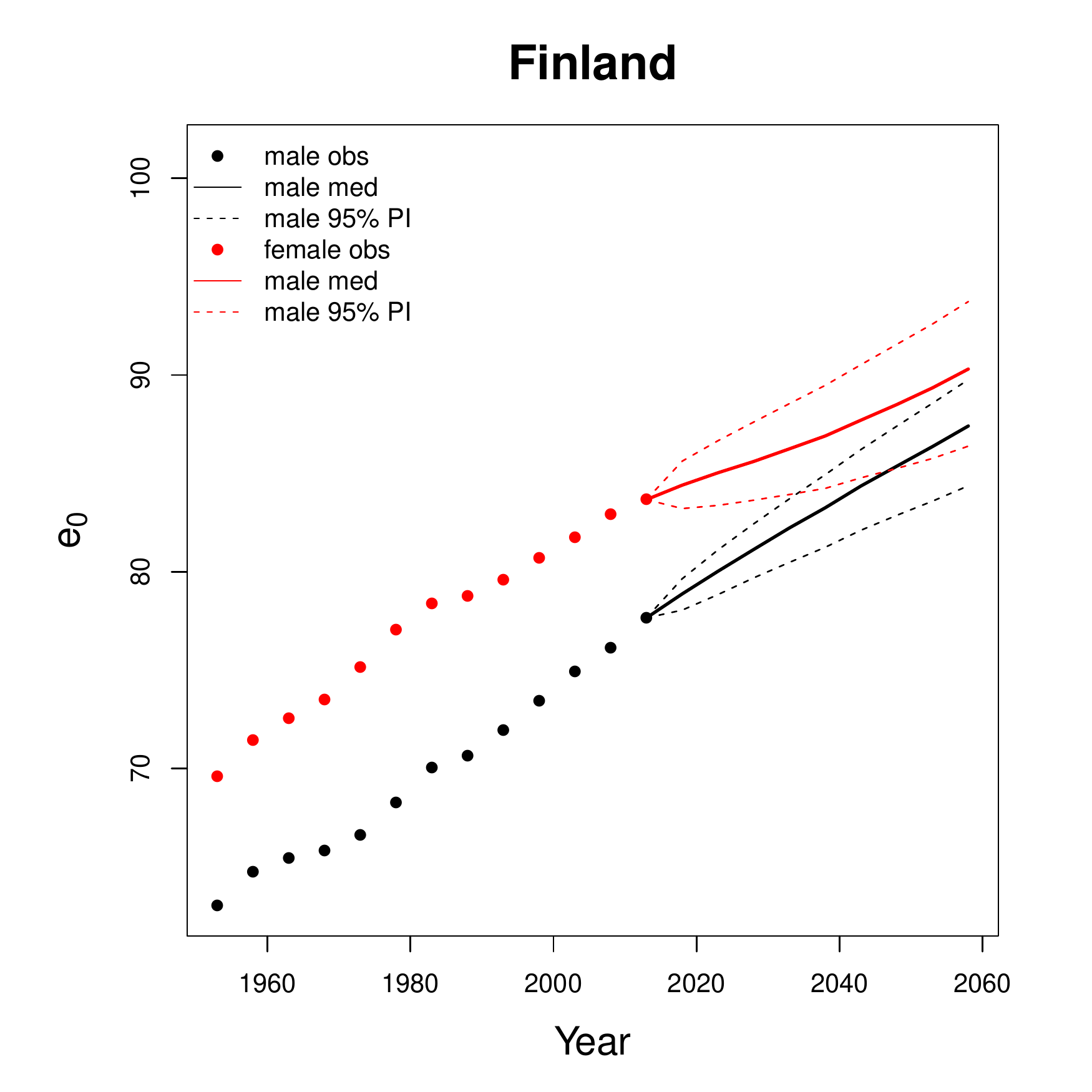}
			\includegraphics[scale=0.43]{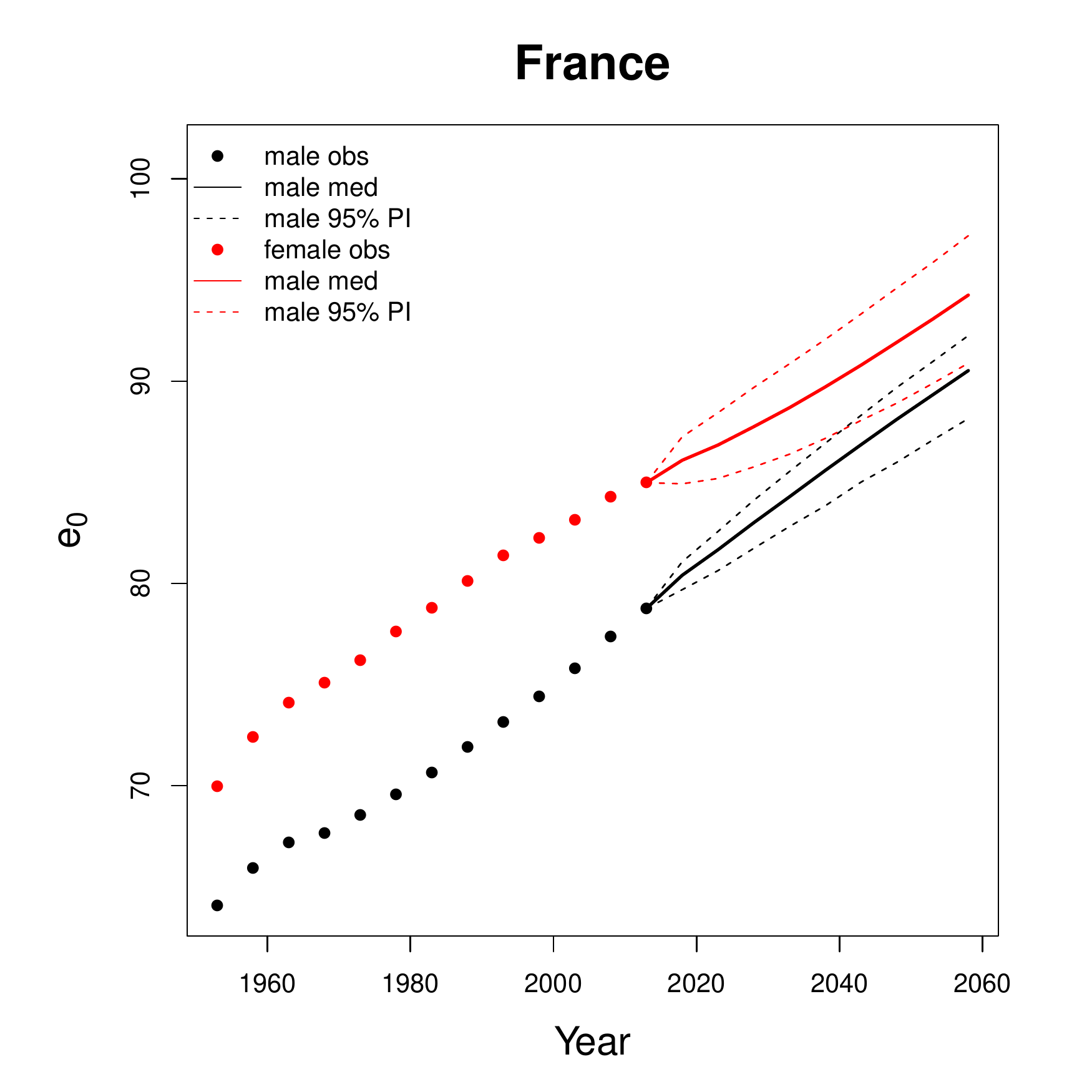}
			\includegraphics[scale=0.43]{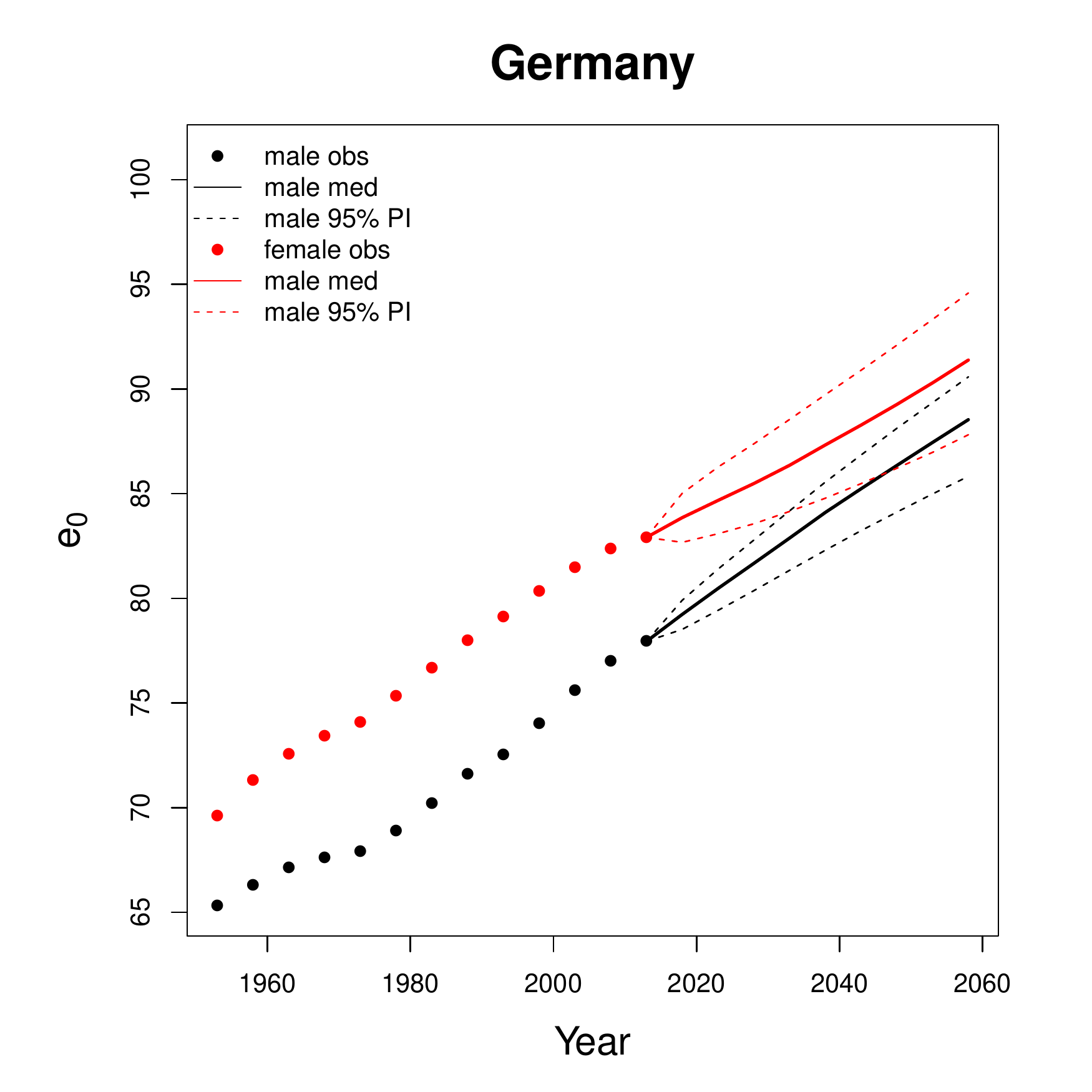}
			\includegraphics[scale=0.43]{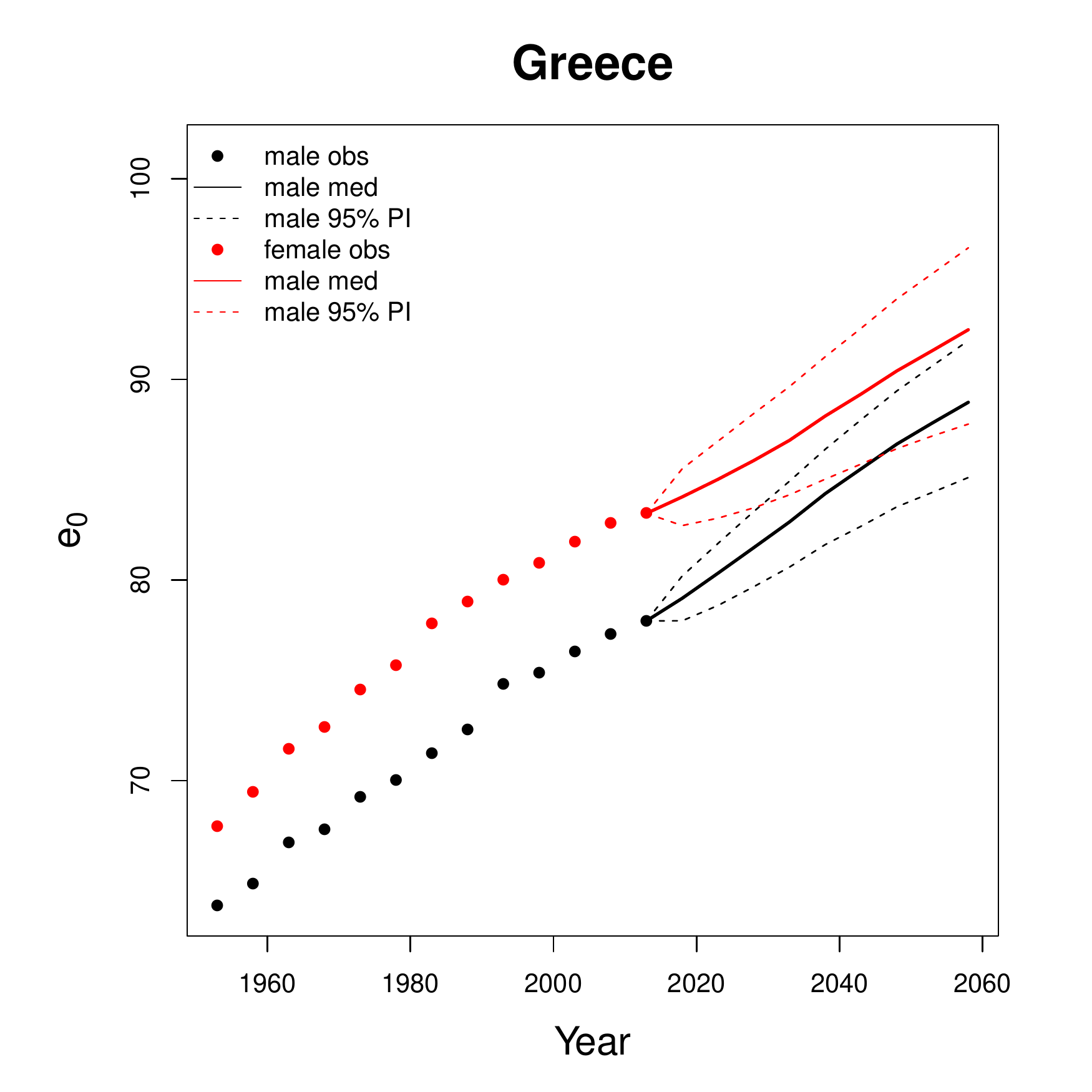}			
		\end{center}
	\end{figure}

	\begin{figure}[H]
		\begin{center}
			\includegraphics[scale=0.43]{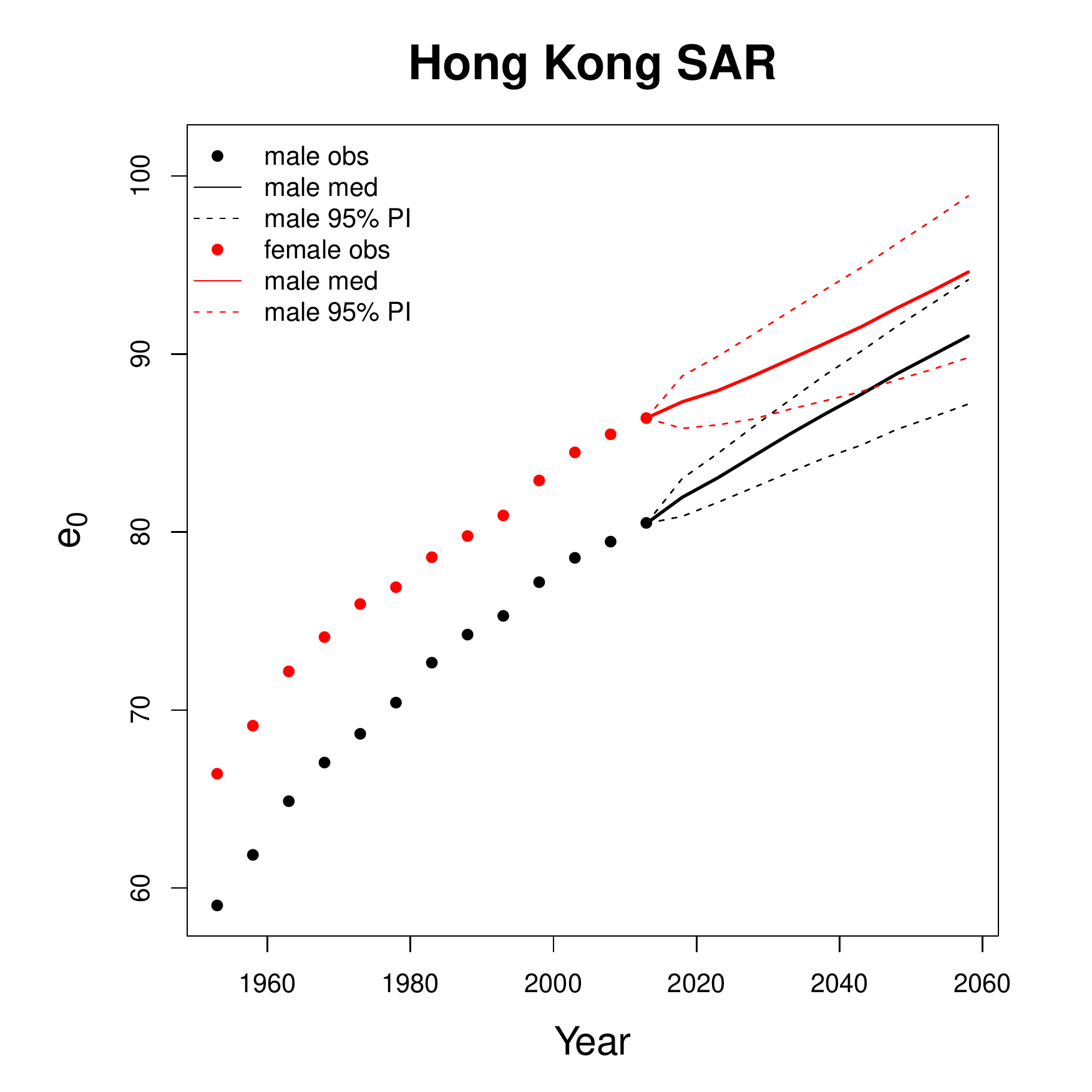}
			\includegraphics[scale=0.43]{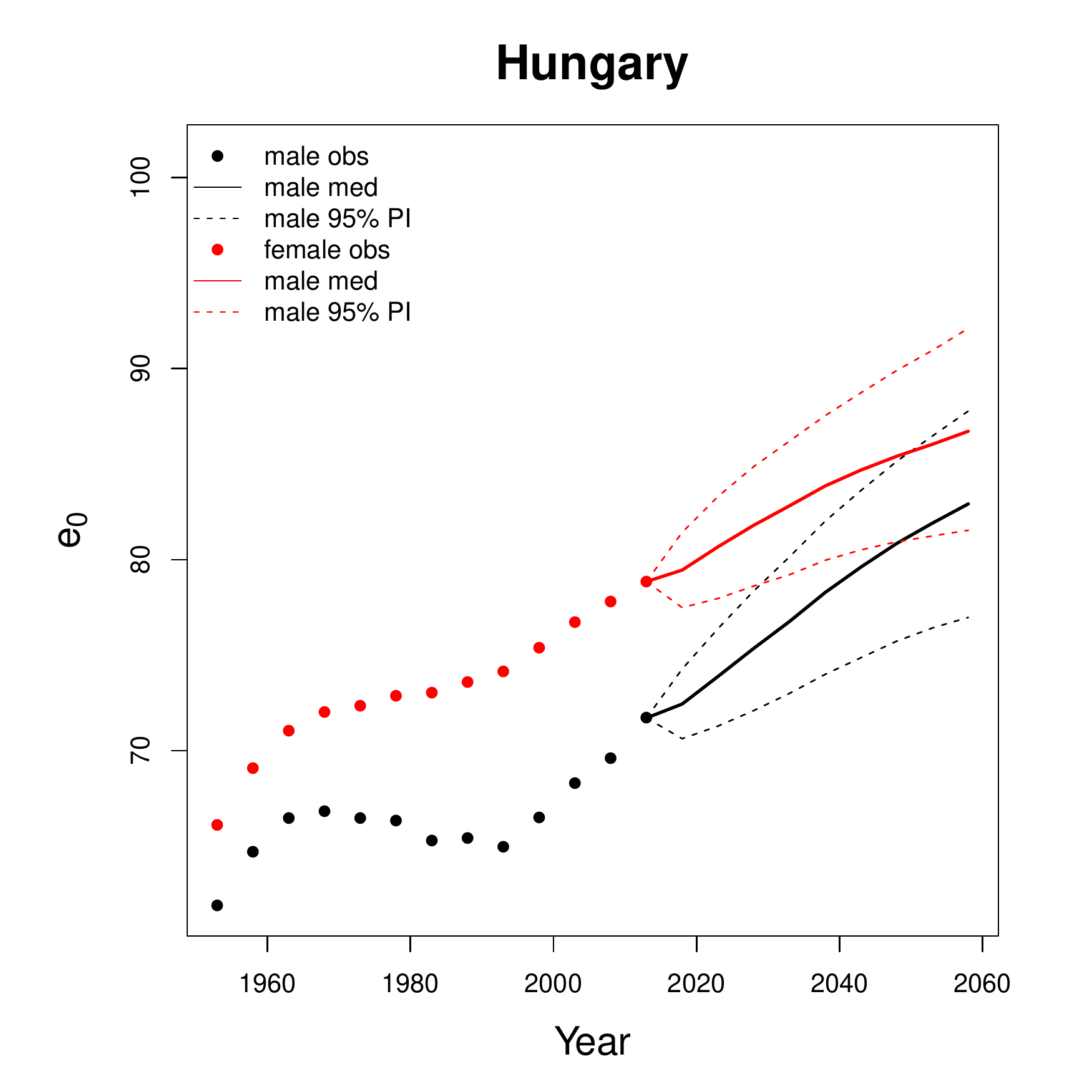}
			\includegraphics[scale=0.43]{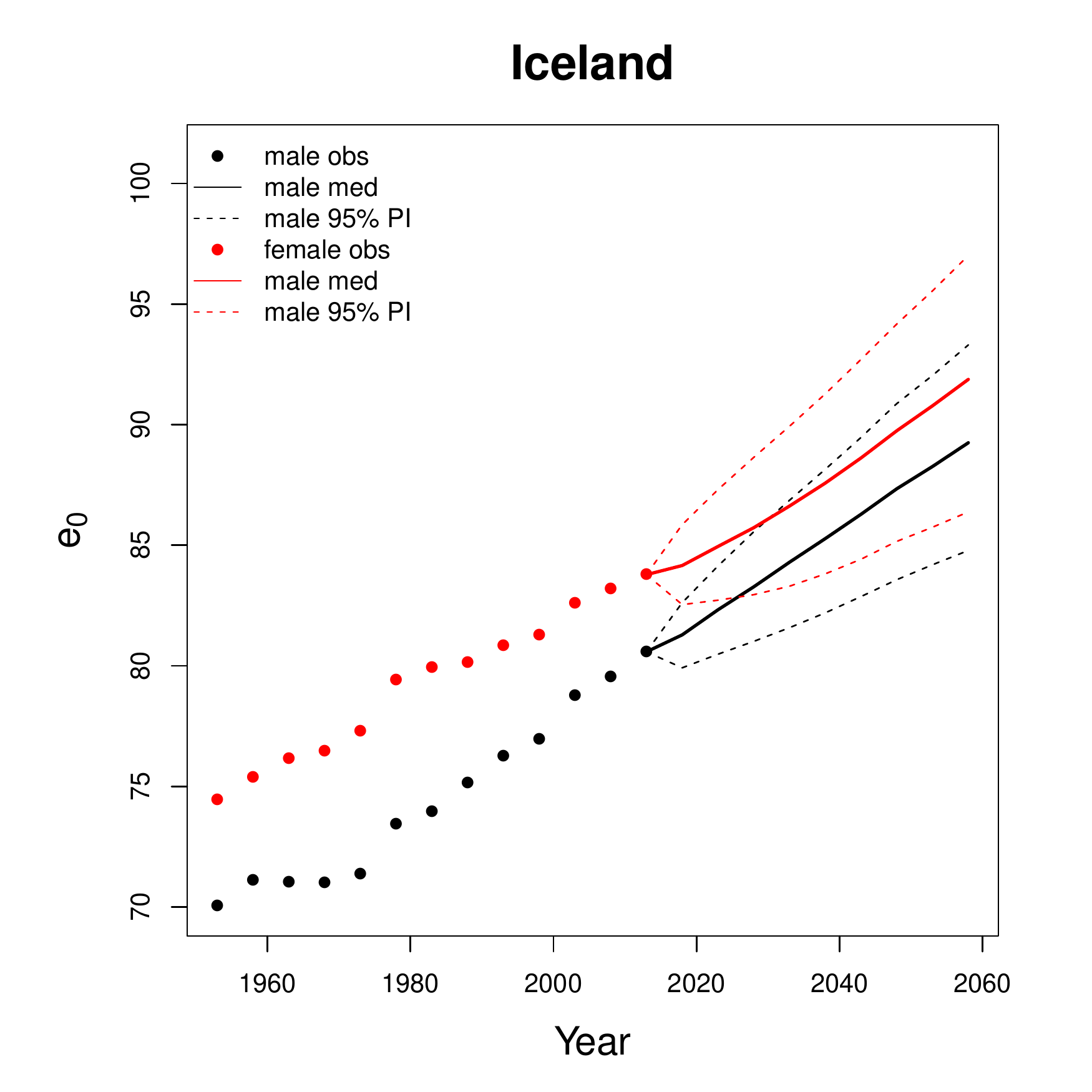}	\includegraphics[scale=0.43]{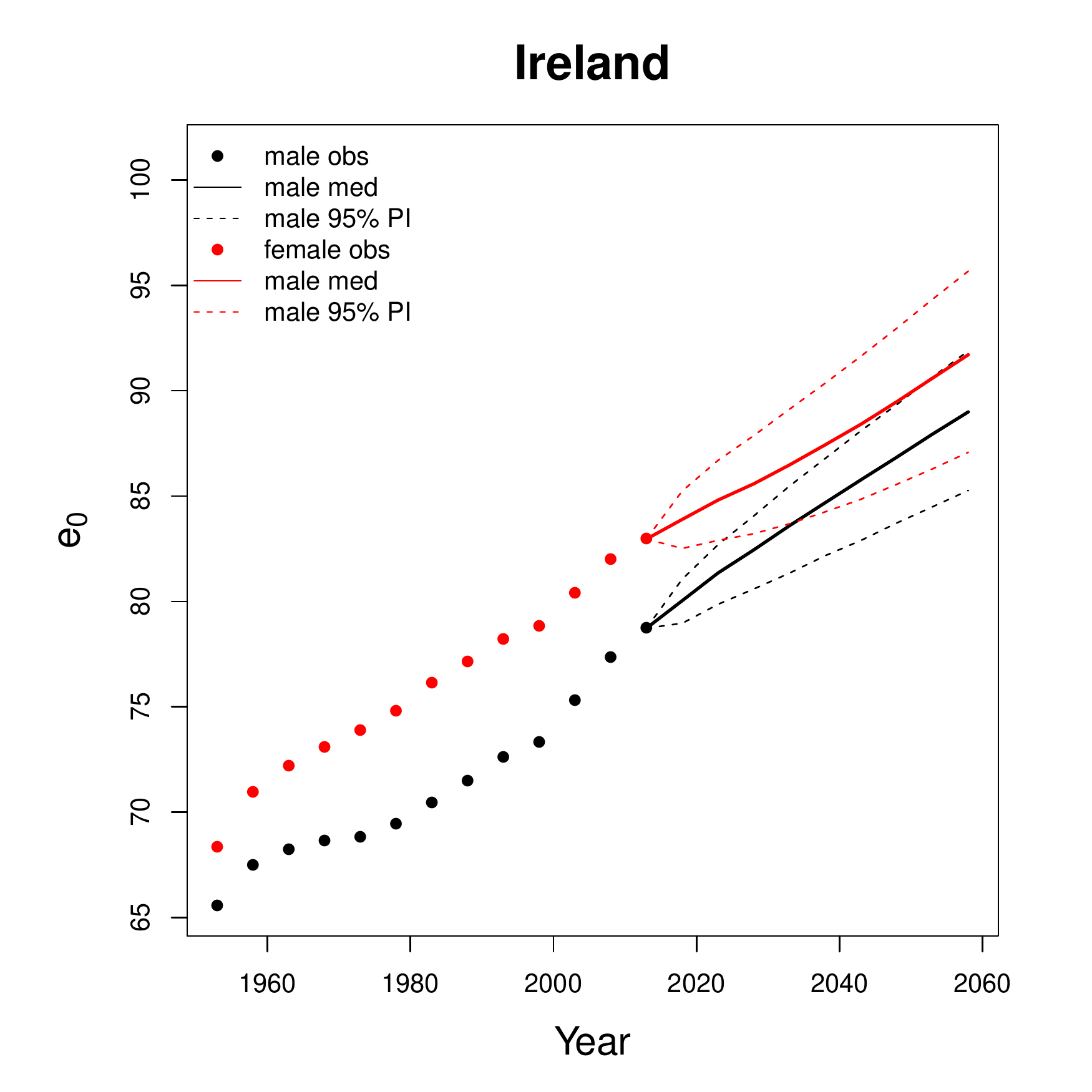}
			\includegraphics[scale=0.43]{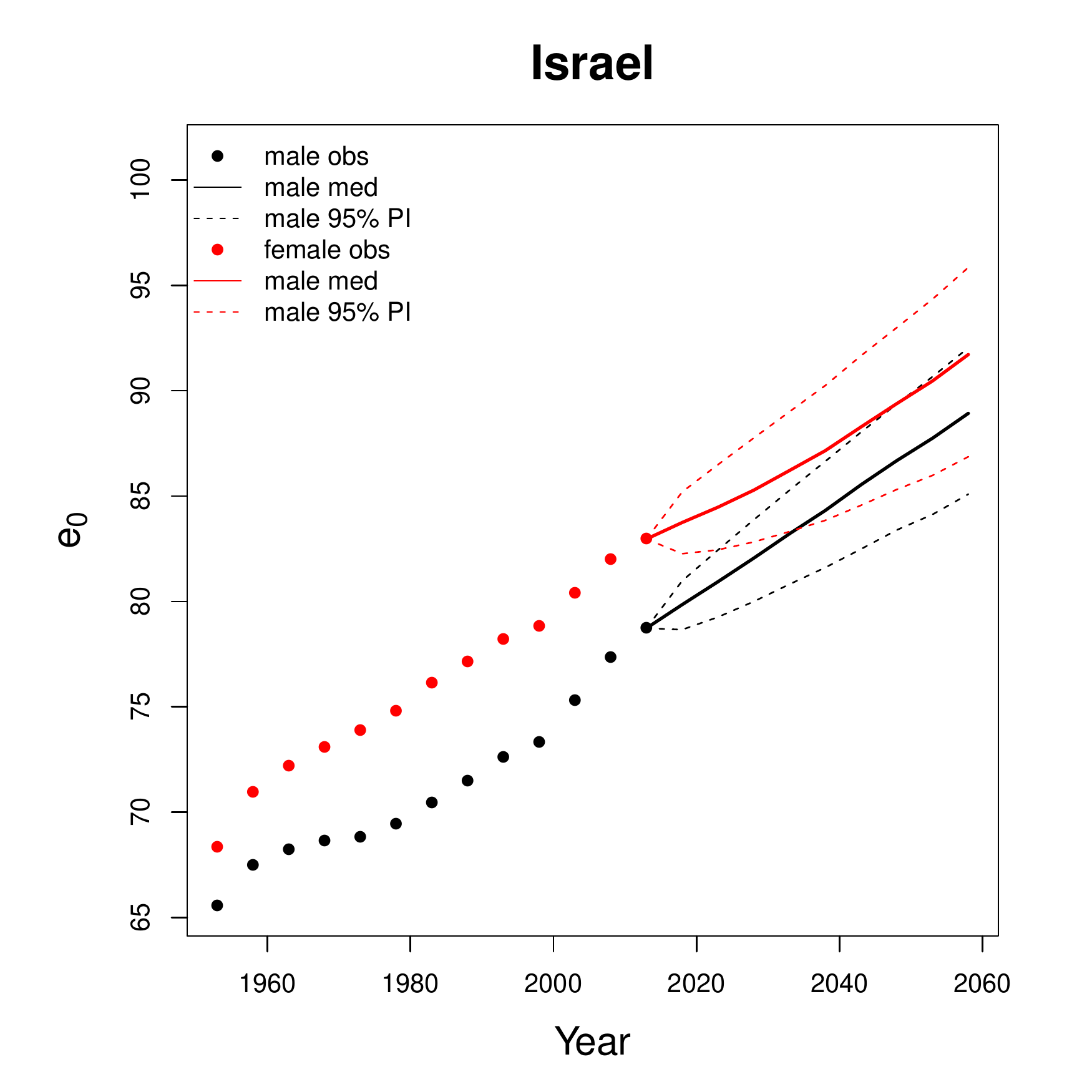}
			\includegraphics[scale=0.43]{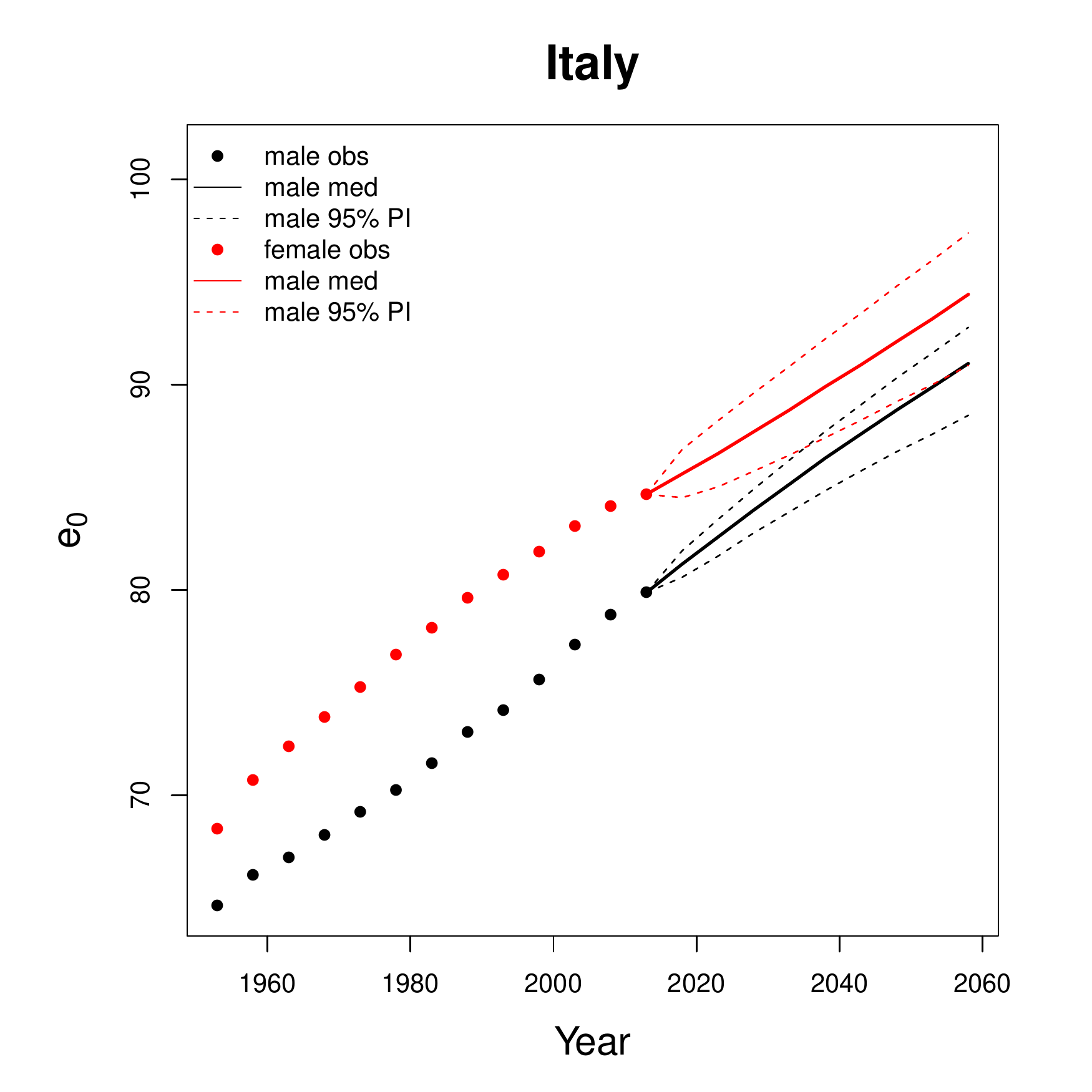}			
		\end{center}
	\end{figure}

	\begin{figure}[H]
		\begin{center}
			\includegraphics[scale=0.43]{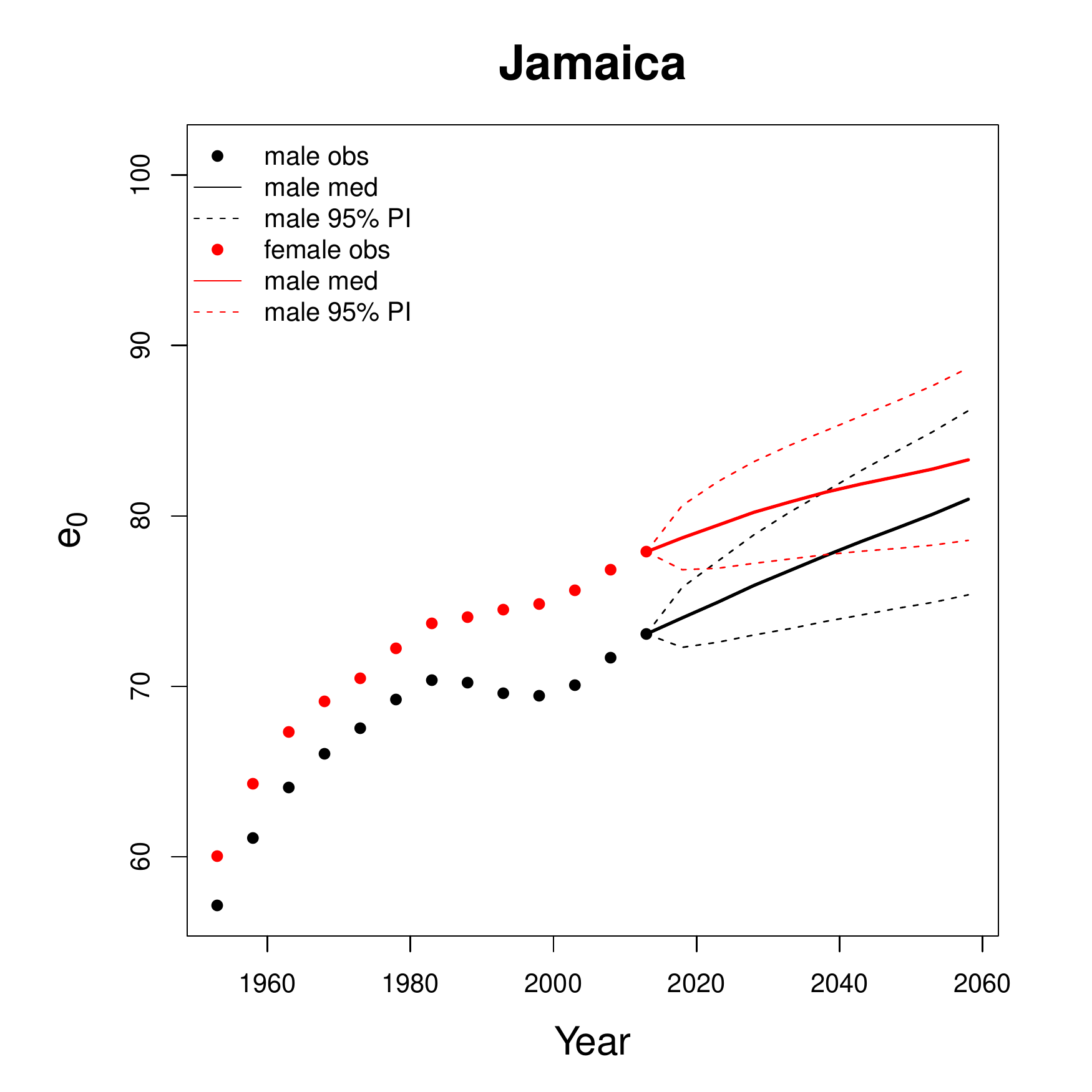}
			\includegraphics[scale=0.43]{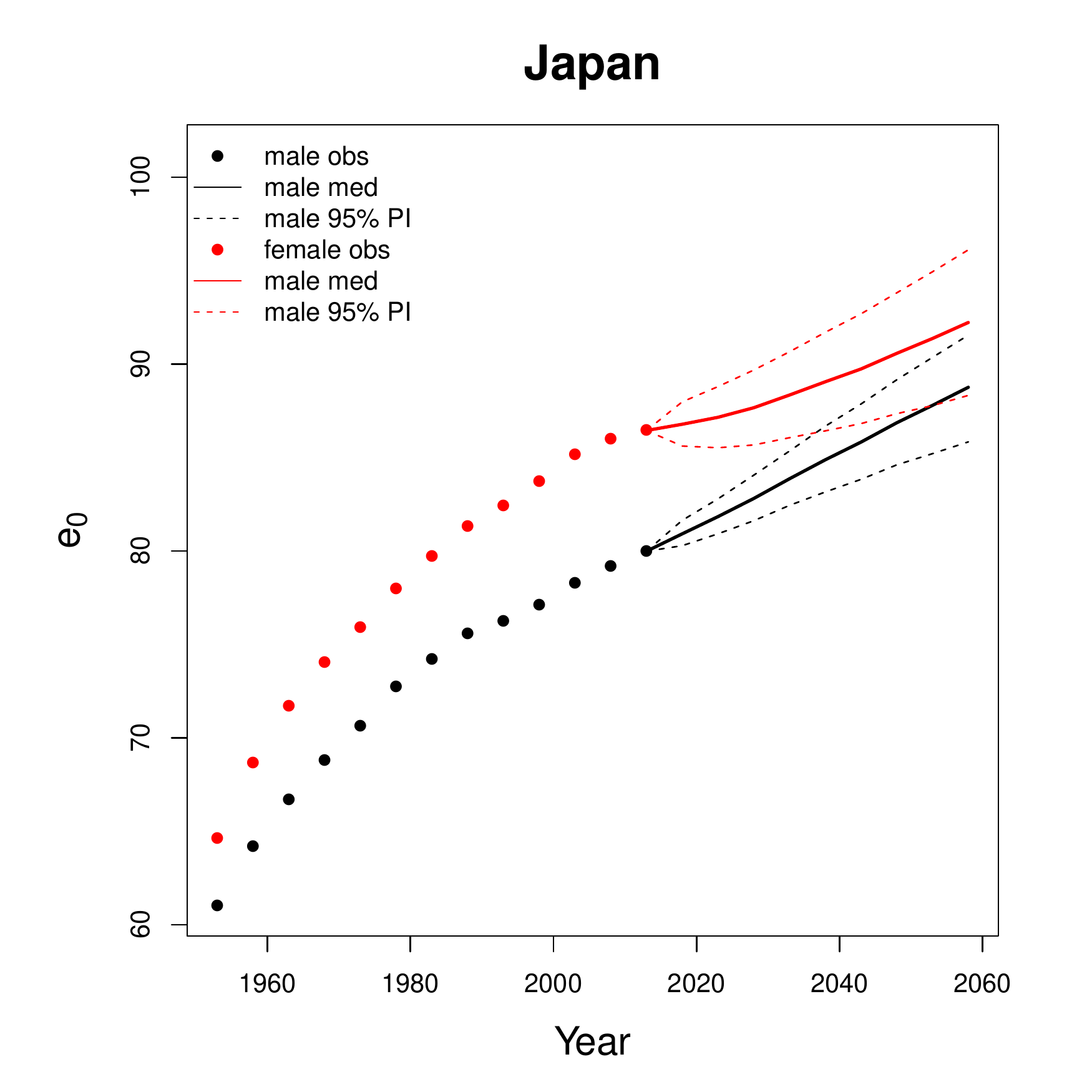}
			\includegraphics[scale=0.43]{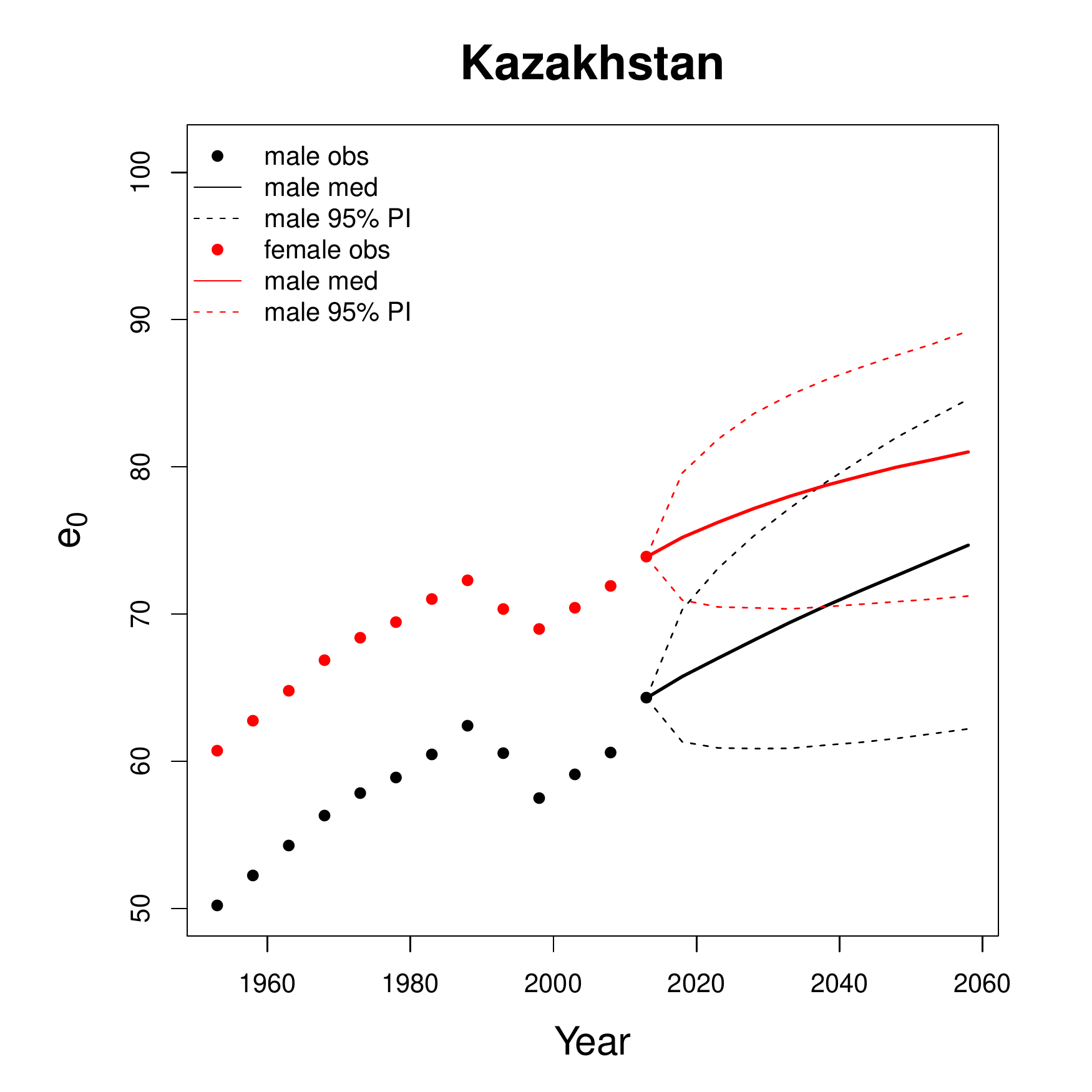}	\includegraphics[scale=0.43]{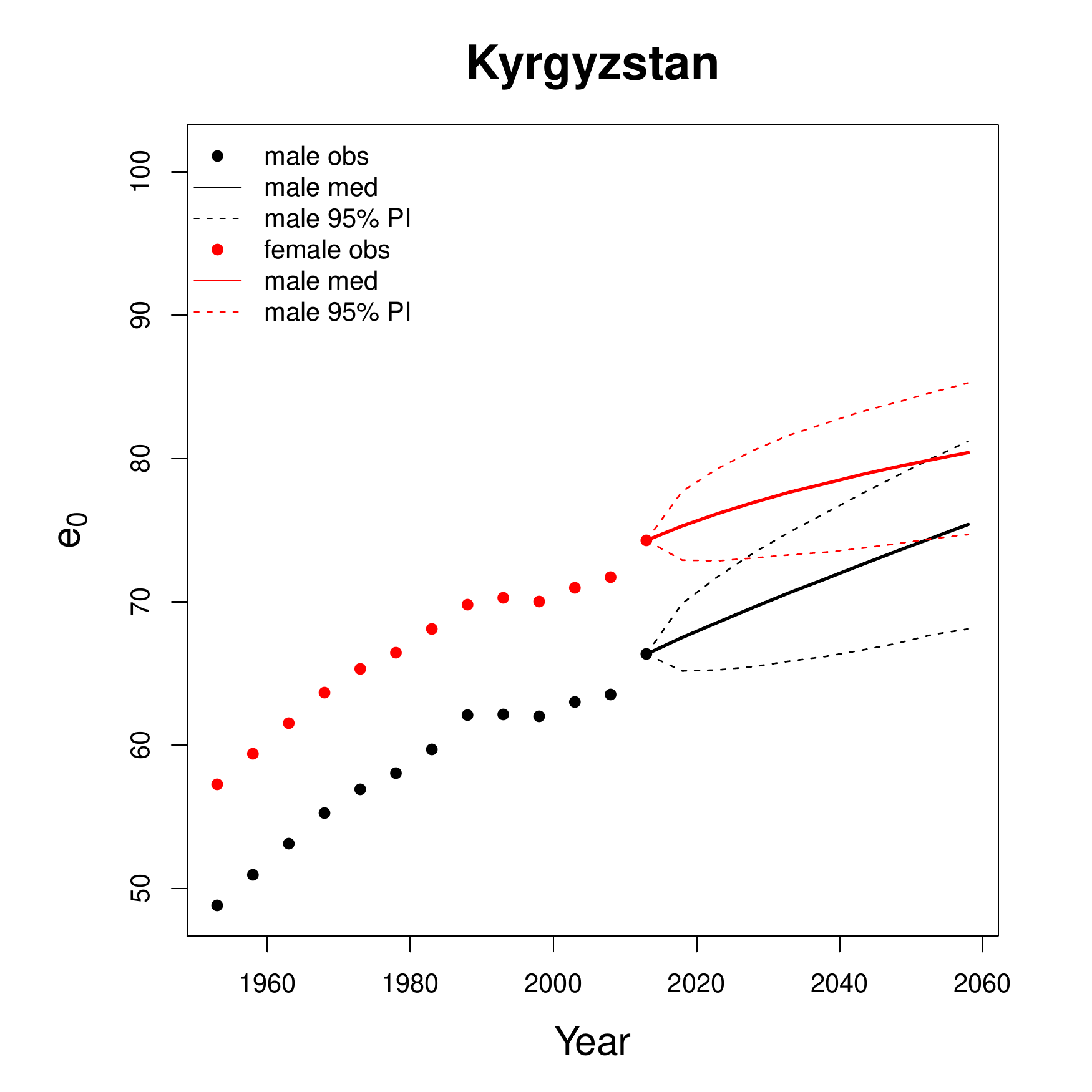}
			\includegraphics[scale=0.43]{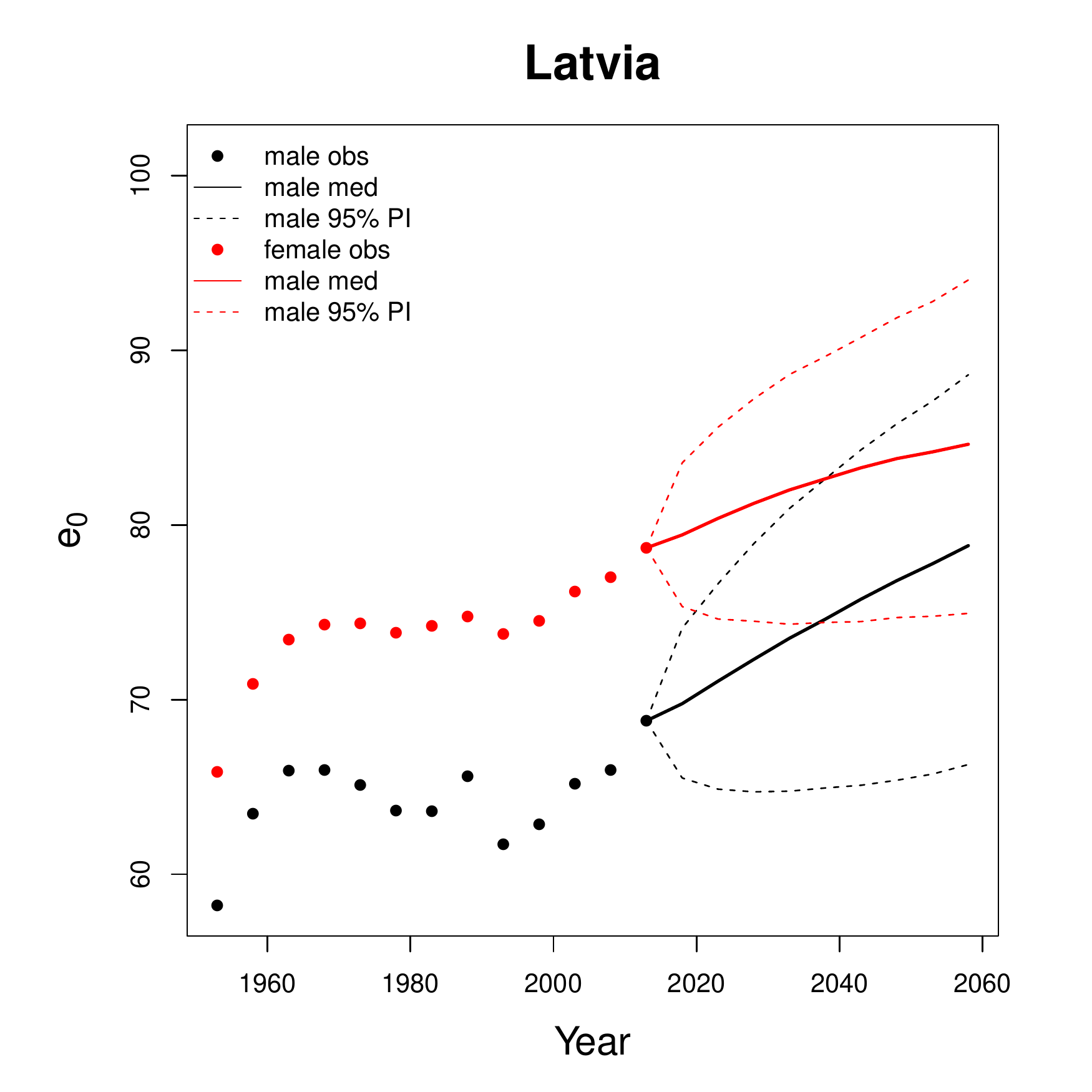}
			\includegraphics[scale=0.43]{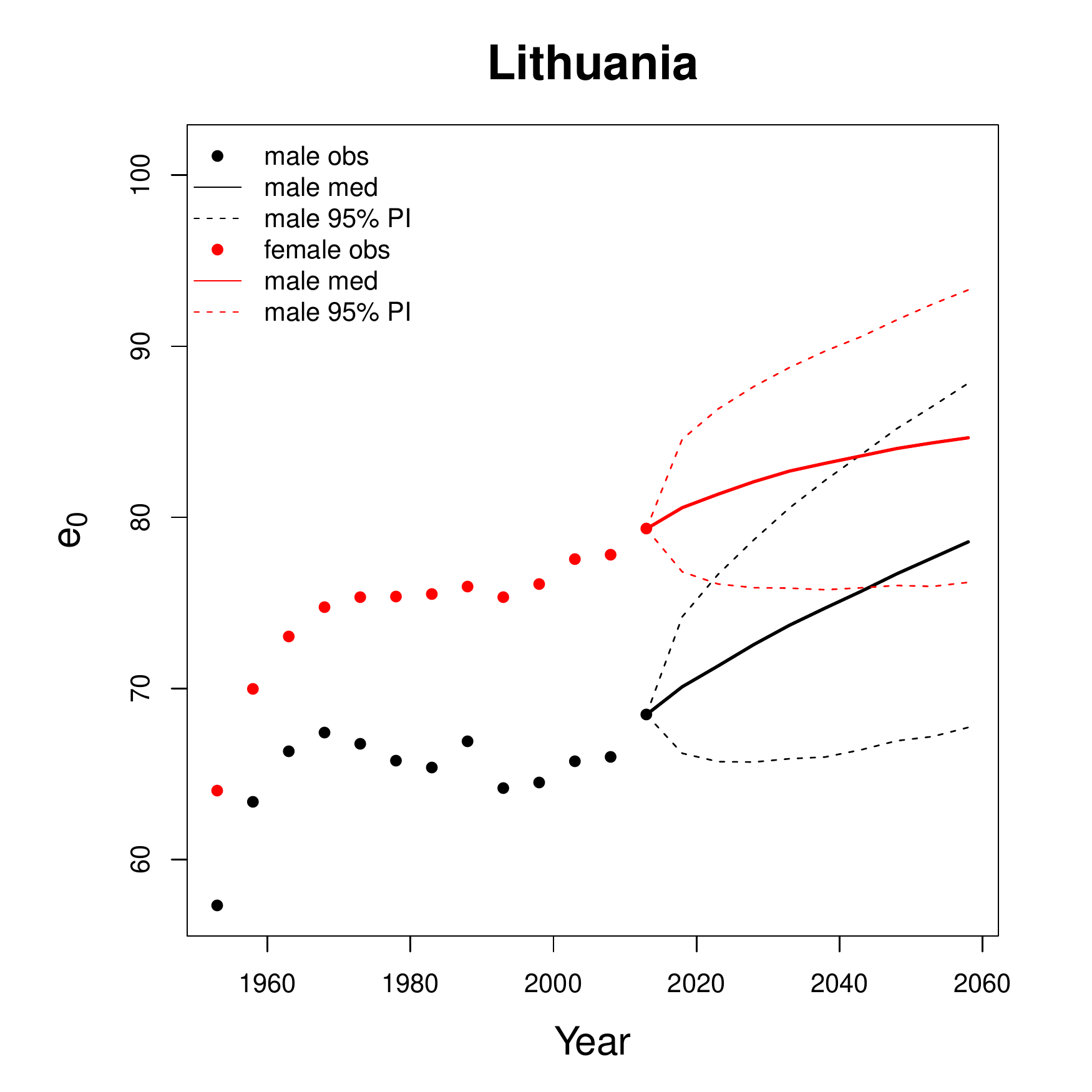}			
		\end{center}
	\end{figure}

	\begin{figure}[H]
		\begin{center}
			\includegraphics[scale=0.43]{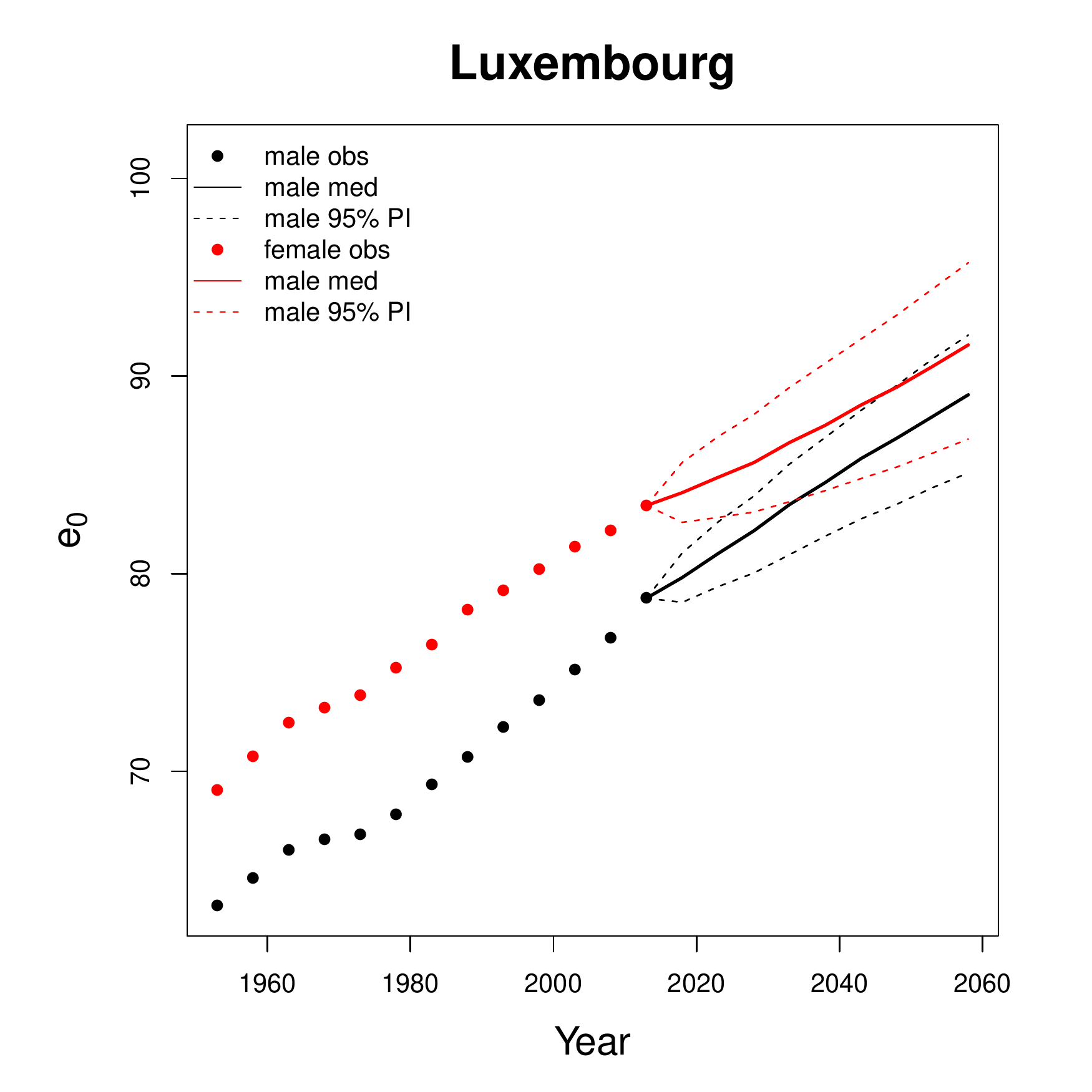}
			\includegraphics[scale=0.43]{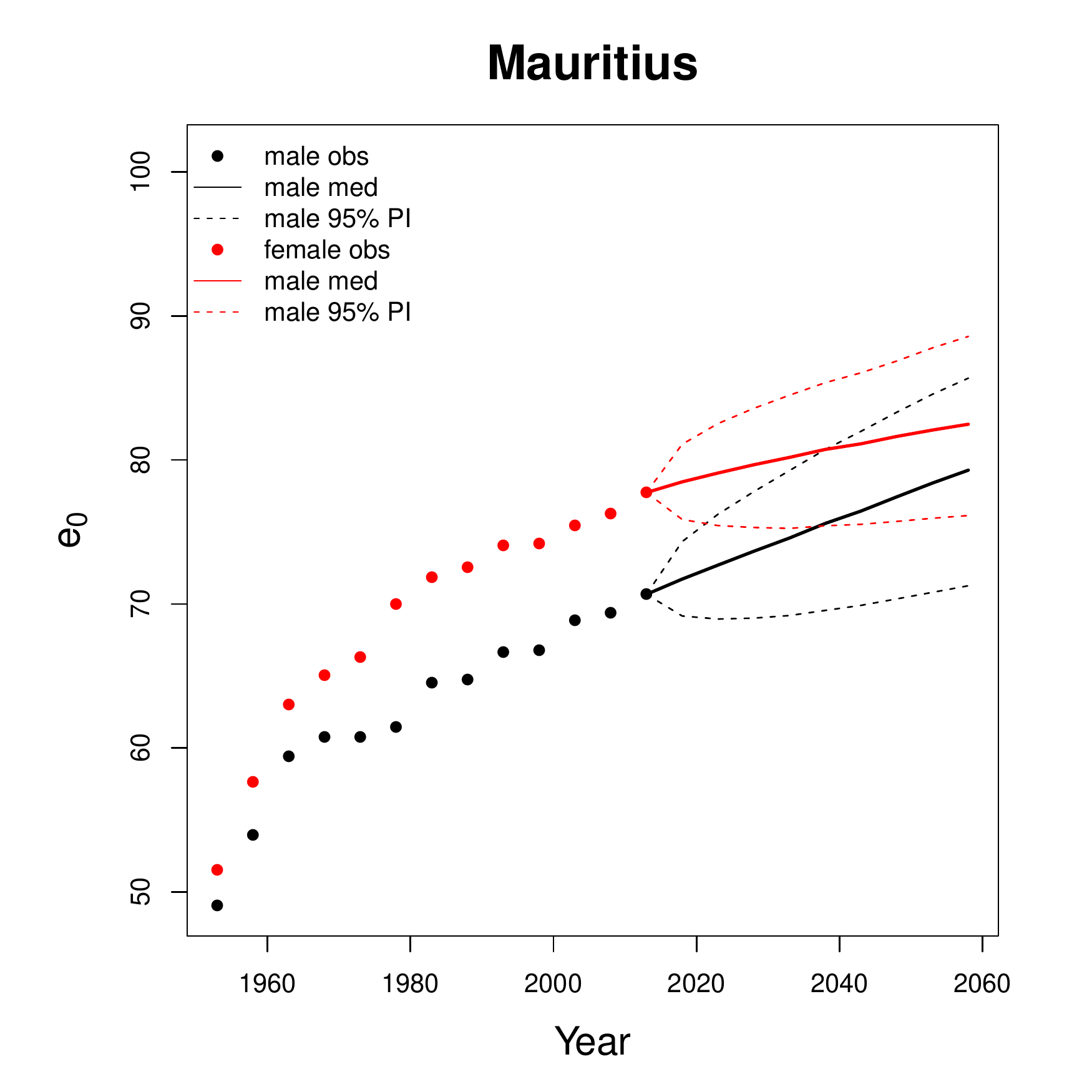}
			\includegraphics[scale=0.43]{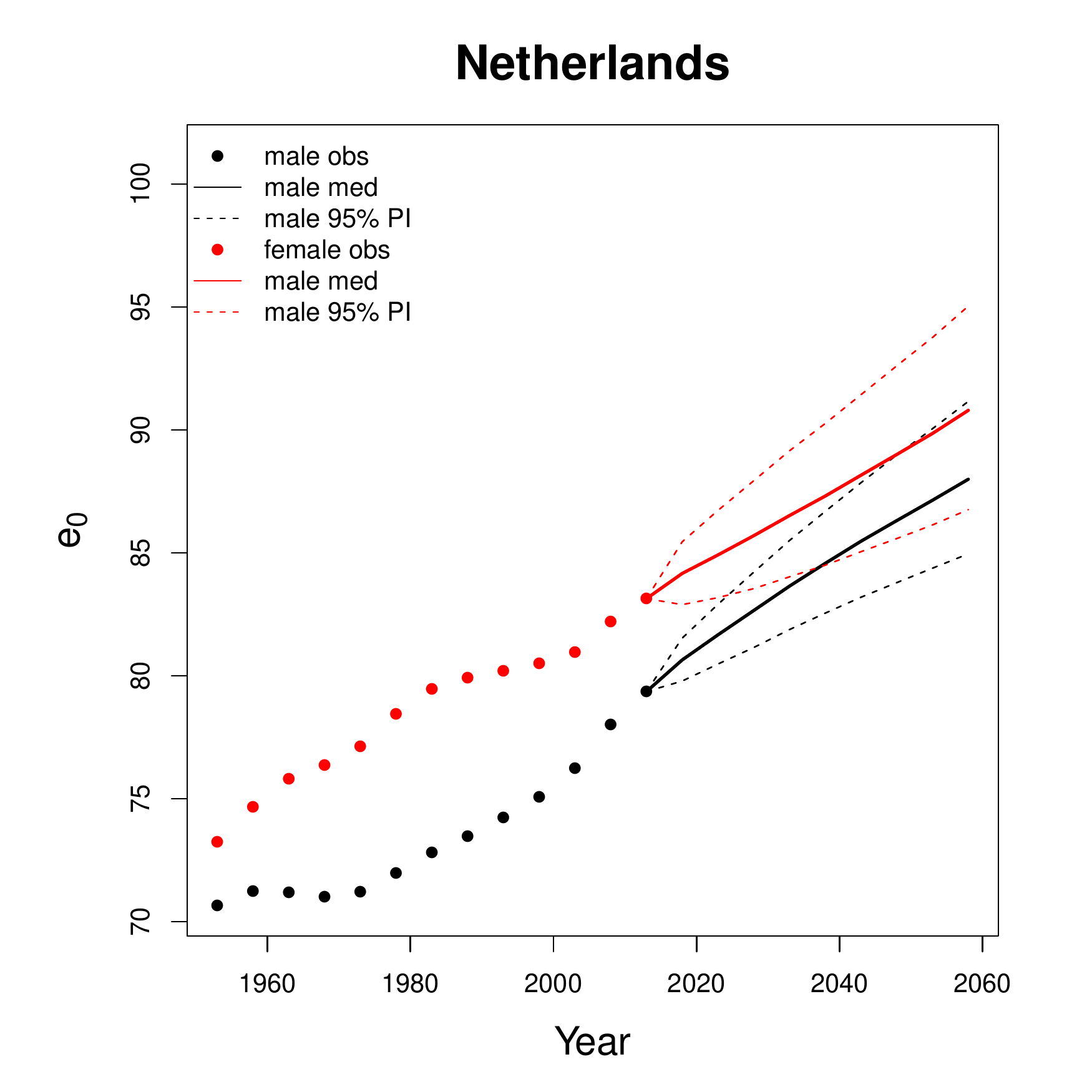}	
			\includegraphics[scale=0.43]{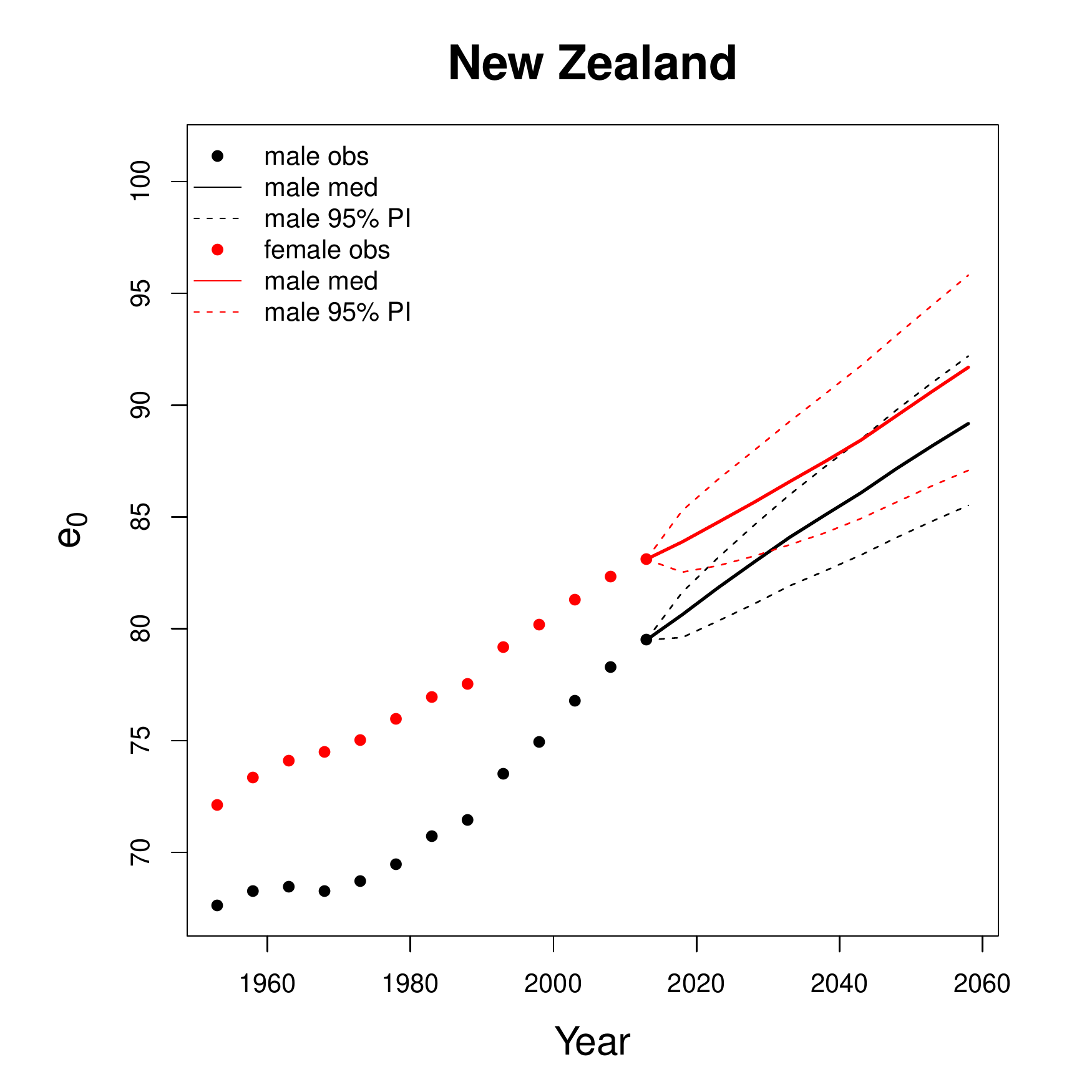}	\includegraphics[scale=0.43]{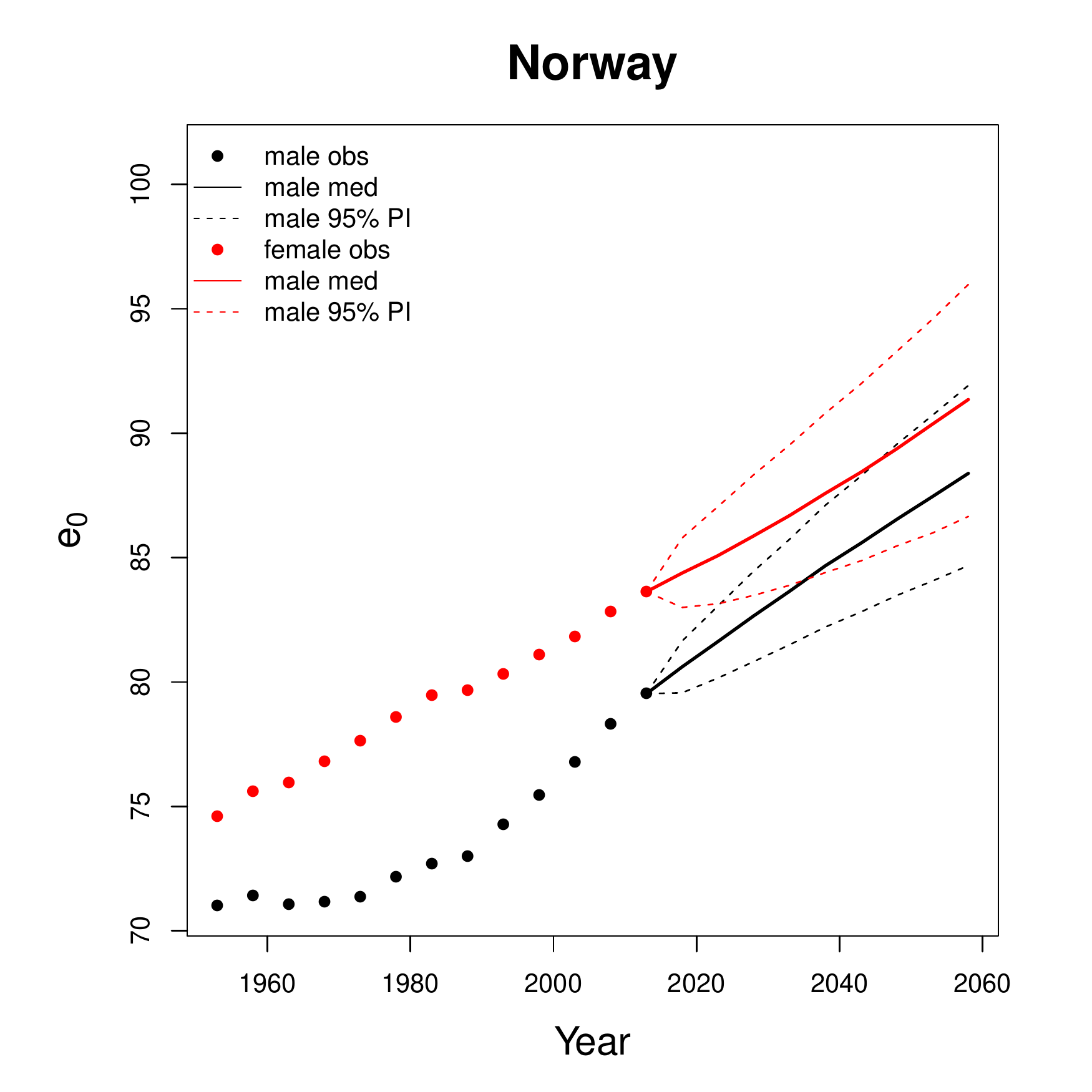}
			\includegraphics[scale=0.43]{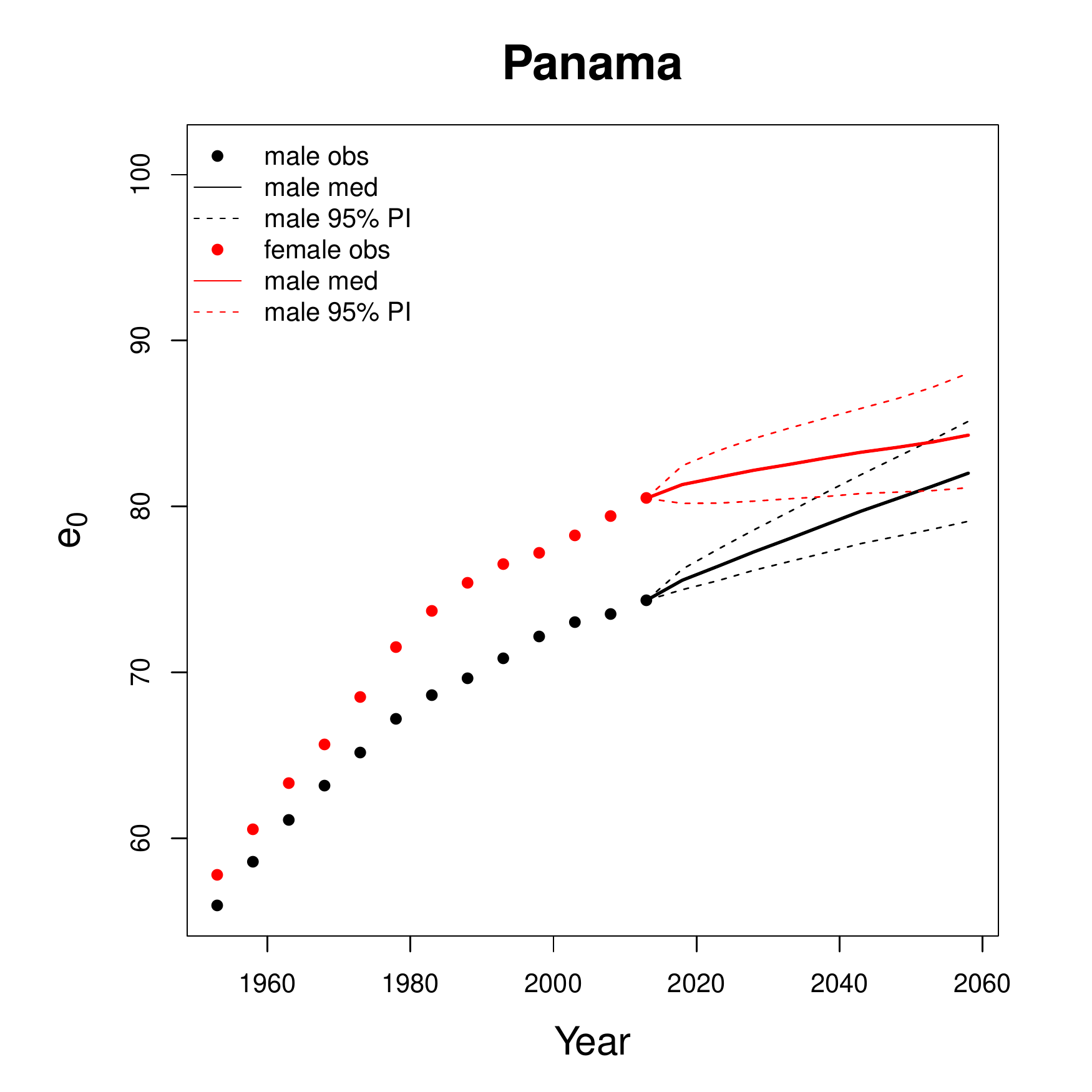}
			
		\end{center}
	\end{figure}

	\begin{figure}[H]
		\begin{center}
			\includegraphics[scale=0.43]{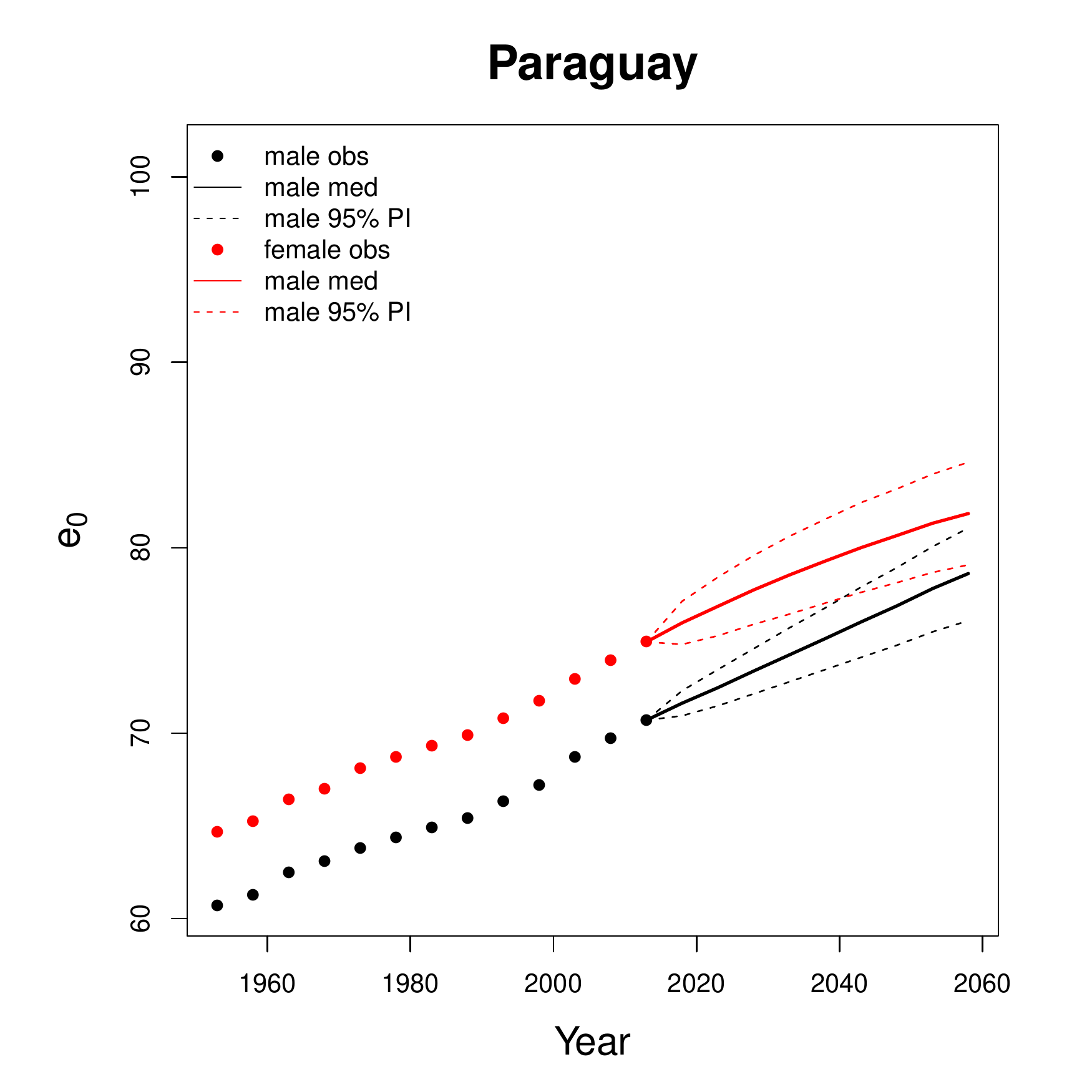}			\includegraphics[scale=0.43]{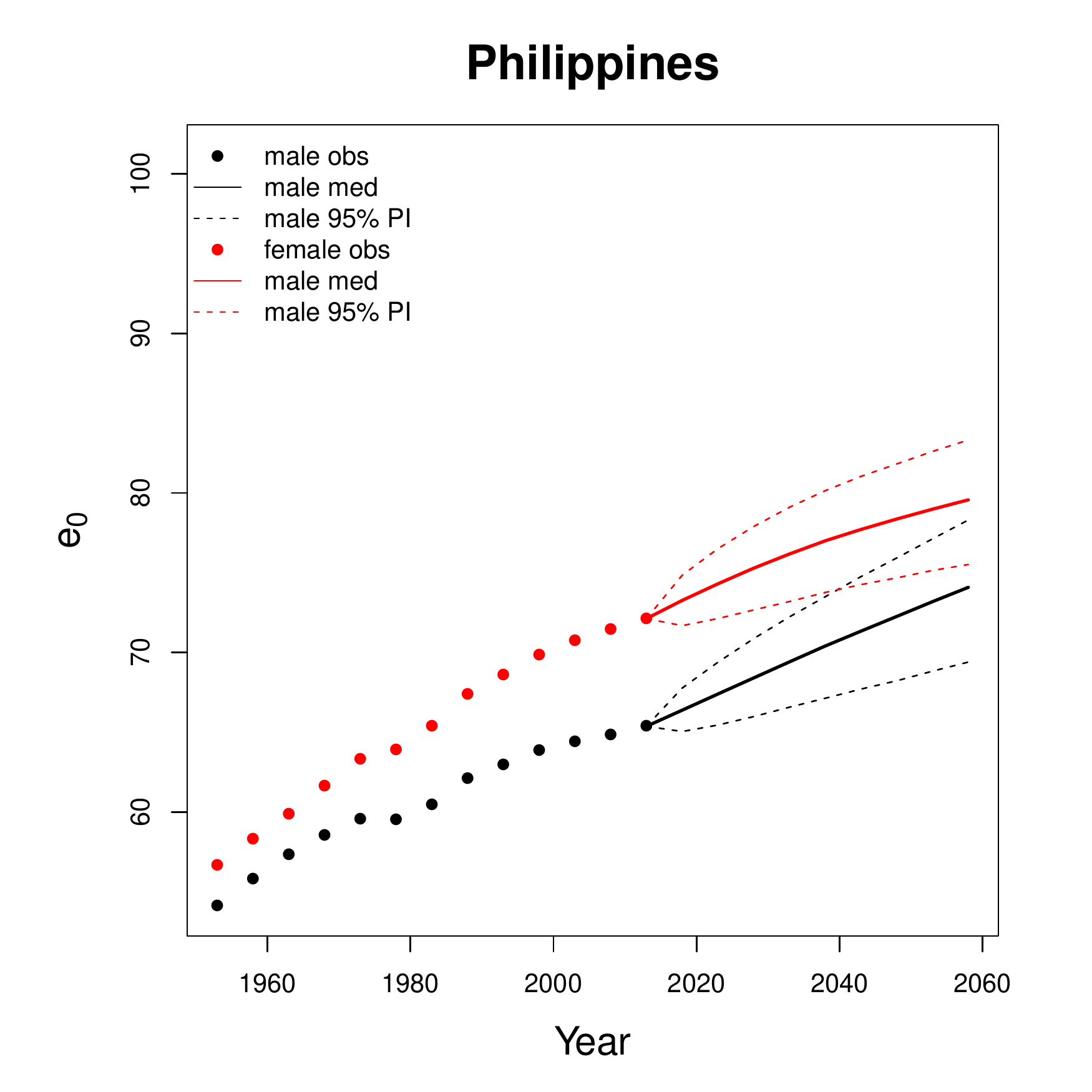}
			\includegraphics[scale=0.43]{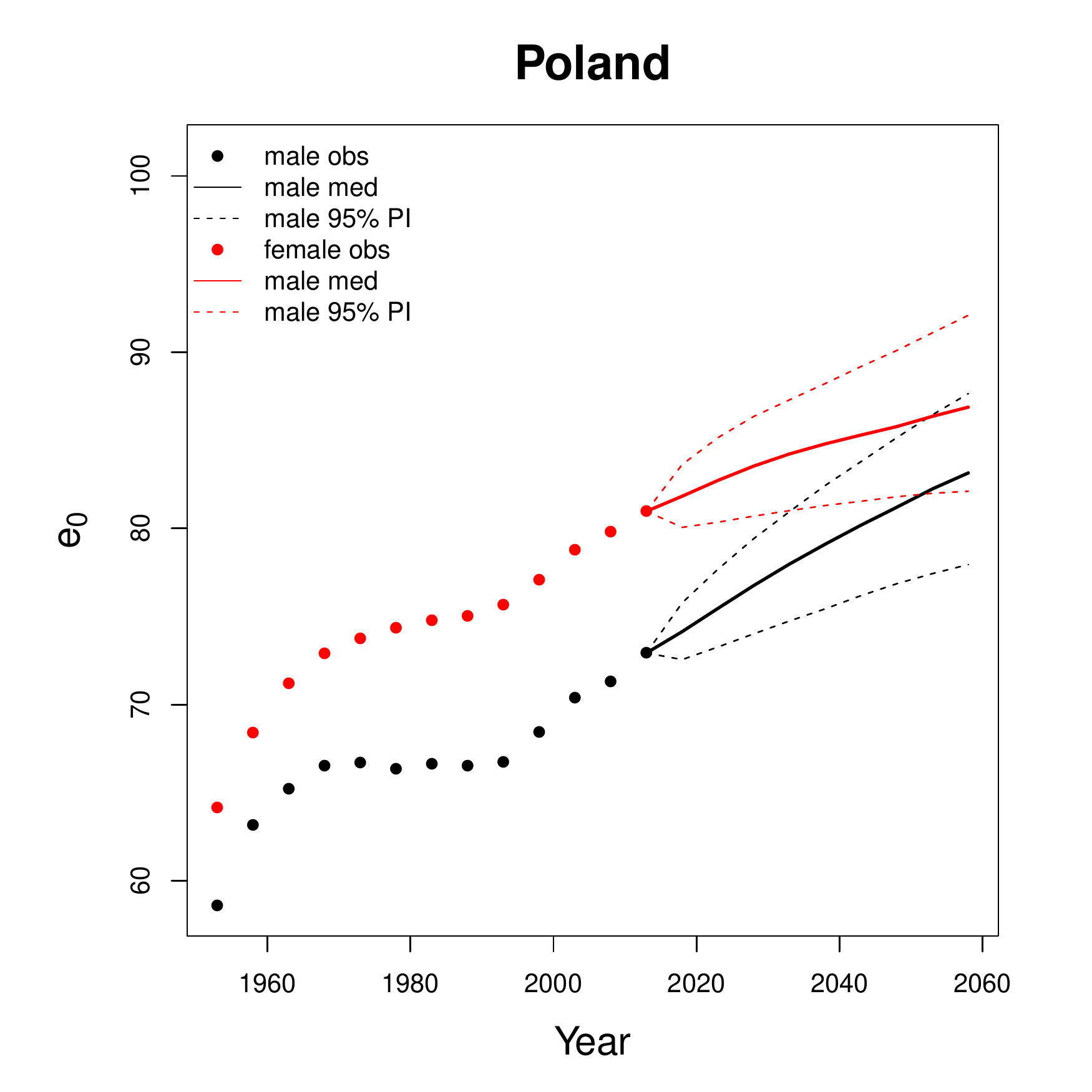}
			\includegraphics[scale=0.43]{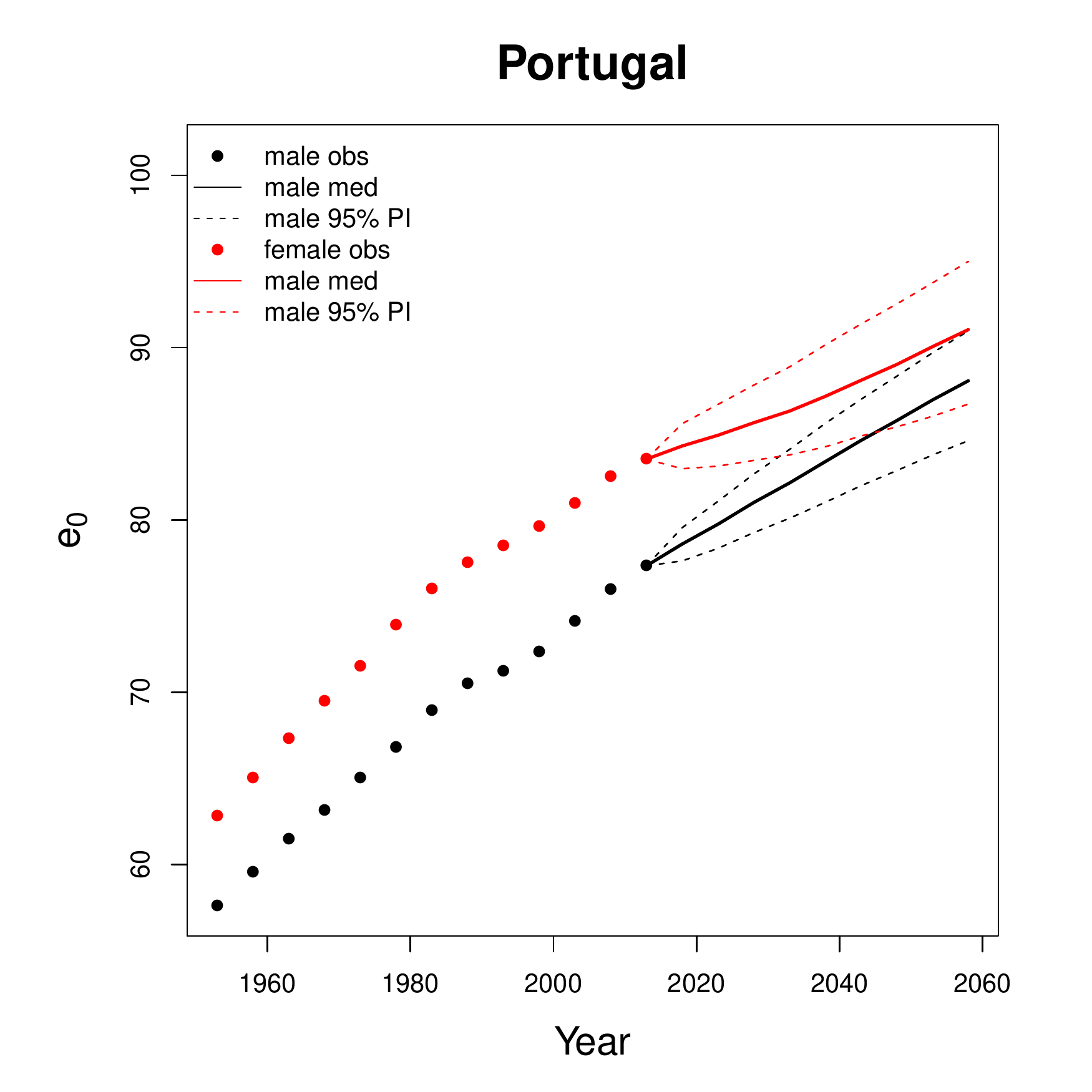}	\includegraphics[scale=0.43]{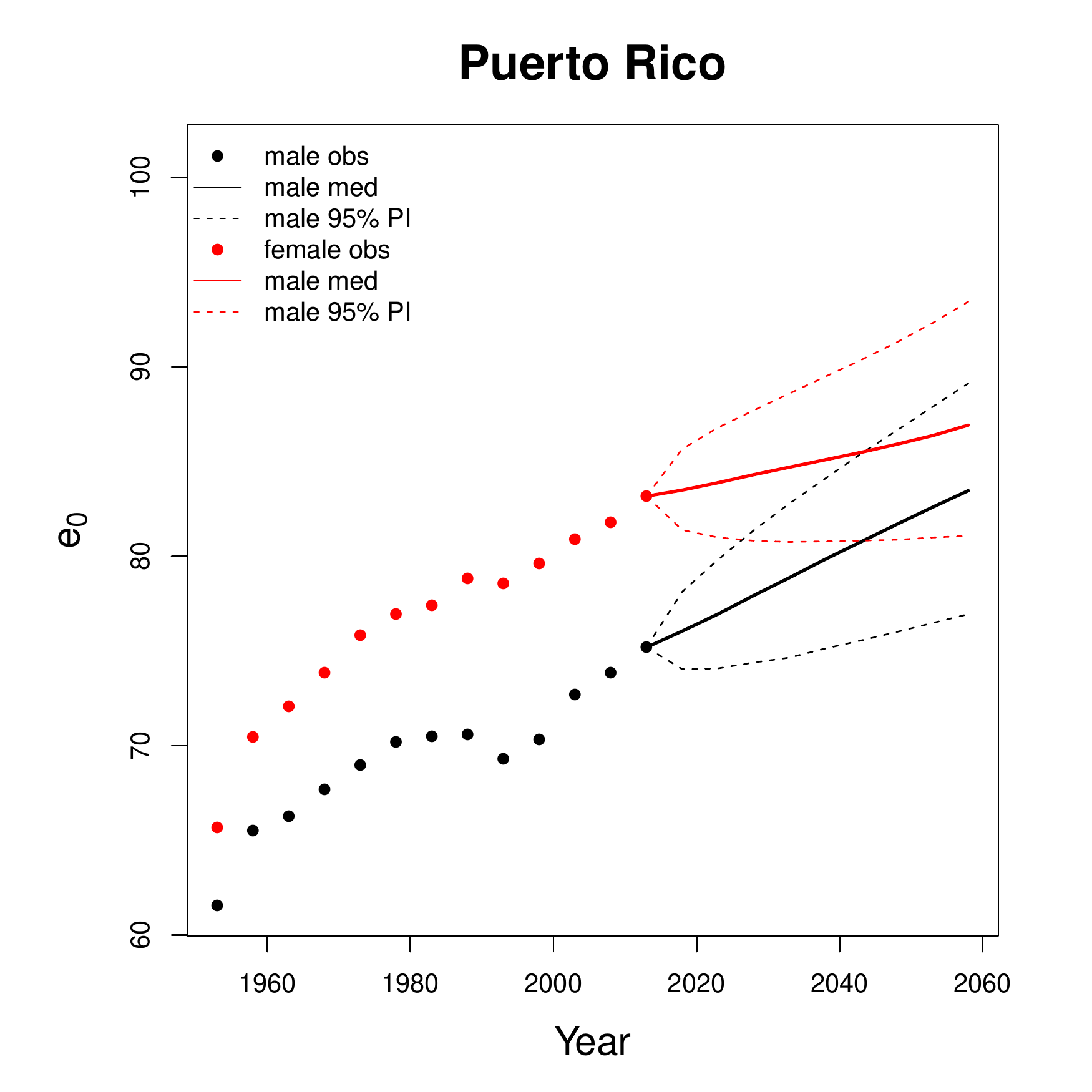}
			\includegraphics[scale=0.43]{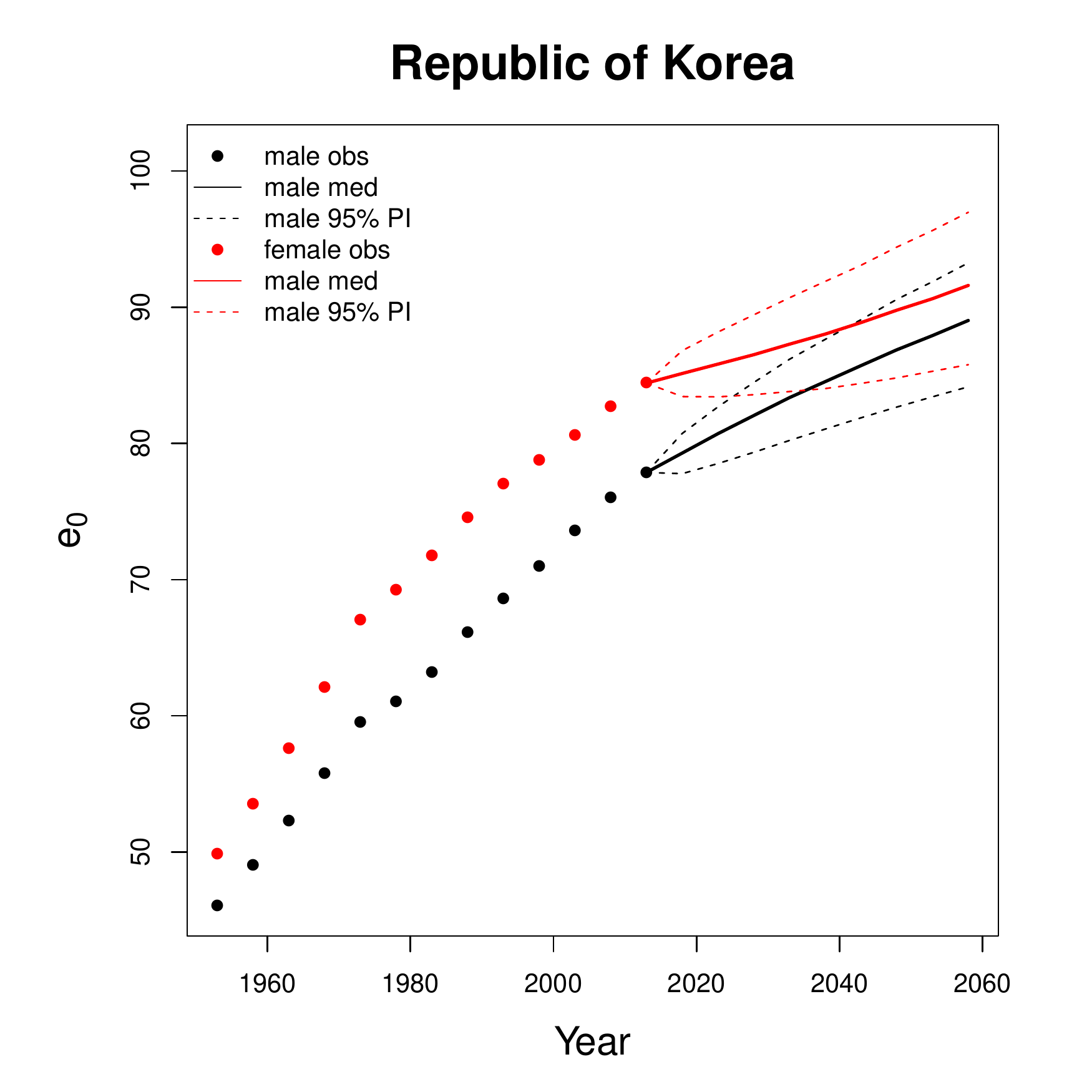}
			
		\end{center}
	\end{figure}

	\begin{figure}[H]
		\begin{center}
			\includegraphics[scale=0.43]{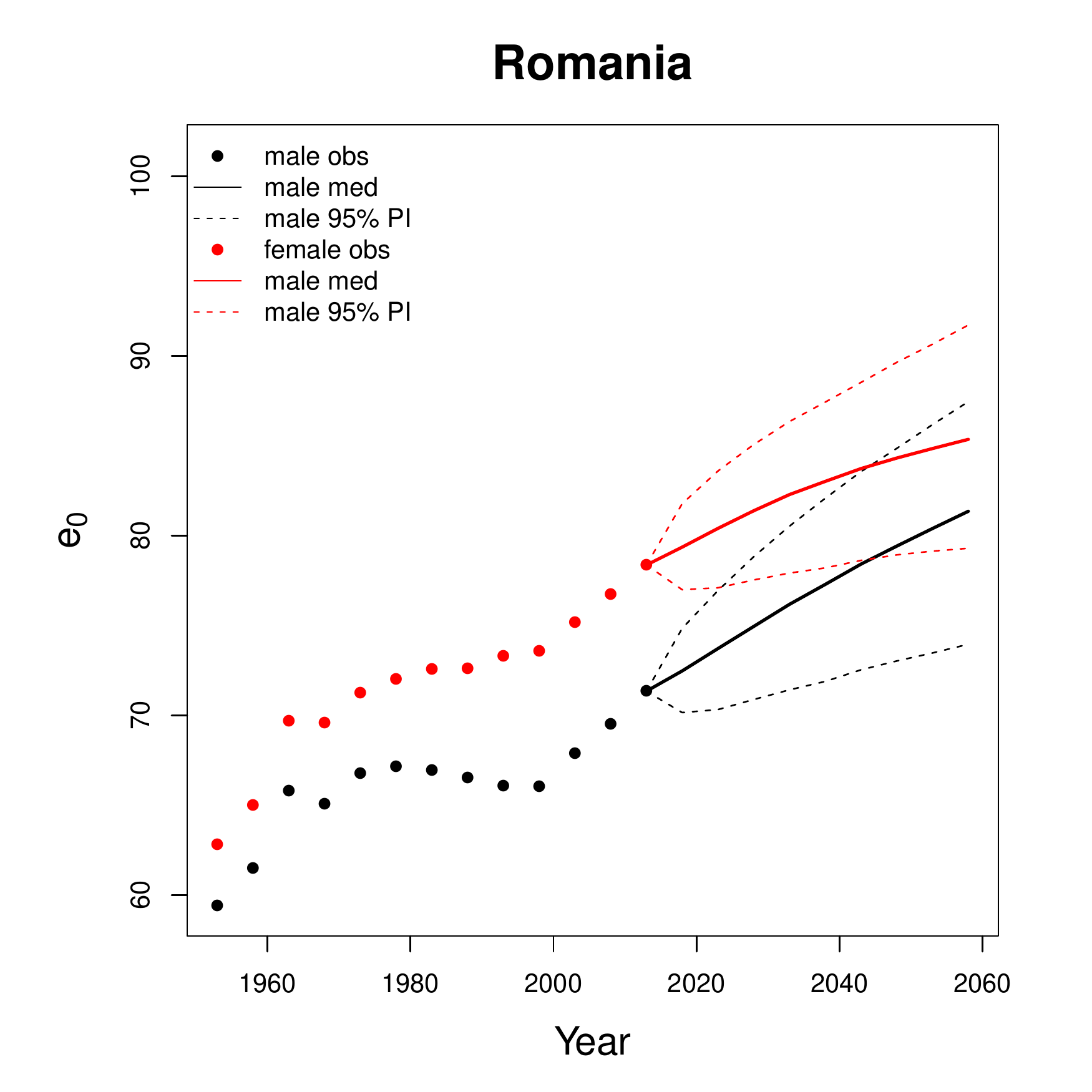}			\includegraphics[scale=0.43]{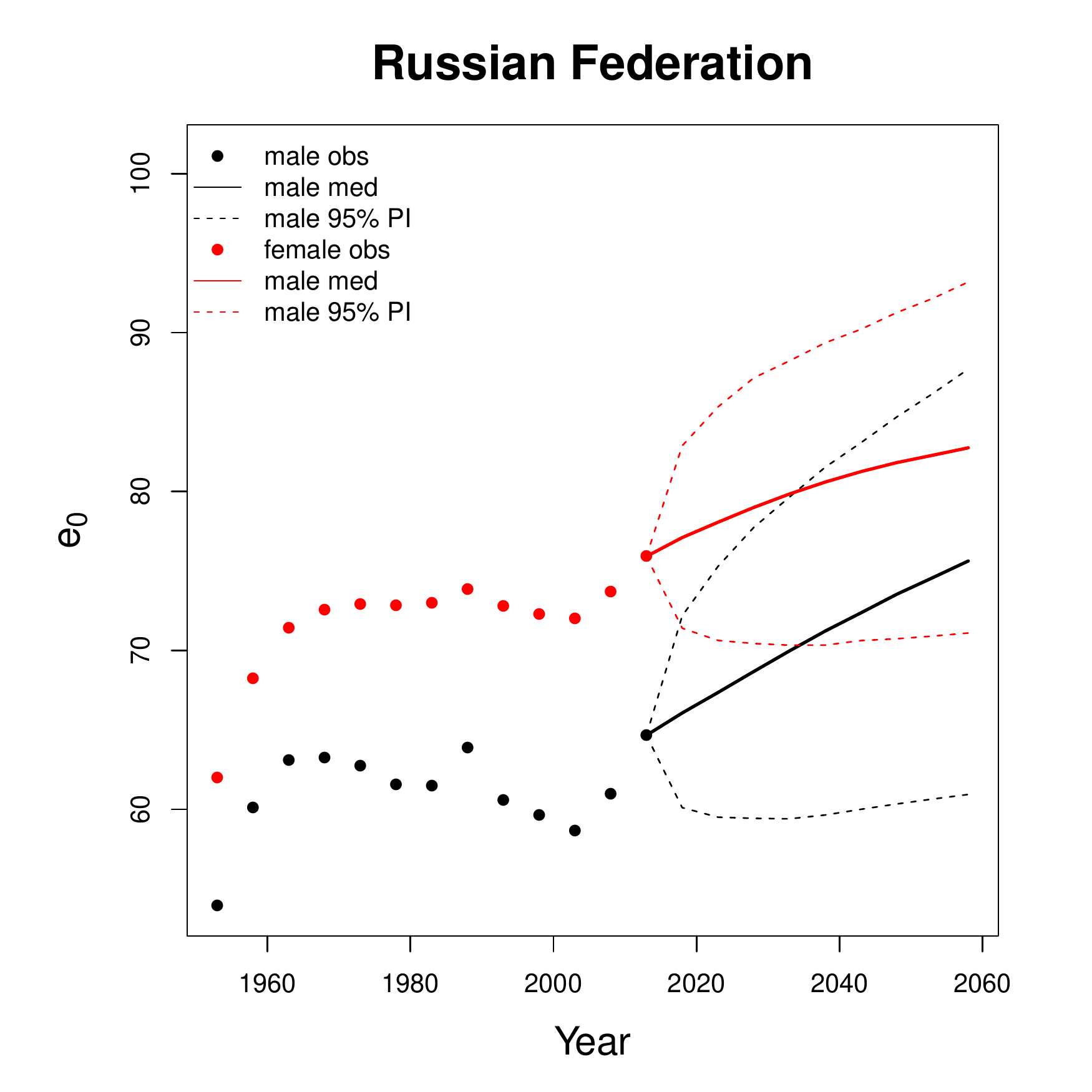}
			\includegraphics[scale=0.43]{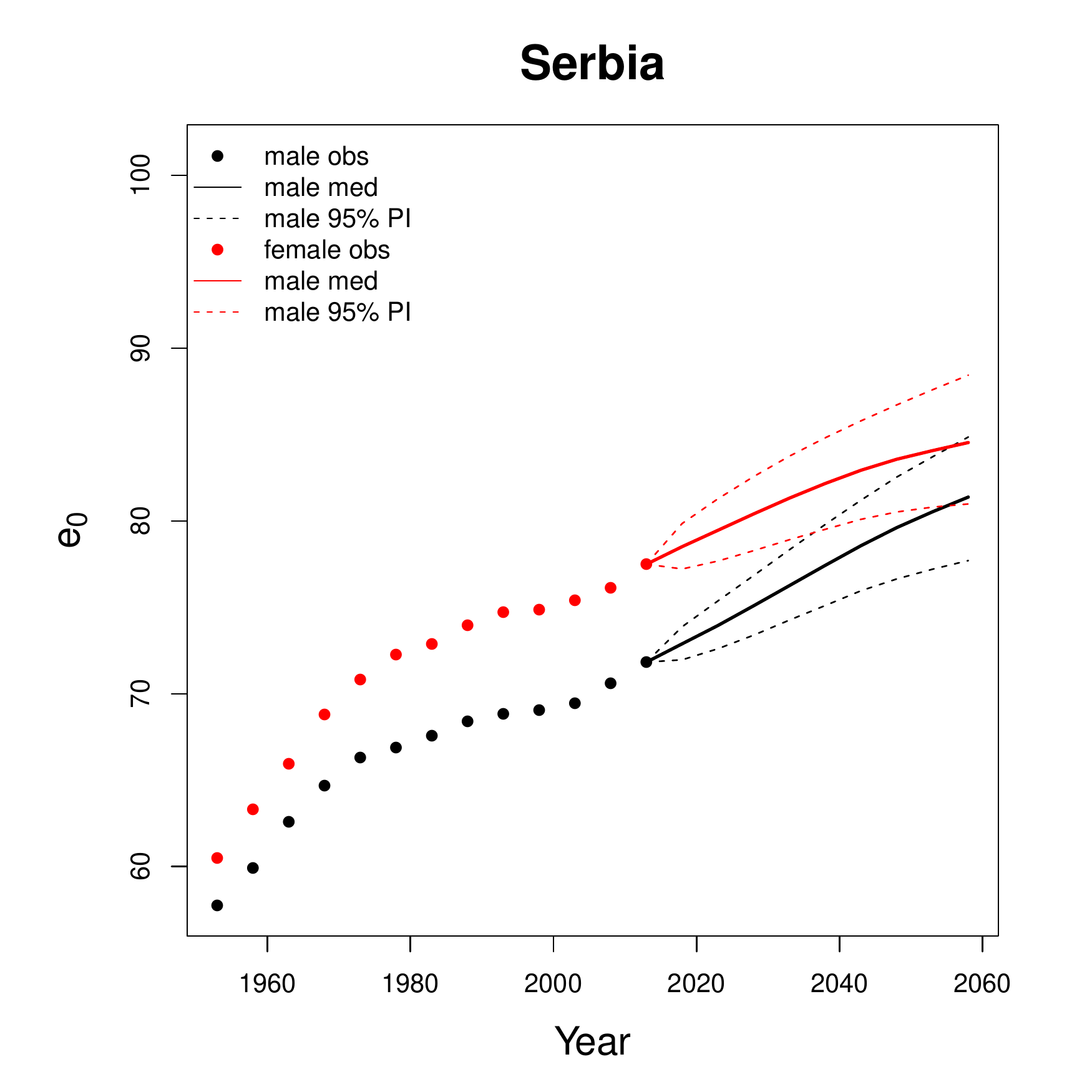}	
			\includegraphics[scale=0.43]{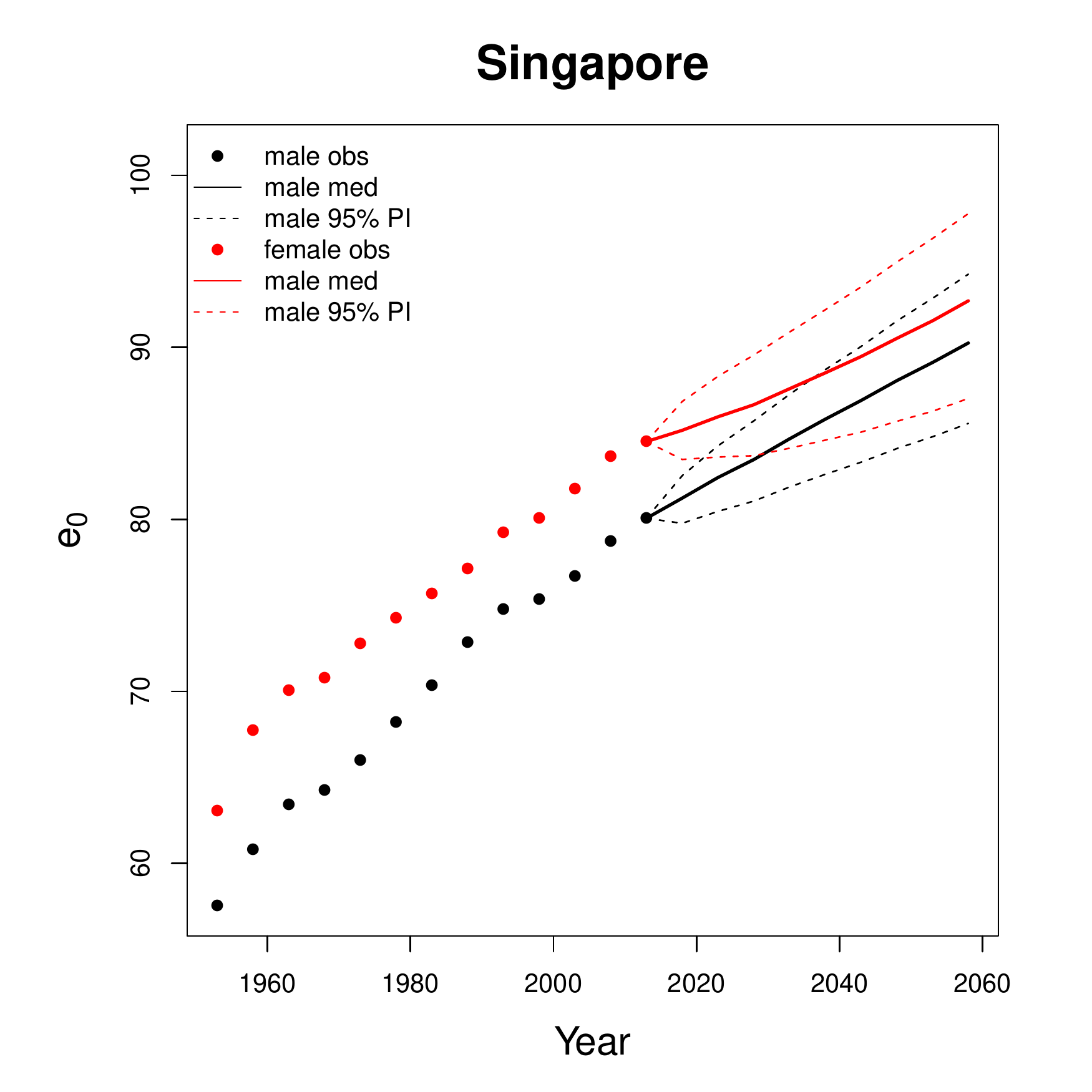}	\includegraphics[scale=0.43]{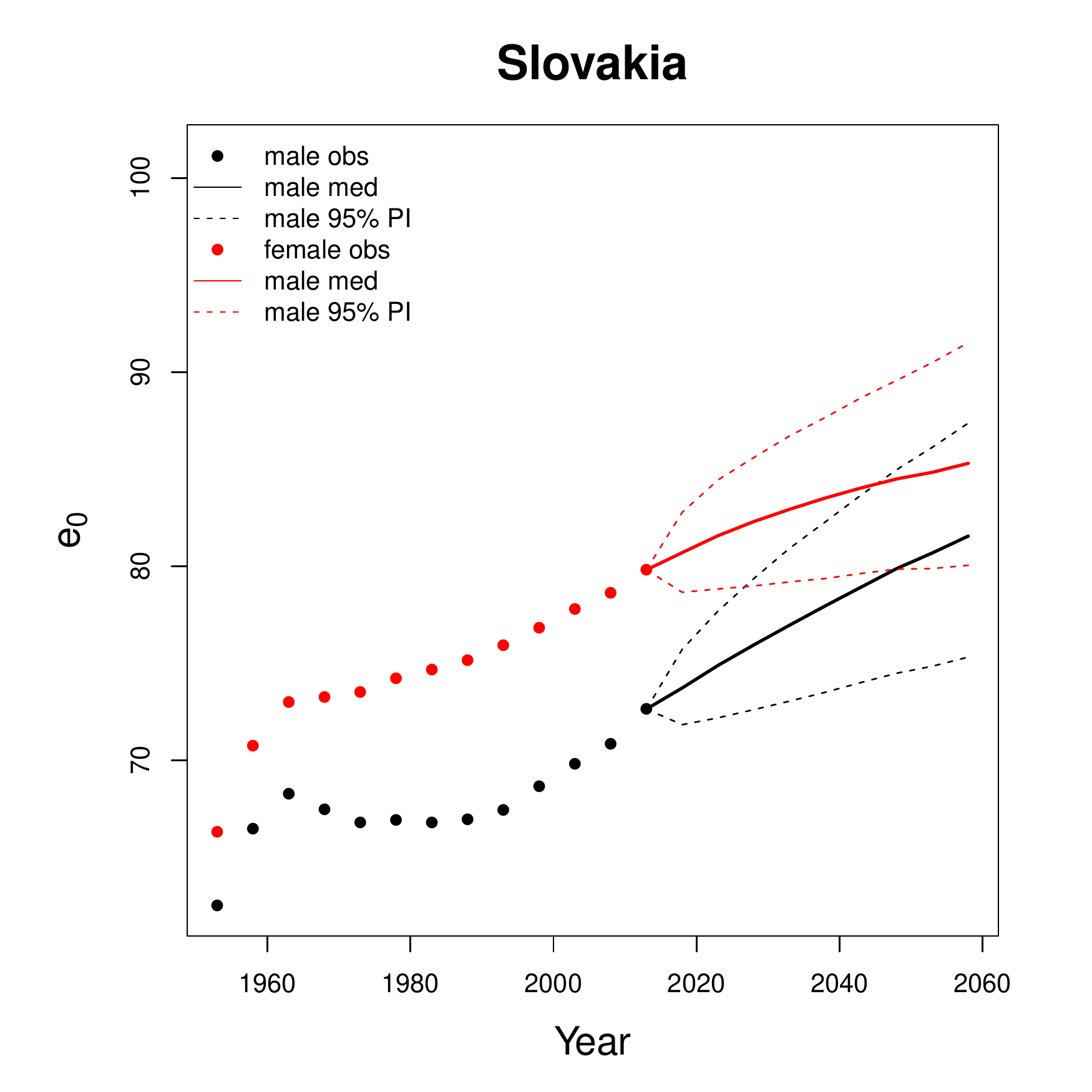}
			\includegraphics[scale=0.43]{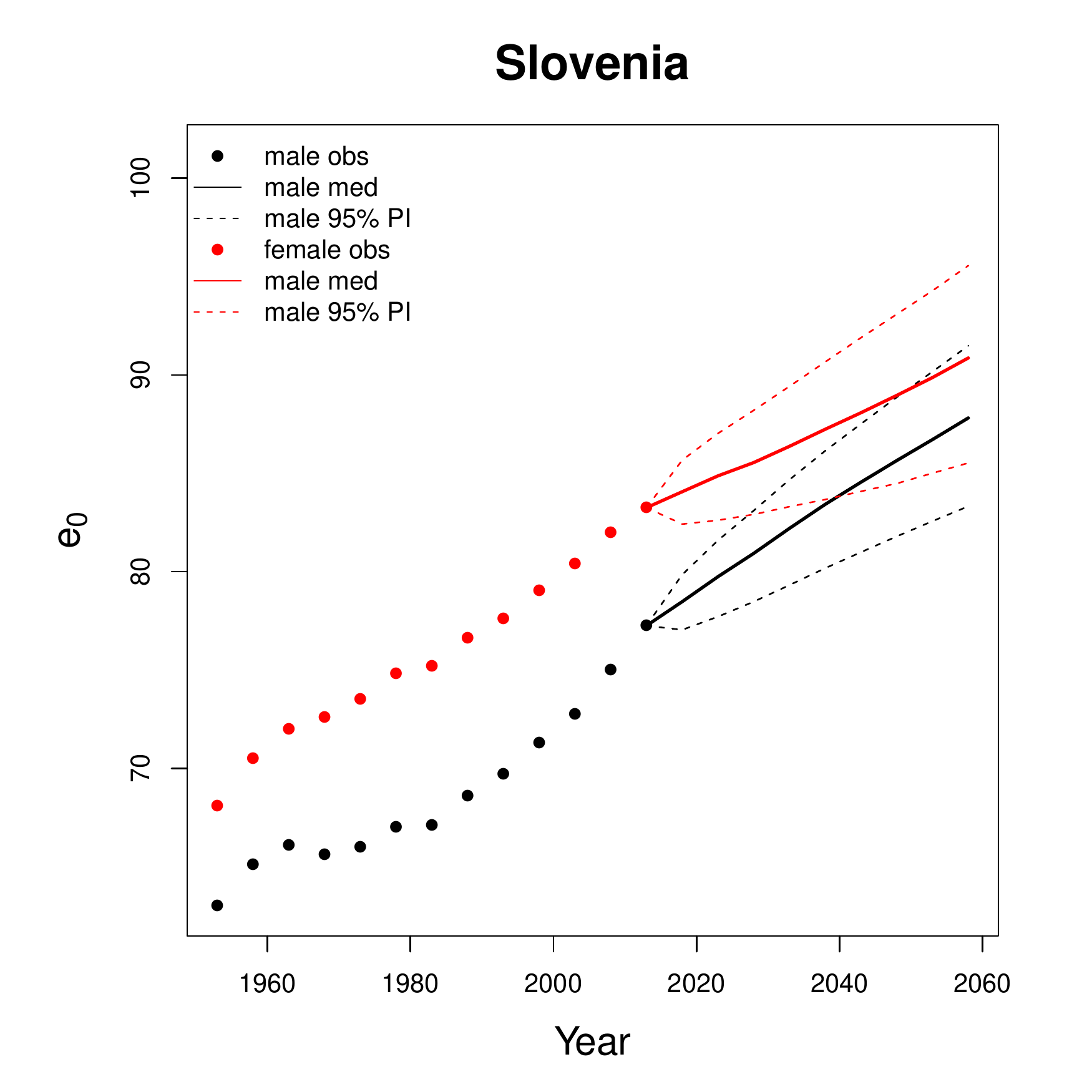}
			
		\end{center}
	\end{figure}

	\begin{figure}[H]
		\begin{center}
			\includegraphics[scale=0.43]{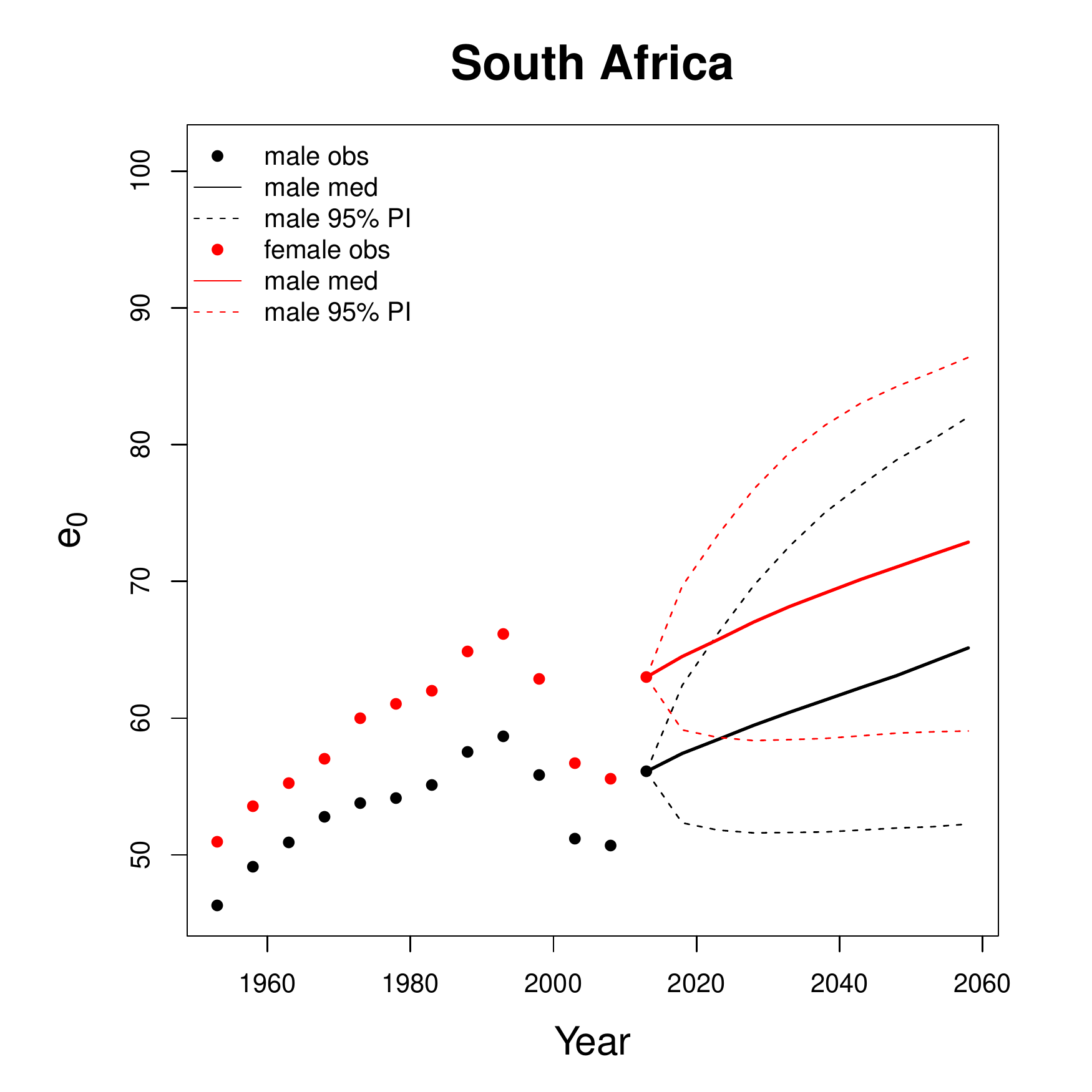}			\includegraphics[scale=0.43]{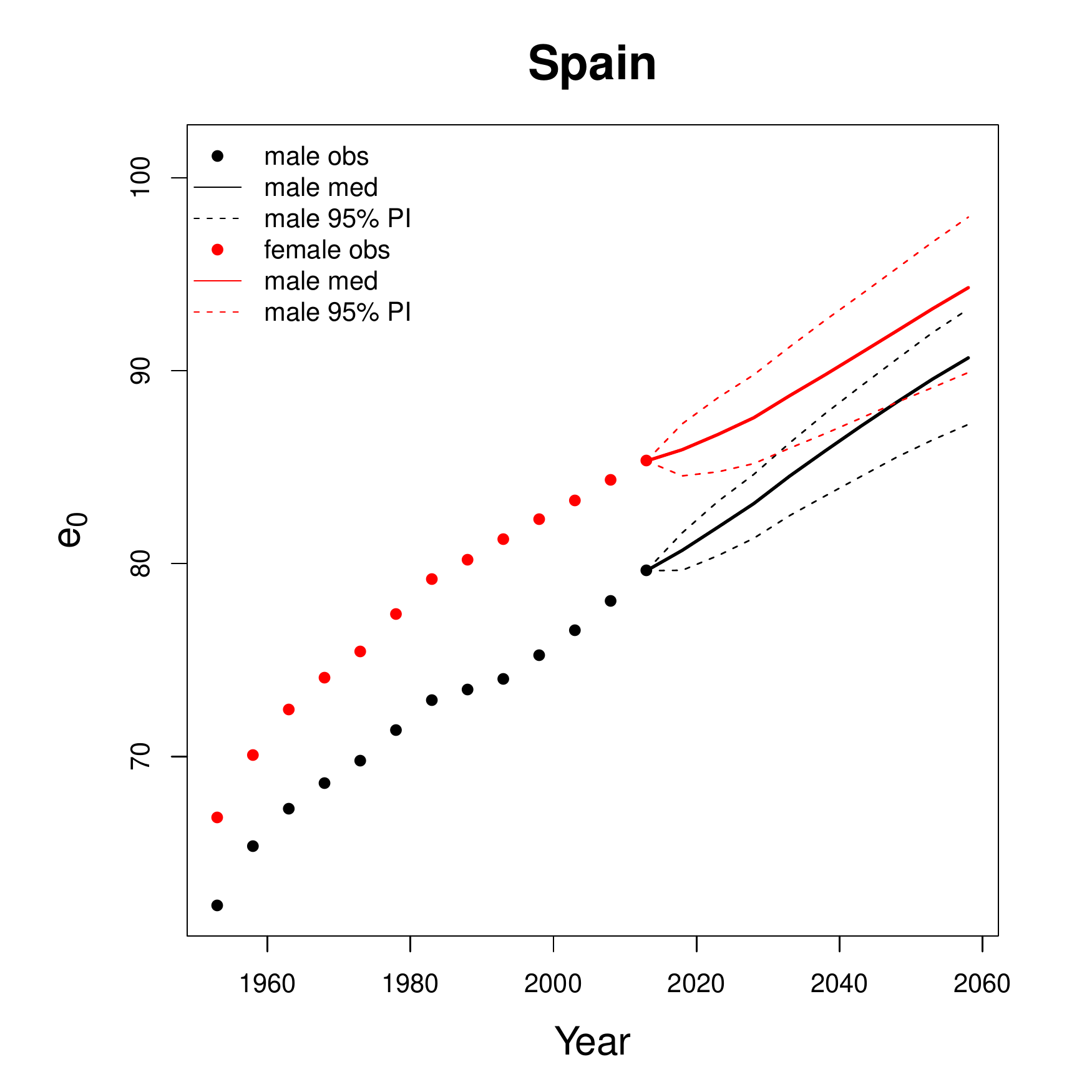}
			\includegraphics[scale=0.43]{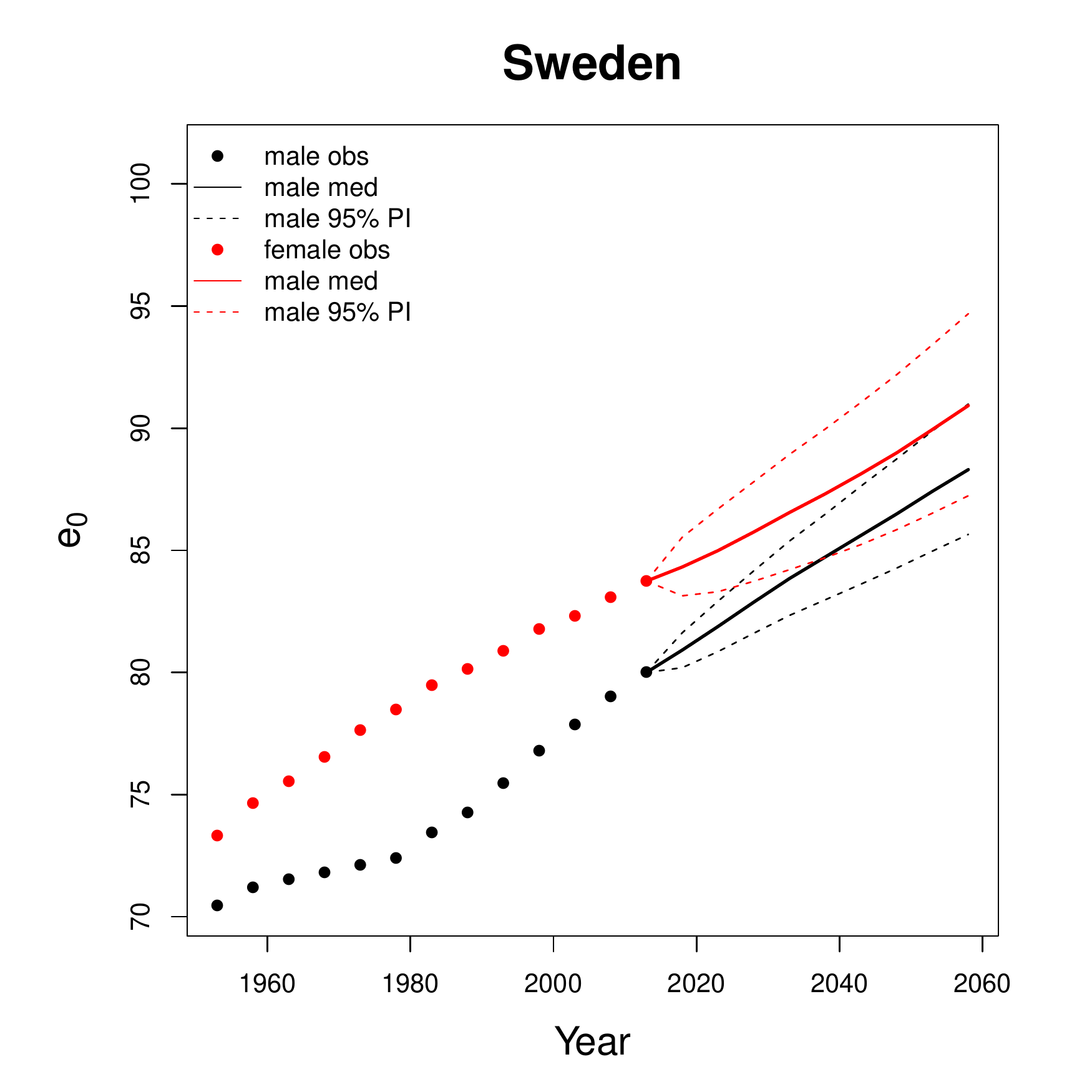}
			\includegraphics[scale=0.43]{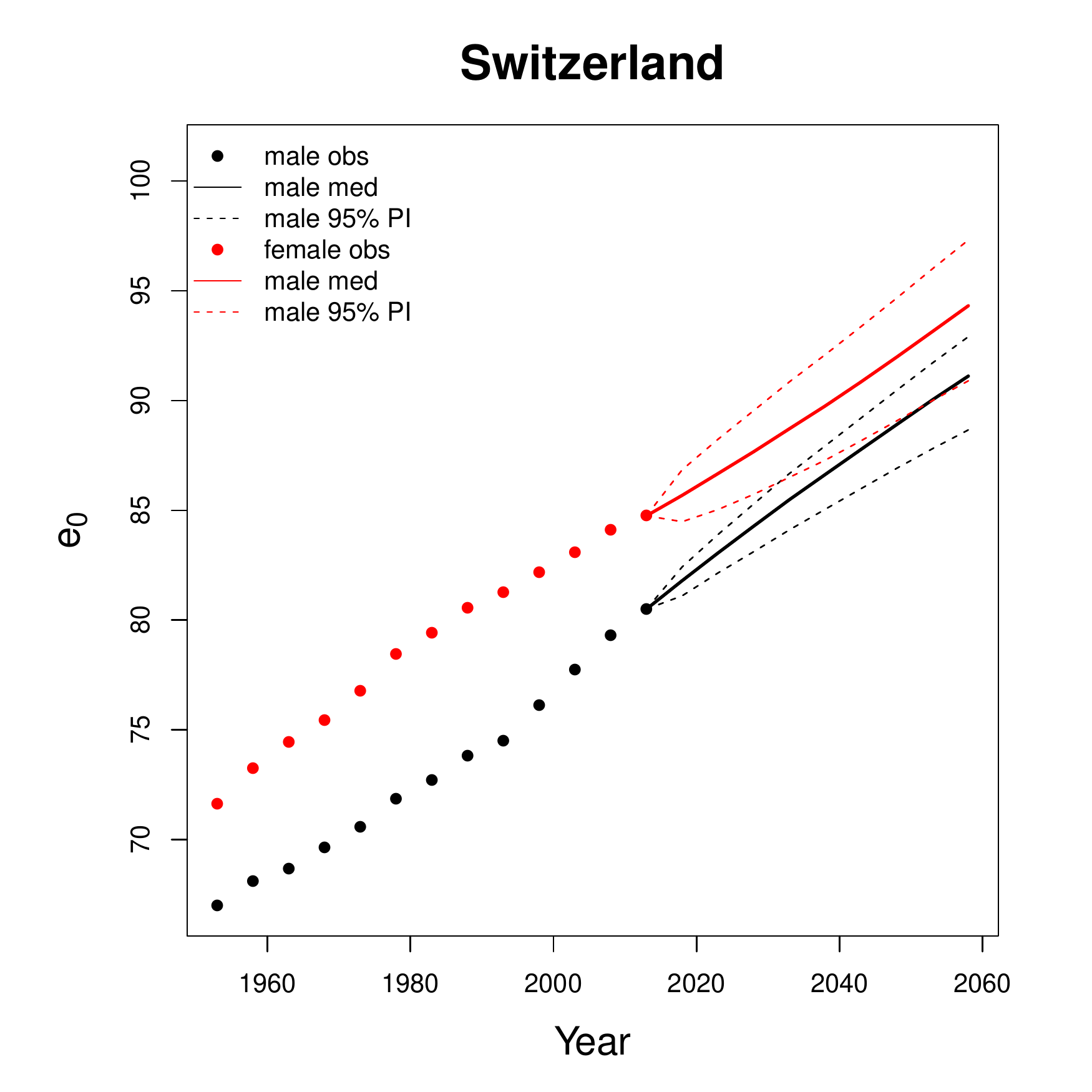}
			\includegraphics[scale=0.43]{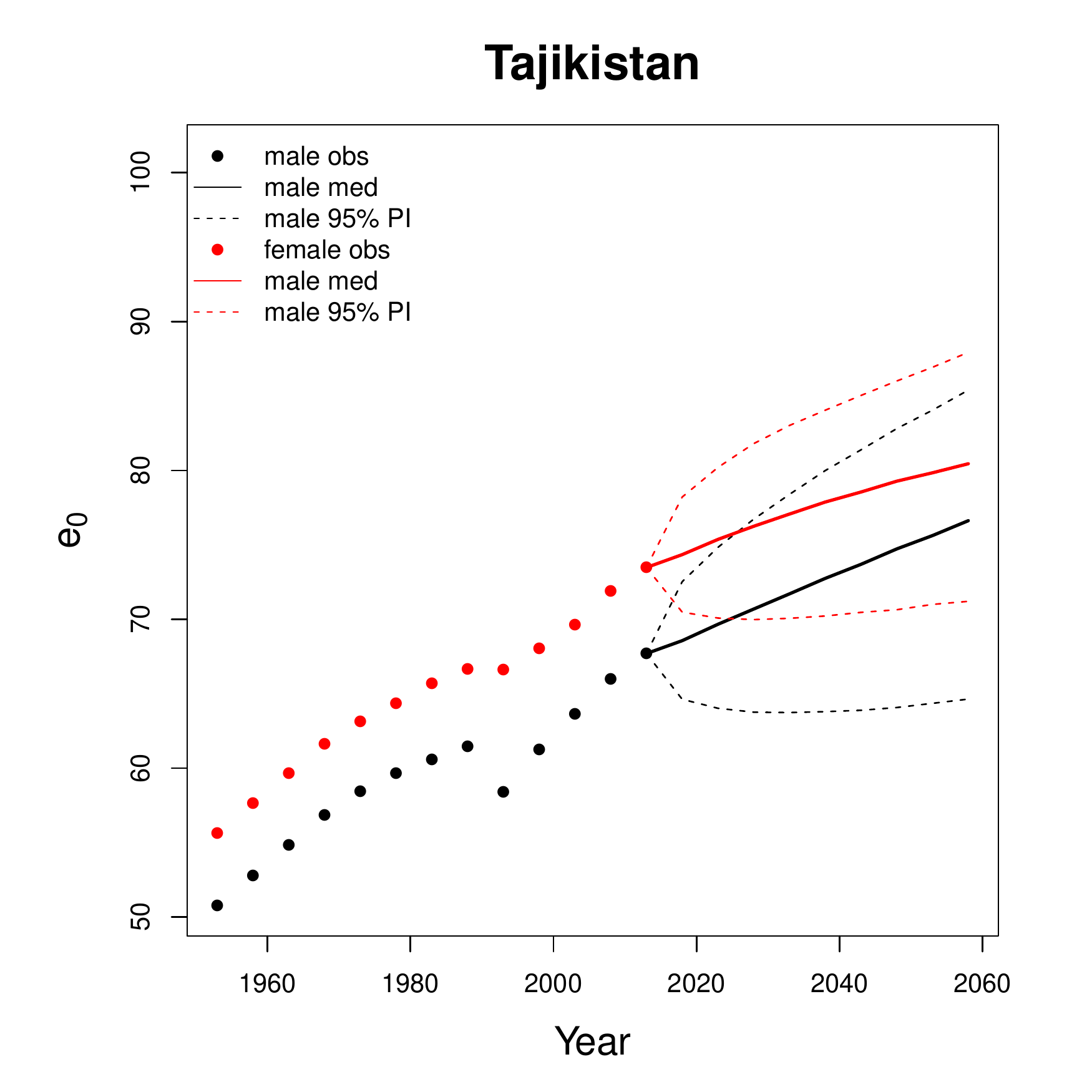}
			\includegraphics[scale=0.43]{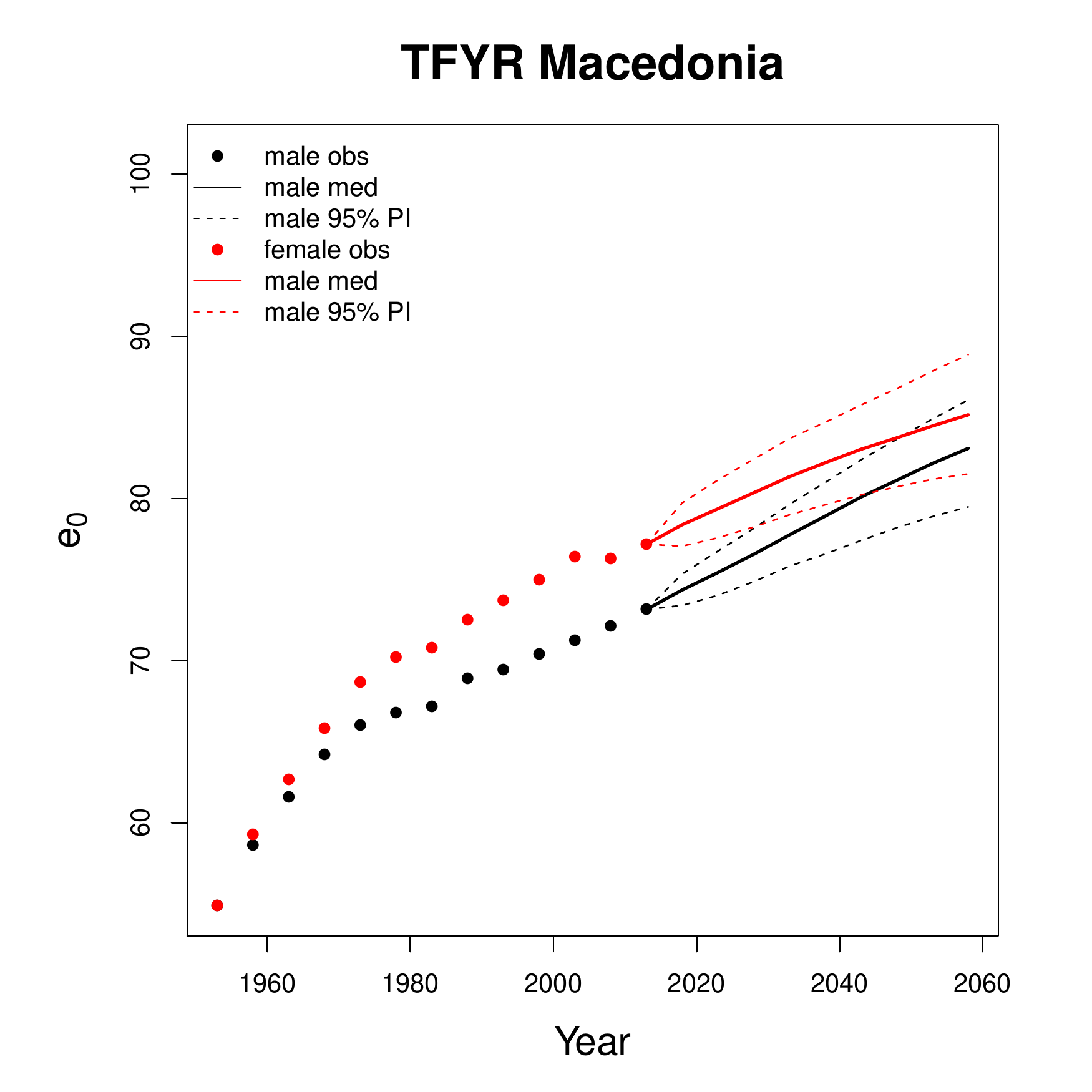}
			
		\end{center}
	\end{figure}

	\begin{figure}[H]
		\begin{center}	
			\includegraphics[scale=0.43]{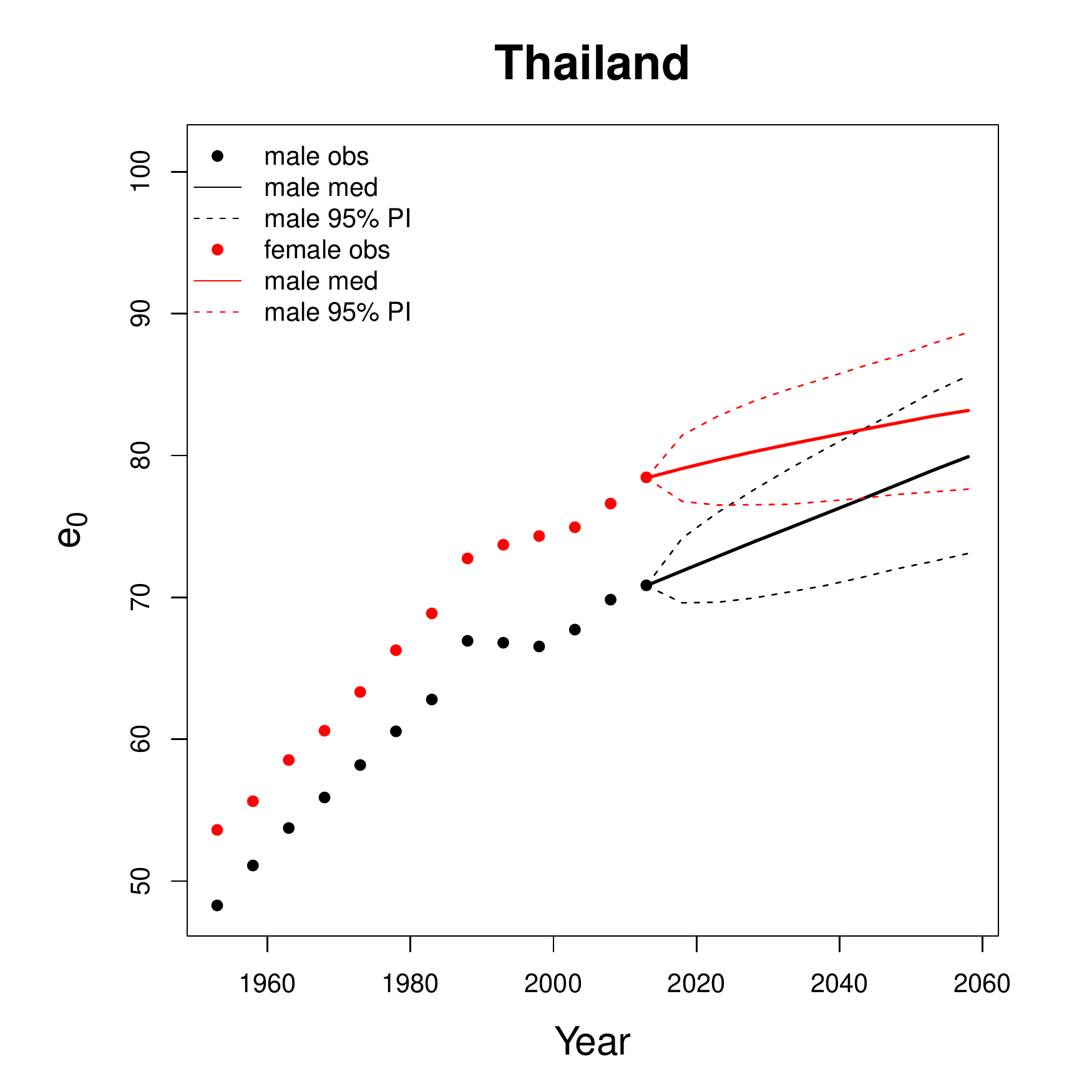}
			\includegraphics[scale=0.43]{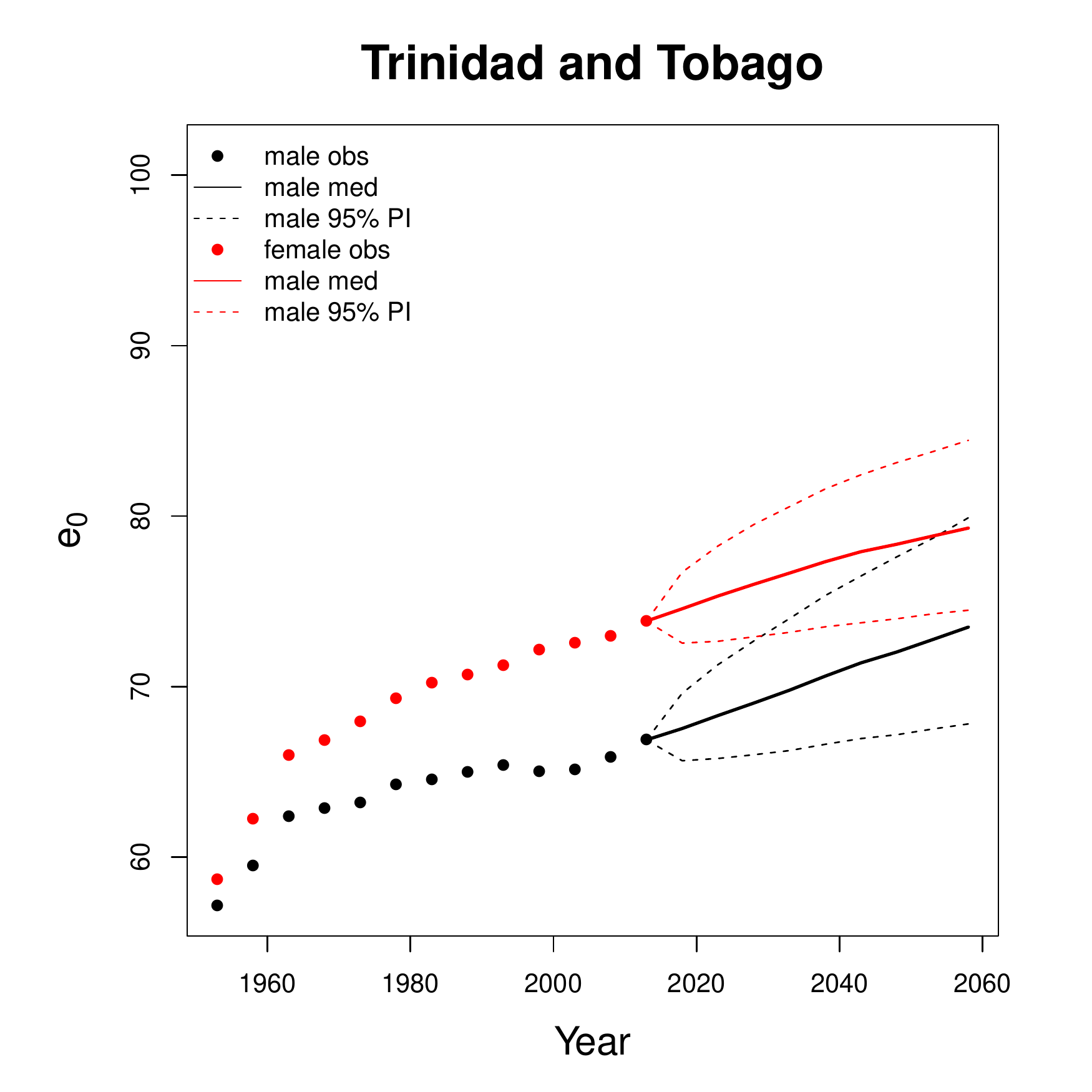}
			\includegraphics[scale=0.43]{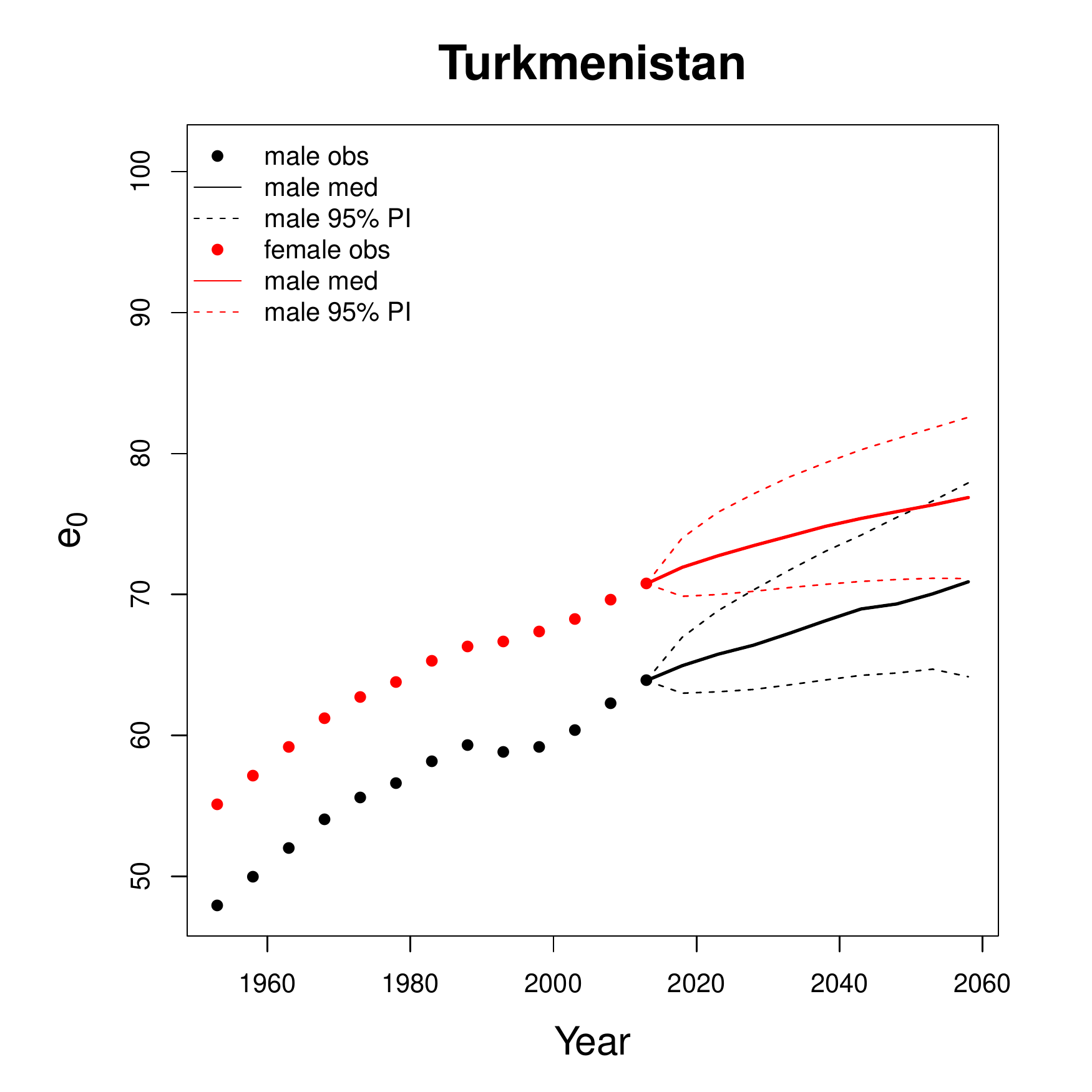}
			\includegraphics[scale=0.43]{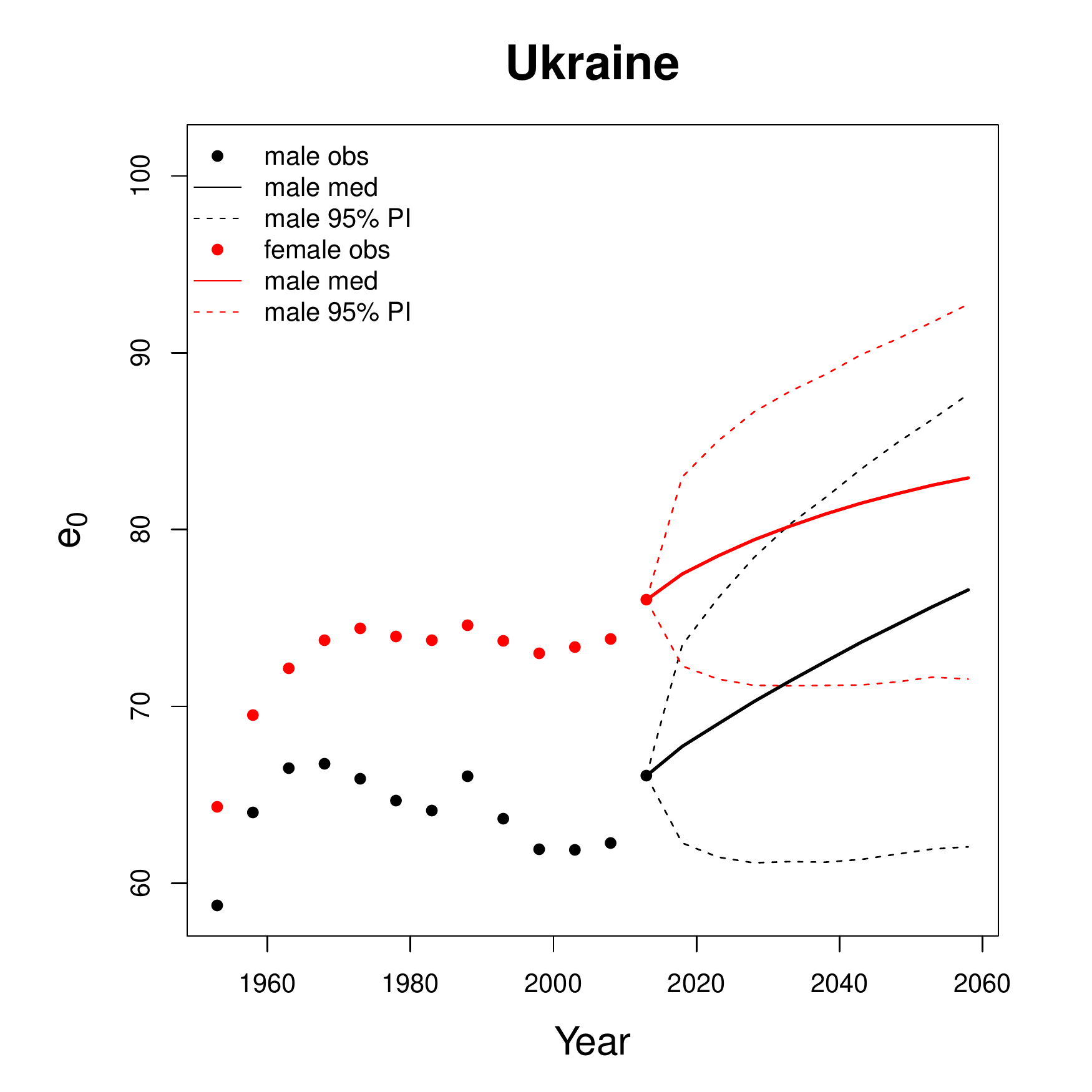}
			\includegraphics[scale=0.43]{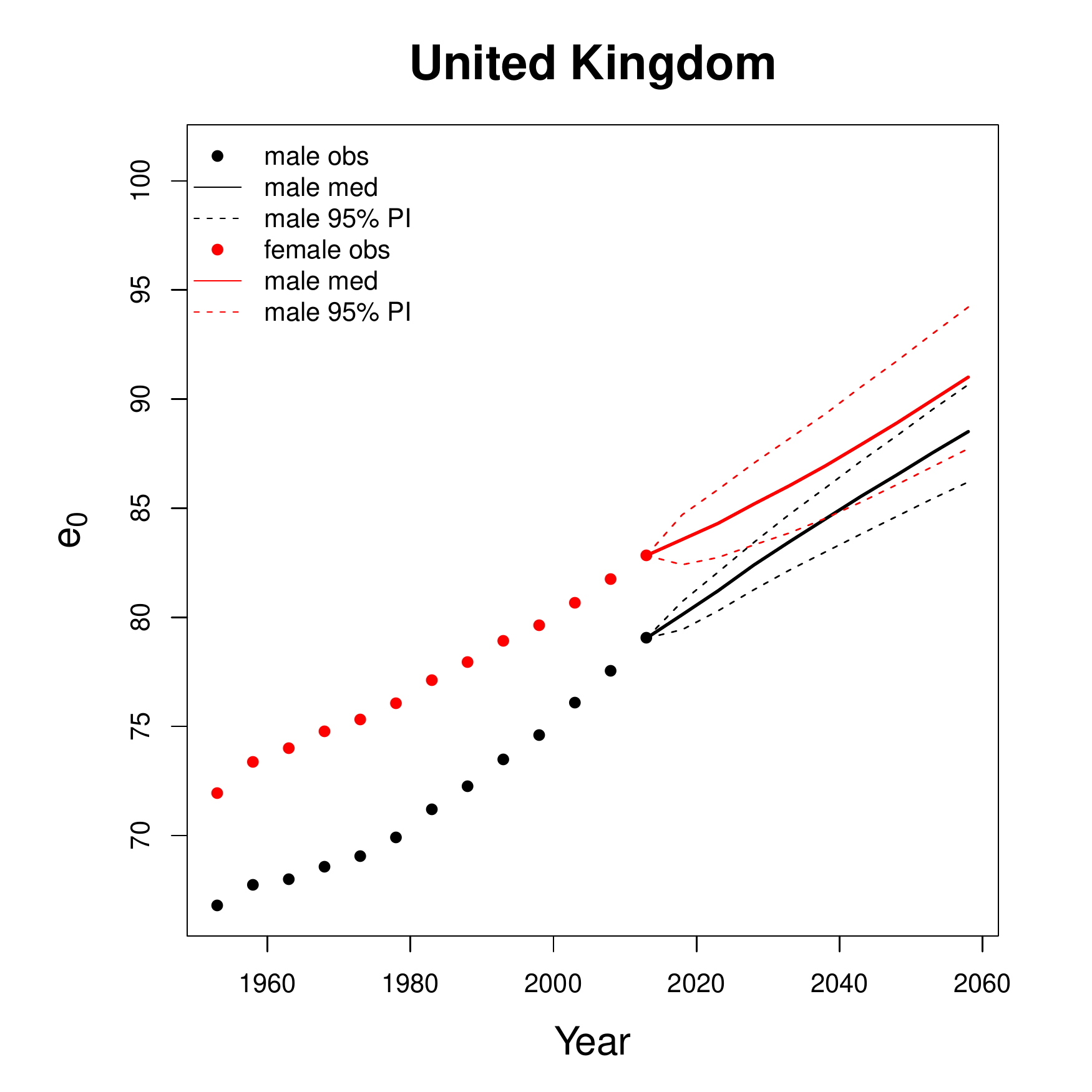}
			\includegraphics[scale=0.43]{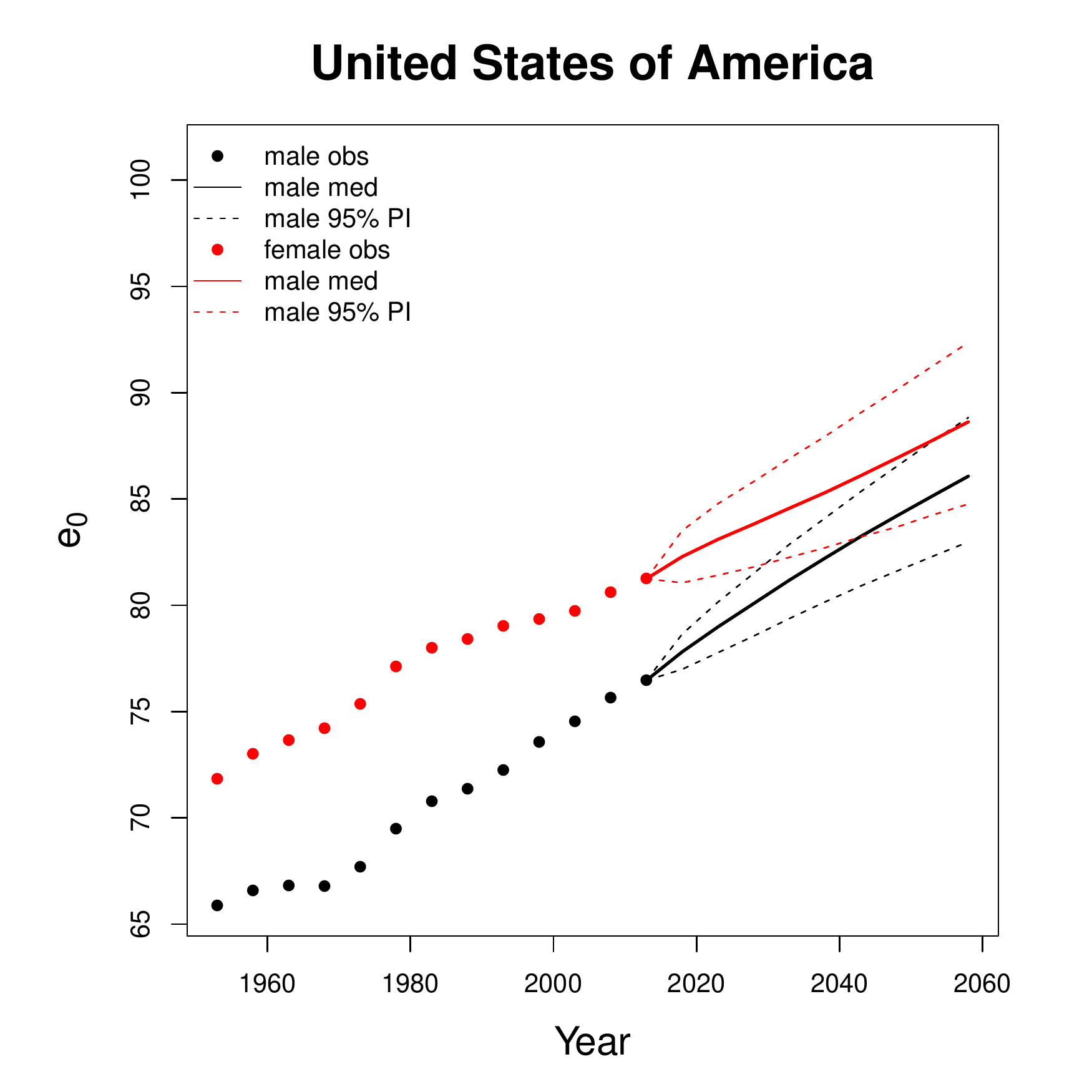}
			
		\end{center}
	\end{figure}

	\begin{figure}[H]
		\begin{center}	
			\includegraphics[scale=0.43]{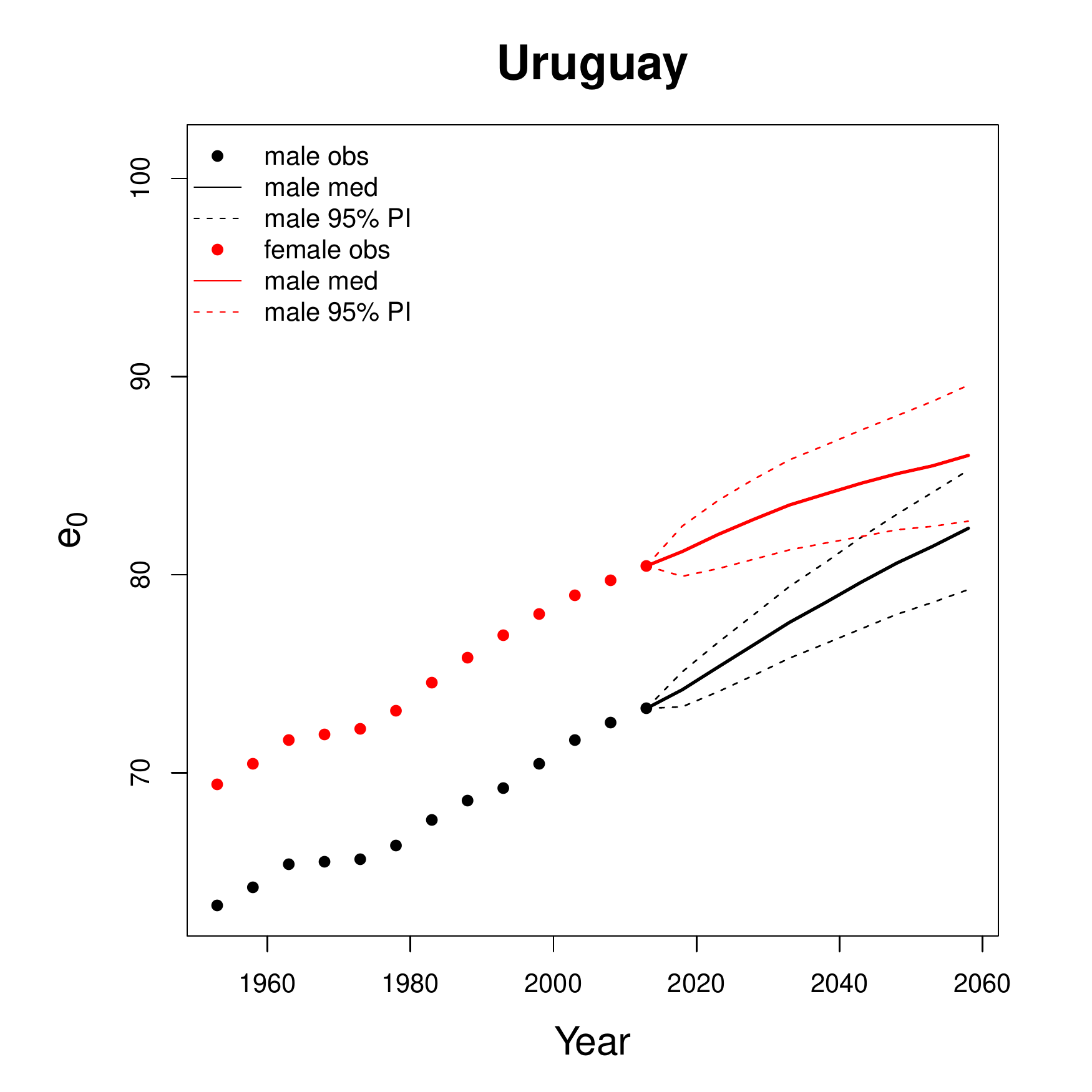}
			\includegraphics[scale=0.43]{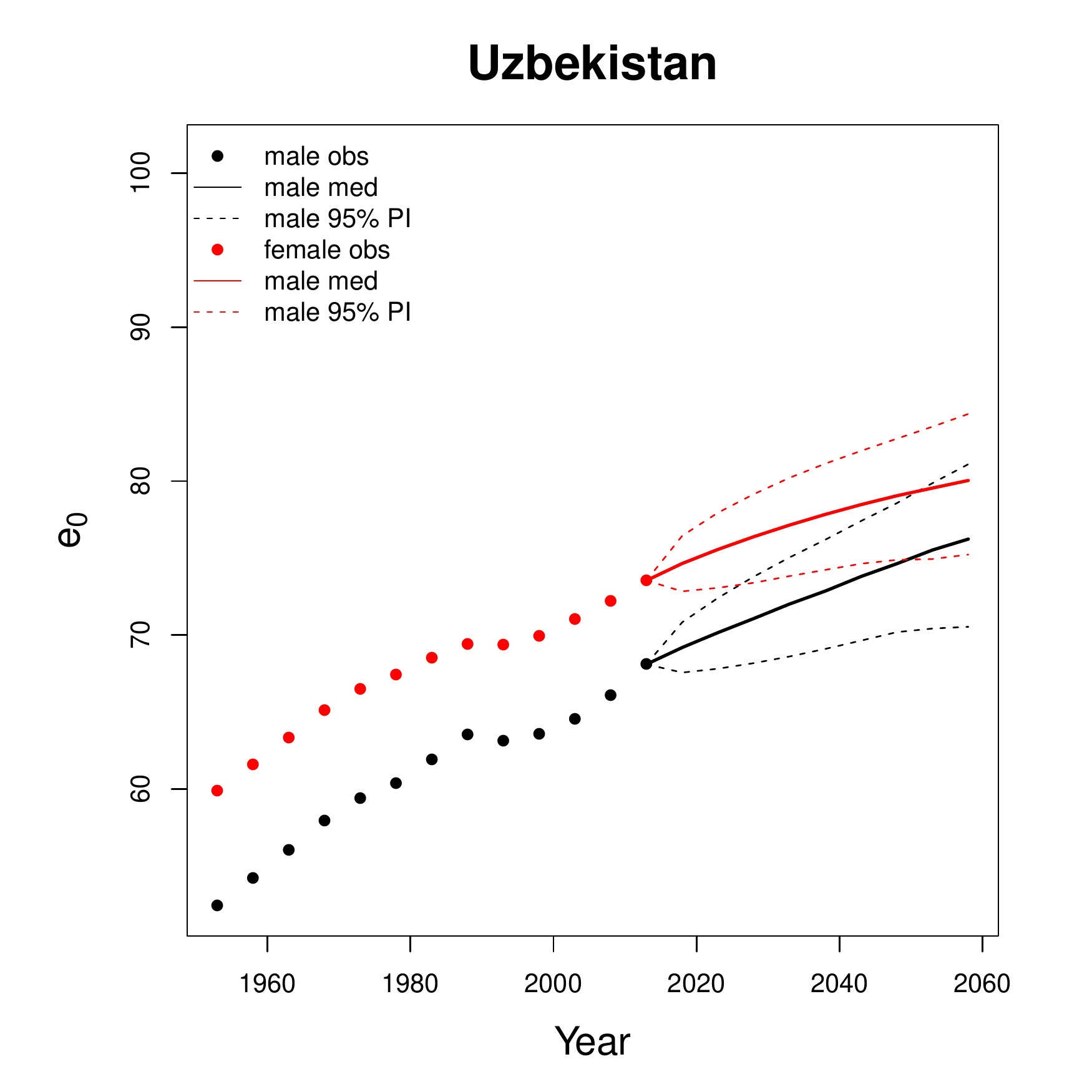}
			\includegraphics[scale=0.43]{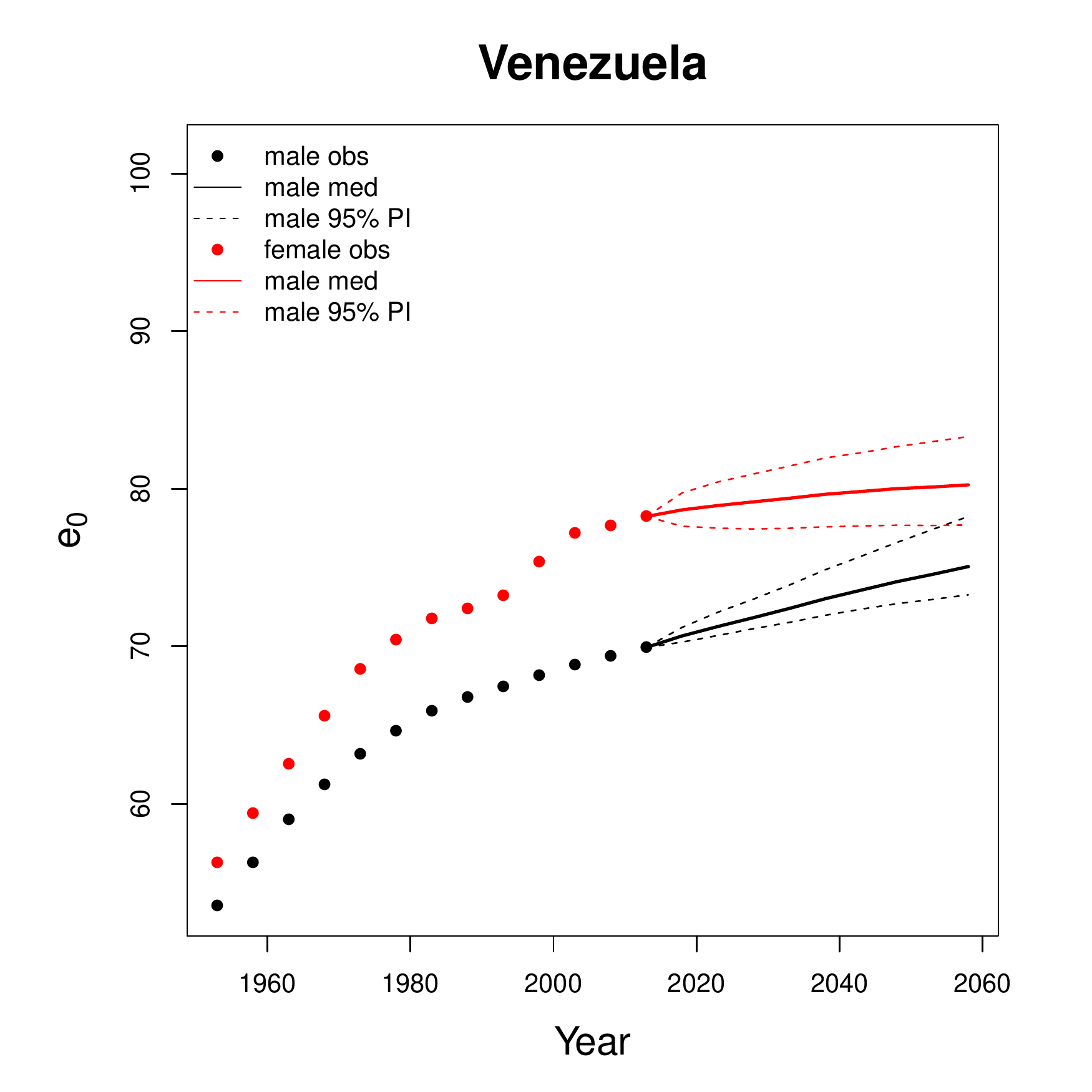}
		\end{center}
	\end{figure}

\end{appendices}

\end{document}